\PassOptionsToPackage{pdfpagelabels=false}{hyperref}
\documentclass[useAMS,usenatbib]{mnras}
\pdfoutput=1

\usepackage[pdftex]{graphicx}
 \usepackage{amsmath}
\usepackage{amssymb}
\usepackage{natbib}
\usepackage{color}
\usepackage{times}

\def \MSUN{{\rm M}_{\odot}}

\def \RSTARS{R_{\rm stars}}
\def \MS{M_{\rm stars}}

\def \VMAX{V_{\rm rot}}
\def\HA{\rm H\alpha}
\def\KMPERS{\rm km~s^{-1}}

\newcommand{\rvvv}[1]{\textcolor{black}{#1}}
\volume{{\rm submitted}}
	
\usepackage{etoolbox}
\makeatletter
\patchcmd\@combinedblfloats{\box\@outputbox}{\unvbox\@outputbox}{}{\errmessage{\noexpand patch failed}}
\makeatother
	
\title[TNG50 galaxies: stellar and gaseous disks]{First Results from the TNG50 Simulation:\\ The evolution of stellar and gaseous disks across cosmic time}

\author[Pillepich et al.] {Annalisa Pillepich$^1$\thanks{E-mail: pillepich@mpia-hd.mpg.de},
Dylan Nelson$^2$, Volker Springel$^2$, R{\"u}diger Pakmor$^2$, 
\newauthor
Paul Torrey$^{3}$,
Rainer Weinberger$^4$, 
Mark Vogelsberger$^5$,
Federico Marinacci$^{5,6}$,
\newauthor
Shy Genel$^7$,
Arjen van der Wel$^{1,8}$,
and Lars Hernquist$^4$
\vspace{2mm}
\\
$^1${Max-Planck-Institut f{\"u}r Astronomie, K{\"o}nigstuhl 17, 69117 Heidelberg, Germany}\\
$^2${Max-Planck-Institut f{\"u}r Astrophysik, Karl-Schwarzschild-Str. 1, 85741 Garching, Germany}\\
$^3$University of Florida, Department of Physics, 2001 Museum Rd., Gainesville, FL 32611, USA\\
$^4${Harvard--Smithsonian Center for Astrophysics, 60 Garden Street, Cambridge, MA 02138}\\
$^5${Kavli Institute for Astrophysics and Space Research, Massachusetts Institute of Technology, Cambridge, MA 02139, USA}\\
$^6${Department of Physics and Astronomy, University of Bologna, via Gobetti 93/2, 40129 Bologna, Italy}\\
$^7${Center for Computational Astrophysics, Flatiron Institute, 162 Fifth Avenue, New York, NY 10010, USA}\\
$^8${Sterrenkundig Observatorium, Universiteit Gent, Krijgslaan 281 S9,B-9000 Gent, Belgium}\\
}

\begin{document}
\maketitle
\begin{abstract}
We present a new cosmological, magnetohydrodynamical simulation for galaxy formation: TNG50, the third and final installment of the IllustrisTNG project. TNG50 evolves $2\times2160^3$  dark-matter particles and gas cells in a volume 50 comoving Mpc across. It hence reaches a numerical resolution typical of zoom-in simulations, with a baryonic element mass of $8.5\times 10^4\MSUN$ and an average cell size of $70-140$ parsecs in the star-forming regions of galaxies. Simultaneously, TNG50 samples $\sim$700 (6,500) galaxies with stellar masses above $10^{10} \,(10^8)\, \MSUN$ at $z=1$. Here we investigate the structural and kinematical evolution of star-forming galaxies across cosmic time ($0 \lesssim z \lesssim 6$). We quantify their sizes, disk heights, 3D shapes, and degree of rotational vs. dispersion-supported motions as traced by rest-frame V-band light (i.e. roughly stellar mass) and by $\HA$ light (i.e. star-forming and dense gas). The unprecedented resolution of TNG50 enables us to model galaxies with sub-kpc half-light radii and with $\lesssim\,$300-pc disk heights. Coupled with the large-volume statistics, we characterize a diverse, redshift- and mass-dependent structural and kinematical morphological mix of galaxies all the way to early epochs. Our model predicts that for star-forming galaxies the fraction of disk-like morphologies, based on 3D stellar shapes, increases with both cosmic time and galaxy stellar mass. Gas kinematics reveal that the vast majority of $10^{9-11.5}\,\MSUN$ star-forming galaxies are rotationally-supported disks for most cosmic epochs ($\VMAX/\sigma>2-3$, $z\lesssim5$), being dynamically hotter at earlier epochs ($z\gtrsim1.5$). Despite large velocity dispersion at high redshift, cold and dense gas in galaxies predominantly arranges in disky or elongated shapes at all times and masses; these gaseous components exhibit rotationally-dominated motions far exceeding the collisionless stellar bodies. 
\end{abstract}



\section{Introduction}
\label{sec:intro}

A fundamental goal of galaxy evolution is to understand the prevalence, origin, and evolution of galaxy morphology across mass scales and as a function of cosmic time.
The structural morphology of a galaxy -- namely, the distribution of mass in three-dimensional space -- is dictated by the kinematics of its components. Morphology provides a snapshot of the orbital mix of constituent stars, gas, and dark matter (DM). In turn, both the structural properties and kinematics of a galaxy's stellar and gaseous bodies are determined by the physical processes governing its formation and subsequent evolution. 

Based on observations as well as theoretical studies of galaxies, a basic picture has become broadly accepted: a disk-like morphology (in stellar light) is usually associated with gas-phase kinematics dominated by rotation. This idea is supported by observations of star-forming, disk galaxies in the Local Universe (starting with the Milky Way \citep{Gunn:1979}, Andromeda, M51, and other nearby galaxies). However, observations of the higher redshift Universe motivate a more complex scenario. The majority of massive, star-forming galaxies detected in stellar light at $z\sim2$ are gas-rich rotating disks \citep[e.g.][]{ForsterSchreiber:2006, Genzel:2006, Genzel:2008}. Yet, these systems have a much larger contribution of random (i.e. disordered, non rotational, or even turbulent) local gas motion \citep[e.g.][]{Weiner:2006, Genzel:2006,Law:2012}. The latter is estimated by the gas velocity dispersion ($\sigma$), as obtained from $\HA$ emission lines for example. The common interpretation of this result is that rotationally-dominated disks are dynamically hotter at high redshift.

According to the canonical theory of cosmological galaxy evolution, galaxy shapes and kinematics are necessarily affected by a number of processes whose effects, signatures, and interplay vary with time and across galaxy masses. These include the accretion of cosmic gas, star formation, gas-rich and gas-poor galaxy mergers and interactions, feedback from intense star formation and super massive black holes, and consequent gas outflows and gas recycling. All such phenomena act upon the foundation provided by the hierarchical growth of structure, tidal forces, and halo-mass assembly as experienced, and dominated by, the underlying DM. 

In fact, it is observationally well established that the galaxy population evidences a wide diversity, if not a dichotomy, of fundamental properties. On one hand, we observe gas-poor galaxies with little or even no ongoing star-formation \citep{Welch:2010, Young:2011, Saintonge:2011}; at the same time, gas-rich galaxies populate a star-forming main sequence \citep{Saintonge:2011}. Different levels of star formation correlate with different morphological properties, at least for the bulk of the galaxy population: quiescent galaxies exhibit on average highly concentrated stellar light distributions typical of bulges while star-forming galaxies are well approximated by exponential disks \citep[e.g.][and references therein, for the local and high-redshift Universe, respectively]{Kauffmann:2003,Wuyts:2011}. Separations of this kind are often made in disjoint observational spaces, mostly for practical reasons and not only for morphological characterizations but also for kinematical descriptions. For example, measures from emission lines trace the gas in galaxies and are typically employed for star-forming, gas-rich galaxies, while measurements from absorption lines trace the stars and so are a viable diagnostic of the kinematics of gas-poor, quiescent galaxies. 

Indeed, the existence of galaxies that have ceased their star formation points towards remarkable physical mechanisms at play. While the processes that bring entire classes of galaxies to quench are highly debated, it is undeniable that different mechanisms have shaped the evolution of star-forming versus quiescent galaxies. It is then quite practical to study morphological evolution of these two galaxy types separately. In particular, we will focus herein on star-forming galaxies only, thereby avoiding the nature of quenching mechanisms and its contributing role to the evolution of galactic structure and kinematics. 

\subsection{Shapes and kinematics of star-forming galaxies in observations}

At intermediate and high redshift ($z \gtrsim 0.5$), the observational analysis of star-forming galaxies encounters a difficulty: galaxy morphologies are typically deduced form multi-wavelength imaging surveys tracing the stellar light, while galaxy kinematics are commonly obtained through $\HA$ spectroscopy. While studies of the stellar kinematics of star-forming galaxies are now common in the local Universe (e.g. with integral field spectroscopy data from SAMI, MANGA, and CALIFA), this is not yet viable at higher redshifts \citep[although see][]{Guerou:2017}.

The observational quantification of galaxy stellar morphologies has a century-long history with methods including visual inspection, light profile fits, and non-parametric measures \citep[see e.g.][and references therein for a review]{Conselice:2014}. In all cases the limitations of projection effects play a role, as data effectively provide two-dimensional structural characterizations. Recently, attempts to infer the three-dimensional geometry of galaxies -- the least ambiguous description of a galaxy's mass distribution -- have emerged \citep[e.g.][with SDSS galaxies]{Padilla:2008}, starting from the ansatz that the shapes of ellipticals and spirals can be well approximated by triaxial ellipsoids \citep{Sandage:1970,Lambas:1992}. For example, at high redshifts, analyses of a large sample of Lyman-break galaxies imply that a non-negligible fraction of star-forming galaxies at $z>1.5$ are intrinsically elongated, or strongly triaxial \citep{Law:2012b, Yuma:2012}. More recently, \citet{VanderWel:2014} and \citet{Zhang:2019} have independently determined the intrinsic shape distributions of star-forming galaxies at $0<z<2.5$ from SDSS and CANDELS. The qualitative picture uncovered is that the overall oblateness (or ``diskiness'') of galaxies increases with time, and that this process starts earlier in higher-mass galaxies. 

In terms of gas kinematics, the most updated measures at intermediate and high redshifts are provided by long-slit and integral field spectrograph (IFS) surveys of gas emission lines, chiefly $\HA$, that traces the $10^4$K gas ionized by young stars in the interstellar medium (ISM) of galaxies. At $z\lesssim1.2$, \citet{Kassin:2012} have analyzed an unbiased sample of  $10^{8-10.7}\MSUN$ stellar mass galaxies from DEEP2 long-slit spectroscopy and distilled the idea that galaxy disks settle with time, increasing in rotational velocity and declining in gas velocity dispersion. A similar analysis by \citet{Simons:2017} included data from the SIGMA survey up to $z\sim2.5$, based on MOSFIRE slit spectra, confirming the qualitative findings: populations of mass-selected galaxies compared across cosmic epochs suggest that galaxies tend toward rotational support with time, and higher-mass systems reach it earlier. More recently, \citet{Price:2019} analyzed the ionized-gas kinematics of more than 700 galaxies at $1.4 \lesssim z \lesssim 3.8$ from the MOSFIRE Deep Evolution Field survey (MOSDEF) finding that the rotational support increases with increasing stellar mass and decreasing specific star formation rate.

In parallel, spatially-resolved kinematic surveys that use IFS instruments at intermediate and high redshifts (MASSIV, SINS/zC-SINF, OSIRIS, AMAZE-LSD, PHIBBS, KMOS$^{\rm 3D}$) have studied the $\HA$ or molecular gas kinematics of samples of massive, star-forming, non-compact galaxies, in samples of a few to several hundreds, and up to $z\sim3.5$. They return a qualitatively consistent picture: at the cosmic noon, the star-forming ``main sequence'' is primarily composed of rotating gaseous disks, with gas velocity dispersions that decline from high to low redshifts \citep[e.g.][and references in Section~\ref{sec:kinematics}]{Swinbank:2012, Wisnioski:2015}. The interpretation is that rapid and efficient gas accretion is required in the first few billion years of cosmic evolution. At the same time, elevated random motions may result from feedback energy during the intense star formation that characterizes galaxies at $z \sim 2$ and/or the dynamics of rapid cosmic gas accretion itself. More recently, \citet{Swinbank:2017} have extended the gas kinematic analysis of star-forming galaxies throughout the $z=0.3-1.7$ range and down to galaxies of $10^8\MSUN$ thanks to MUSE observations.
Finally, \citet{Girard:2018} have further extended the investigated mass range down to $\sim4\times 10^9\MSUN$ galaxies at $1.4 < z < 3.5$ thanks to gravitational lensing.

\subsection{Structures and kinematics of galaxies from hydrodynamical simulations}
A complete theory for the formation of galaxies must reproduce the final, $z=0$, global and statistical properties of the observed galaxy population as a whole. It should also provide insights into the evolution and internal details of individual galaxies' structures and kinematics. This can be achieved in complete generality only with numerical simulations and, in particular, with hydrodynamical simulations of galaxies in the full cosmological context.

There has been significant progress in recovering the morphological diversity of the observed galaxy populations in recent years; particularly in large-volume hydrodynamical simulations like Illustris \citep{Vogelsberger:2014a, Vogelsberger:2014b, Genel:2014, Sijacki:2015} and EAGLE \citep{Schaye:2015, Crain:2015}, at roughly kpc and $10^6\MSUN$ spatial and mass resolution. With Illustris, \citet{Snyder:2015, Bottrell:2017a, Bottrell:2017b} have demonstrated, via realistic mock observations of Illustris galaxies, that the observed connections among mass, star formation, and galaxy structure can arise naturally from models matching global star formation and halo occupation functions, albeit with a deficit of bulge-dominated galaxies. In parallel, \citet{Correa:2017} and \citet{Thob:2019} have demonstrated, for example, that the EAGLE model also naturally produces a galaxy population for which stellar morphology is tightly correlated with the location in the color-mass diagram, with the red sequence mostly composed of elliptical galaxies and the blue cloud of disk galaxies. 

More recently, the connection between galaxy stellar morphologies and galaxy colors has been recovered also in the successor of Illustris, the IllustrisTNG simulations \citep{Naiman:2018, Nelson:2018, Marinacci:2018, Pillepich:2018, Springel:2018}, as demonstrated qualitatively by \citet{Nelson:2018}. The realism of the galaxies produced by currently-available hydrodynamical simulations is reaching unprecedented quantitative levels indeed with the TNG100 simulation, as explored by \cite{RodriguezGomez:2019}. Through synthetic images obtained by post-processing the simulated galaxies with SKIRT and including the effects of dust, we have demonstrated that the average $\MS \gtrsim 10^{10}\MSUN$ TNG100 galaxies exhibit values of optical morphological estimators like Gini-M20, concentration-asymmetry-smoothness statistics, and 2D Sersic indexes and sizes that are in remarkable agreement with Pan-STARRS data at $z\sim 0.05$.

However, in closely comparing structural measurements of galaxies to those available in observations, numerical works have generally neglected to quantify the most intrinsic and less ambiguous characterization of a galaxy's mass distribution: its three-dimensional shape, as for example approximated by an ellipsoid. Notable studies include \citet{Trayford:2018} and \citet{Thob:2019} of EAGLE galaxies, and \citet{Zhang:2019} together with \citet{Ceverino:2015} and \citet{Tomassetti:2016} who qualitatively contrasted the shape fractions inferred from CANDELS data to $\sim\,$30 high-redshift VELA zoom-in galaxies with stellar masses in the $10^{9-10}\MSUN$ range, finding broadly similar trends. The bulk of the currently-available analyses  of galaxy structures from simulations have focused almost exclusively on the low-redshift Universe.

In general, zoom-in simulations provide trailblazing insights into the formation and evolution of e.g. Milky Way-like galaxies and their morphological components \citep[e.g.][with Eris and the Auriga simulations, respectively]{Guedes:2013, Grand:2017}. However, they have been less useful in broadly testing the outcome of their underlying physical models against {\it population-wide} morphological observed estimators, being also more prone to be affected by subtle phenomena, like the butterfly effect \citep{Genel:2019}. This is because such projects are designed to sample only a few, or up to a few tens, galaxies at the time. These are often chosen to represent specific classes of galaxies, and typically favor better numerical resolution over statistics in comparison to $\sim\,$100 Mpc cosmological simulations as mentioned above. 

On the other hand, zoom-in simulations from the last five to eight years are the first actually suitable to provide predictions for the internal kinematics of galaxies. In most cases, kinematical analysis has focused on the measurement of the rotation velocities for Milky Way-like local disk galaxies \citep[e.g.][]{Guedes:2011}. In a few cases, velocity dispersion has been measured, particularly for disk galaxies at $z=0$. For example, this has been done for the stellar component with the purposes of understanding radial migration \citep{Roskar:2013}; and for the gas component to highlight the direct connection to the star formation-feedback loop \citep{Agertz:2013}. Thinking about higher redshifts, \citet{Kassin:2014} have analyzed four Milky Way-like galaxies and shown that they follow similar disk-settling trends as the observations suggest, with increasing (decreasing) rotational velocity (gas velocity dispersion) with time. However, those galaxies at high redshifts represent objects chosen, by construction, to be cold disks at $z=0$. More recently, \citet{Hung:2019} have measured the velocity dispersion of star-forming gas in four FIRE galaxies from $z=0$ to $z=4$ and found that $\sigma$ increases steeply from $z=0$ to $z\sim1.5$, with 100-Myr time variation that is connected to the evolution of the star formation rates and gas-mass inflows. However, also in that work, the simulated galaxies cover a narrow mass range, and with lower masses than those probed by current IFS surveys at intermediate and high redshift.

No studies have yet provided an extensive and conclusive kinematics analysis of populations of simulated star-forming galaxies that can be broadly contrasted to the results from slit-like and IFS-like current observations. So far, large uniform-volume simulations have been analyzed in order to shed light on the dichotomy and formation of fast and slow rotators \citep[e.g.][with Illustris and Magneticum]{Penoyre:2017, Li:2018, Schulze:2018} and to quantify the connection between galaxy morphology and spin on the one side and AGN feedback, environment and/or mergers on the other \citep[][with Illustris, Horizon-AGN \rvvv{and EAGLE}]{Rodriguez-Gomez:2015, Dubois:2016, Choi:2018, Lagos:2018, Lagos:2018b}. Recently, \citet{VanDeSande:2019} have compared structural and kinematic properties of simulated galaxies from EAGLE, Hydrangea \citep{Bahe:2017}, Horizon-AGN \citep{Dubois:2016}, and Magneticum (\textcolor{blue}{Dolag et al. in prep}) with observed galaxies in the SAMI, ATLAS$^{\rm 3D}$, CALIFA and MASSIVE surveys. However, \rvvv{in most cases} these works have \rvvv{principally} focused on massive early-type galaxies at $z\sim0$, which are expected to be mostly quenched. This choice has been dictated by the limited numerical resolution of such models, which is generally thought to be insufficient to properly capture even the kiloparsec-scale kinematics of intermediate and low-mass galaxies that are observationally accessible, especially in the low-redshift Universe. Other analyses have investigated the relation between stellar shapes and stellar kinematics with EAGLE and across galaxy types \citep{Thob:2019}. Furthermore, there is a rich body of works on the problem of the evolution and conservation of galaxies angular momentum, also informed by the results of large uniform-volume simulations and across galaxy types \citep[e.g. more recently][with Illustris and EAGLE]{Genel:2015, Lagos:2017}. However, in most cases so far, the focus has been placed on the final $z=0$ outcome, or on the stellar components, or the analyses have been carried out via theoretical characterizations of galaxies properties (e.g. resolution element-based angular momenta) that cannot be easily contrasted to observations. In fact, so far, no quantitative analysis of the spatially-averaged or map-based internal kinematics of star-forming galaxies within large uniform-volume simulations exists. 

{\renewcommand{\arraystretch}{1.2}
\begin{table*}
  \caption{Physical and numerical parameters of TNG50 and the other two flagship runs of the IllustrisTNG series, TNG100 and TNG300. The parameters are: the box side-length, the initial number of gas cells and dark matter particles, the target baryon mass, roughly equal to the average initial stellar particle mass, the dark matter particle mass, the $z$\,=\,0 Plummer equivalent gravitational softening of the collisionless component, the minimum comoving value of the adaptive gas gravitational softenings, and the average cell size of star-forming gas across the simulated volumes. Lastly, the total run time including substructure identification in CPU core hours, and the number of compute cores used. For further details on the adaptive mass and spatial resolution of the gas component see Section~\ref{sec:res}.}
  \label{tab:sims}
  \begin{center}
    \begin{tabular}{lcccccccccc}
     \hline
     
 Run Name & $L_{\rm box}$ & $N_{\rm GAS}$ & $N_{\rm DM}$ & $m_{\rm baryon}$ & $m_{\rm DM}$ & $\epsilon_{\rm DM,stars}^{z=0}$ & $\epsilon_{\rm gas,min}$ & $\bar{r}_{\rm cell, SF}$& CPU Time & $N_{\rm cores}$ \\
  & [\,com Mpc\,] & - & - & [\,M$_\odot$\,] & [\,M$_\odot$\,] & [\,pc\,] & [\,phys pc\,] & [\, com pc\,] & [\,Mh\,] & - \\ \hline
\textbf{TNG50}(-1) & $51.7$ & $2160^3$ & $2160^3$ & $8.5\times10^4$  & $4.5\times10^5$   & 288 & 72 & 140 & $\sim130$    & 16\,320 \\
 \textbf{TNG100}(-1) & $110.7$ & $1820^3$ & $1820^3$ & $1.4\times10^6$  & $7.5\times 10^6$   & 738 & 190 &355& 18.0    & 10\,752 \\
 \textbf{TNG300}(-1) & $302.6$ & $2500^3$ & $2500^3$ & $1.1\times10^7$   & $5.9\times10^7$    & 1477 & 370 &715& 34.9    & 24\,000 \\
 \hline
 
    \end{tabular}
  \end{center}
\end{table*}}

\subsection{The current work and TNG50}

The new TNG50 simulation that we introduce here provides a transformational step forward in uncovering the structural and kinematics properties of simulated galaxies across cosmic time. Specifically, we model and analyze here for the first time the 3D shapes and internal kiloparsec-scale structure and dynamics of thousands star-forming galaxies, including hundreds of $\geq10^{10}\MSUN$ at $z\gtrsim1$, all modeled with a physically-motivated, though necessarily simplified, numerical treatment of star formation and feedback that acts below $\sim\,$100 parsec scales in the ISM. 

Our overarching goal is to present theoretical predictions from TNG50 for both $\HA$ and stellar-light tracers. We then place them into the context of the findings of currently available long-slit and IFS observations (see above), in anticipation of future, highly detailed high-redshift galaxy observations. Our approach is supported by a numerical model that includes the dominant mechanisms that are expected to influence the evolution of the galaxy properties of interest (see above and Section~\ref{sec:tng50}). In particular, we measure the sizes, disk heights, intrinsic 3D shapes, rotational velocities and velocity dispersions of stars and star-forming i.e. $\HA$-emitting gas in the inner regions of galaxies, with two objectives. First, to uncover outcomes of TNG50 for which the model has not been in any way calibrated and is thus predictive. Second, to contrast structural versus kinematical features, as well as the properties of the stellar versus gaseous components of galaxies. We hence focus on a redshift regime ($z\gtrsim 0.5$) where such comparisons are currently prohibitive in observations, though soon to emerge: in this way, we maximize the predictive return of this work and fill a gap in the current landscape of theoretical studies based on hydrodynamical models. In this first paper on the topic, we focus on the evolution of galaxy populations on the star-forming main sequence, postponing an analysis of quenched galaxies. We connect galaxy populations across redshift at fixed stellar mass, postponing the study of individual galaxy tracks. 
 
The structure of the paper is as follows. In Section~\ref{sec:tng50}, we describe the numerical and technical details of the new TNG50 simulation. We provide the analysis methods and operational definitions of galaxy properties in Section~\ref{sec:methods}. Statistics of the star-forming galaxy population of TNG50 are given in Section~\ref{sec:sample}. The results are quantified and discussed in two phases: Section~\ref{sec:results_1} for the structural morphologies and Section~\ref{sec:results_2} for the kinematics, respectively. A discussion that brings together and critically addresses all the presented results is given in Section~\ref{sec:discussion}, and we summarize and conclude in Section~\ref{sec:summary}.


\section{The TNG50 Simulation}
\label{sec:tng50}

TNG50 is the most computationally-demanding and highest-resolution realization of the IllustrisTNG simulation project\footnote{\url{www.tng-project.org}}. The first two simulations of the suite, TNG100 and TNG300, have already been introduced and analyzed for a variety of galaxy-physics and cosmology topics \citep{Nelson:2018, Naiman:2018, Marinacci:2018, Pillepich:2018, Springel:2018}. Here, together with the companion paper by \textcolor{blue}{Nelson et al. 2019}, we present the third and final volume of the project.

TNG50 provides an unprecedented combination of volume and resolution. It evolves dark-matter, gas, stars, black holes, and magnetic fields within a uniform periodic-boundary cube of 51.7 comoving Mpc on a side (hence the name), containing 2$\times 2160^3$ total initial resolution elements: half dark-matter particles and half Voronoi gas cells, in addition to an equal number of Monte Carlo tracer particles. Table~\ref{tab:sims} provides the most important details and a comparison to the other flagship IllustrisTNG runs. As with all flagship TNG simulations, TNG50 (aka TNG50-1) is also equipped with a series of lower-resolution realizations of the same volume (to control for resolution effects), each accompanied by its dark-matter only counterpart (to uncover baryonic effects). 
The second level, TNG50-2, has a comparable, but slightly better, resolution to TNG100, while the third level, TNG50-3, compares closely to TNG300. TNG100 (TNG300) has a particle resolution that is a factor of 15 (120) worse than TNG50(-1) -- see Appendix~\ref{sec:app_res}. 

\subsection{The initial conditions}

In order to simulate a representative volume and minimize sample variance, we have selected our simulation from a set of sixty random realizations of the initial density field that have been evolved at low resolution to the current epoch. From these sixty realizations we have chosen the most average cumulative dark matter halo mass function, considering haloes more massive than $10^{10}\MSUN$. The volume selected for TNG50 includes at $z=0$ one object exceeding $1.82\times10^{14}\MSUN$ (a Virgo-cluster analog), about sixty massive galaxies residing in halos of $10^{12.5}\MSUN$ and above (massive ellipticals), $\sim\,$ 130 galaxies as massive as the Milky Way, including a priori both disk and elliptical galaxies, and thousands of dwarf galaxies. It is therefore a fully representative patch of the Universe, encompassing a diverse range of environments and large-scale overdensities across cosmic time. The inclusion of the cosmologically expected average abundance of objects, in particular the more massive halos, is reflected in the computational expense to execute such a run all the way to redshift zero. 

The initial conditions of all simulations of the series have been created at $z=127$ using the Zeldovich approximation and the {\sc N-GenIC} code (\textcolor{blue}{Springel, priv. comm.}). The adopted cosmological parameters are consistent with the \citet{PlanckXIII:2015} results, with matter density $\Omega_{\rm m} =\Omega_{\rm dm} + \Omega_{\rm b} = 0.3089$, baryonic density $\Omega_{\rm b} = 0.0486$, cosmological constant $\Omega_\Lambda=0.6911$, Hubble constant $H_0 = 100\,h\, {\rm km\, s^{-1}Mpc^{-1}}$ with $h=0.6774$, normalisation $\sigma_8 = 0.8159$ and spectral index $n_s=0.9667$.

\subsection{The model}

TNG50, as with the other IllustrisTNG simulations, has been evolved with {\sc arepo}  \citep{Springel:2010}, a massively-parallel simulation code optimized for large runs on distributed memory machines. It solves the coupled equations of ideal magnetohydrodynamics (MHD) and self-gravity. For the (magneto)hydrodynamics, the code uses a finite volume method on an unstructured, moving, Voronoi tessellation of the simulation domain. This provides a spatial discretization for Godunov's method with a directionally unsplit second order scheme \citep{Pakmor:2016a}. A fundamental strength of the code is that the generating points of the Voronoi mesh are allowed to move in time, through the assignment of a velocity close to the local fluid velocity field. Consequently, the mesh has no preferred directions or Cartesian grid structure. Ideal MHD is solved in a single-ion plasma approximation, with the divergence constraint of the magnetic field handled by the eight-wave Powell cleaning scheme \citep{Pakmor:2011, Pakmor:2013}; a tiny, uniform primordial magnetic-field seed ($10^{-14}$ comoving Gauss at $z\sim127$) is imposed at the initial conditions. Finally, Poisson's equations for gravity are solved via a tree-particle-mesh (TreePM) algorithm, with Voronoi gas cells treated as point masses at the position of their centres of mass and all other matter components appropriately sampled as particles. 

The model for galaxy formation includes stochastic, gas density-threshold based star formation; evolution of mono-age stellar populations represented by star particles; chemical enrichment of the interstellar medium (ISM) and the tracking of nine different chemical elements (H, He, C, N, O, Ne, Mg, Si, Fe) in addition to the total gas metallicity and Europium; gas cooling and heating; feedback from supernovae in the form of galactic winds; seeding and growth of supermassive black holes and the injection of energy and momentum from them into the surrounding gas. All physics aspects of the model, including parameter values and the simulation code, are described in the two IllustrisTNG method papers \citep{Weinberger:2017, Pillepich:2018Method} and remain entirely unchanged for our production simulations, including TNG50. One numerical update has been applied to execute the TNG50 simulation, namely a change to the shape of the effective equation of state of the star-forming gas: this has been done in order to promptly convert the highest-density gas into stars and hence avoid time steps as small as {\it 10 years} at the resolution of TNG50. In practice, this change was motivated by numerical reasons alone and has no impact on galactic properties or statistics at any cosmic time (see Section 2.2 of \textcolor{blue}{Nelson et al. 2019} for details). 

\begin{figure*}
\centering      
\includegraphics[width=15.5cm]{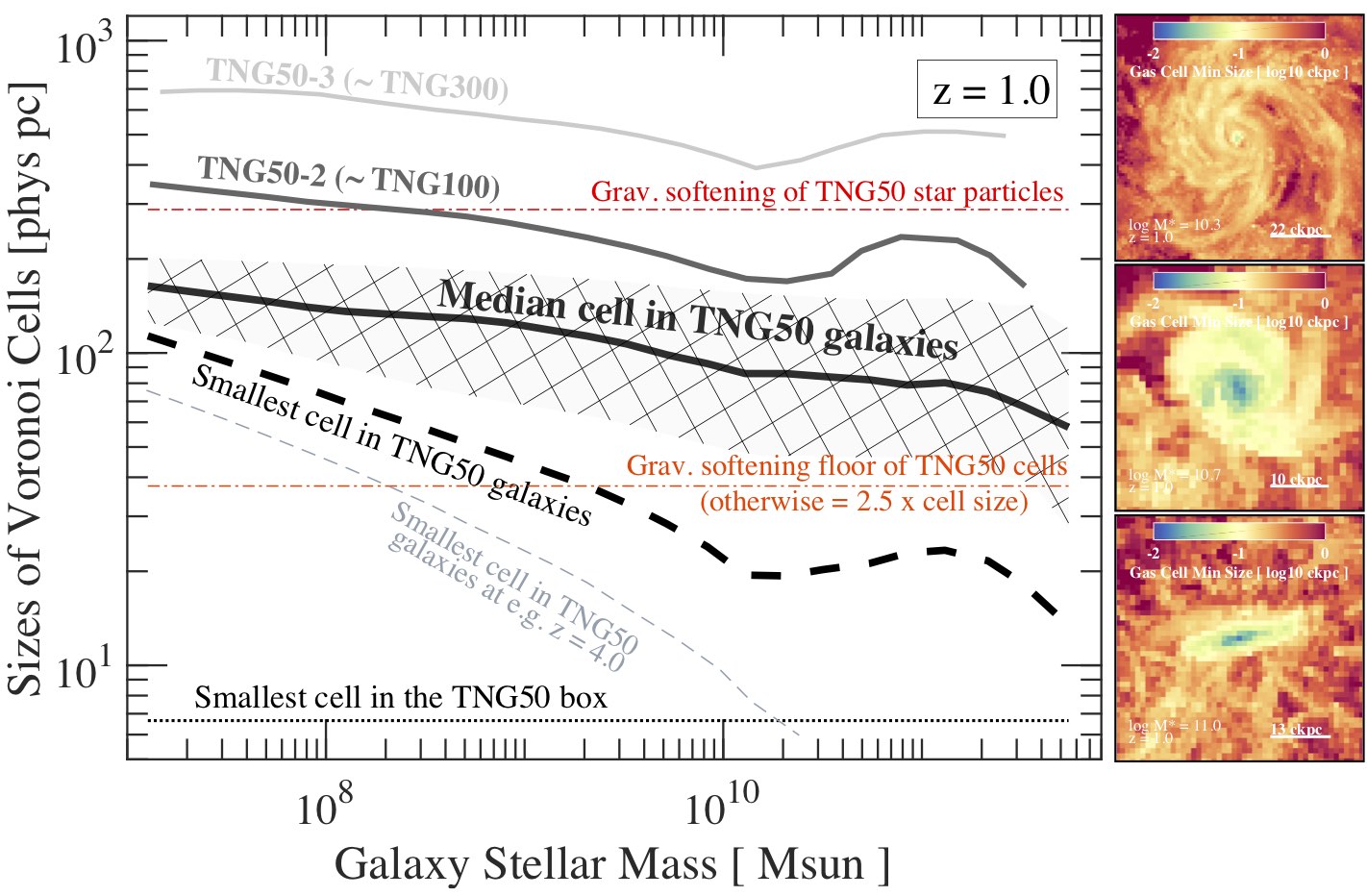}
\caption{\label{fig:cellsizes} Spatial resolution of the baryonic elements in TNG50. Left panel: distribution of the Voronoi cell sizes in TNG50 galaxies, taken as the spherical volume equivalent radii and representing the spatial resolution over which the equations of magnetohydrodynamics are solved. The black solid curve represents the average across the galaxy population of the {\it median} sizes of star-forming gas cells within galaxies at $z=1$ (within twice the stellar-half mass radius); the hatched region denotes the 5th to 95th percentiles. The smallest cells within galaxies can have sizes that are several times smaller than the median: the galaxy-population average at $z=1$ is shown with the black dashed curve, and compared to the $z=4$ measurement (thin gray dashed curve). For reference, the smallest gas cell across the whole volume measures 6.5 phys pc. These sizes are compared to the gravitational softening lengths of both stars and gas, in red and orange respectively. Analog measurements at the lower-resolution levels of TNG50-2 and TNG50-3 are also included (dark and light gray solid curves). Right stamps: spatial distribution of the Voronoi cell sizes in three random galaxies, showing the smallest cell size along the line of sight, binned in pixels of size 1 ckpc. While central regions typically reach resolutions of a few tens of parsecs, the majority of the star-forming disk is resolved at $\sim\,$100 parsec scales.}      
\end{figure*}

\subsection{Notes on numerical resolution}
\label{sec:res}

A defining feature of TNG50 in comparison to previous cosmological uniform-volume simulations is its high numerical resolution. We therefore provide some important comments on this aspect.

As summarized in Table~\ref{tab:sims}, in TNG50 the DM particle mass is $4.5\times10^5\MSUN$, while the mean baryonic gas mass resolution is $8.5\times10^4 \MSUN$, maintained within a factor of two of this target through cell (de-)refinement. Stellar particles inherit a similar initial mass but subsequently lose mass via stellar evolution. 

The Plummer equivalent gravitational softening of the collisionless components (dark matter, stars and wind particles) is 575 comoving pc until $z=1$, after which it is fixed to its physical value of 288 pc at $z=1$, down to the current epoch. The gravitational softening of the gas component is, instead, adaptive and proportional to the effective cell radius, $\epsilon_{\rm gas} = 2.5 r_{\rm cell}$. This value is calculated as the radius of the sphere with volume equal to the Voronoi gas cell volume, a good approximation for the cell size since mesh regularization maintains roughly spherical cell shapes. A minimum softening is enforced: here it corresponds to 72 comoving pc at all times, so that it does not become too much smaller than the collisionless softening. 

The adaptivity of the Voronoi mesh translates into a continuous distribution of possible cell sizes that drop to very small values in the highest density environments. As a result, no single number can robustly describe the spatial resolution over which the magnetohydrodynamics is resolved. In Fig.~\ref{fig:cellsizes}, we therefore illustrate and quantify the distribution of cell sizes ($r_{\rm cell}$) in TNG50 galaxies at $z=1$. The black solid curve shows the average across the galaxy population of the {\it median} size of star-forming gas cells within galaxies (within twice the stellar-half mass radius). The smallest cells within galaxies can have sizes that are many factors smaller than the median size: the galaxy-population average at $z=1$ is denoted by the black dashed curve, and compared to the analog at $z=4$ (thin gray dashed curve). The smallest gas cell across the whole TNG50 volume at $z=1$ is 6.5 physical pc. These sizes can be compared to the gravitational softening lengths of both stars and gas, in red and orange annotations respectively. The median cell sizes in typical galaxies at two lower-resolution levels are also shown: TNG50-2 and TNG50-3 (dark and light gray solid curves), which are similar to TNG100 and TNG300, respectively.

On the right column of Fig.~\ref{fig:cellsizes} we illustrate the spatial distribution of  Voronoi cell size within three random galaxies at $z=1$. The colorbar indicates the smallest Voronoi cell along the line of sight, binning in pixels of 1 comoving kpc, and ranges from 10 comoving parsecs (in blue) to 1 comoving kpc (in red). By construction the cell sizes trace the gas density, with cells as small as a few tens of parsecs in the central, high-density regions of galaxies. Overall, we see that Voronoi sizes of about 100 pc are typical of the star-forming regions of galaxies, e.g. in the spiral arms: our numerical prescriptions -- e.g. the conversion of gas into stars and the launching of stellar winds -- are subgrid only below this $\sim$100-parsec scale in the ISM.


\section{Analysis Methods and operational definitions}
\label{sec:methods}

The goal of this paper is to provide an atlas of the structural and kinematical properties of star-forming galaxies and their evolution across time in the TNG50 simulation. To begin, we first describe how galaxies are identified and their properties characterized.

\subsection{Galaxy identification}

We identify haloes, and the subhaloes which inhabit them, via the Friends-of-Friends \citep[FoF][]{Davis:1985} and {\sc subfind} \citep{Springel:2001} algorithms, respectively. Within each FoF halo, subhaloes identified by the {\sc subfind} algorithm are made up of all the resolution elements (gas, stars, dark matter, and black holes) which are gravitationally bound to the subhalo. By definition these elements are disjoint from the material in other subhaloes, including satellite galaxies. In our framework, galaxies are subhaloes with non vanishing stellar mass (see definition below), excluding those with less than 20 per cent of their mass in dark matter \citep[see Section 5.2 of][]{Nelson:2019a}. The sample includes both central and satellite galaxies; for each, its position is taken as the location of the particle with the minimum gravitational potential energy. A central galaxy is generally the most massive subhalo in its halo, and may be associated with one or more satellites. Here, satellites are {\sc subfind}-identified galaxies that are members of their parent FoF group regardless of their distance from its centre. Unless otherwise stated, we study both central and satellite galaxies without distinction.

\subsection{Galaxy descriptors}
\label{sec:descriptors}

We consider galaxy bodies holistically, and quantify the phase-space properties of their stellar and gaseous components. Key aspects are schematically summarized in Table~\ref{tab:tracers} and detailed here.

We describe galaxies both in terms of stellar mass and rest-frame V-band light. Stellar mass is simply the sum of the current mass of stellar particles in a given volume, hence taking into account mass loss due to supernovae, AGB stellar winds and general stellar evolution. A Chabrier IMF is adopted throughout \citep{Chabrier:2003}. The V-band light (Johnson-V broad filter centered at $\lambda_{\rm eff} = 5506.9$\AA, Vega magnitudes) is calculated for every stellar particle via the \citet{BruzualANDCharlot:2003} stellar population synthesis model given a particle's stellar age and metallicity. The redshifted rest-frame V band falls within the NIR-Cam filters on board the James Webb Space Telescope (JWST), hence providing a useful, observationally accessible descriptor of galaxy populations from intermediate to very high redshifts ($1\lesssim z \lesssim9$). 

In terms of the gaseous component, we are interested in (i) the total gas mass, without distinctions based on density, temperature or metallicity, as well as in (ii) star-forming gas, together with its observational proxies. In the simulation, when a gas cell exceeds a density threshold ($n_{\rm H} = 0.13\,$cm$^{-3}$), it is dubbed {\it star forming}. In practice, in the IllustrisTNG model, cold high-density gas is placed on an equation of state between temperature and density \citep{Springel:2003}. As a result, our ability to physically describe the small-scale properties of the inherently multi-phase ISM is limited in this regime.

For this paper, star-forming gas in the simulation traces the sites of star formation, and can be described in terms of its mass and star formation rate (SFR, instantaneous). To connect to a widely used and accessible observational tracer, we associate an $\HA$ luminosity to each star-forming gas cell using the canonical calibration of \citet{Kennicutt:1998}:

\begin{equation}
\label{eq:halpha}
  {\rm SFR}(\MSUN~ {\rm year}^{-1}) = 7.9 \times 10^{-42} ~L (\HA) ~ [ergs~ s^{-1}]
\end{equation}
 
\noindent While different measurements of SFRs can be used in this relation (see next subsection), here we adopt the instantaneous, theoretical star formation rate of the gas cells. For the present analysis, we consider only the spatial distribution of $\HA$ luminosity, and not its quantitative magnitude. Still, we note that $\HA$-emitting gas in reality is likely more directly affected in its motions by stellar feedback than in TNG, where we invoke a wind particle scheme which does not directly disturb the star-forming gas from which it originates \citep[see][and \textcolor{blue}{Nelson et al. 2019} for details on the implementation and a characterization of the properties of galactic outflows in the TNG model, respectively]{Pillepich:2018Method}.

Occasionally, we will refer to neutral hydrogen as the fraction of the mass of each gas cell in hydrogen, as self-consistently tracked, which is also neutral, according to the fiducial self-shielding prescription. We combine together atomic and molecular phases and ignore complications therein (see \citet{Diemer:2018, Stevens:2019} and \textcolor{blue}{Popping et al. 2019} for further details on the treatments of HI and H$_2$). Finally, we neglect any effect of dust on the considered observables.

{\renewcommand{\arraystretch}{1.2}
\begin{table*}
  \caption{\label{tab:tracers}Schematic overview of the galaxy components or tracers considered in this paper:}
\begin{tabular}{|p{3cm}|p{6cm}|p{7.3cm}}
\hline
Galaxy Descriptor & Technical Description & Notes \\
\hline
Stellar Mass & Sum of the mass of all individual stellar particles of a galaxy, assuming a \citet{Chabrier:2003} IMF. & Not directly observable, but traces the distribution and dynamics of mono-age stellar populations. \\
V-band Stellar Light & Integrated light from star particles at $0.55 \mu$m (rest-frame), using  \citet{BruzualANDCharlot:2003}. & Traces stellar particles according to their age and metallicity, neglecting dust attenuation effects. \\
Gas Mass & Sum of the mass of all gas cells particles of a galaxy. & Not observable, as it includes all gaseous phases regardless of temperature, density, ionization state, chemical composition. \\
Neutral Hydrogen Mass & Sum of the fraction of the hydrogen cell mass (or density) in neutral phase. & Includes both molecular and atomic phases without attempting to separate the two. \\
$\HA$ Light & Sum of the light associated to star formation within a galaxy, i.e. proportional to the instantaneous SFR of gas cells, as per Eq.~\ref{eq:halpha}. & In TNG traces gas that is star forming (i.e. cold and dense). Given our ISM model, this $\HA$-emitting gas would trace multiple observational phases including molecular and ionized gas without necessarily distinguishing between them. \\
\hline
\end{tabular}
\end{table*}
}

\subsection{Galaxy properties} 
\label{sec:props}

The operational definitions and hence measurements of galaxy properties in this paper serve two simultaneous purposes. First, we want to describe our simulated galaxies based on properties that are broadly similar to available observables. At the same time, we want to uncover intrinsic characteristics and their interrelations without the limitations and biases encountered in observations. In practice we do not create `mock' observations, and our measurements do not exactly replicate any observational choice available in the literature. As a result, comparisons to observational datasets should be undertaken with care -- detailed quantitative observational comparisons are postponed to future analyses. We measure:

\paragraph*{Galaxy center and bulk velocity.} The center and bulk (or systemic) velocity define the coordinate system with respect to which positions and velocities of all galaxy components are measured. Although the simulation halo finder places a galaxy center at the position of its most-gravitationally bound resolution element, this does not necessarily coincide with e.g. the peak of the stellar light distribution. 
We therefore take the average 3D position and 3D velocity of the 5 per cent most gravitationally bound resolution elements. We use stars for stellar mass (or light) measurements, and likewise gas cells for gas mass, $\HA$ or neutral hydrogen measurements.

\paragraph*{Stellar mass.} Throughout this paper, a galaxy's stellar mass is the sum of the current mass of all stellar particles within 30 physical kpc from the galaxy center, excising satellites. 

\paragraph*{Star formation rate (SFR).} Unless otherwise stated, SFRs are estimated from the instantaneous gas properties, which we regard as the true SFR of the galaxies. Alternatively, we could measure the SFRs in galaxies averaged over different time scales, i.e. 10, 50, 200 Myrs: the relationship between these measurements has been quantified in TNG by \citet{Donnari:2019}. 
We always sum over resolution elements within an aperture of 30 physical kpc. 
The specific star formation rate (sSFR) is the ratio between SFR and stellar mass, the latter as defined above. 

\paragraph*{Size ($r_{1/2}$).} Every galaxy is characterized by a 3D stellar and 3D total gas half-mass radius and 2D half-light radii in V band, $\HA$ or neutral hydrogen mass. In all cases, to determine the total mass or light, we account for all gravitationally-bound material without mimicking any surface-brightness or signal-to-noise limits. As  a result our size estimates may be larger than those typically inferred observationally. Importantly, for the 2D projected, circularized sizes, we first rotate galaxies in their face-on projections (see below). The symbol $R_{\rm stars}$ always denotes the 3D stellar half-mass radius of a galaxy. 

\paragraph*{Galaxy `up vector' and edge-on vs. face-on orientation.} We define the `up vector' of a galaxy as the shortest eigenvector (minor axis) of the mass tensor, accounting for all the gravitationally-bound mass within the 3D half-mass radius. In a coordinate system that is aligned with the eigenvectors (principal axes) of the mass tensor, many edge-on projections are possible, and we choose one at random. In general, we determine the mass tensor (and projections) of a galaxy independently for stellar or gas measurements. The diagonalization of the mass tensor provides us with a definition of the structural major and minor axes of a galaxy. 

\paragraph*{Disk height ($h_{1/2}$).} We measure a ``thickness'' or ``disk height'' for every galaxy by rotating it into an edge-on projection and determining how rapidly the mass (or light) profiles decrease as a function of distance from the mid-plane. The disk thickness is taken as the half-mass or half-light height ($h_{1/2}$) accounting for all the mass or light (of stars or gas) within $2R_{\rm stars}$ in both the major and minor axes directions, i.e. we place a squared slit aligned with the galaxy's major axis.\footnote{We have also measured the disk heights by considering all material within $1R_{\rm stars}$ or excluding the central regions, i.e. in the $1-2R_{\rm stars}$ range, in order to exclude the possible contribution of a bulge. In fact, stellar heights excluding the innermost regions are up to a factor of two larger, consistently across masses and times, indicating disk flaring. The qualitative conclusions and trends we describe in the next sections are unaffected by these choices and the quantitative effects are negligible for the scope of this analysis.} We do not characterize disk heights based on exponential-profile fitting because we require a measure that applies to all types of galaxies, not just disk-like galaxies.

\paragraph*{Shape.} We characterize the structural major and minor axes of a galaxy based on the 3D intrinsic shapes of galaxies, diagonalizing the stellar (or $\HA$-emitting gas) mass tensor. The shape of a galaxy is summarized with two parameters: the minor-to-major axis ratio ($s$ or sphericity) and the middle-to-major axis ratio (the $q$ parameter). To estimate a galaxy's shape we: a) define shapes in elliptical shells; b) iteratively diagonalize the mass tensor and redefine the elliptical shells for the calculation until the direction and modulo of the eigenvectors of the mass tensor stabilize. This is done by keeping fixed the direction of the major axis and keeping the major-axis length fixed to $2R_{\rm stars}$ (as in \citealt{Chua:2019}). Our fiducial choice is to measure 3D shapes at $2R_{\rm stars}$ and within an elliptical shell with a thickness of $0.4 R_{\rm stars}$.\footnote{Due to shape changes as a function of galactocentric distance within galaxies, quantitative results depend on the elliptical aperture. We will generally comment on the comparison between shape measurements at $R_{\rm stars}$ versus $2R_{\rm stars}$. We also note that non-iterative or non-shell determination of axis ratios systematically biases the results, labeling galaxies more spherical than implied by their iso-density contours.}

\paragraph*{Kinematics.} 
From the kinematics view point, we characterize our simulated galaxies with summary statistics of the velocity fields of the V-band and $\HA$-emitting components within apertures of $2R_{\rm stars}$ in projection. Our measurements of velocity dispersion and rotational velocity capture intrinsic dynamics, and deliberately take advantage of optimal view directions, unlike in observations.\\

 \noindent {\it Velocity dispersion} ($\sigma$). We measure the velocity dispersion of V-band stellar light and $\HA$-emitting gas by projecting galaxies face-on and including particles/cells within a cube of side-length $8R_{\rm stars}$. We thus calculate the average line-of-sight unweighted velocity dispersion in pixels of 0.5 comoving kpc\footnote{Our pixel sizes are chosen to fall between what our numerical resolution enables and what is observationally possible at present (spatial resolutions in IFS observations ranging from a fraction to a few physical kpc, from low redshift to $z\sim 2-3$). Average velocity dispersion values do depend on pixel size, especially in the cases where the velocity field is highly inhomogeneous (larger pixels imply larger $\sigma$).
Detailed comparison with observational results requires a more closely matched analysis.}, as defined in Appendix~\ref{sec:app_sigma}. Next, we employ a virtual ``slit'' aligned with the structural major axis of the galaxy that extends from $[-2 \RSTARS, + 2 \RSTARS] $ and $[-1/5 \RSTARS, + 1/5 \RSTARS]$ along the major and middle axes, respectively. We measure one line-of-sight velocity dispersion per galaxy: the mean dispersion of the pixels within $1-2\times R_{\rm stars}$ along the slit. We measure the velocity dispersion in face-on projection in order to remove the contribution from possibly ordered motions. We also exclude the central regions because we are interested in the kinematics of galaxy disks, excluding for example contributions from stellar bulges. This procedure is visualized in Figs.~\ref{fig:kinematics1} and \ref{fig:kinematics2} for the stellar and $\HA$ components, respectively. Our estimates for velocity dispersion naturally and deliberately include the effects caused by feedback on the motions of the gaseous components (and indirectly the stellar components). They also include contributions from extraplanar gas, depending on the 3D distribution of the $\HA$-emitting gas. \rvvv{Throughout the analysis, we quantify the effects of the thermal motions of the gas: unless otherwise explicitly noted, our results do not take into account the contributions from thermal motions.} This choice is dictated by the limitations of the underlying ISM model, whereby it is not easy to associate a physically-meaningful value to the internal energy of star-forming (i.e. $\HA$-emitting) gas cells: \rvvv{in practice, in our model, to include a thermal component to the $\HA$ velocity dispersions entails some degree of uncertainty (see Appendix~\ref{sec:app_sigma} for more details)}. \rvvv{On the other hand, from an observational perspective, it is often hard, if not impossible, to obtain an intrinsic measure of $\sigma_{\rm gas}$ subtracted of the contribution from thermal motions: therefore in order to provide results that are closer to what typically obtained observationally, we also provide TNG50 estimates of the total gas velocity dispersion. In any case, our} measures correspond to ``vertical'' velocity dispersions if disks are in place. They are a priori different from estimates of the tangential or ``in-plane'' velocity dispersions but can be more directly related to the vertical heights of galaxies, as we do in the Discussion. \\ 

 \noindent {\it Rotational velocity} ($\VMAX$). We measure rotational velocity as the maximum of the rotation curve within $2R_{\rm stars}$ in projection. In particular, we: 1) project galaxies edge-on, including particles/cells within a cube of side-length $8R_{\rm stars}$, 2) derive the mean line-of-sight unweighted velocity of the stars (or gas) in pixels of 0.5 ckpc; 3) place a virtual ``slit'' aligned with the structural major axis of the galaxy that extends from $[-2 \RSTARS, + 2 \RSTARS] $ and $[-1/5 \RSTARS, + 1/5 \RSTARS]$ along the major and minor axes, respectively; 4) derive the 1D profile of mean line-of-sight velocity versus distance along the slit; 5) record the maximum of the absolute value of this rotation curve within $2R_{\rm stars}$, labeling it $\VMAX$. Given the statistical nature of our simulated sample, we deliberately neglect possible asymmetries in the rotation curves and the possibility that the de-facto kinematical major axis is mis-aligned with respect to the structural major axis. However, we explore a number of alternative definitions and comment on these outcomes below. Examples of rotation curves from TNG50 galaxies are shown in Figs.~\ref{fig:kinematics1} and \ref{fig:kinematics2}, for the stellar and $\HA$ components, respectively. Our measures correspond to ``azimuthal'' velocities within the disk plane, if a disk is in place, and reflect more prominently the velocities of the material along the structural major axis, by construction.

\begin{figure*}
\centering                                      
\includegraphics[height=5.7cm]{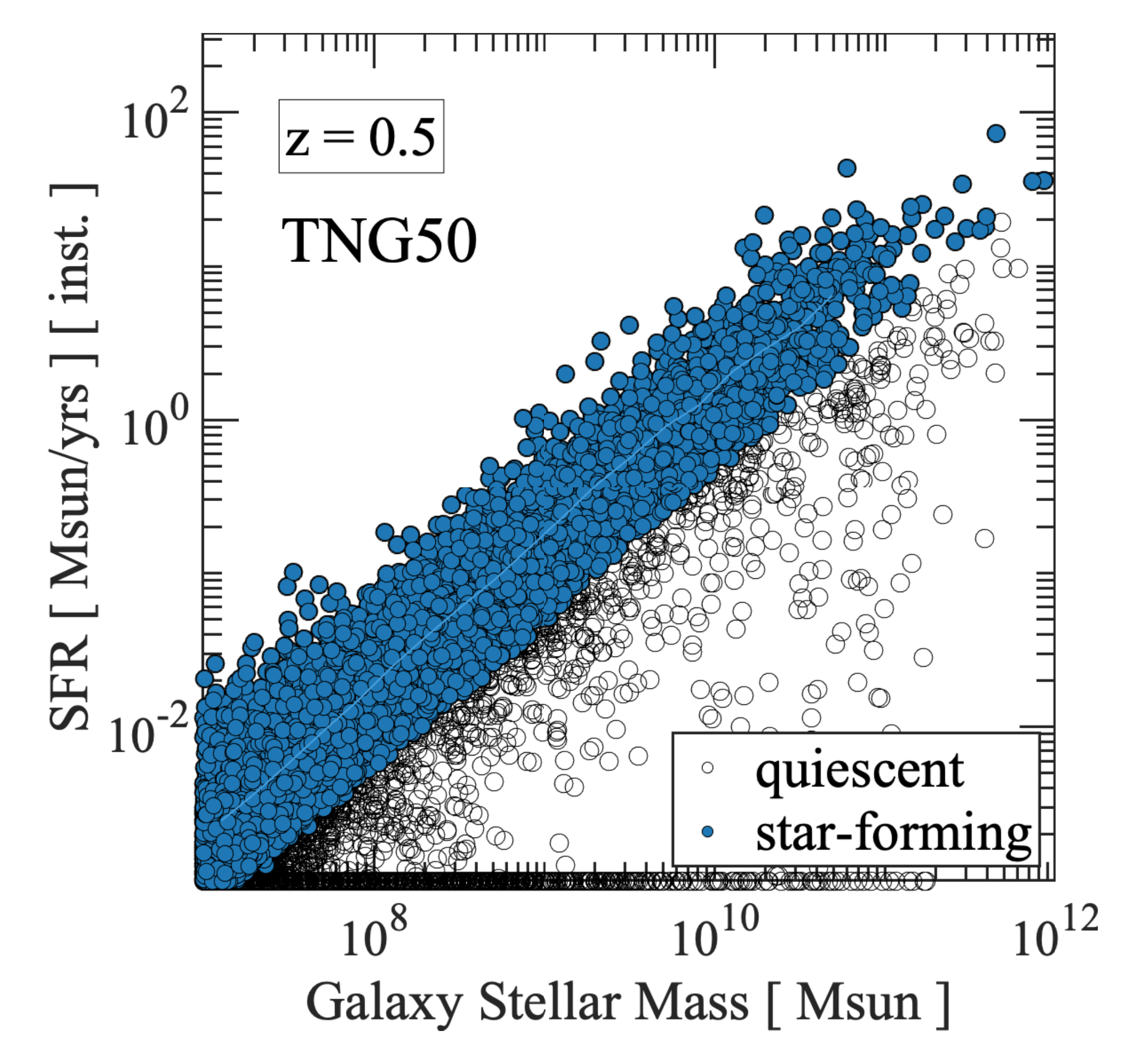}
\includegraphics[trim={1.5cm 0 0 0}, height=5.7cm]{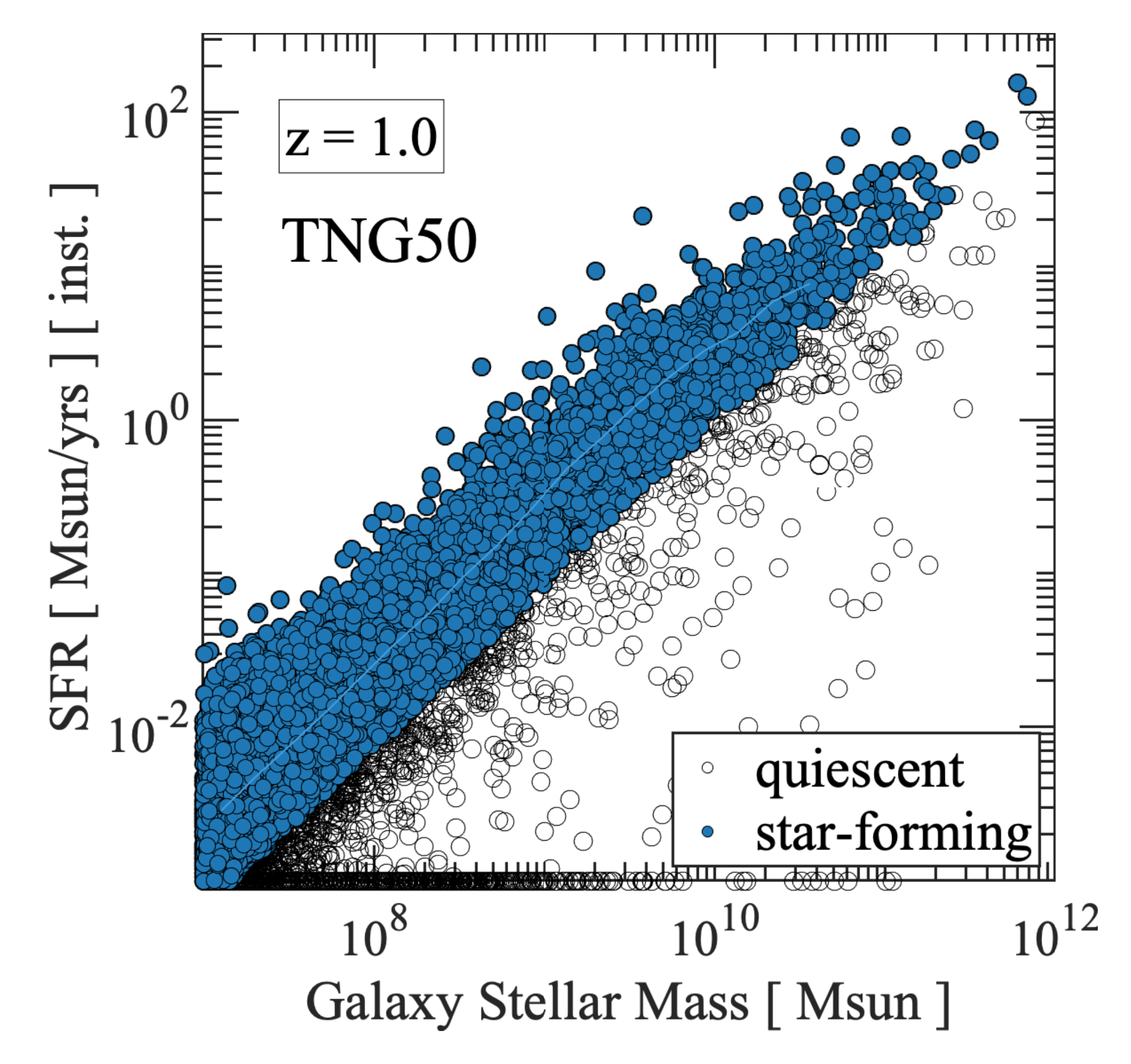}
\includegraphics[trim={1.5cm 0 0 0}, height=5.7cm]{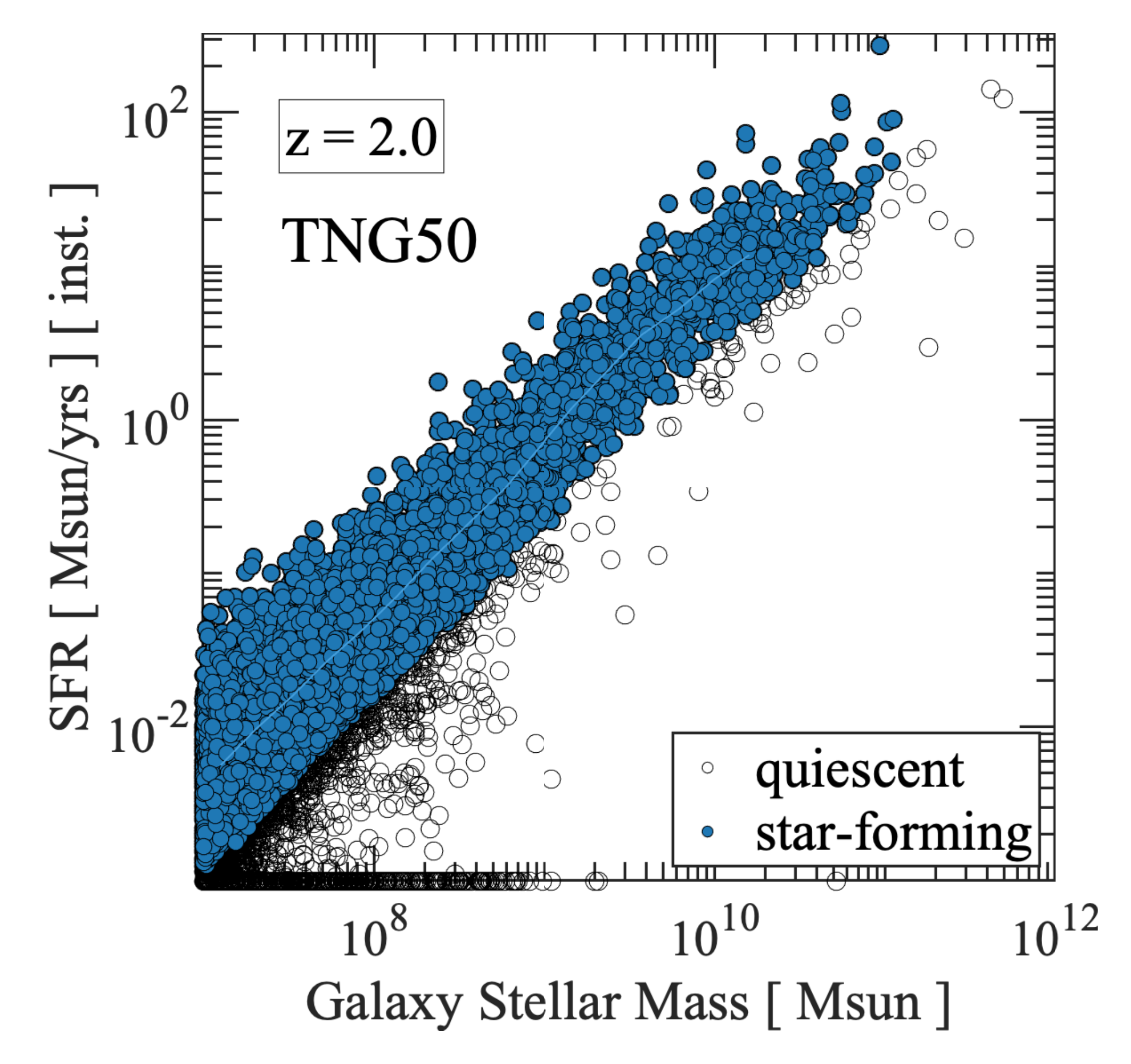}
\includegraphics[height=5.7cm]{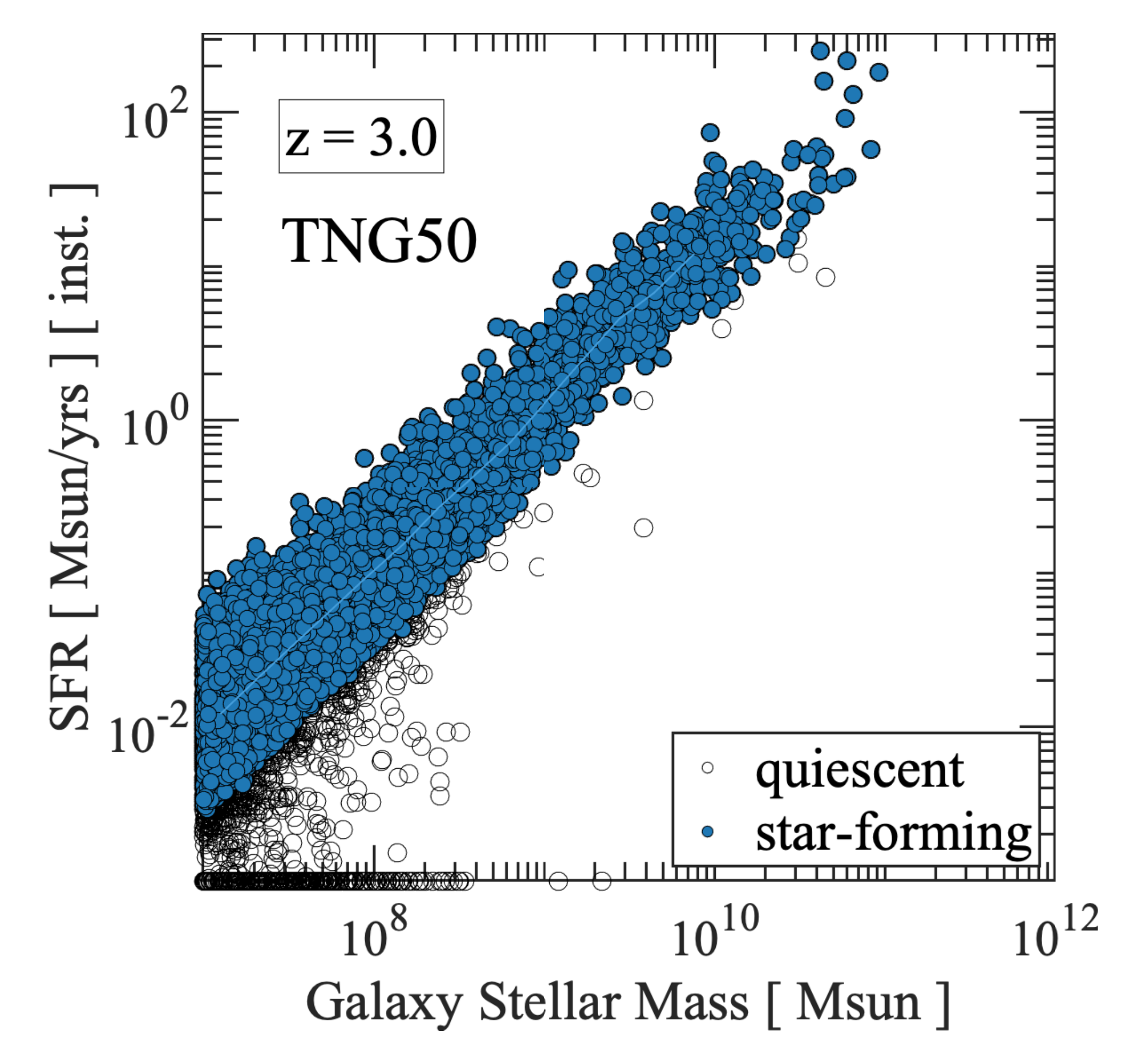}
\includegraphics[trim={1.5cm 0 0 0}, height=5.7cm]{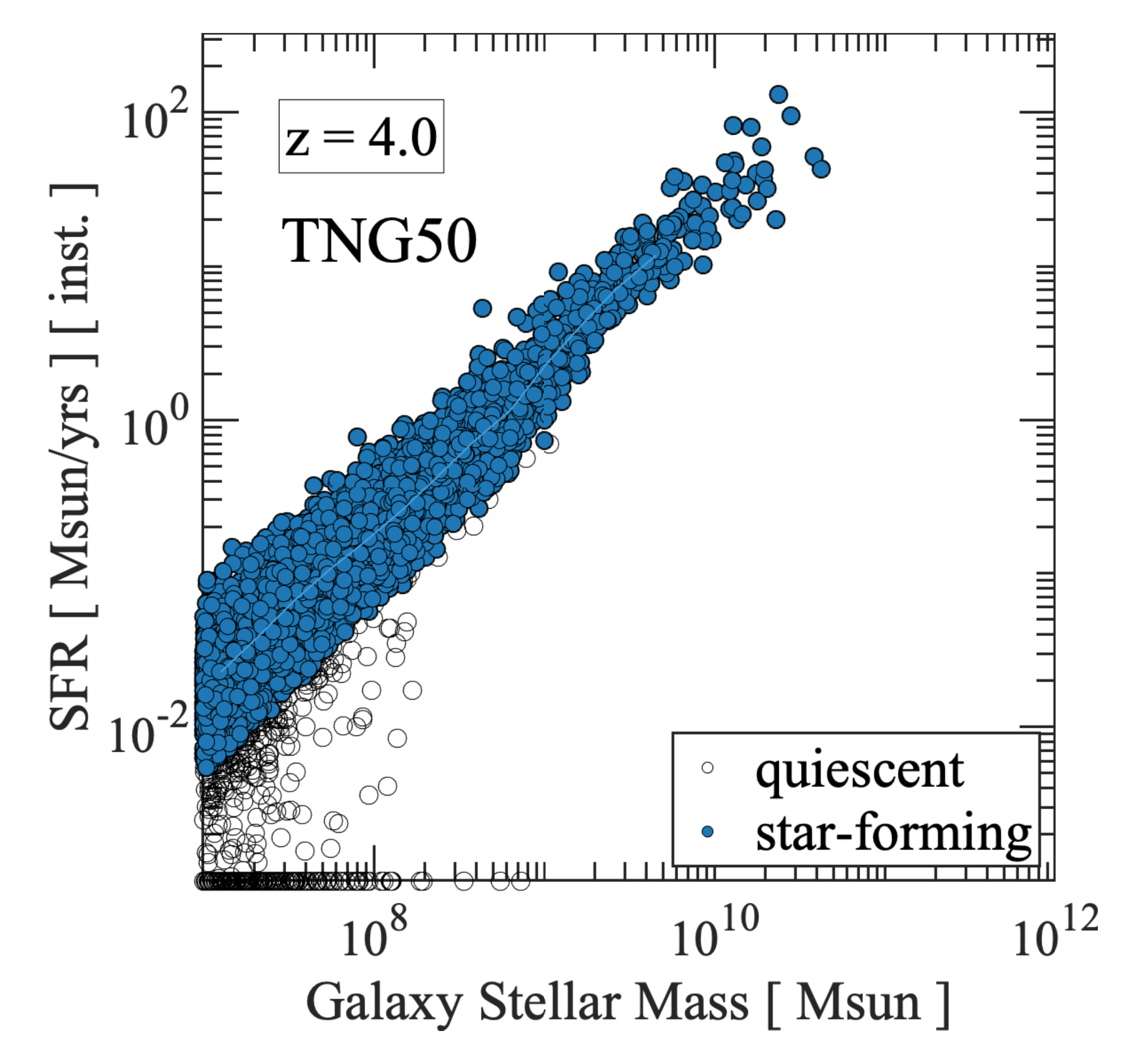}
\includegraphics[trim={1.5cm 0 0 0}, height=5.7cm]{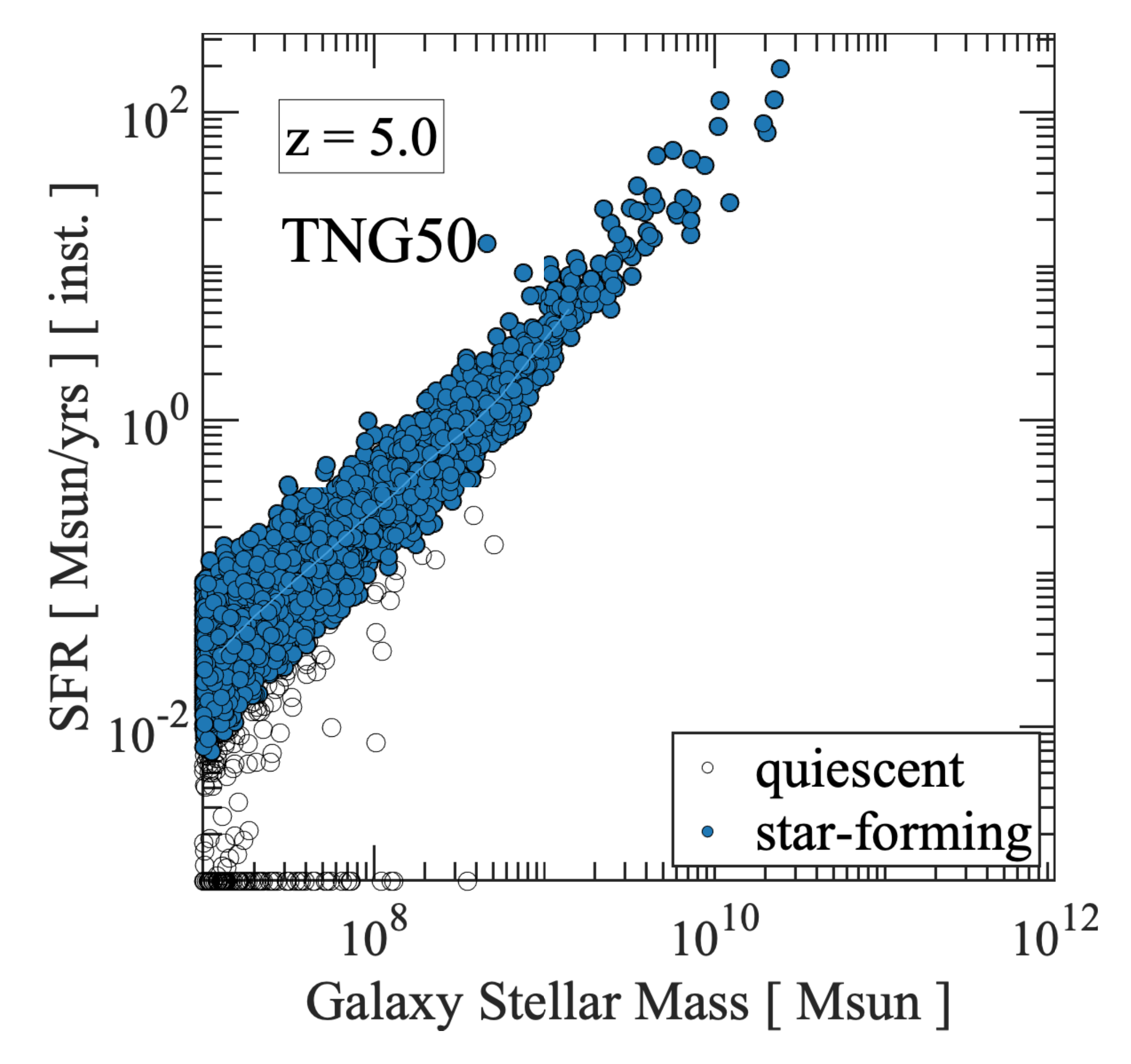}
\caption{\label{fig:ssfr} The star formation rate vs. galaxy stellar mass plane in the TNG50 simulation at different redshifts, for all galaxies in the box (centrals and satellites). In this analysis we study exclusively star-forming galaxies, here denoted as blue filled circles. The distinction between star-forming (blue filled circles) and green-valley or quiescent galaxies (black open circles) is made based on a recursive refinement of the star-forming main sequence (SFMS; see Section~\ref{sec:sample} for details). A SFMS naturally emerges in the simulations and is present down to at least $10^{7}\MSUN$ in stellar mass, as well as already at high redshift.} 
\end{figure*}

\subsection{Notes on plots and units}
Relations across galaxy properties are shown by drawing running medians, i.e. calculating median properties as a function of e.g. galaxy stellar mass in bins of 0.2-0.3 dex. Median curves are plotted only for bins that contain at least 10 galaxies each, unless otherwise noted. The abbreviation ``pkpc'' denotes ``physical kpc'' to distinguish from comoving units (``ckpc''). All velocities are in physical i.e. proper km s$^{-1}$.


\section{The TNG50 sample of star-forming galaxies}
\label{sec:sample}

In this paper, we are interested in the widest possible range of galaxy stellar mass across cosmic epochs. Unless otherwise stated, we consider all simulated {\it star-forming} galaxies with $10^7 \MSUN$ in stellar mass and above. In this Section, we expand on these choices and operational definitions.

The stellar-mass minimum enforces at least 140 stellar particles per galaxy and in practice, in the $1-2 \times 10^7 \MSUN$ stellar mass range, a median of about 300 stellar particles and about 20\,000 total resolution elements among DM, gas and stars. For comparison, galaxies at $\sim10^{10}\MSUN$ are resolved with $10^5$ stellar particles and more than 1.5 million resolution elements in total. In the subsequent sections, we elaborate to which level TNG50 results are converged across this mass range on a property-by-property basis. For now, it should be noted that this mass selection, paired with the ($\sim$50 Mpc)$^3$ comoving volume of the simulation, provides a sample of about 20,000, 21,000, 10,500, and 1,400 galaxies at $z=1, 3, 5$, and 8, respectively, with a high-mass end ($\MS \ge 10^{10}\MSUN$) sampled with $\sim\,$700, 140, and 20 galaxies at $z=1, 3$ and 5, respectively. 

In Fig.~\ref{fig:ssfr}, we show how TNG50 galaxies populate the SFR vs. stellar mass plane, from low ($z=0.5$) to high ($z=5$) redshift, making no distinction between centrals and satellites. Galaxies with SFR$<10^{-3}\MSUN$ yrs$^{-1}$ -- roughly the TNG50 limit imposed by numerical resolution -- are placed by hand at SFR $\equiv 10^{-3}\MSUN$ yr$^{-1}$. 

At all epochs, the TNG50 simulation returns a densely populated star-forming main sequence (SFMS), here approximately linear in logarithmic space across the sampled mass range, with galaxies dropping down in smaller numbers from an average of e.g. a few $10^{-10}$ yr$^{-1}$ ($z\lesssim 1.5$). Following common practice in the observational community, we label our galaxies as star-forming (blue filled circles in Fig.~\ref{fig:ssfr}) vs. green-valley or quenched (black open circles in Fig.~\ref{fig:ssfr}, with no distinction) by recursively searching for the ridge line of the SFMS (defined as the median sSFR of the star-forming galaxies). In practice, we use 0.2-dex bins of stellar mass and ``reject'' green-valley and quenched galaxies as those whose logarithmic distance from the SFMS ridge line at their respective stellar mass lie in the range $-1.0 < \Delta {\rm log}_{10}$sSFR $ < -0.5$ and $\Delta {\rm log}_{10}$sSFR $\leq -1.0$, respectively. 
We assume the SFMS holds up to a few $10^{10}\MSUN$, above which the reference locus of the SFMS is linearly extrapolated. Star-forming galaxies are those with $\Delta {\rm log}_{10}$sSFR $ > -0.5$. This sSFR-based classification has been shown to be consistent at the 1 to 10 percent level with commonly-adopted selections in the UVJ-diagram: this is the case for e.g. CANDELS galaxies \citep{Fang:2018} but also, and more importantly for the analysis at hand, for TNG galaxies \citep{Donnari:2019}.

Fig.~\ref{fig:ssfr} illuminates the distinction between star-forming and quiescent galaxies. We see that a SFMS emerges naturally in IllustrisTNG, being a property of the simulated galaxy population that was not searched for during the IllustrisTNG model development. This was also the case for the Illustris model \citep{Torrey:2014a, Sparre:2015}. From the median relation (solid blue curve) it is clear that the SFMS moves to higher SFR values at higher redshifts, qualitatively consistent with observational trends, and exists all the way up to high redshift (at least to $z\sim 8$, not shown here) and down to low mass ($10^7\MSUN$). The 1$-\sigma$ logarithmic scatter of the SFMS in TNG50 in the $10^{8-10}\MSUN$ mass range is about 0.2 dex or smaller. Such {\it intrinsic} scatter decreases towards higher redshifts (a trend that is confirmed also with the larger volumes and hence richer statistics of TNG100 and TNG300) and is approximately $ 0.2$ dex regardless of the way SFR is measured (e.g. instantaneous or averaged over 10 or 100 Myrs). These estimates are broadly consistent with observational findings (see e.g. \citealt{Speagle:2014} and \citealt{Donnari:2019} for discussions about observations and the comparison between the latter and the TNG results, respectively).

From Fig.~\ref{fig:ssfr} we also see the co-existence of star-forming, green-valley, and quenched galaxies in TNG50 at almost all mass scales at $z\lesssim2$. For example, at $z=0.5$, at the high mass end ($\MS \ge 10^{11}\MSUN$), about 35 per cent of galaxies are star-forming and the remaining ones are either transitioning or quiescent. However, the fraction of high-mass quiescent galaxies declines quickly at earlier epochs: in the TNG50 volume there are no quenched massive galaxies at $z=3$, due in part to their relative scarcity. In practice, star-forming galaxies dominate the TNG50 galaxy population across the majority of the mass and redshift range studied here.


We adopt the selection of star-forming TNG50 galaxies visualized in Fig.~\ref{fig:ssfr} throughout this analysis.


\section{Results on structural properties}
\label{sec:results_1}

\subsection{Galaxy sizes and the extent of the different galactic components}
\label{sec:sizes}

The most fundamental characterization of galaxy structure is size, which we use as the foundation of the spatial scales of interest for the remainder of the analysis. Stellar mass and r-band projected sizes of TNG100 galaxies at $\MS\ge 10^9\MSUN$ and $z\leq 2$ have been quantified and compared to observations by \citet{Genel:2018}. There we have shown that star-forming and quiescent TNG100 galaxies exhibit different evolutionary tracks and size-mass relations; overall, TNG galaxy sizes were found to be consistent with observations typically at the 0.1$-$0.2 dex level, including uncertainties. Here we further consider lower-mass galaxies and earlier epochs, and by comparing the extent of stellar bodies to that of the gaseous and star-forming components.

\begin{figure}
\centering                                      
\includegraphics[width=8.7cm]{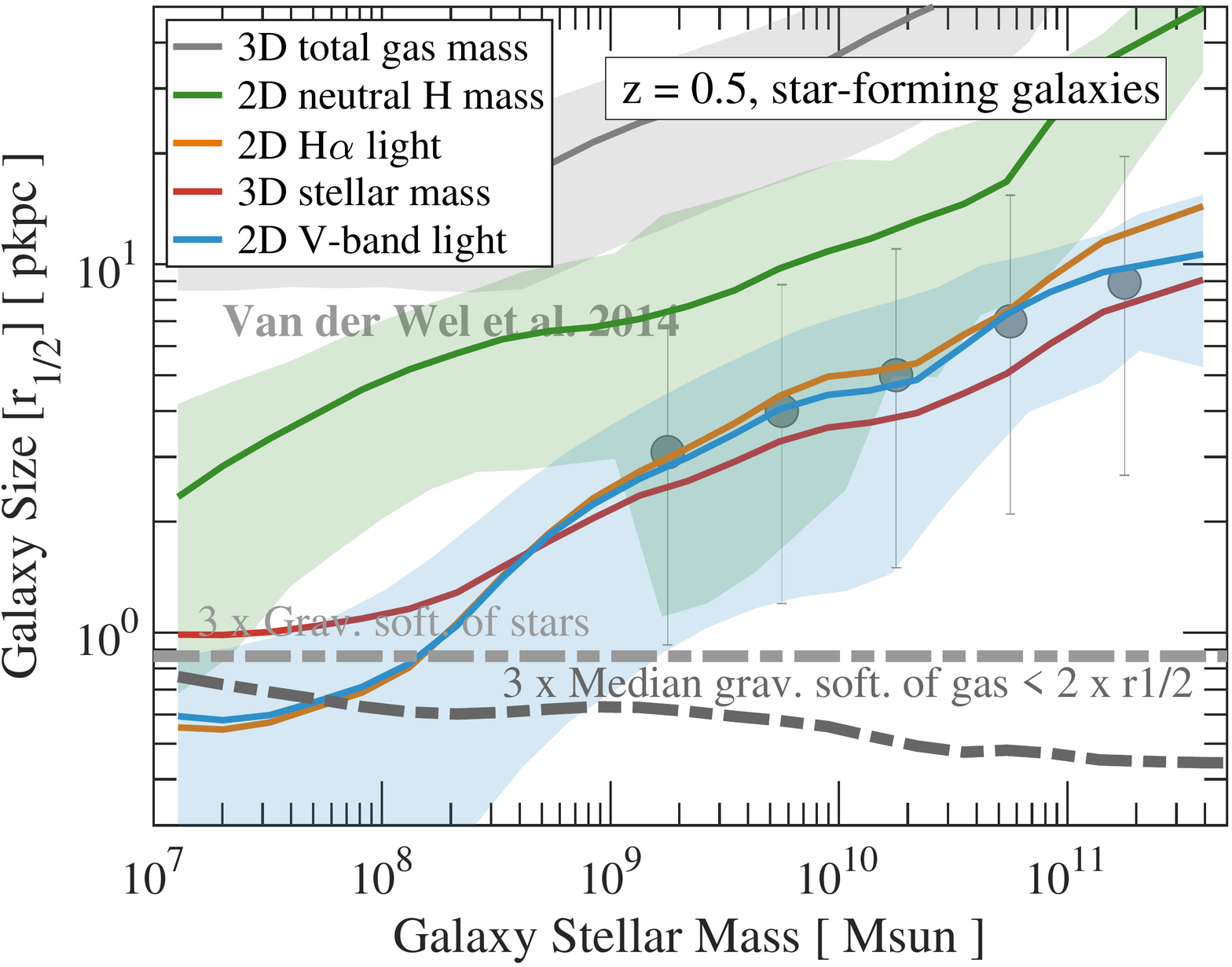}
\includegraphics[width=8.7cm]{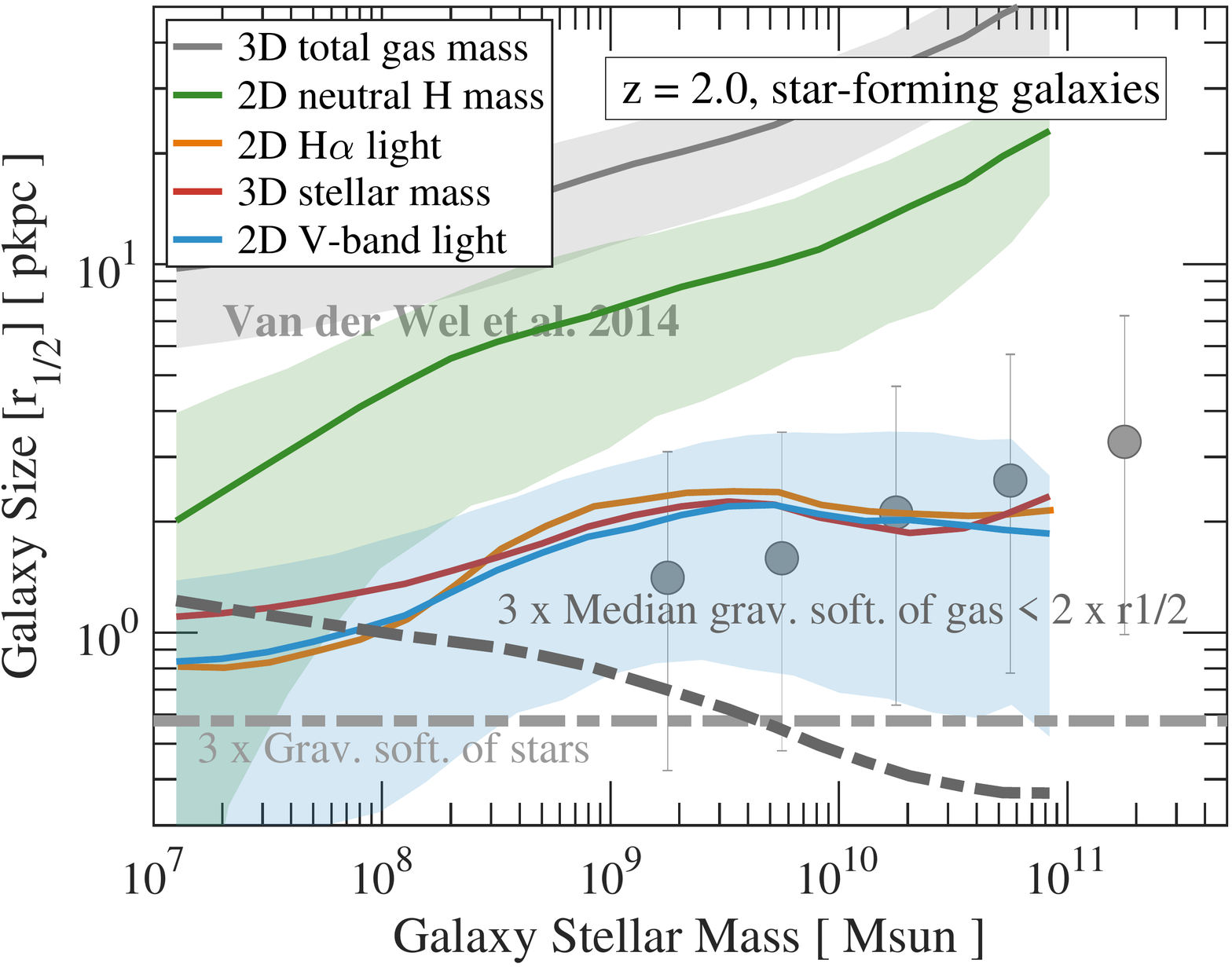}
\includegraphics[width=8.7cm]{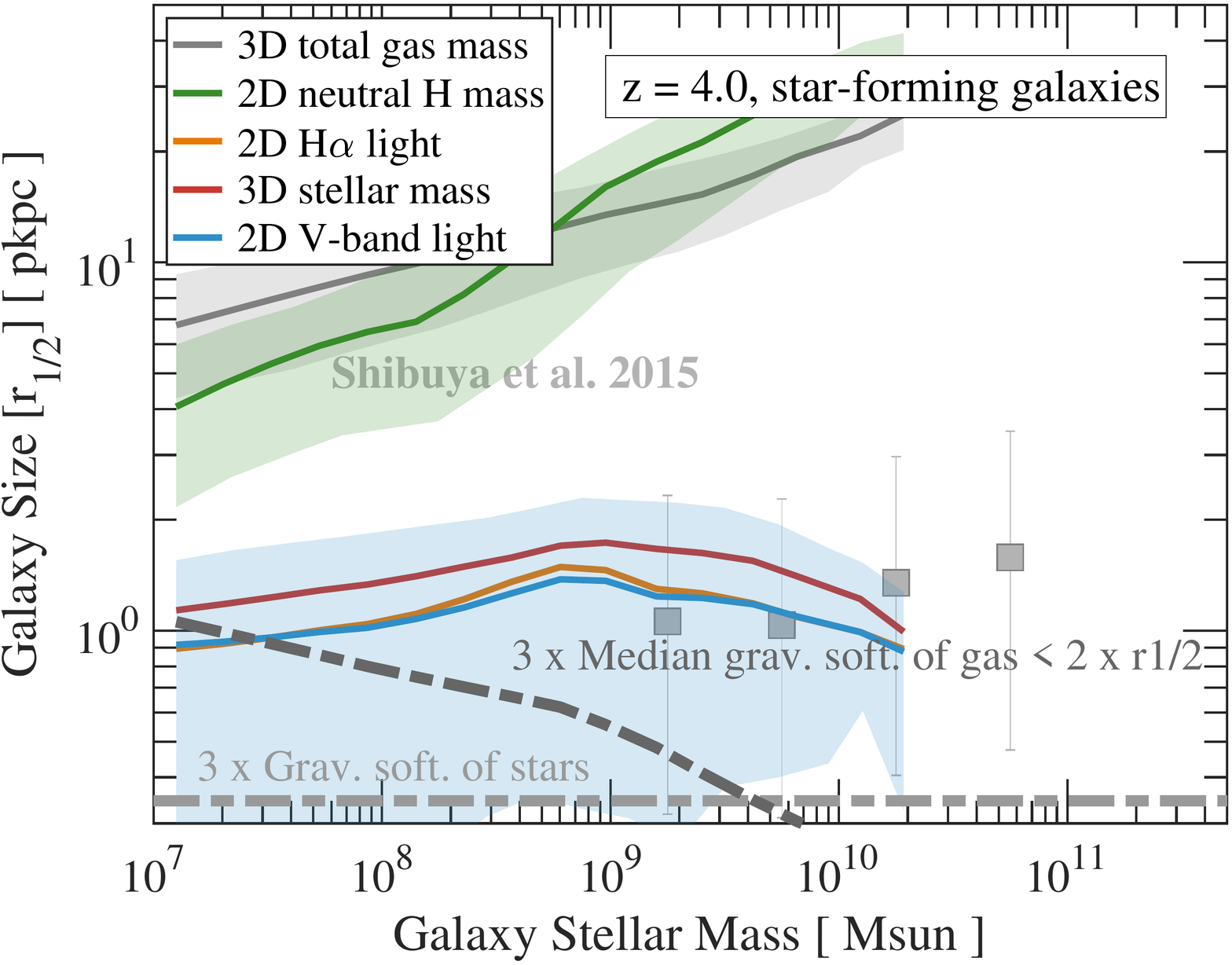}
\caption{\label{fig:sizes} Median galaxy sizes of TNG50 star-forming galaxies as a function of galaxy stellar mass, from low (top) to high (bottom) redshift. Different colors denote 3D or 2D face-on circularized half-mass or half-light radii of different galaxy components. We include both central and satellite galaxies. Shaded areas denote the $\pm1\sigma$ dispersions of the size distributions at fixed stellar mass (omitted for several curves to avoid overcrowding the plot). Gray annotations mark the locus of the typical gravitational softening of the stellar and gaseous resolution elements, for reference (see Appendix~\ref{sec:app_res} for a quantification of the level of convergence across masses, redshift, and matter components).} 
\end{figure}

The sizes of TNG50 galaxies, measured independently for different matter components and tracers, are shown in Fig.~\ref{fig:sizes} as a function of galaxy stellar mass, from $z=0.5$ to $z=4.0$. Solid thick curves show medians across the galaxy population in 0.2-dex bins of stellar mass; shaded regions denote $\pm$ 1$-\sigma$ galaxy-to-galaxy variations. We show, from the largest to the smallest, 3D total gas (gray), 2D projected neutral Hydrogen (green) and 3D stellar (red) half-mass radii, together with 2D, face-on projected, circularized V-band (blue) and $\HA$ (orange) half-light radii, all in physical kpc. As a reminder, here $\HA$ light is a proxy for SFR (see Section~\ref{sec:descriptors}) and traces the location of dense, cool gas.

Noticeably and not surprisingly, the extent of the total gas is much larger than any stellar proxy, at all times, this being the case also for the radii of neutral hydrogen, at least at $z\ge 0.7$. Specifically, neutral-hydrogen sizes are about 2-10 times larger than the extent of the stellar and star-forming bodies, and more so at higher redshifts and for more massive galaxies. Stellar- and $\HA$-light instead trace one another within a factor of $\sim1.5-2$ for all redshifts and masses we consider here: in particular, $\HA$ 2D sizes are larger than V-band ones for more massive galaxies at more recent times. This is not a trivial accord: while the spatial distribution of $\HA$ light traces the sites of stars {\it at birth}, and can be further affected by radiative processes, a number of physical mechanisms such as galaxy mergers or migration can redistribute stars over their lifetime, affecting the overall stellar sizes differently than those of the cold and dense gas out of which they form. The similarity of optical and $\HA$ sizes of galaxies is, indeed, broadly consistent with observational findings at $z\sim1$ \citep[e.g.][]{NelsonE:2013, NelsonE:2016} and provides guidance for future observations at higher redshifts, e.g. with JWST.

Importantly, TNG50 galaxies below $10^{9.5}\MSUN$ have half-light radii of 0.5-2 physical kpc on average, with little to no redshift trend. The weak redshift evolution between $z=4$ and $z=0.5$ and below the $10^{10}\MSUN$ scale is consistent with observational findings on stellar effective radii, e.g. by \citet{VanDerWel:2014s} and \citet{Shibuya:2015} in restframe optical wavelengths. For context, these observations are shown in Fig.~\ref{fig:sizes} (gray symbols, with scatter bars): our 2D face-on circularized radii (blue curves) should be compared to projected long-axis sizes. These are readily available from \citet{VanDerWel:2014s} at $z\leq3$ (gray circles). At higher redshifts, we convert the values from \citet{Shibuya:2015} from circularized to long-axis projected sizes by adding 0.15 dex, reflecting the typical projected shape of b/a=0.5 at this stellar mass and redshift range (square symbols in the lower panel). Observations indicate that low-mass galaxies are at least as small as the TNG50 expectation and the latter are overall in the observational ballpark both in the median and galaxy-to-galaxy variation.

Indeed, in TNG50 galaxies are as small as a few hundred physical pc, and this can occur also at the high-mass end in the compact tails of the size distribution. It is possible to resolve such sizes only with the high numerical resolution of TNG50. For reference, the dot-dashed light gray lines in Fig.~\ref{fig:sizes} denote three times the gravitational softening of the stars. This never exceeds 576 comoving pc at any redshift, and can be as small as e.g. $\sim192$ physical pc at $z=2$. On average, the median gravitational softening of the gas cells within galaxies have similar extent, but the densest cells can be evolved with even smaller softening lengths (see Fig.~\ref{fig:cellsizes}). Here, the dot-dashed dark gray curves in the panels of Fig.~\ref{fig:sizes} denote three times the median gravitational softening of the gas which is contained within twice the stellar half-mass radius. It is manifest from these comparisons that the physical extent of TNG50 galaxies can be smaller than any nominal measure of the gravitational softenings of the various components. As we show in Appendix~\ref{sec:app_res} and comment on in Section~\ref{sec:betterres}, in our numerical and physical model the softening per se' is {\it not} responsible for setting the size of simulated galaxies. While at $z\lesssim1$ the extent of galaxies below $\lesssim 10^{8}\MSUN$ may be somewhat overestimated because of the still limited numerical resolution, at higher redshifts and masses TNG50 median sizes are convergently resolved.


\begin{figure*}
\centering 
\includegraphics[trim={0 0 0 0.7cm},clip,width=4cm]{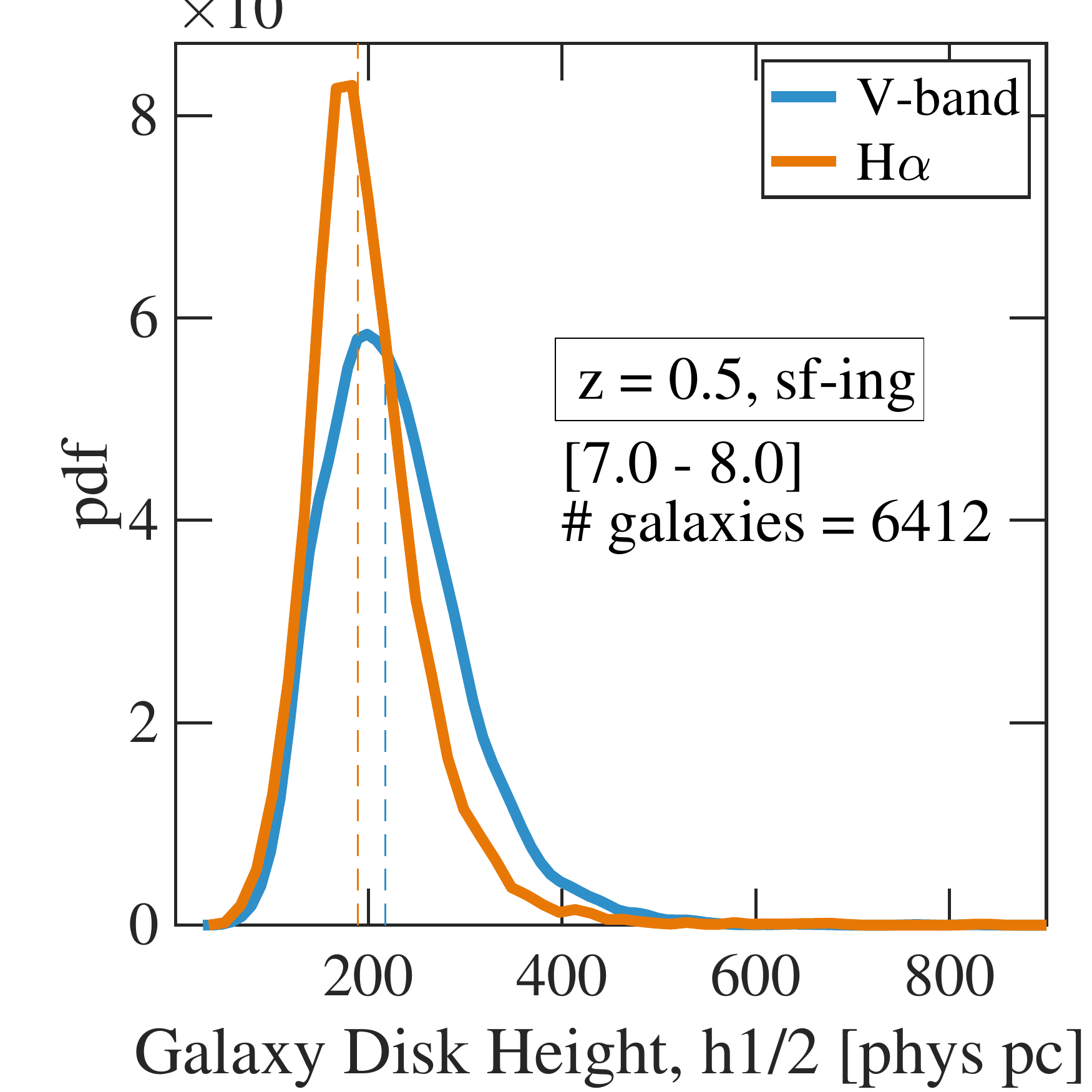}
\includegraphics[trim={0 0 0 0.7cm},clip,width=4cm]{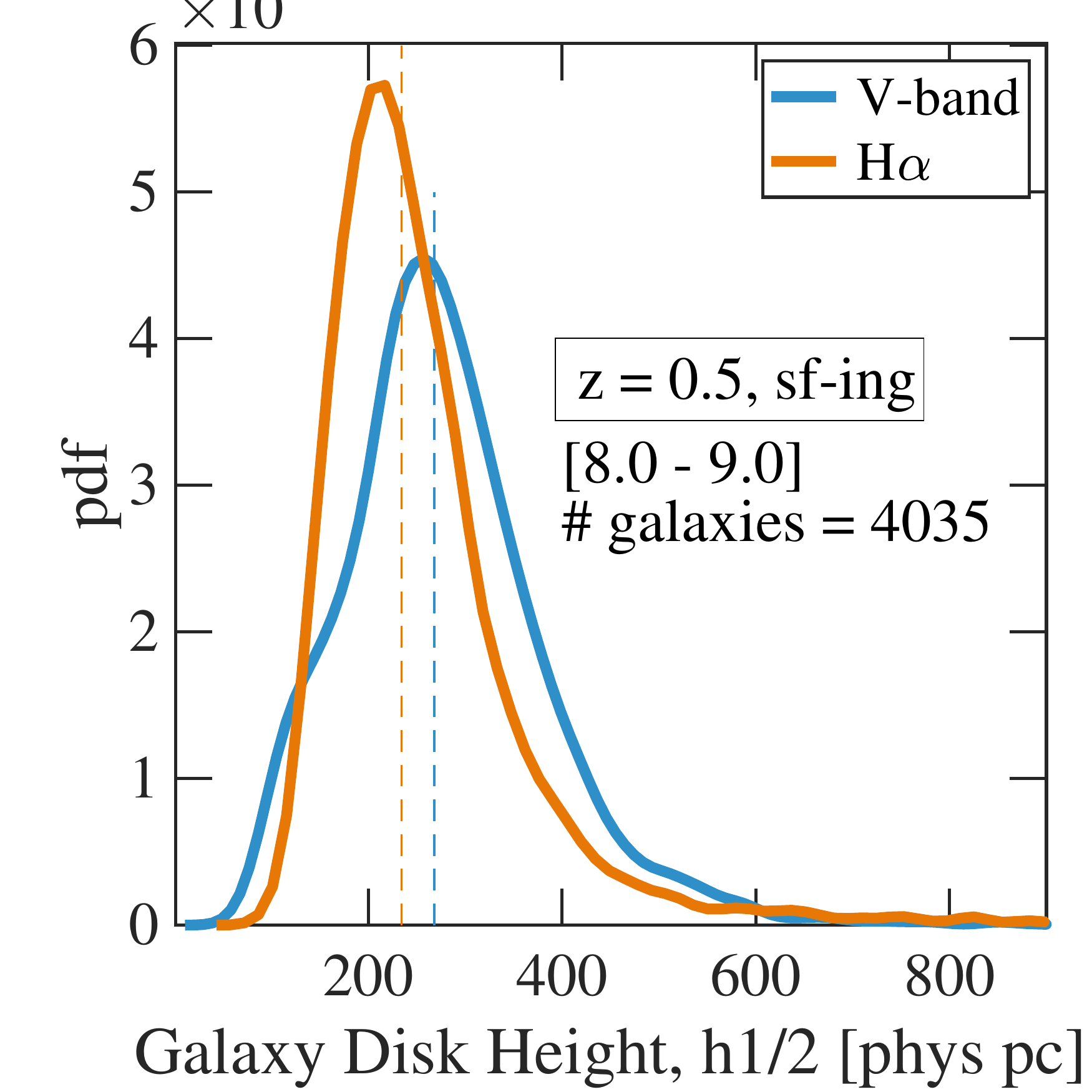}
\includegraphics[trim={0 0 0 0.7cm},clip,width=4cm]{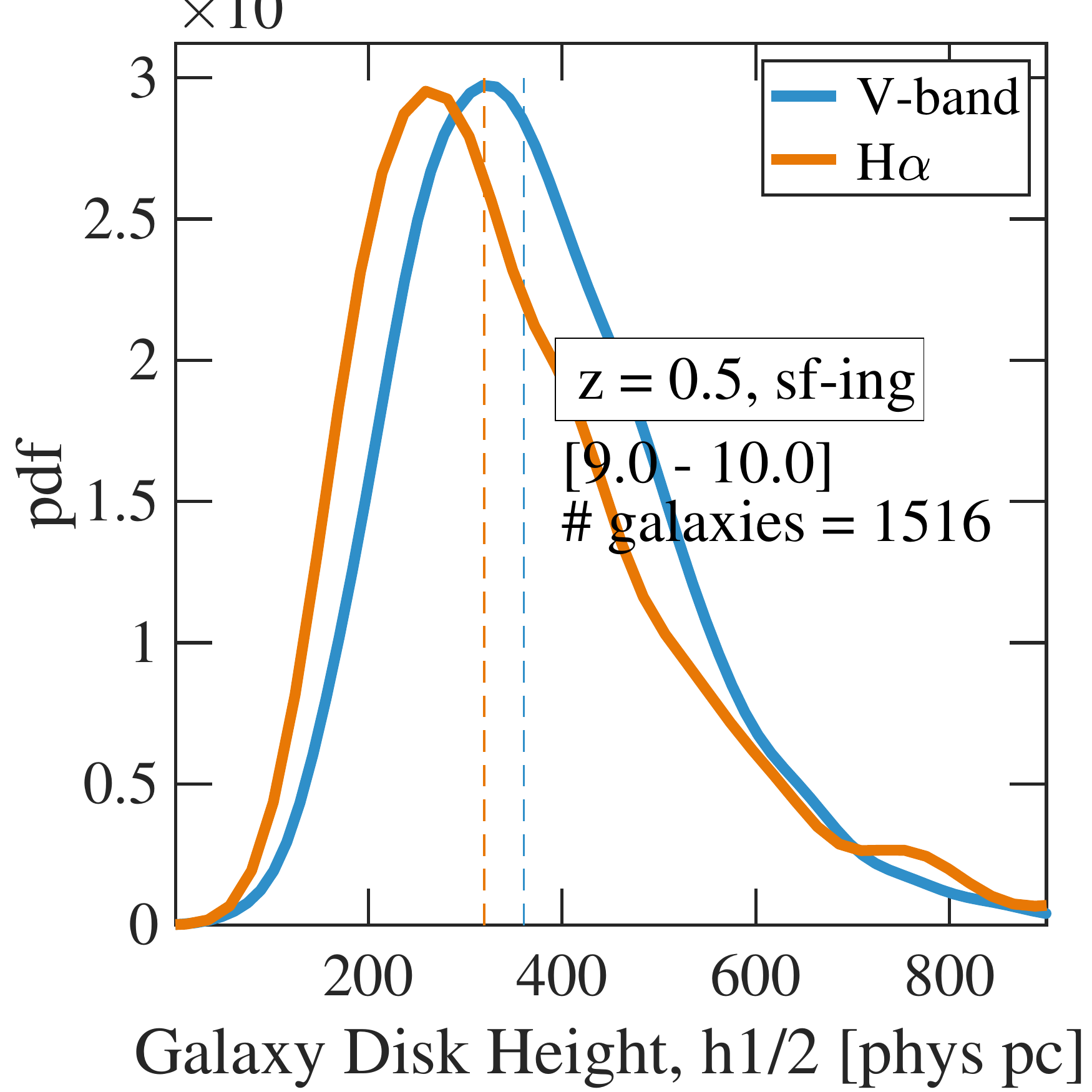}
\includegraphics[trim={0 0 0 0.7cm},clip,width=4cm]{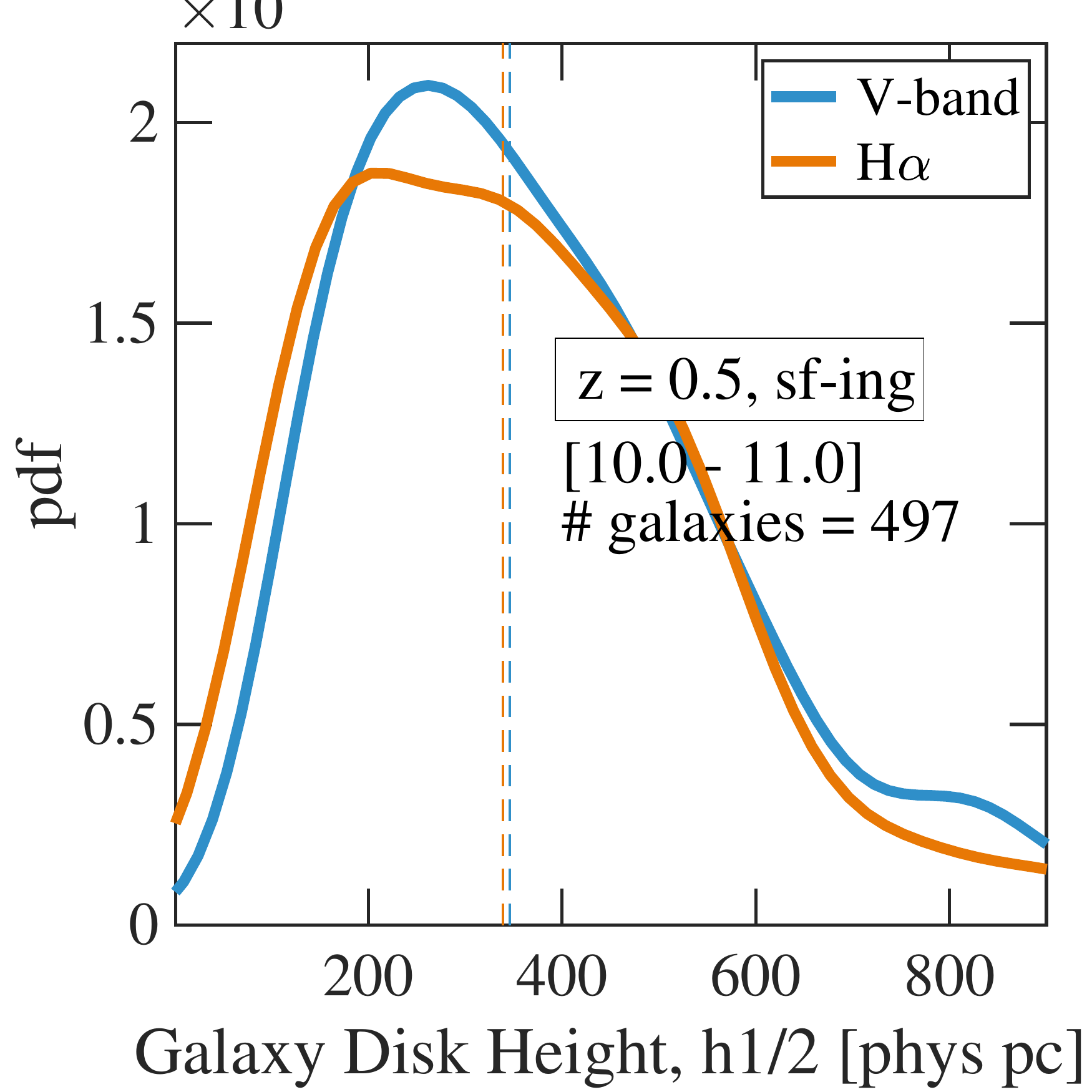}

\includegraphics[trim={0 0 0 0.7cm},clip, width=4cm]{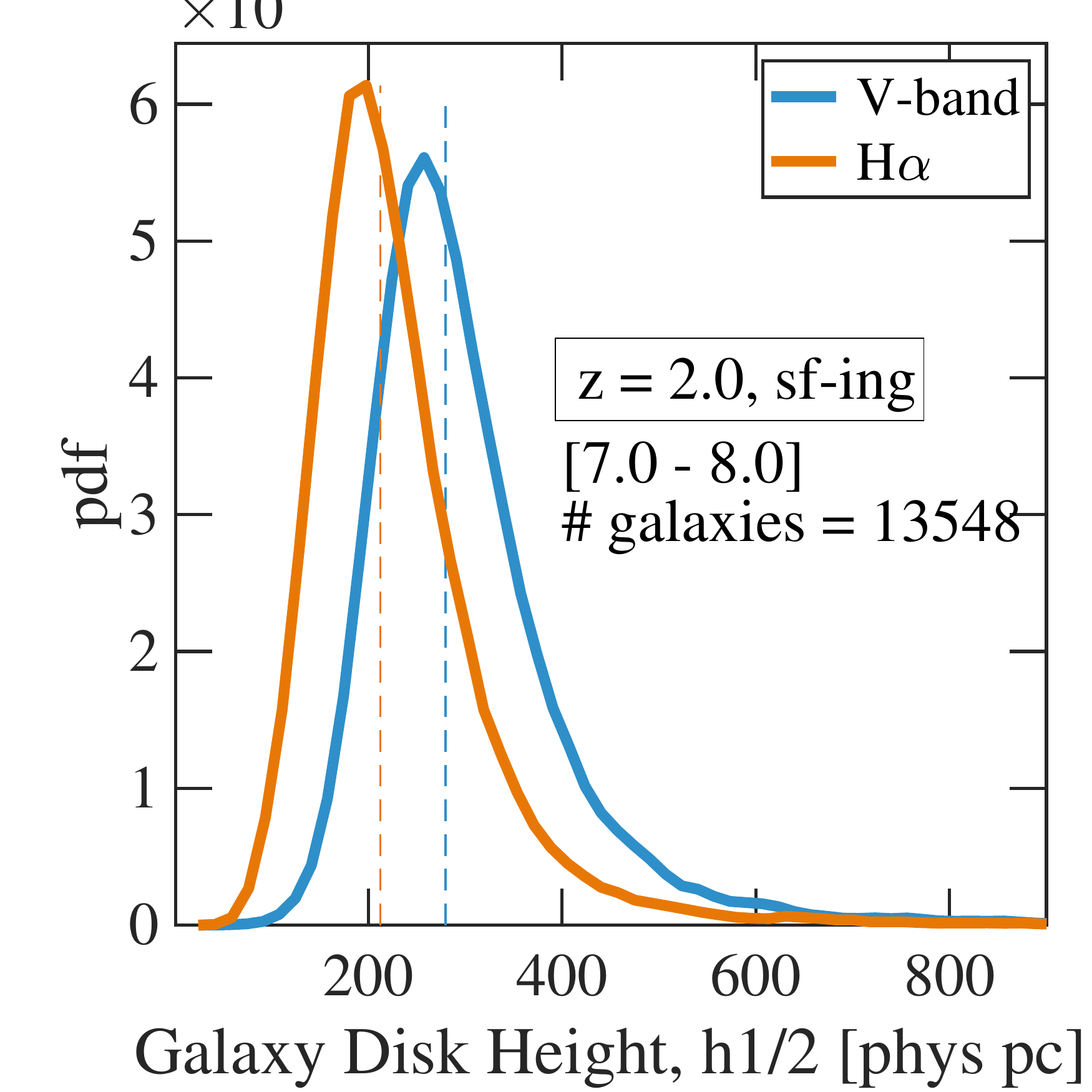}
\includegraphics[trim={0 0 0 0.7cm},clip,width=4cm]{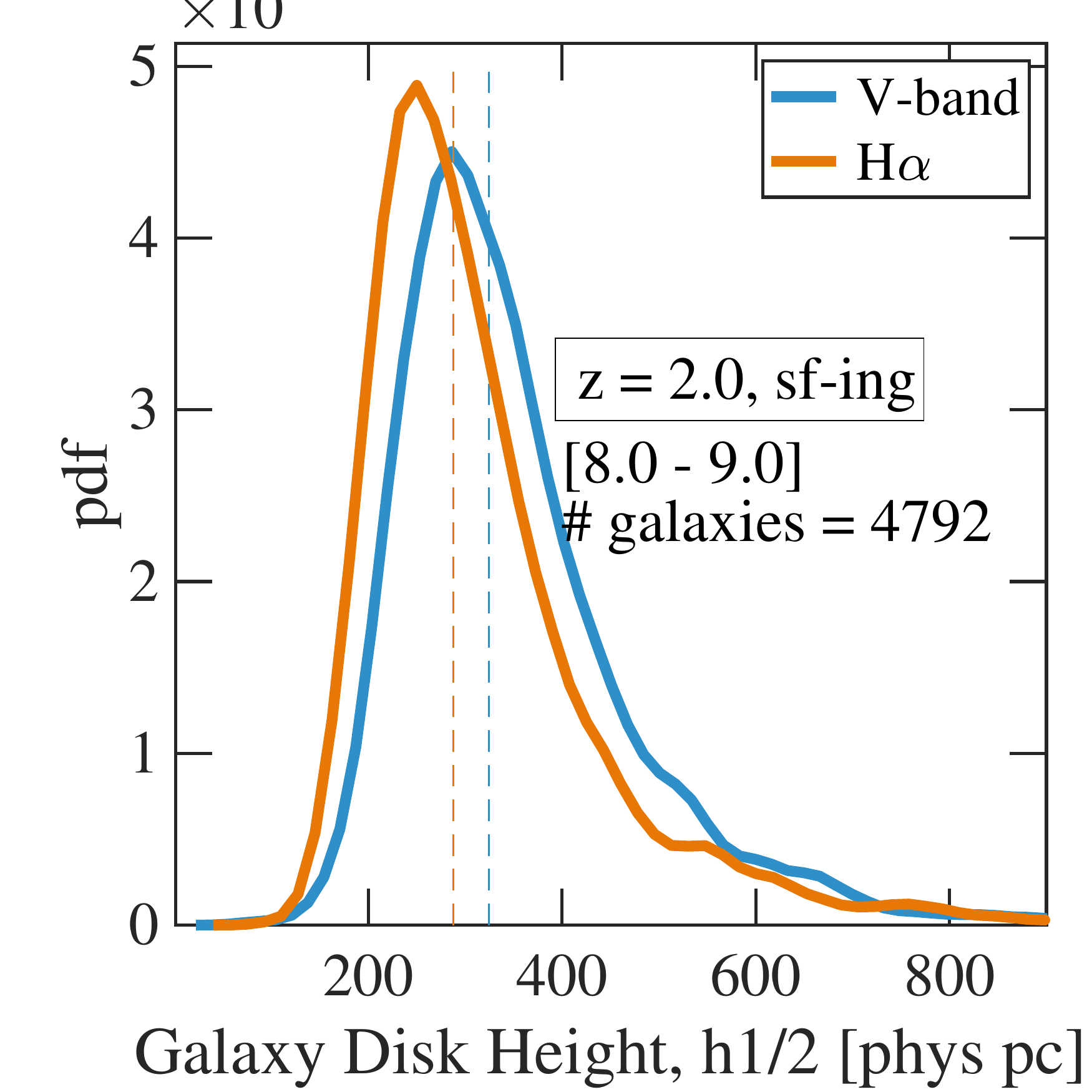}
\includegraphics[trim={0 0 0 0.7cm},clip,width=4cm]{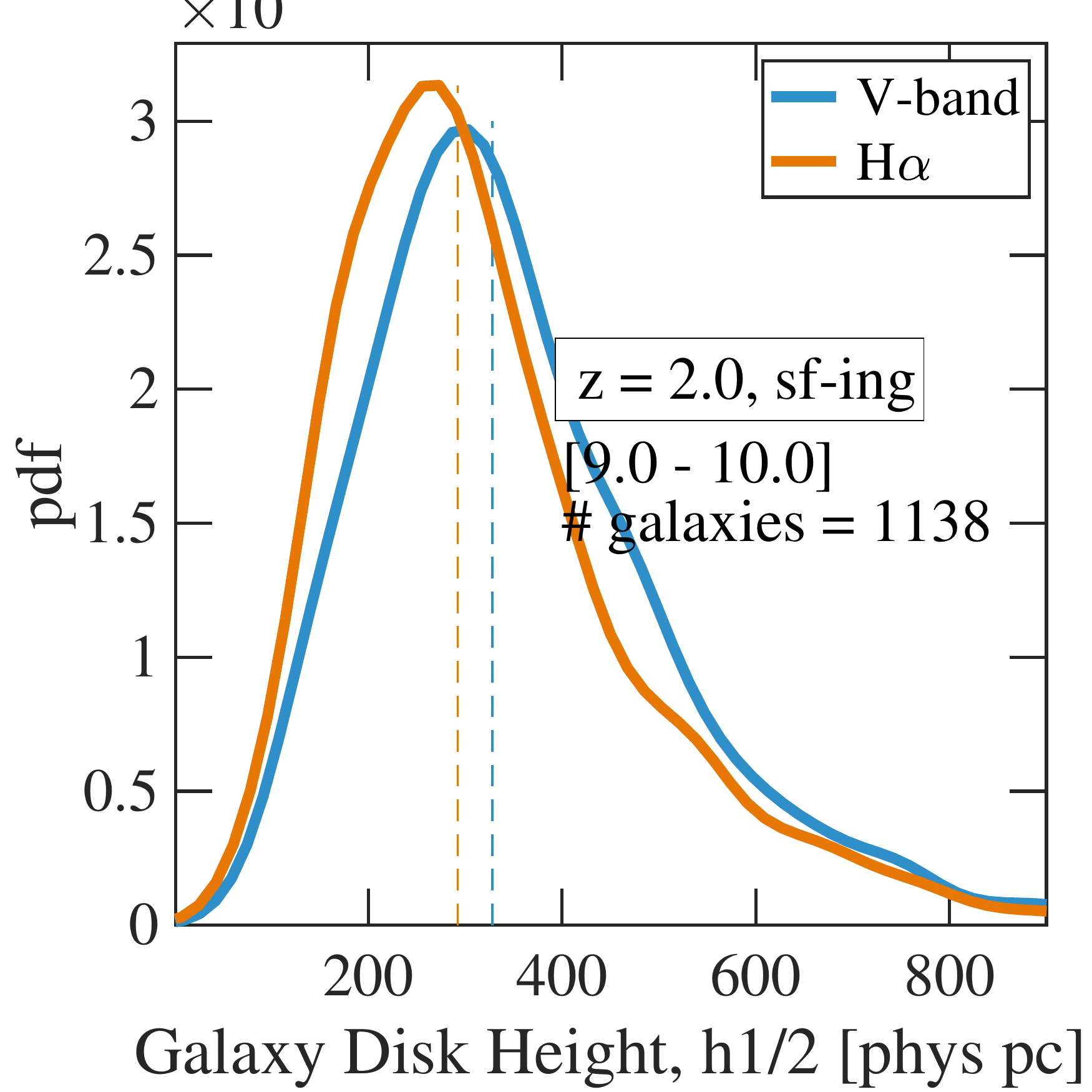}
\includegraphics[trim={0 0 0 0.7cm},clip,width=4cm]{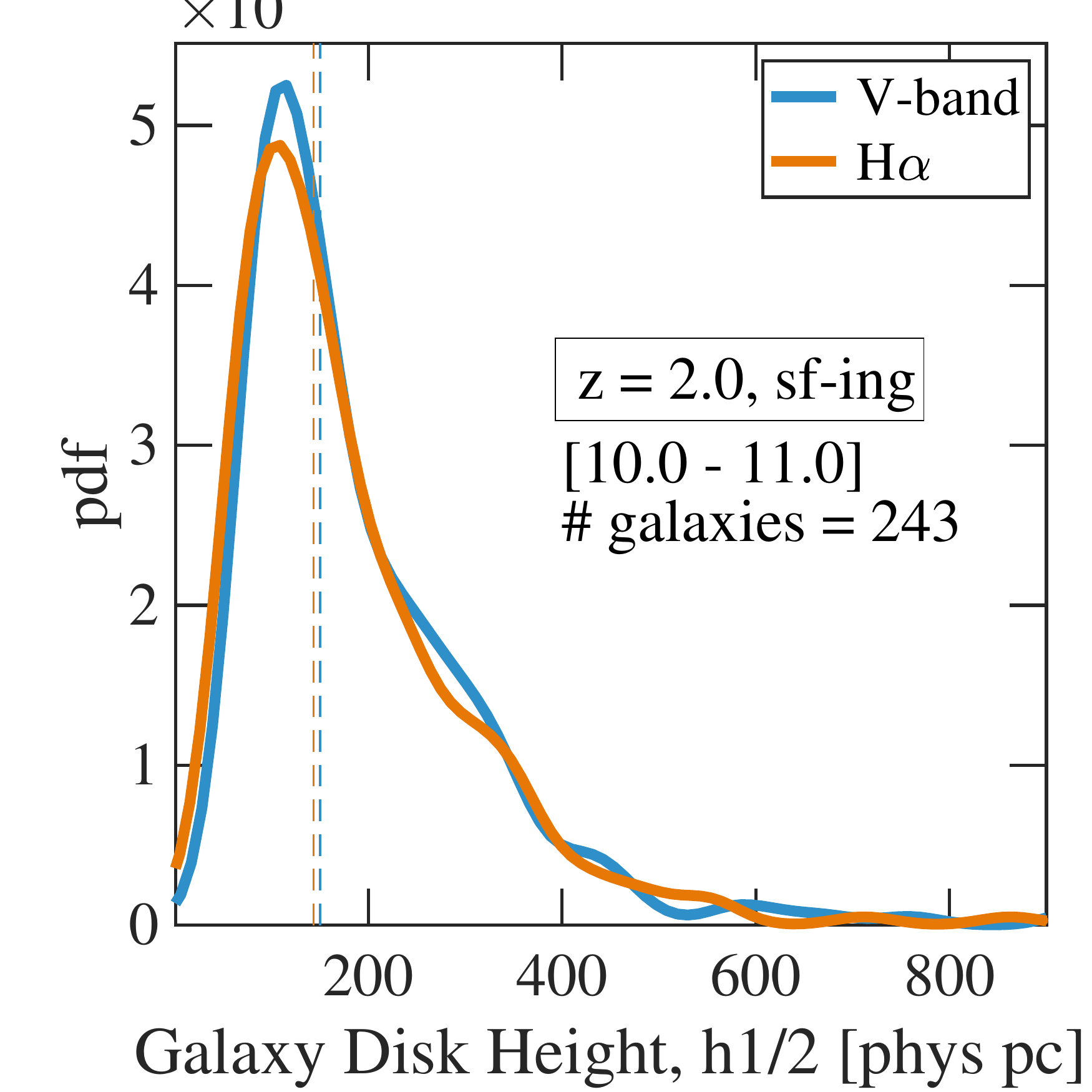}

\includegraphics[trim={0 0 0 0.7cm},clip,width=4cm]{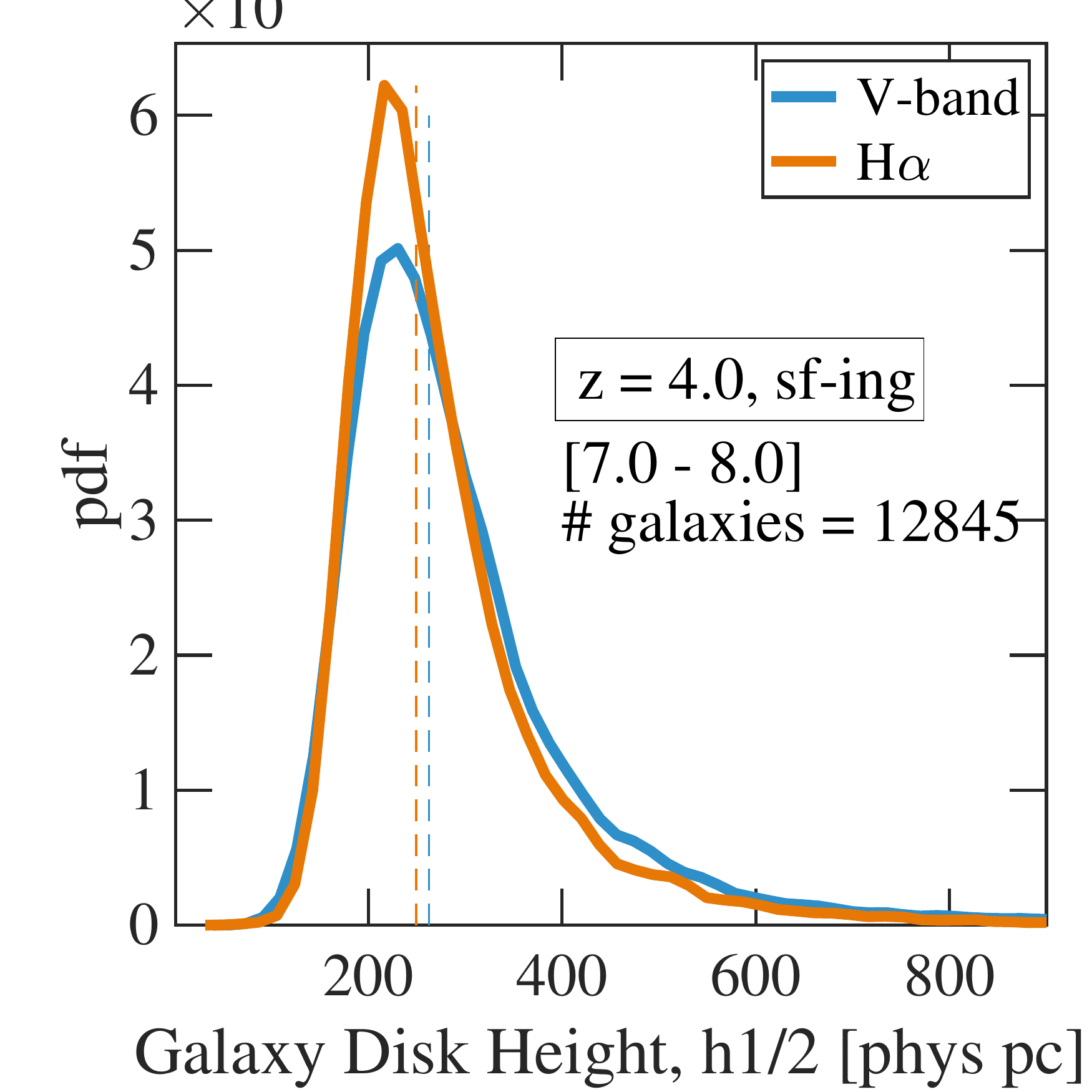}
\includegraphics[trim={0 0 0 0.7cm},clip,width=4cm]{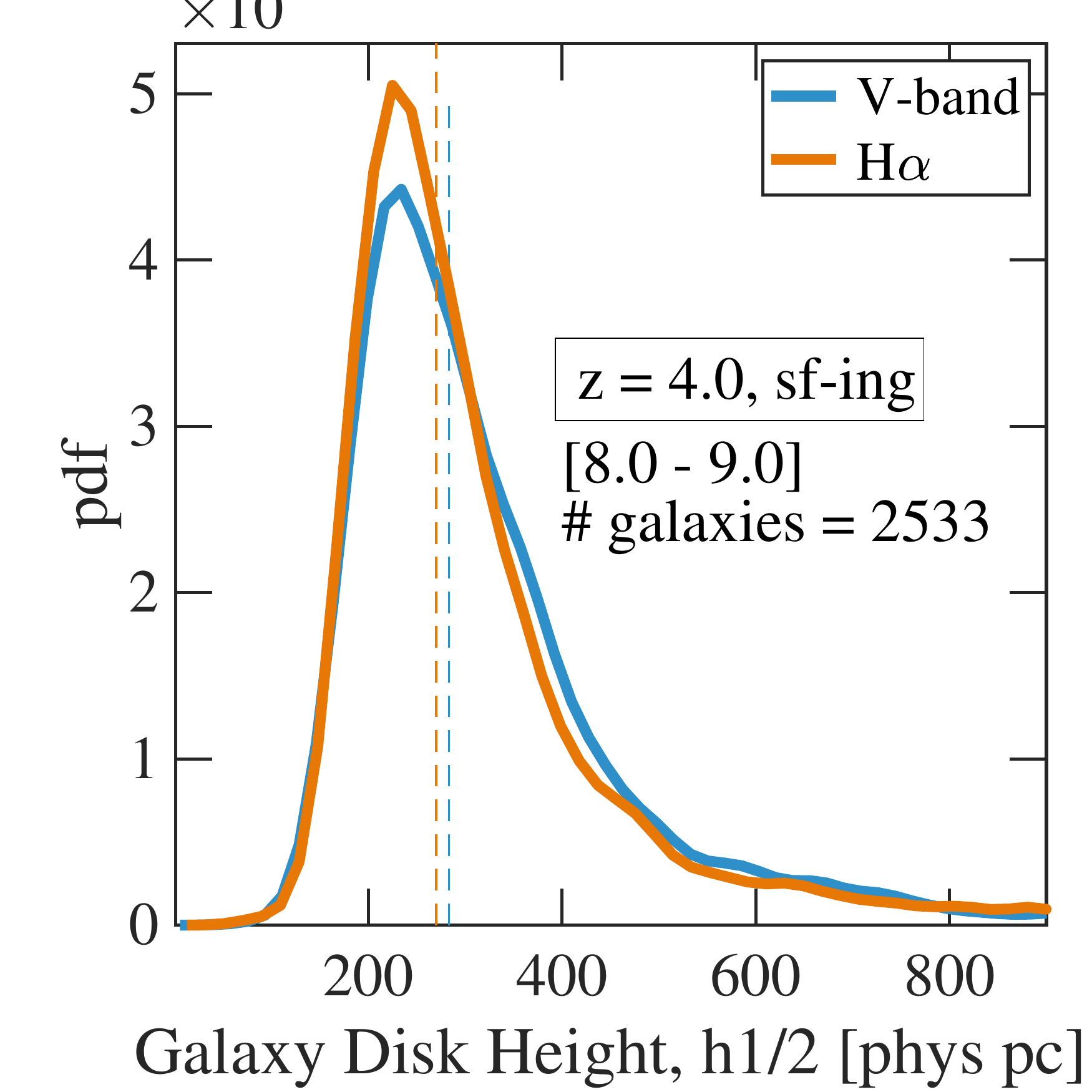}
\includegraphics[trim={0 0 0 0.7cm},clip,width=4cm]{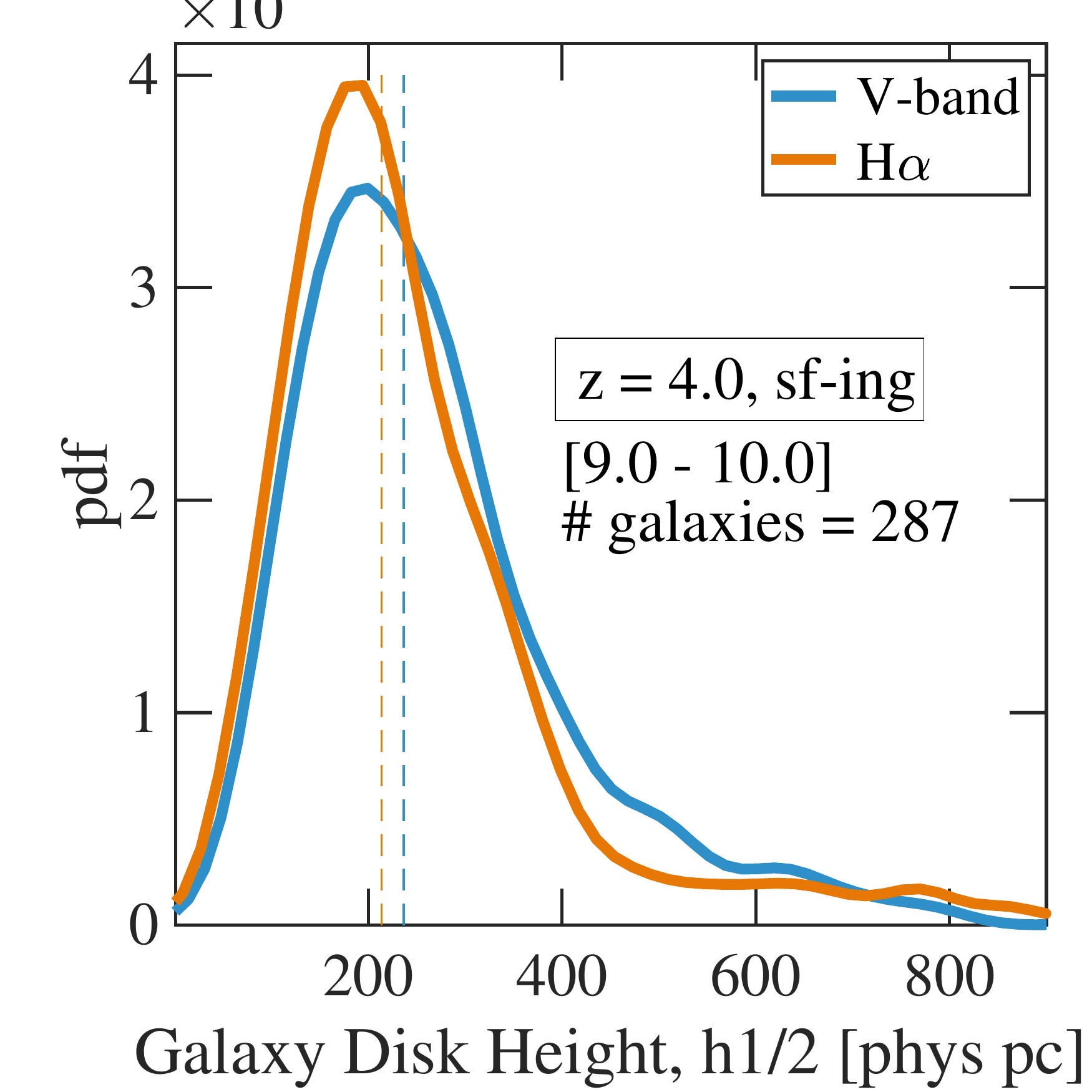}
\includegraphics[trim={0 0 0 0.7cm},clip,width=4cm]{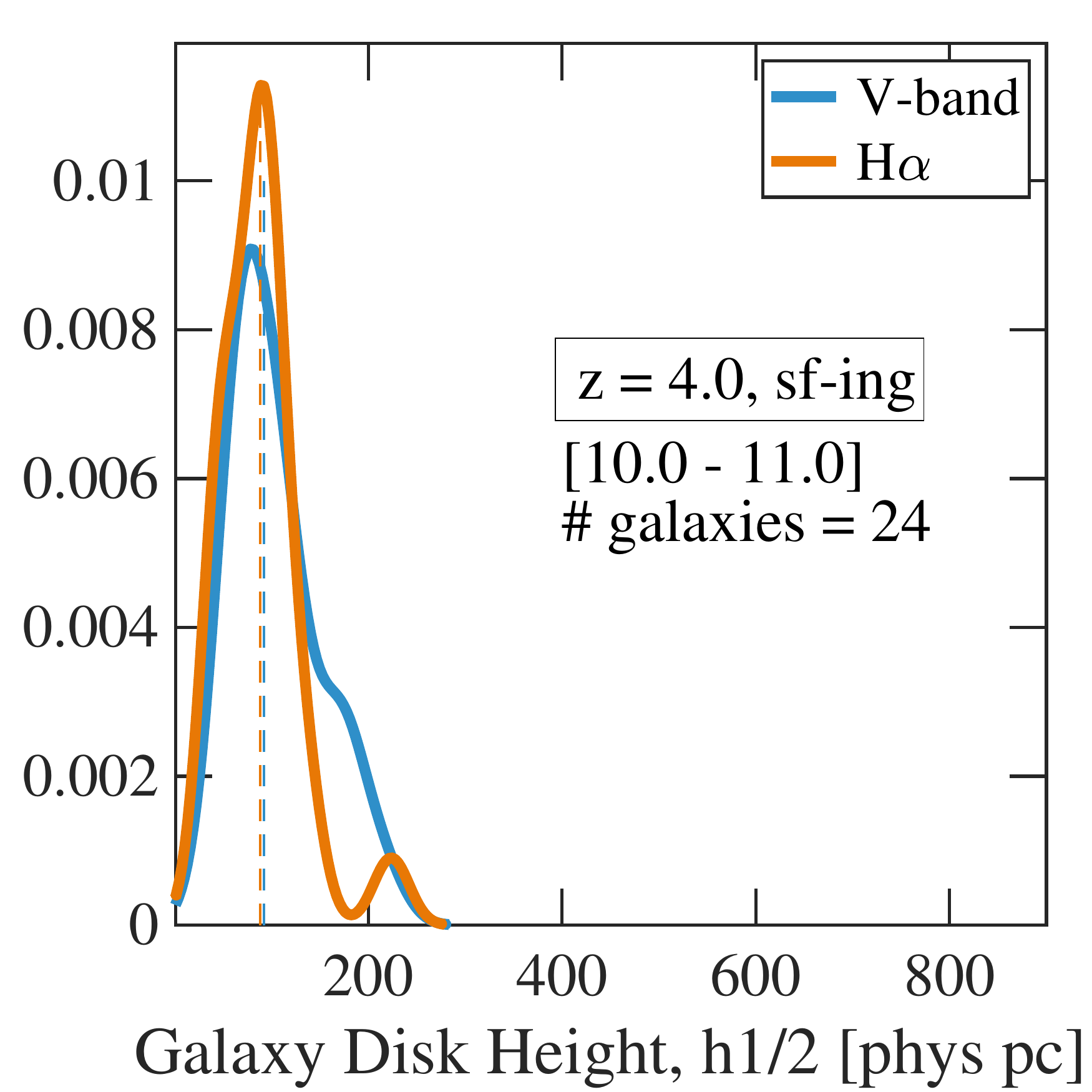}

\caption{\label{fig:diskheights_distribs} Distributions of disk heights in physical parsecs for TNG50 star-forming galaxies. We show three discrete redshifts, from $z=0.5$ (top) to $z=4$ (bottom), and in several bins of increasing galaxy stellar mass (left to right). Galaxies' thicknesses are estimated as described in Section~\ref{sec:props} from both the V-band (blue) and $\HA$ (orange) light distributions in edge-on projections. Vertical dashed lines denote medians of the distributions. The TNG50 calculation can return (star-forming) galaxies with averages ``disk'' heights of 100-400 physical parsecs in general, and as thin as 50-80 parsecs. In Appendix~\ref{sec:app_res}, we quantify the level of convergence of TNG50 heights across masses, redshift, and matter components.}
\end{figure*}

\begin{figure*}
\centering 
\includegraphics[width=8cm]{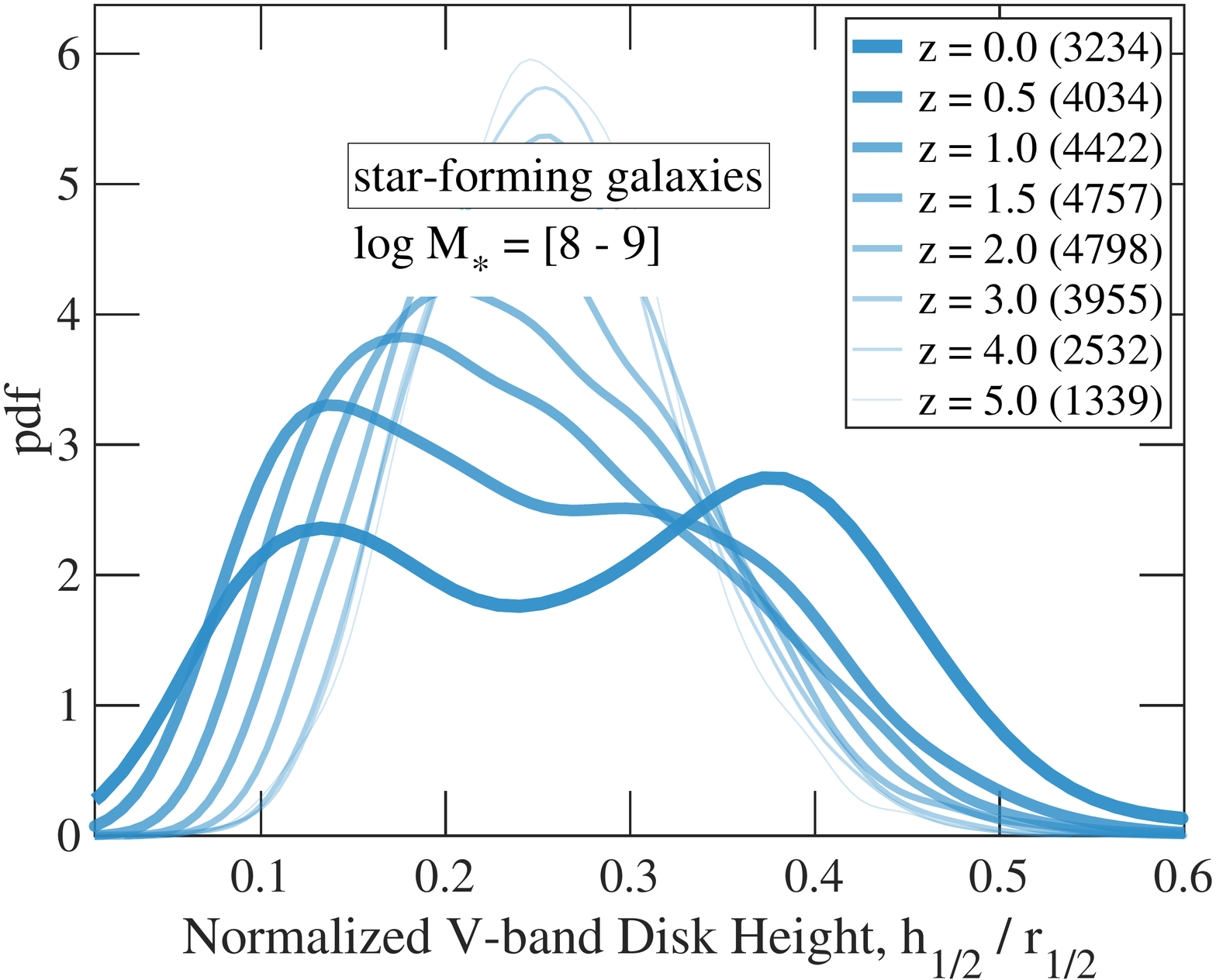}
\includegraphics[width=8cm]{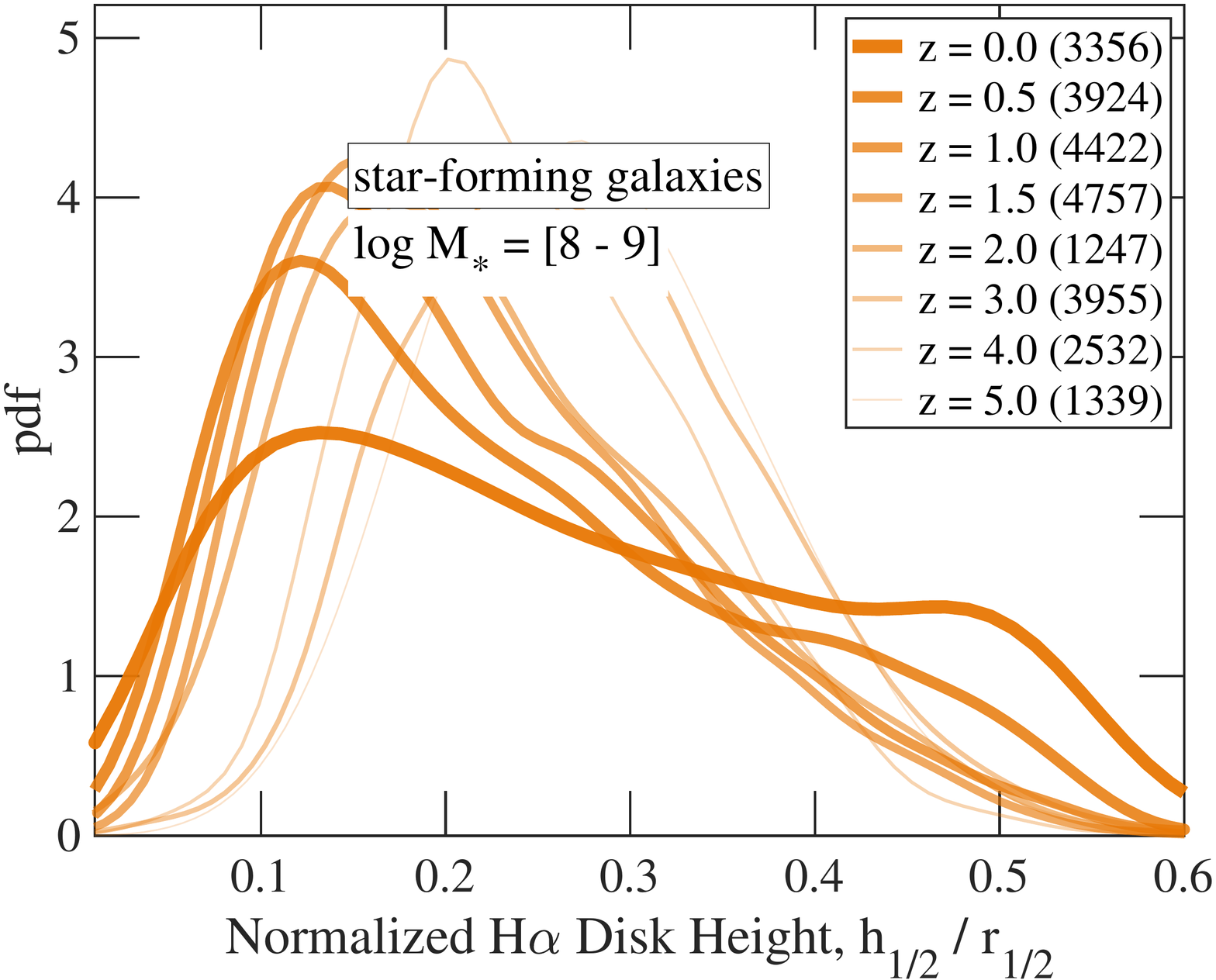}
\includegraphics[width=8cm]{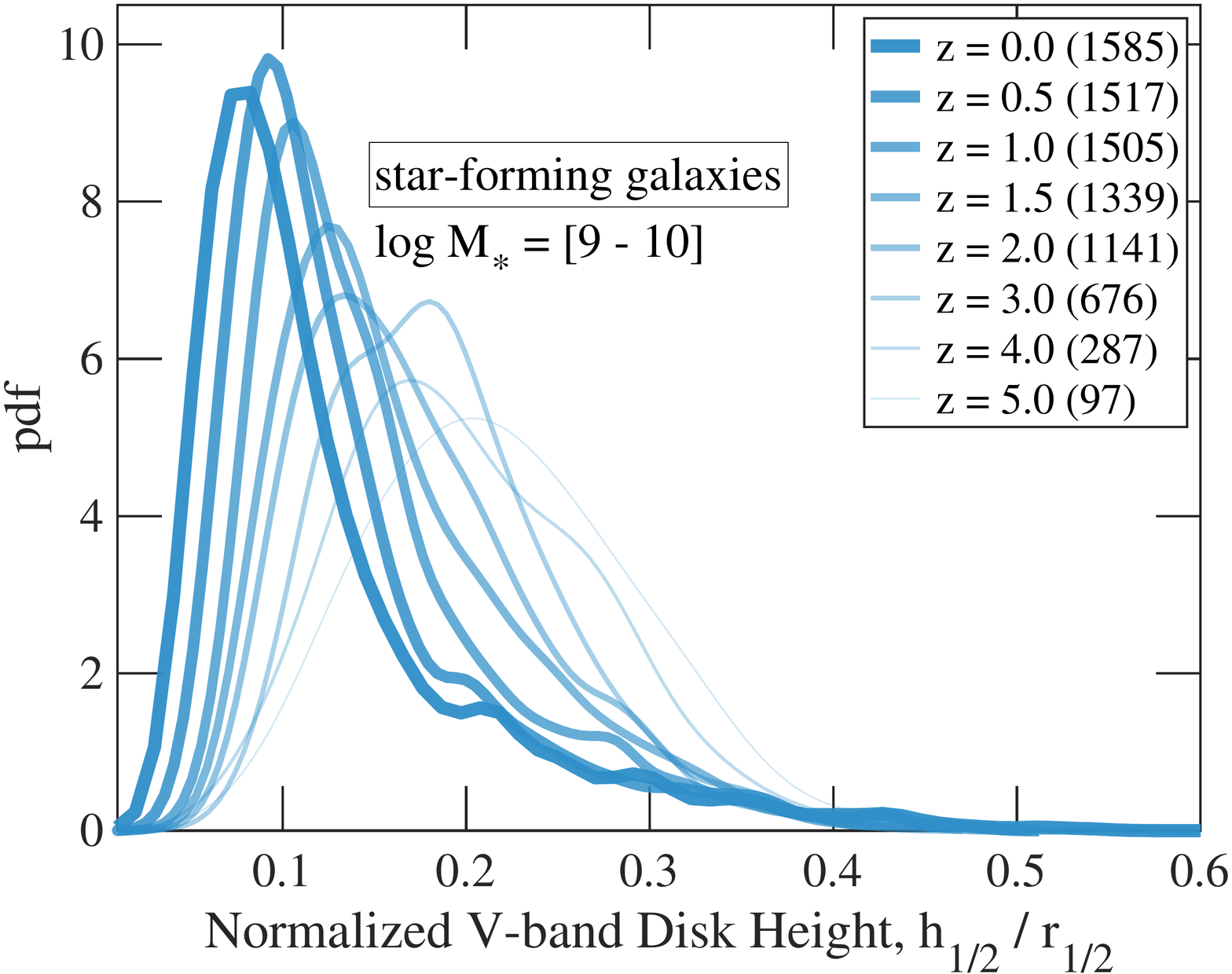}
\includegraphics[width=8cm]{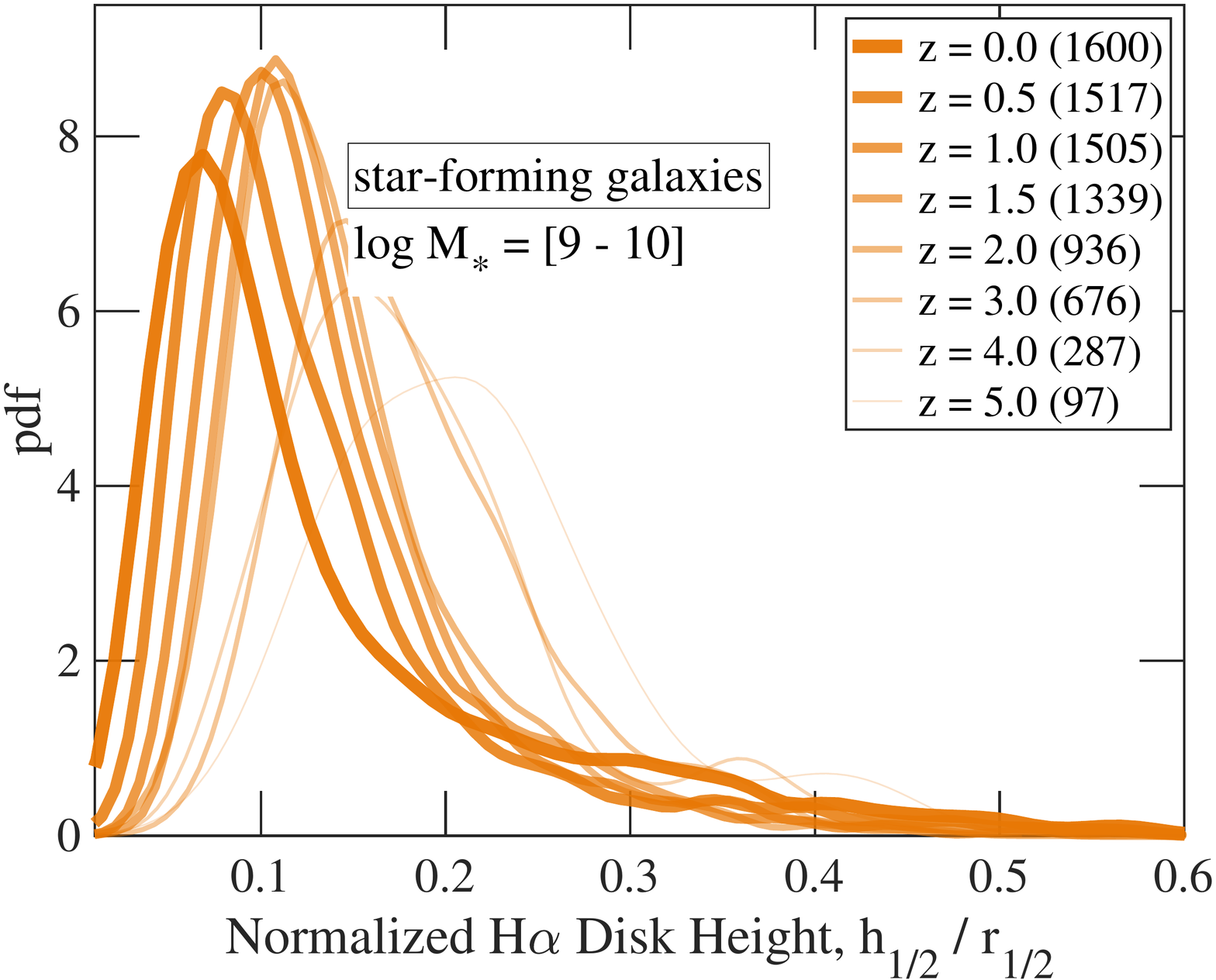}
\includegraphics[width=8cm]{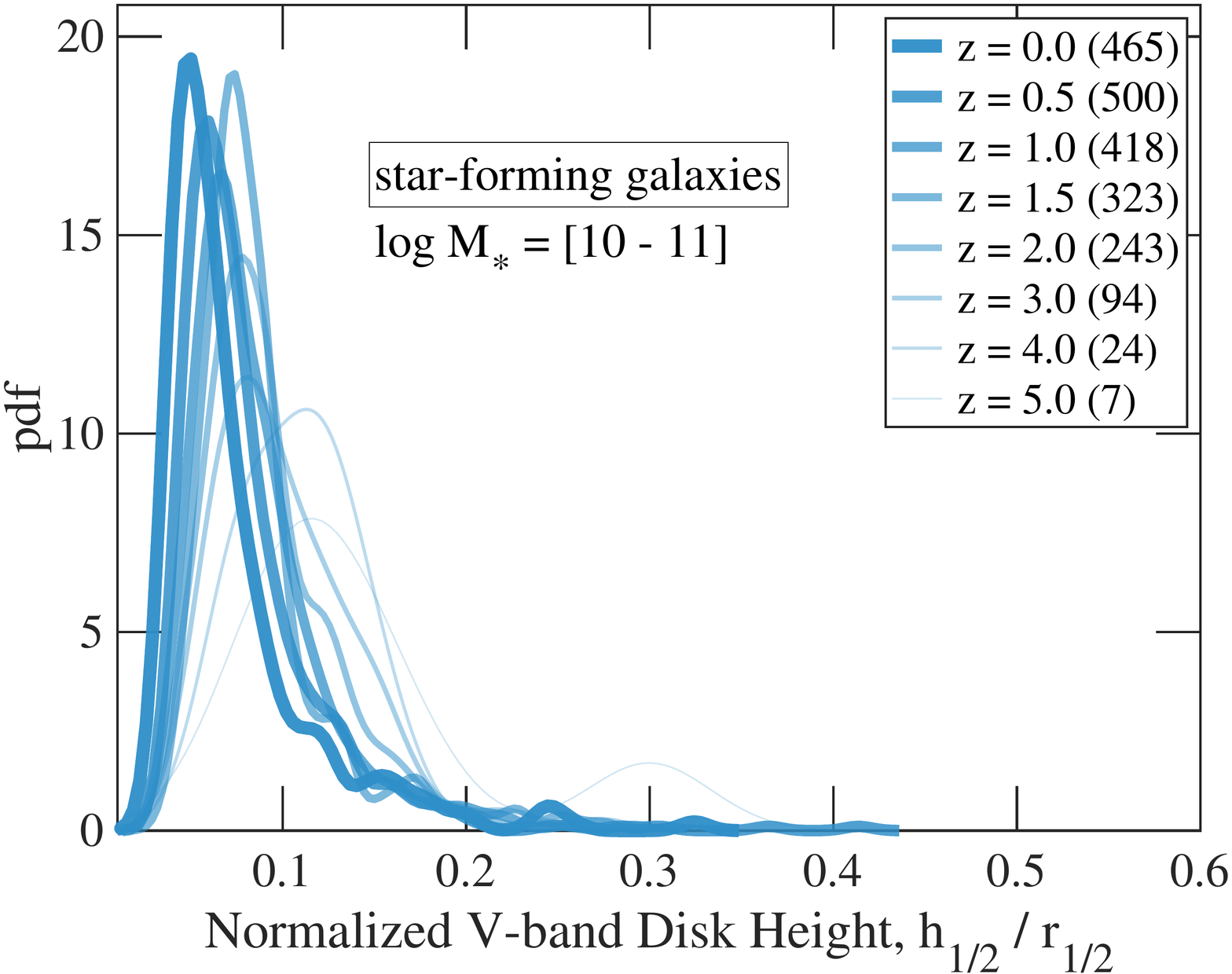}
\includegraphics[width=8cm]{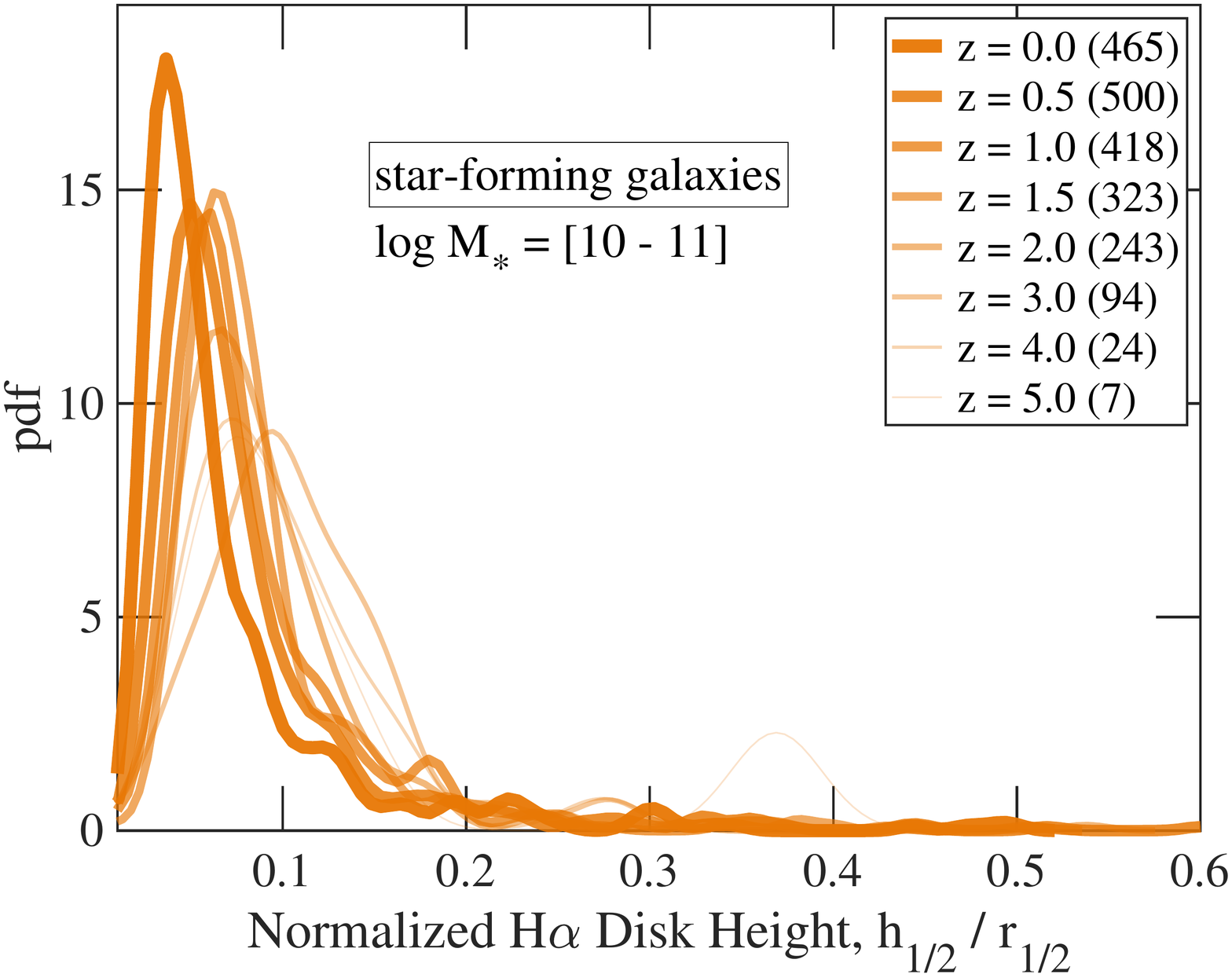}
\caption{\label{fig:diskheights_ev}   Time evolution of the distributions of {\it normalized} disk heights, for TNG50 star-forming galaxies. The distributions represent the relative thickness or flatness of galaxies, in bins of galaxy stellar mass (from top to bottom), derived from either V-band (left) or $\HA$ (right) light. In each panel, thicker and brighter curves depict lower redshifts. Galaxy populations exhibit flatter or ``diskier'' morphologies at more recent times.}
\end{figure*}

\subsection{Thickness of galaxy disks, and flatness}
\label{sec:heights}

Moving on from two and three dimensional sizes we proceed to investigate how TNG50 galaxies are distributed in terms of their edge-on extent, i.e. their thickness. 
\rvvv{In an upcoming paper (\textcolor{blue}{Donnari et al. 2020}), we will explicitly show the vertical stellar mass and light profiles of TNG50 disky galaxies. Here, we directly provide estimates of their ``heights''.}
As defined in Section~\ref{sec:props}, we label our galaxies with half-mass and half-light heights ($h_{1/2}$) using edge-on projections: in the case of actually flattened and disky galaxies, $h_{1/2}$ is a proxy for exponential disk heights.

Fig.~\ref{fig:diskheights_distribs} shows the independently-normalized distributions of galaxy disk heights ($h_{\rm 1/2}$) from the rest-frame V band (blue curves) and $\HA$ light (orange curves) profiles, from low to high redshifts (top to bottom) and from low to high stellar masses (left to right). The thicknesses or disk heights are given in physical parsecs. The richness of the simulated sample is indicated with the number of galaxies per mass bin and redshift: we have a maximum of about 13500 galaxies at $z=2$ in the $10^{7-8}\MSUN$ bin and a minimum of 24 galaxies in the $10^{10-11}\MSUN$ bin at $z=4$.

First, the two different proxies (rest-frame V band and $\HA$) return overall similar distributions, but for the $\HA$ disks being somewhat thinner below $10^{10}\MSUN$, by up to 45 per cent in the lowest mass bins at the cosmic noon. In both stellar and gaseous cases, the TNG model naturally returns populations of (massive) galaxies that vary in thickness by up to one order of magnitude at $z\lesssim1.5$: for example, at $z=1$ (not shown), heights range from about 80 to 600 parsecs between the 5th and 95th percentiles of star-forming $10^{10-11}\MSUN$ galaxies. We point out that the disk-height distributions of the total gas (cold or hot, dense or not: not shown) would be many times broader and peaking at values that are many times larger than those of the $\HA$-emitting gas, the latter representing the densest star-forming gas in the simulation. Stellar-mass (not shown) and V-band disk-height distributions, in contrast, are quite similar. Finally, the $\HA$-light vertical distribution is ``thinner'' than that of the gas mass that emits it, with vertical thicknesses up to a factor of two smaller at the high-mass end (also not shown): in other words, the vertical profiles of the SFRs or $\HA$ light are steeper than those of the star-forming gas mass. 

Second, the average galaxy thickness is between 100-400 physical parsecs for most of the distributions, which broaden for larger-mass galaxies. As with galaxy sizes, such small heights are only resolved with the improved numerical resolution of TNG50, namely in comparison to the other runs of the TNG series and other cosmological simulation projects (see Appendix~\ref{sec:app_res}). Importantly, the bulk of $10^{8-9}\MSUN$ galaxies exhibits disk heights of roughly 300 pc at all times above $z\sim1$, even at earlier epochs (e.g. up to $z\sim8$, not shown) when the stellar softening length (or three times its value) is many factors smaller. This hints that, within our model, disk heights are not a trivial consequence of the choices for the numerical gravitational softenings, as we demonstrate in Appendix~\ref{sec:app_res}. There, we conclude that the galaxy thicknesses of Fig.~\ref{fig:diskheights_distribs} are already converged to the 20-40 per cent level across all studied masses and redshifts.

Finally, the median disk heights increase for more massive galaxies, and exhibit a non monotonic trend with redshift. Massive galaxies ($\gtrsim10^9\MSUN$) reach larger thicknesses at the cosmic noon, and all disks become thinner with time after $z\sim1$.

As galaxies themselves are larger (more extended) at larger stellar masses (see Fig.~\ref{fig:sizes}) and galaxy populations exhibit larger sizes in both comoving and physical units towards lower redshifts (at least at $\MS \gtrsim 10^{10}\MSUN$), a normalization of the disk heights is required in order to better capture the mass and redshift trends of galaxy shapes. In Fig.~\ref{fig:diskheights_ev} we show the distributions of the {\it normalized} disk heights ($h_{1/2} / r_{1/2}$), dividing by the 2D face-on projected circularized half-light radii of the corresponding tracer, whether V-band or $\HA$ light. In practice, these normalized disk heights are somewhat similar to the $q$ parameter often used in observations (minor to major axis of the ellipses fit to a given isophote in projection), with the benefit that our measure avoids any undesirable projection effects. Alternatively, we can view the normalized disk height as an estimate of the intrinsic flatness of a galaxy, with smaller values denoting flatter and ``diskier'' galaxies.

The distributions of Fig.~\ref{fig:diskheights_ev} encompass galaxies from low (top) to high (bottom) stellar masses, and for V-band (left) vs. $\HA$ (right) heights. We omit the smallest $10^{7-8}\MSUN$ bin to avoid results possibly affected by resolution issues in the sizes (see Appendix~\ref{sec:app_res}). In each panel, thicker and darker curves refer to lower redshifts, in the range $0.5 \lesssim z \lesssim 5$. Two main trends are clearly manifest. First, at fixed galaxy stellar mass, the populations of star-forming galaxies are progressively thinner, flatter (i.e. diskier) at more recent times than at early cosmic epochs, with the bulk of the $10^{9-11}\MSUN$ galaxy population exhibiting flatness indices as small as 10 per cent by $z \lesssim 1$. Second, at any given time, more massive galaxies are relatively more disky than lower-mass counterparts, as can be appreciated by comparing across the three rows. V-band and $\HA$ light measurements both return a very consistent picture, albeit with some quantitative differences (see Discussion). Given Figures \ref{fig:sizes} and \ref{fig:diskheights_distribs}, these trends appear to be the result of the redshift evolution of 2D sizes and physical ``thicknesses'' alike.

While it is well-known from observations of the local Universe that massive star-forming galaxies have disk-like shapes (e.g. the Milky Way, Andromeda, M51 \citep{Schinnerer:2013}, the galaxies of the PHANGS survey (\textcolor{blue}{Leroy et al. 2019}) and from the surveys cited in Section 6.2), the precise quantification and a robust understanding of the evolution of stellar and gaseous shapes across cosmic time and masses remains a topic of ongoing investigation. The trends provided by  Fig.~\ref{fig:diskheights_ev} provide a prediction from the TNG50 simulation which can be quantitatively tested against observations in the future.

\begin{figure*}
\centering 
\includegraphics[width=17cm]{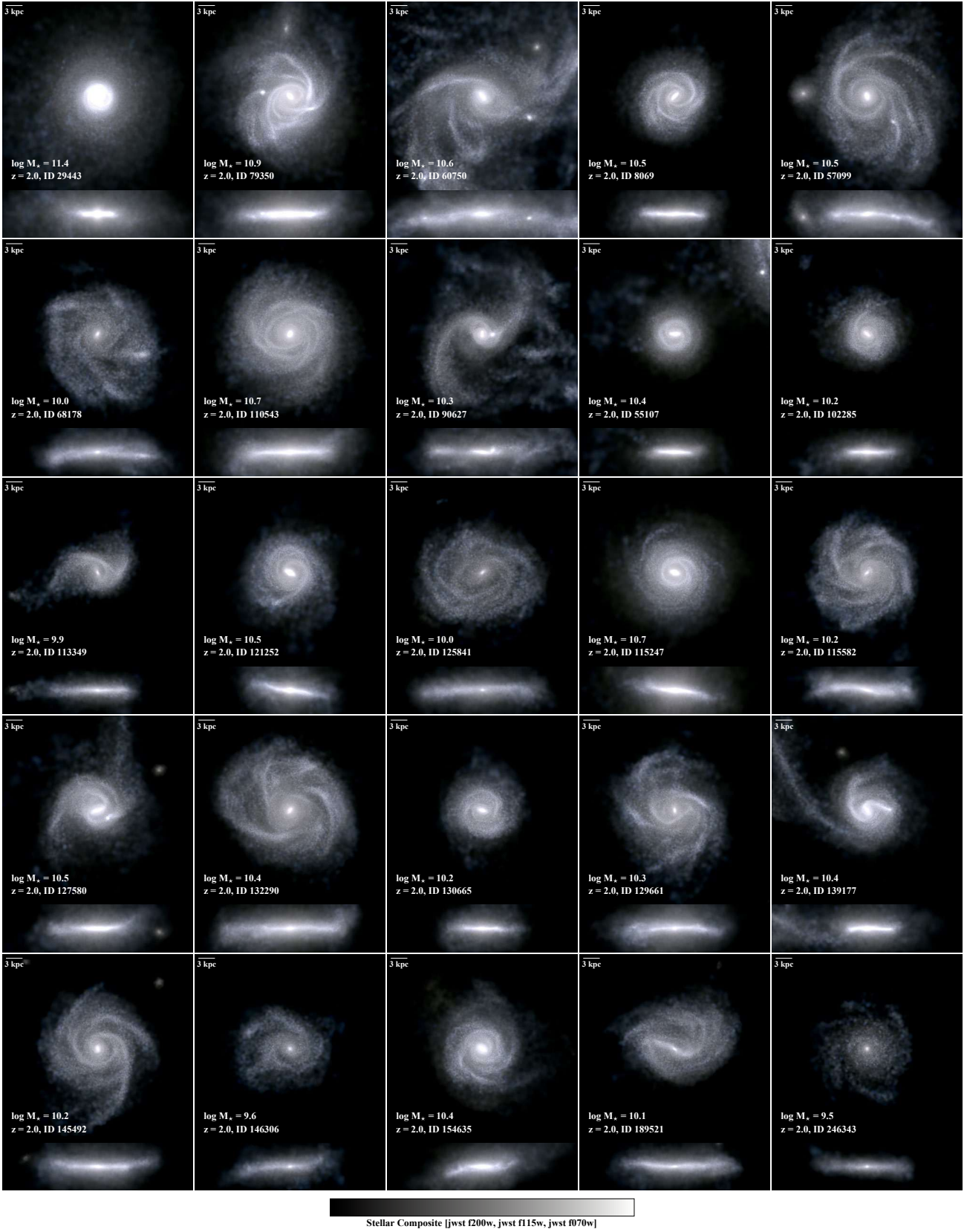}
\caption{\label{fig:images_1} Stellar light composites of a random sample of $z=2$ TNG50 star-forming galaxies, selected \rvvv{to be thin i.e. to} have $h_{1/2} \leq 0.1 \times r_{1/2}$ in the V-band and $\MS \ge 10^{9}\MSUN$. Images are for the JWST NIRCam f200W, f115W, and F070W filters (rest-frame), neglecting any dust effects and including all stellar light within the given projection. Every panel is 40 physical kpc on a side. The rich diversity of stellar structure is captured and quantified by the metrics explored herein.}
\end{figure*}

\begin{figure*}
\centering 
\includegraphics[width=17cm]{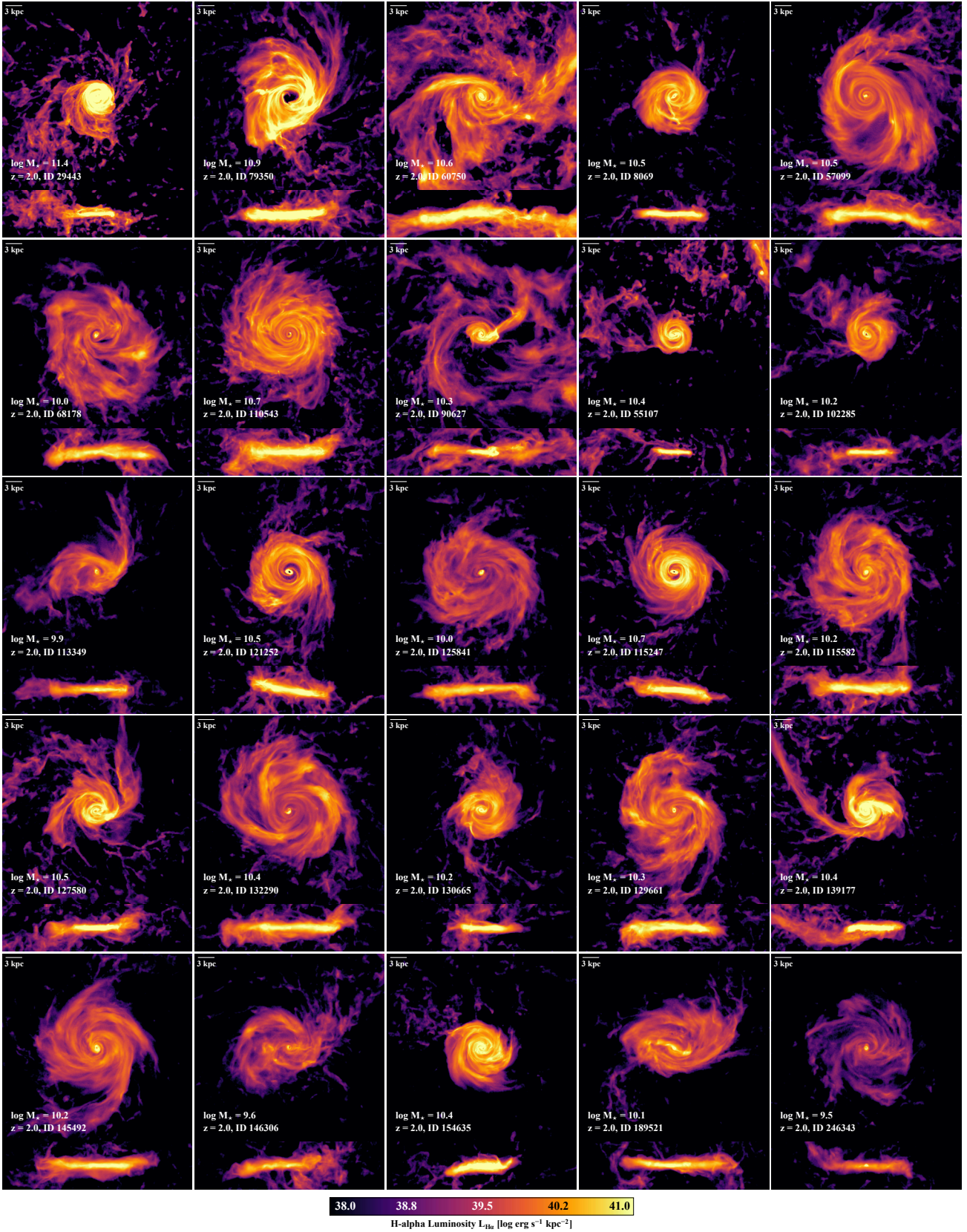}
\caption{\label{fig:images_2} Same as in Fig.~\ref{fig:images_1} but for $\HA$ light, i.e. the light from the star-forming gas within simulated galaxies \rvvv{selected to have small relative disk heights}. The complex gas-phase structure of $z=2$ star-forming galaxies and deviations from symmetric disk-like configurations, including the presence of warps and large-scale asymmetries, highlight the challenge of quantifying galactic structure using $\HA$ as a morphological tracer.}
\end{figure*}

To give a visual impression of how such galaxies simulated in TNG50 are structured, we show in Figs.~\ref{fig:images_1}, \ref{fig:images_2} a random sample of 25 ``thin'' star-forming galaxies at $z=2$, selected to have very thin disks, namely with V-band half-light height smaller than one tenth of the V-band 2D galaxy radius ($h_{1/2} \leq 0.1 \times r_{1/2}$). Fig.~\ref{fig:images_1} shows the face-on and edge-on projections imaged in JWST NIRCam f200W, f115W, and F070W filters (no dust, rest-frame) and Fig.~\ref{fig:images_2} the corresponding $\HA$ luminosity maps. The stellar structures span a wide diversity, from compact spirals to large extended disks, and from massive stellar bulges, to prominent bars, to nearly bulgeless surface brightness profiles. There is frequently complex structure evident in the star-forming gas, particularly e.g. in deviations from perfect disk-like morphologies in the form of warps and large-scale asymmetries due to gas inflows and outflows common at this epoch.

\begin{figure*}
\centering   
\includegraphics[height=5.6cm]{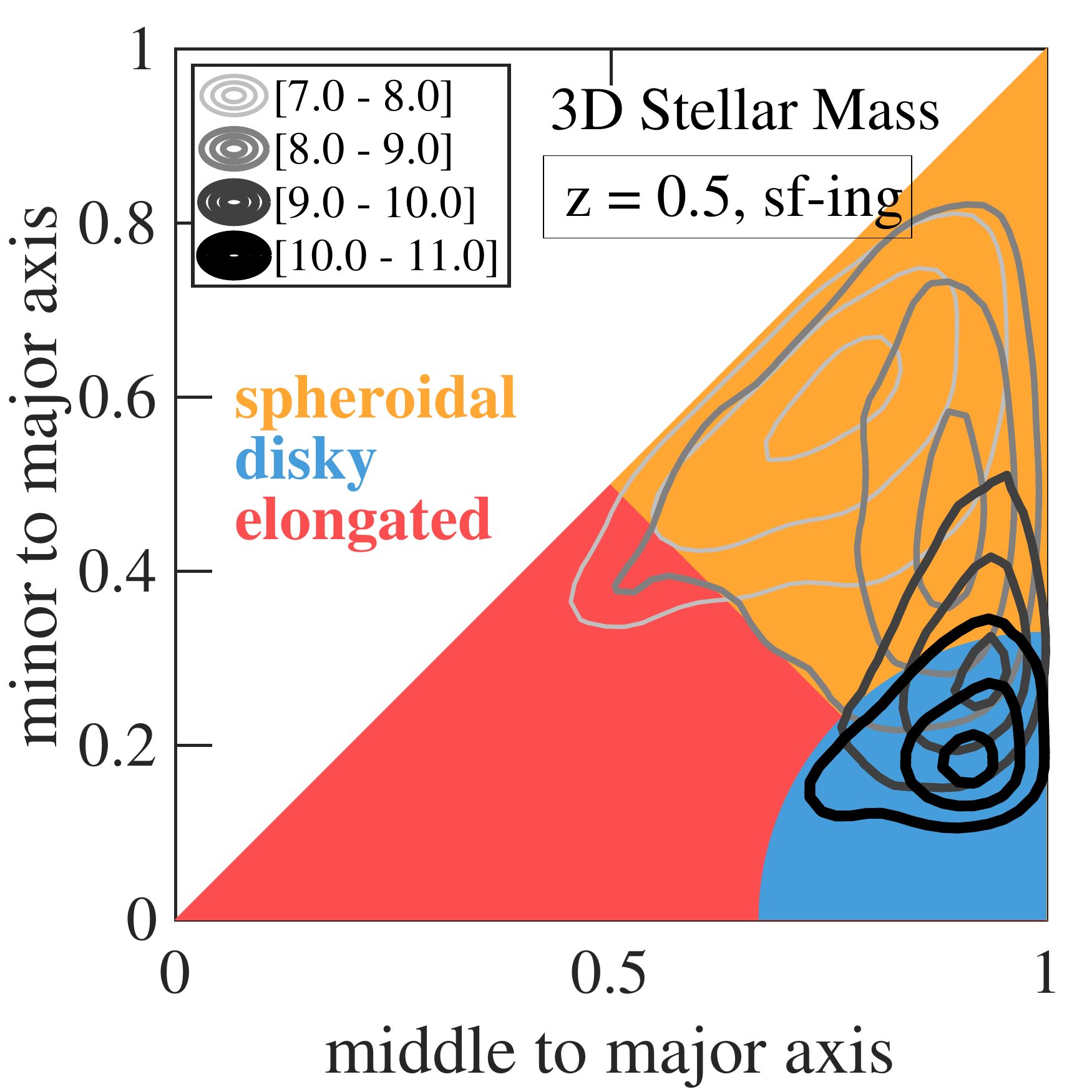}
\includegraphics[trim={1cm 0 0 0}, height=5.6cm]{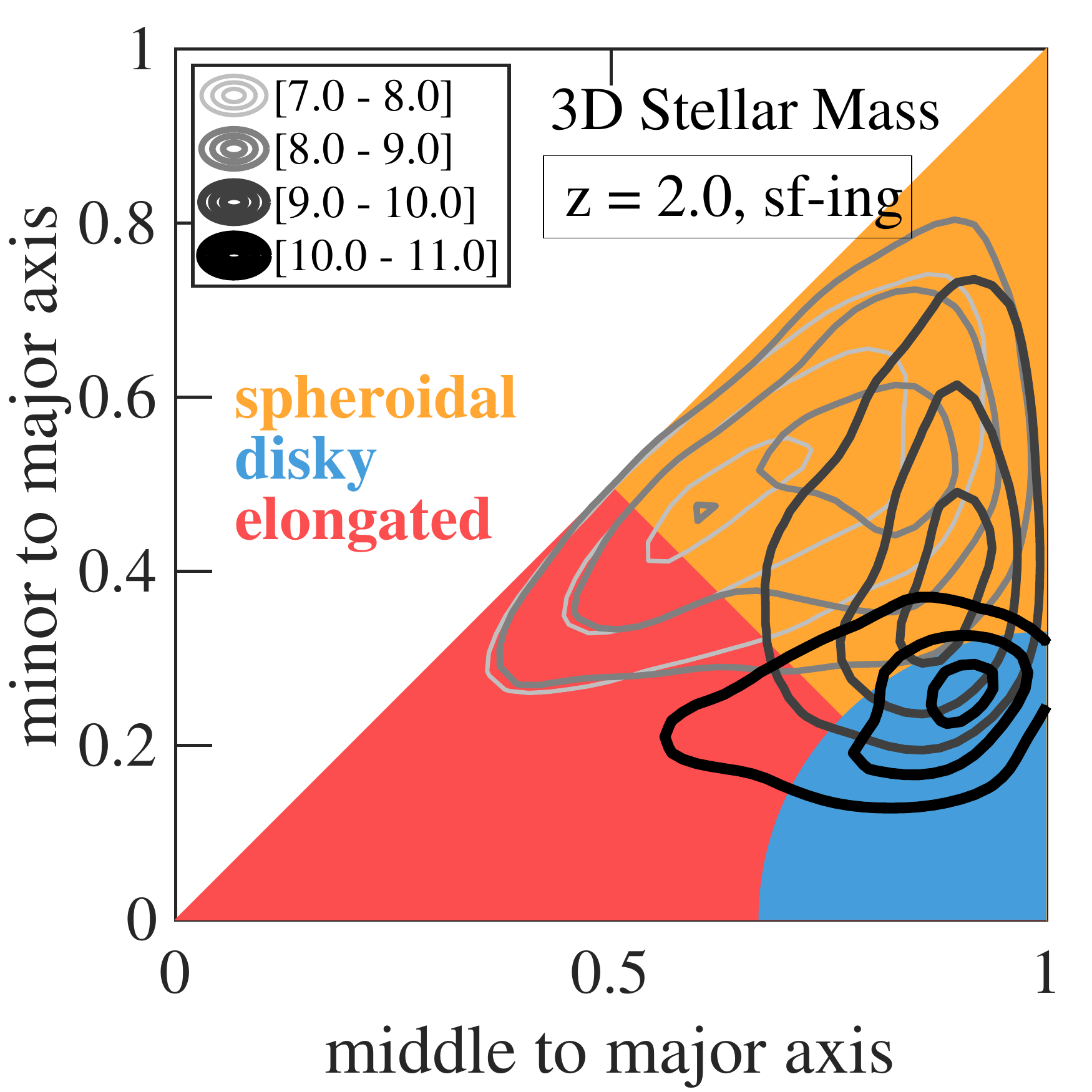}
\includegraphics[trim={1cm 0 0 0}, height=5.6cm]{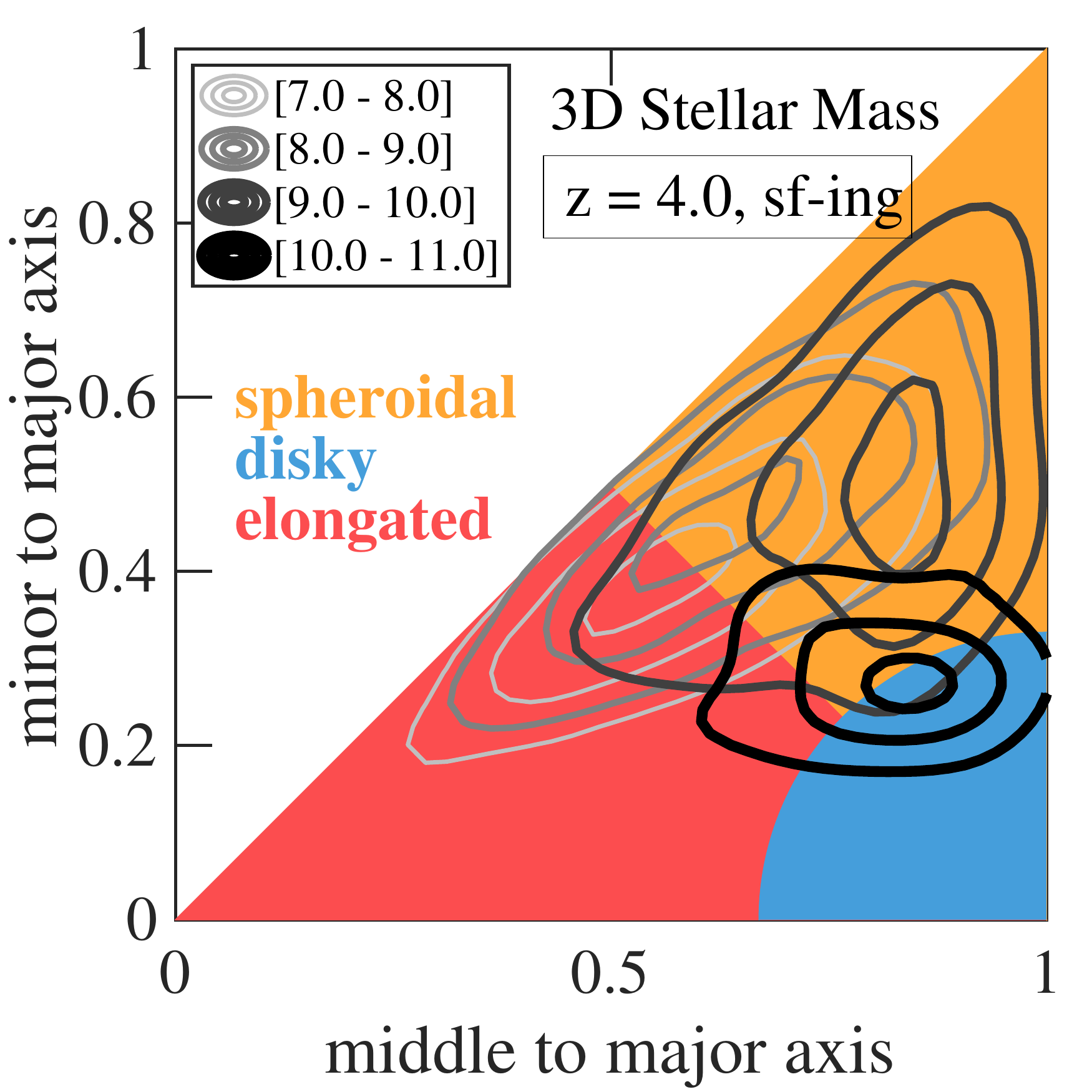}

\includegraphics[height=5.6cm]{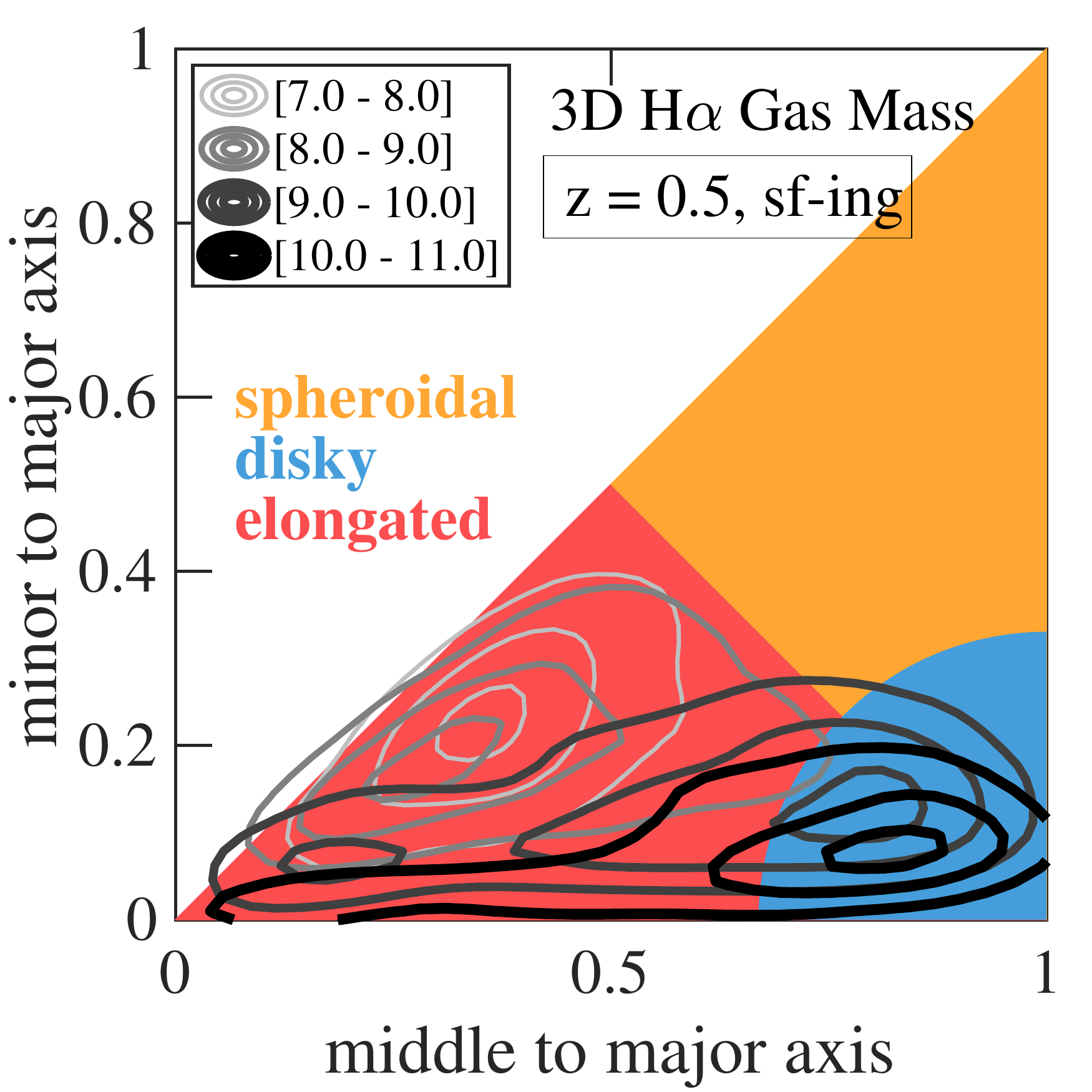}
\includegraphics[trim={1cm 0 0 0}, height=5.6cm]{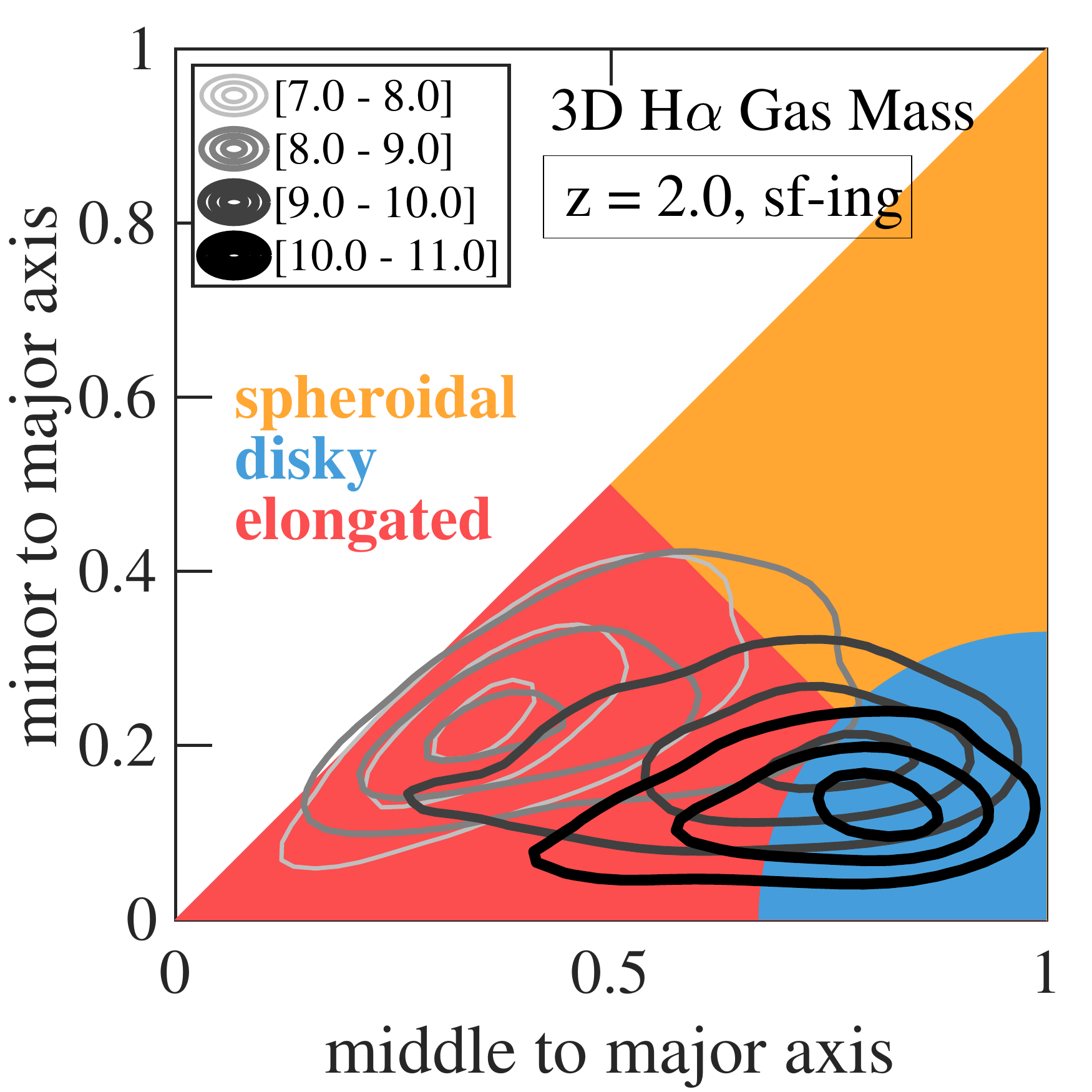}
\includegraphics[trim={1cm 0 0 0}, height=5.6cm]{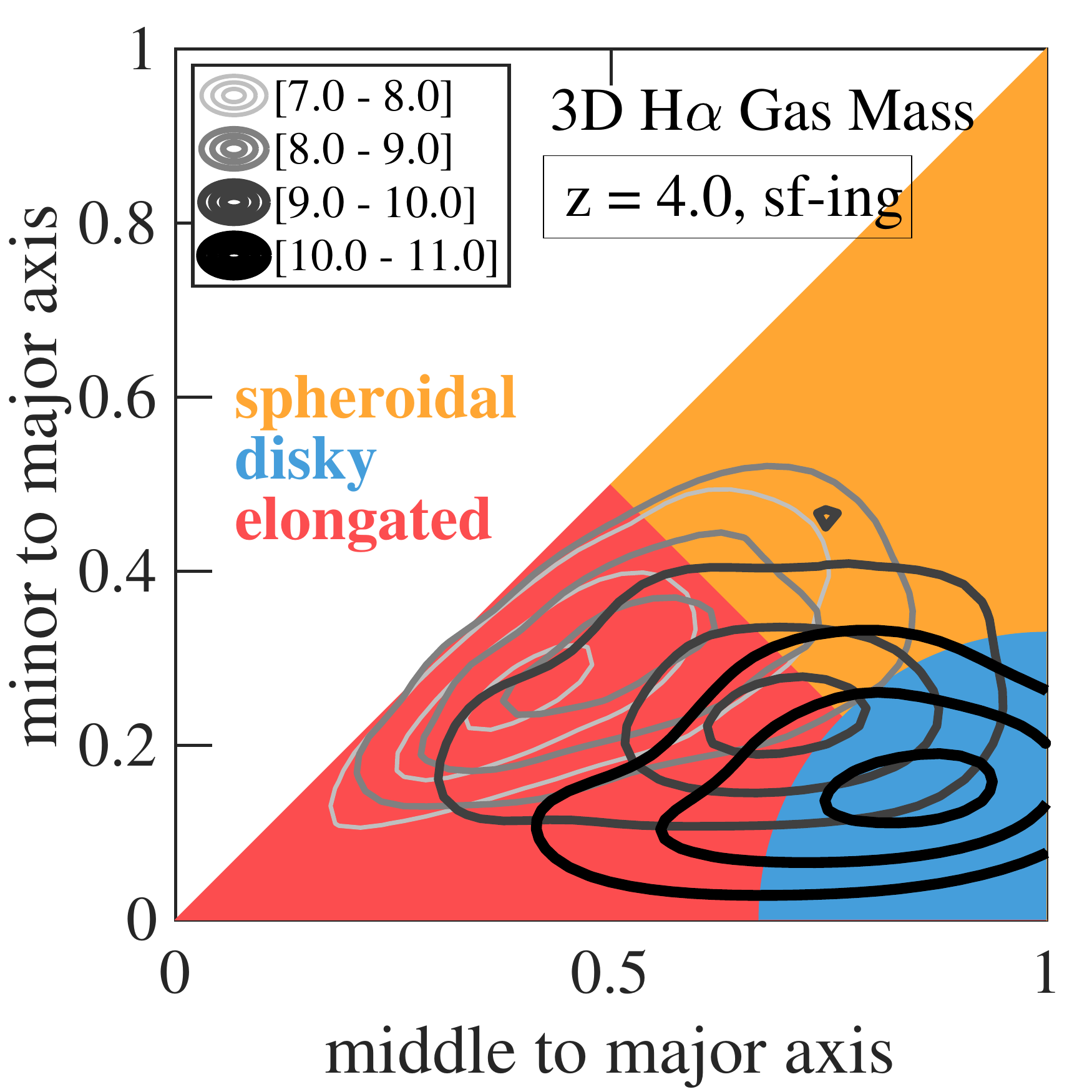}
\caption{\label{fig:shapes} The structural i.e. morphological mix of TNG50 star-forming galaxies as a function of mass and time, according to the 3D shapes of their stellar-mass (top) and $\HA$-emitting gas mass (bottom). We measure the minor to major and middle to major axis ratios at twice the stellar-half mass radius. Contours are the results of a bivariate kernel density estimation, and denote levels at 90, 60, and 30 per cent of the galaxy number density distribution maxima. Different shades of black/gray denote different mass bins. We plot the shape distributions only for the mass and redshift bins that contain at least 20 galaxies. Gaseous structures are systematically more disky or elongated (lower minor to major axis) than stellar bodies.} 
\end{figure*}

\subsection{Three-dimensional intrinsic shapes}
\label{sec:3Dshapes}

A direct assessment of the structural morphology of galaxies can be obtained by identifying the three-dimensional ellipsoidal shape that best describes their mass spatial distribution. While the 3D shape is certainly directly accessible with simulated galaxies, recently the 3D shapes of galaxies have been estimated also from IR observations of e.g. CANDELS galaxies through de-projection methods \citep[e.g.][]{Zhang:2019, VanderWel:2014}. Effectively, the 3D shape of a galaxy is the most intrinsic quantification of how its material is distributed in space, and hence provides a less ambiguous characterization in comparison to e.g. any mass or light-profile based indexes, like Sersic or concentration parameters. 

In Fig.~\ref{fig:shapes}, we show the number density distribution of TNG50 galaxies in the minor-to-major axis ratio ($s$) vs. middle-to-major axis ratio ($q$) plane at several redshifts. Here we describe the ellipsoidal shape of galaxies using a shell centered around twice the stellar half-mass radius, as described in Section~\ref{sec:props}. The shape is estimated for the stellar mass (top panels) and the $\HA$-emitting gas (bottom panels). Contours of the galaxy number density in different colors, from black to light gray, denote different stellar mass bins. Colored regions in the shape plane visualize the morphological classification proposed by \citet{VanderWel:2014} and that we adopt throughout: disky vs. spheroidal vs. elongated galaxies, in close analogy to the oblate vs. spherical vs. prolate characterizations of dark-matter haloes based on the triaxiality parameter.

From Fig.~\ref{fig:shapes}, three main results quickly arise: 1) stellar-mass based disky galaxies are more frequent at lower redshifts than at higher redshifts and at larger masses than lower masses; 2) elongated i.e. `cigar-like' stellar mass distributions are more frequent at higher redshifts ($z\gtrsim1$) and lower masses ($\MS \lesssim 10^9\MSUN$, depending on redshift); 3) the spatial distributions of star-forming gas are markedly biased towards disky and elongated shapes. The first two statements are qualitatively consistent with recent observational findings at intermediate redshifts by \citet[][]{Zhang:2019, VanderWel:2014} (see also below). The $\HA$-gas results are an absolute novelty and a confirmation that, despite the unavoidable simplifications within the underlying galaxy-physics model, in TNG star-formation occurs in relatively thin disk-like configurations of dense gas, in accordance with the basics principles of galaxy formation theory.

\begin{figure*}
\centering   
\includegraphics[trim={0 0 0 0}, clip,height=3.8cm]{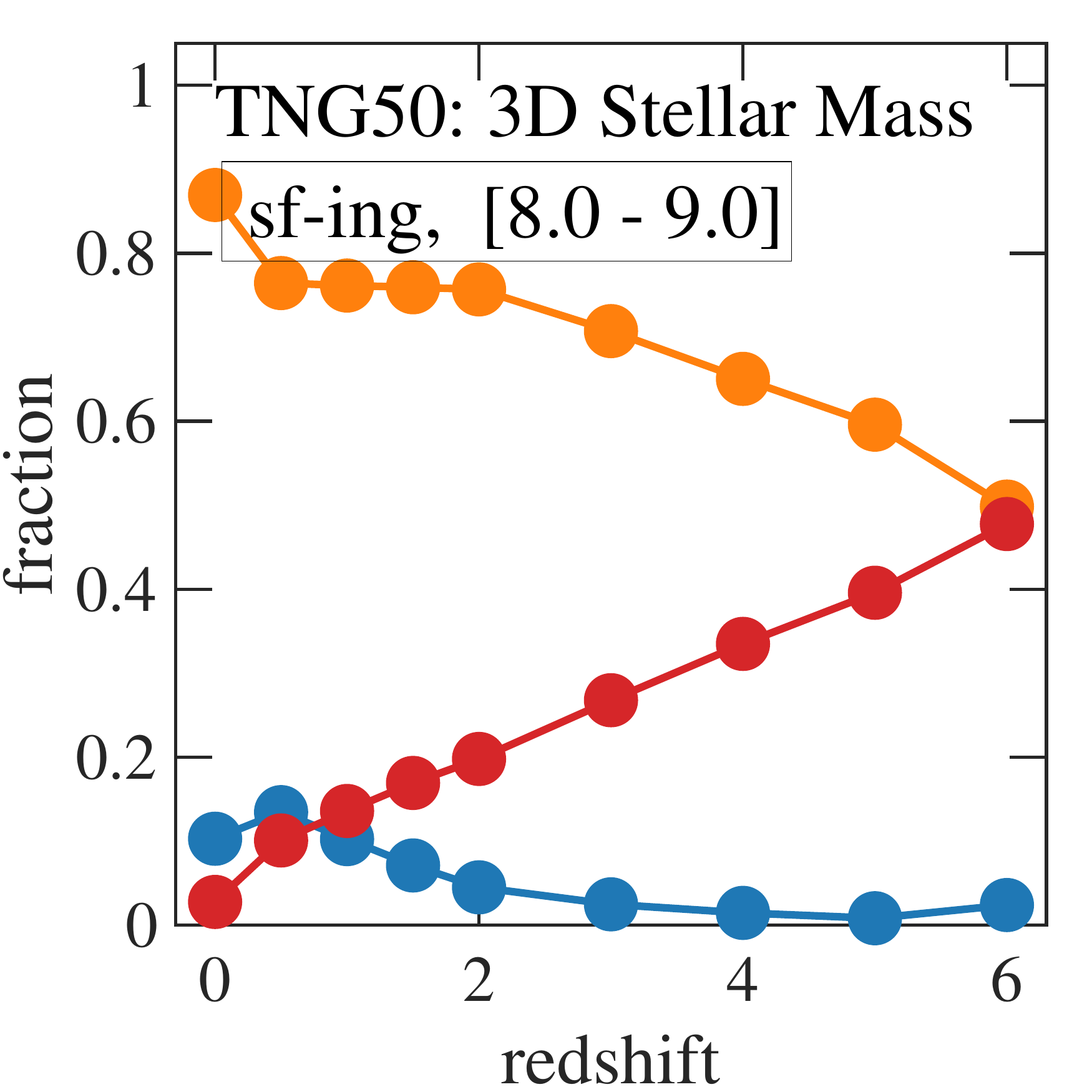}
\includegraphics[trim={2.6cm 0 0 0}, clip,height=3.8cm]{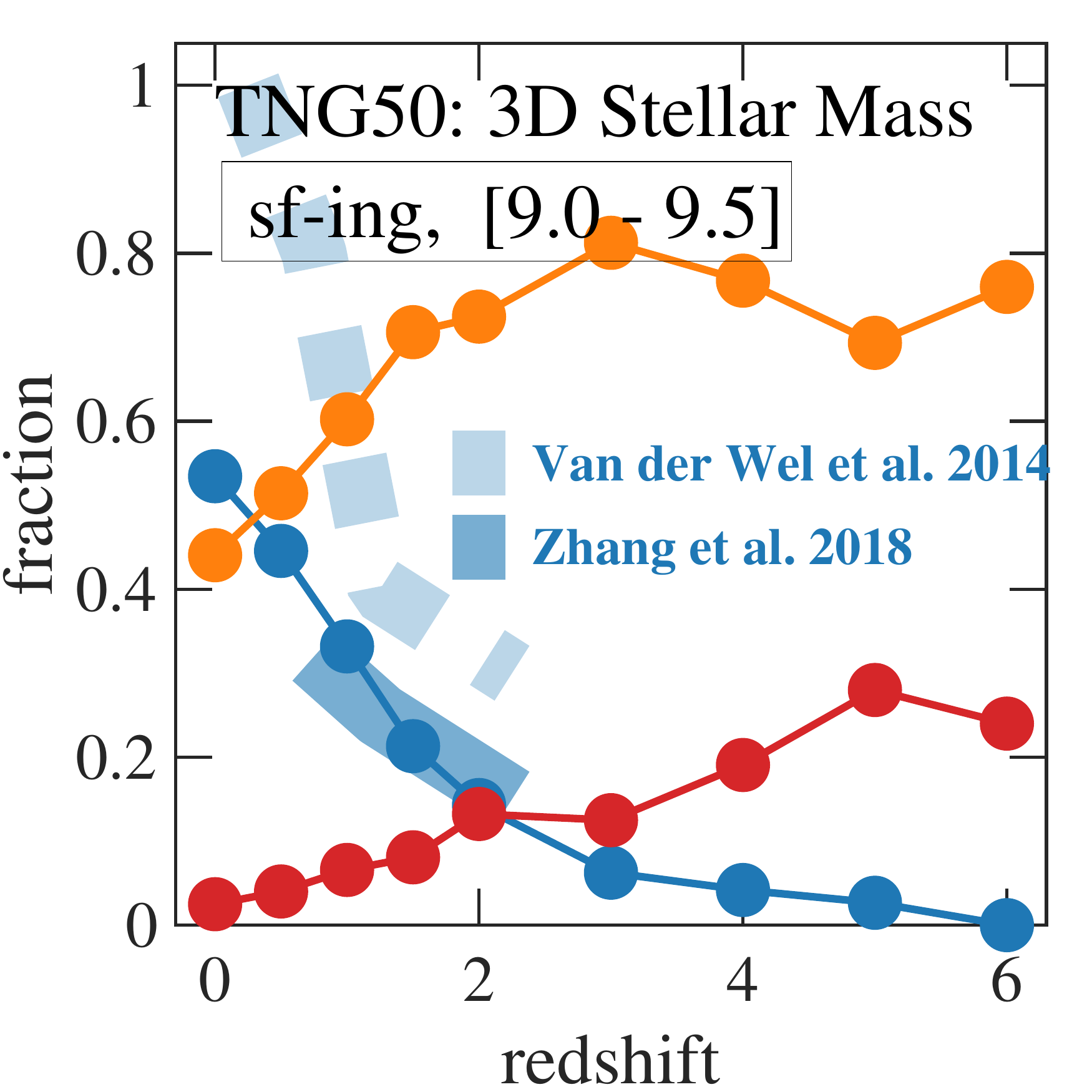}
\includegraphics[trim={2.5cm 0 0 0}, clip,height=3.8cm]{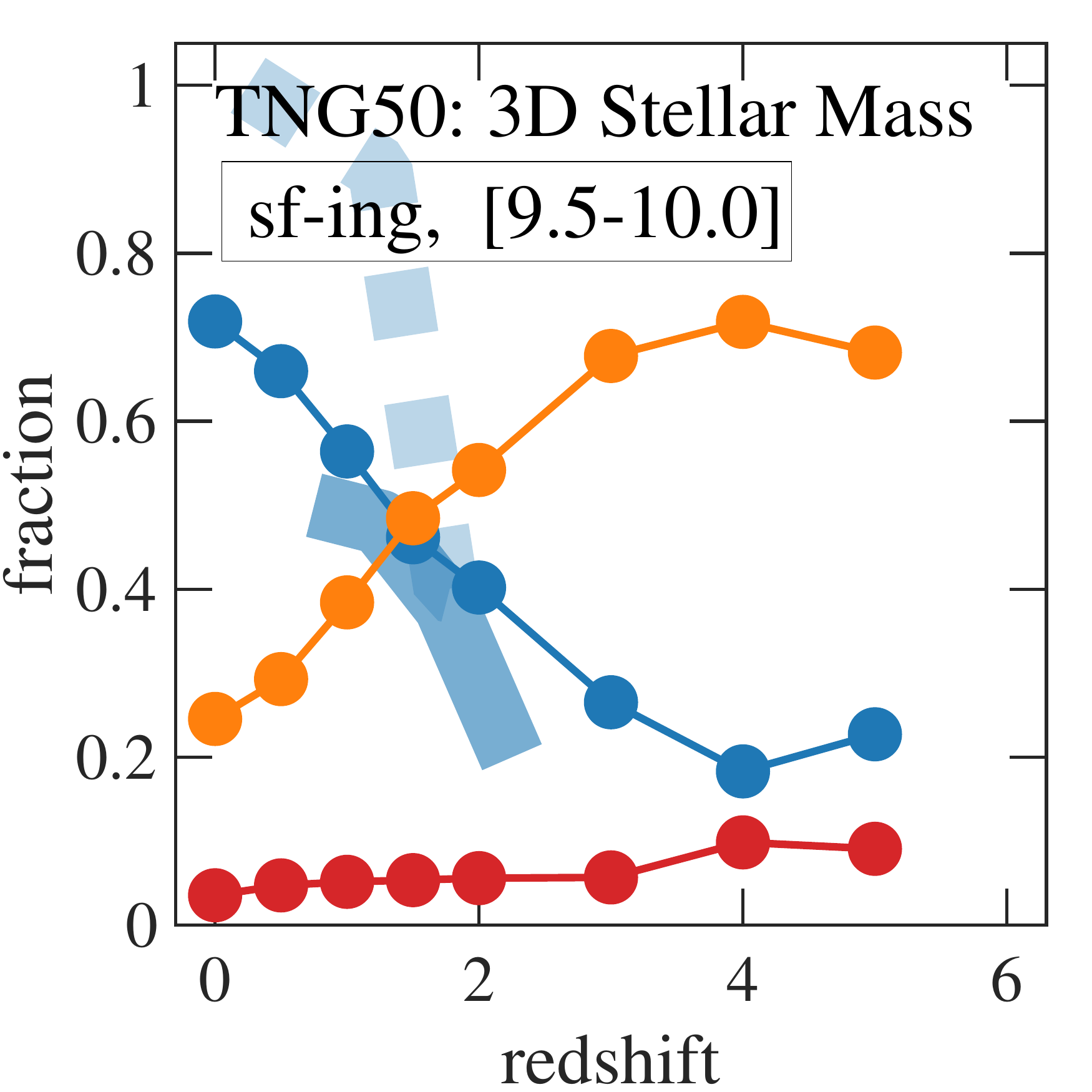}
\includegraphics[trim={2.6cm 0 0 0}, clip,height=3.8cm]{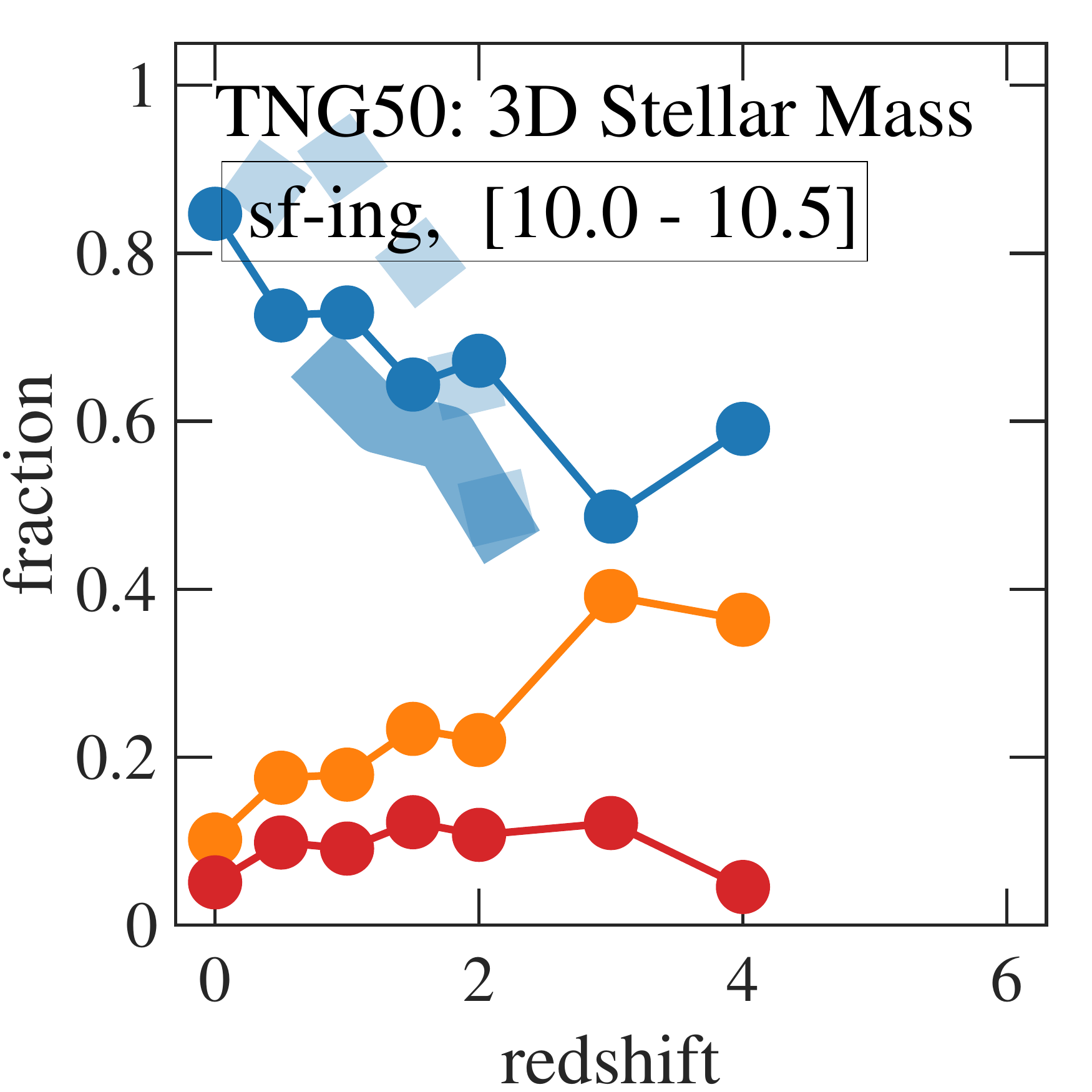}
\includegraphics[trim={2.5cm 0 0 0}, clip,height=3.8cm]{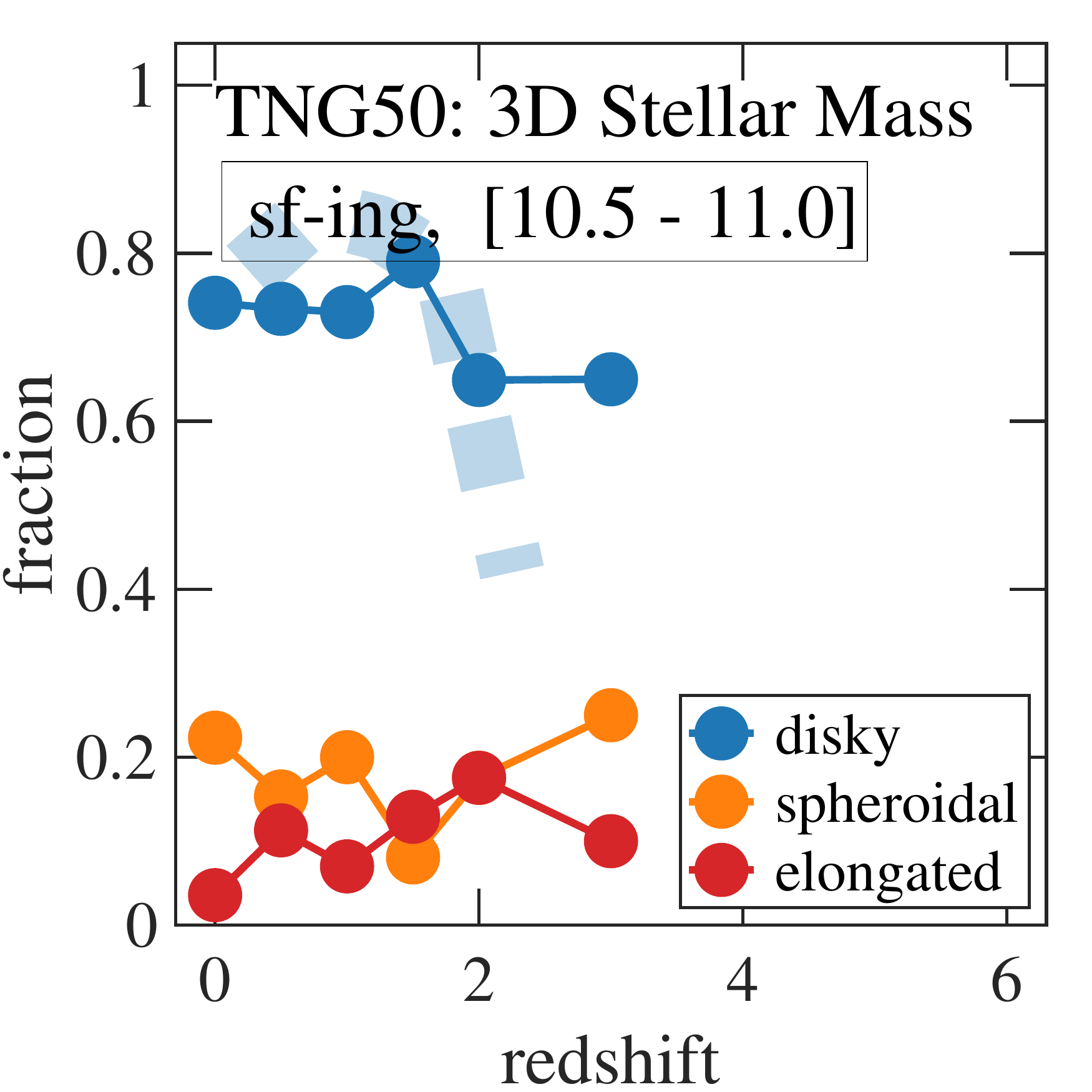}

\includegraphics[trim={0 0 0 0 0}, clip,height=3.8cm]{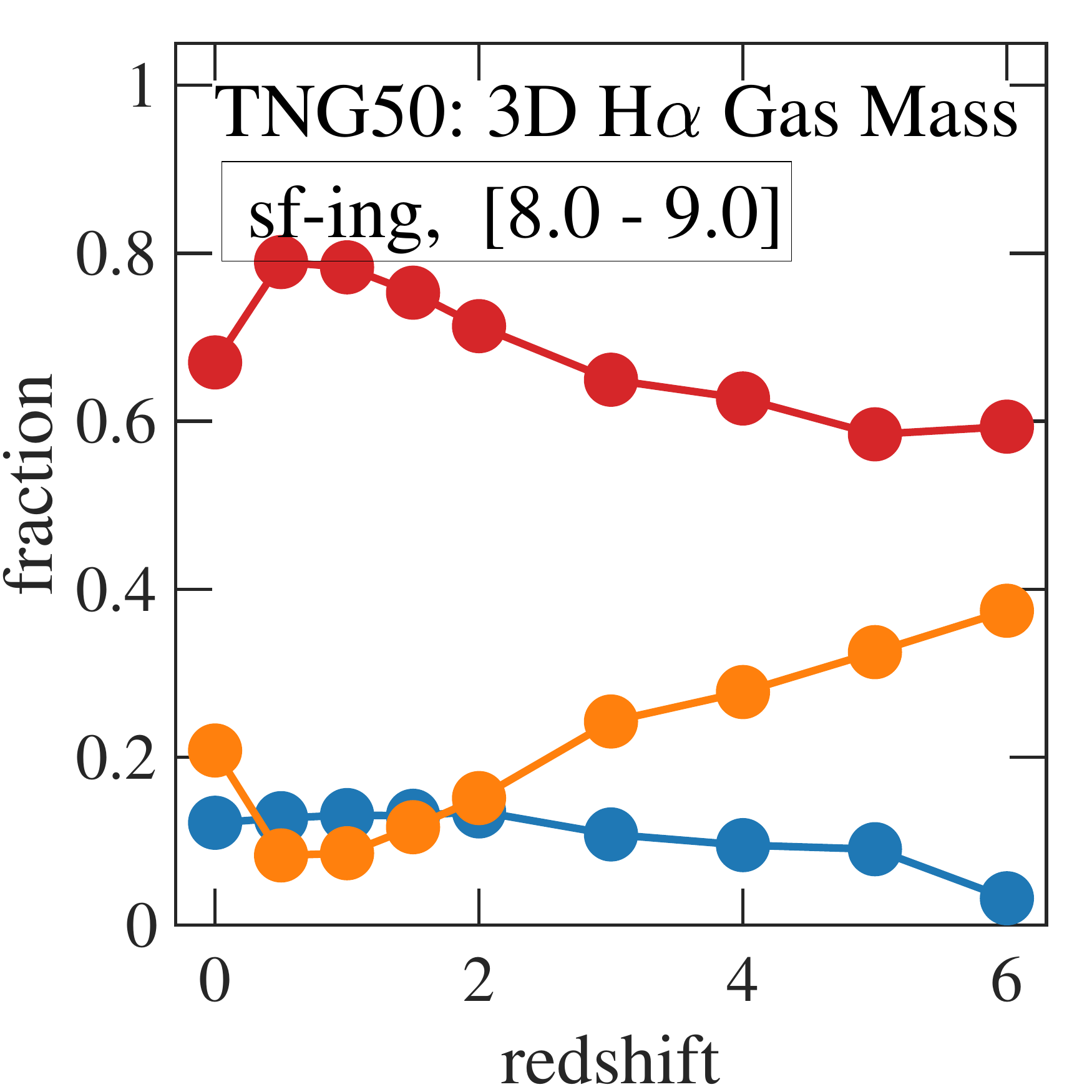}
\includegraphics[trim={2.6cm 0 0 0}, clip,height=3.8cm]{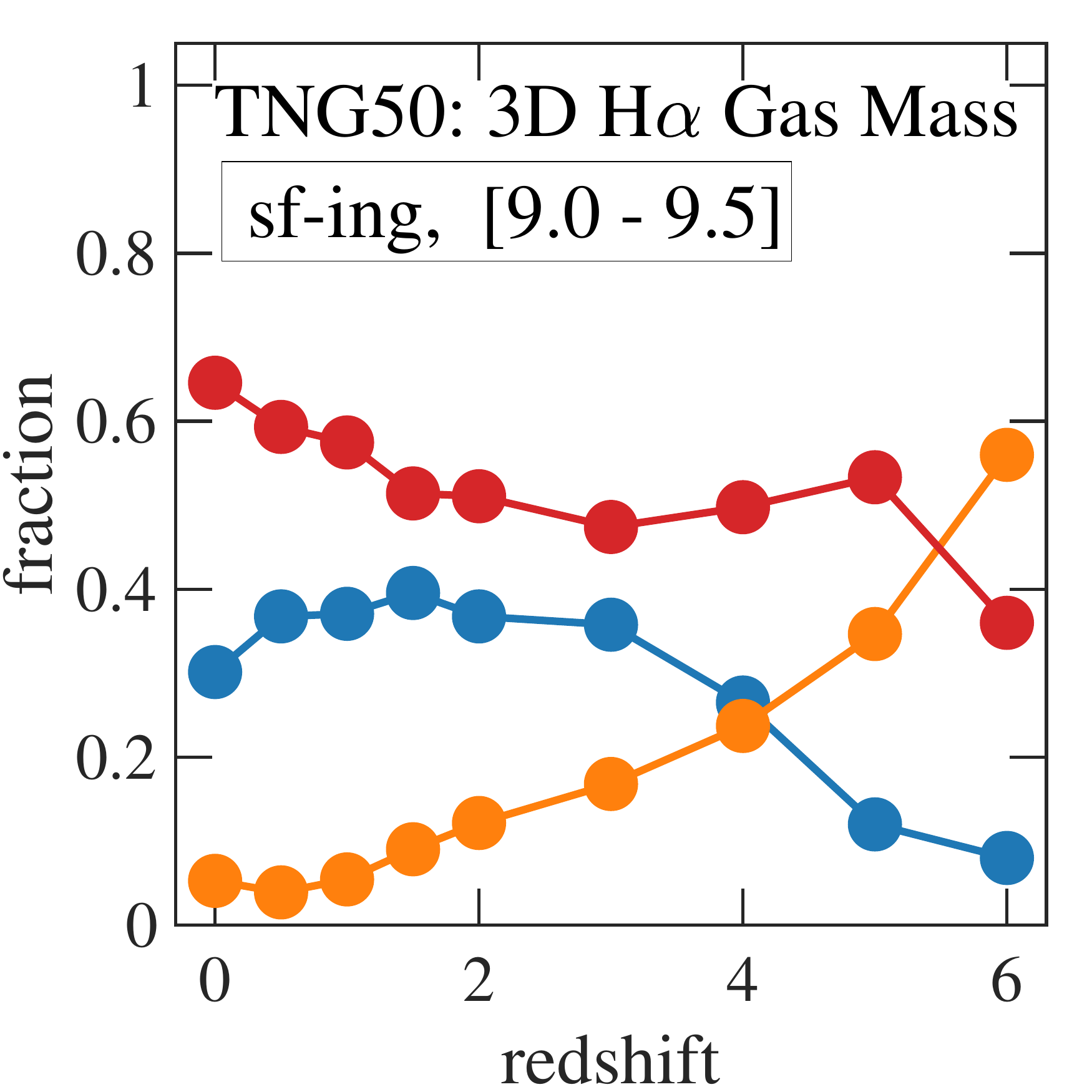}
\includegraphics[trim={2.5cm 0 0 0}, clip,height=3.8cm]{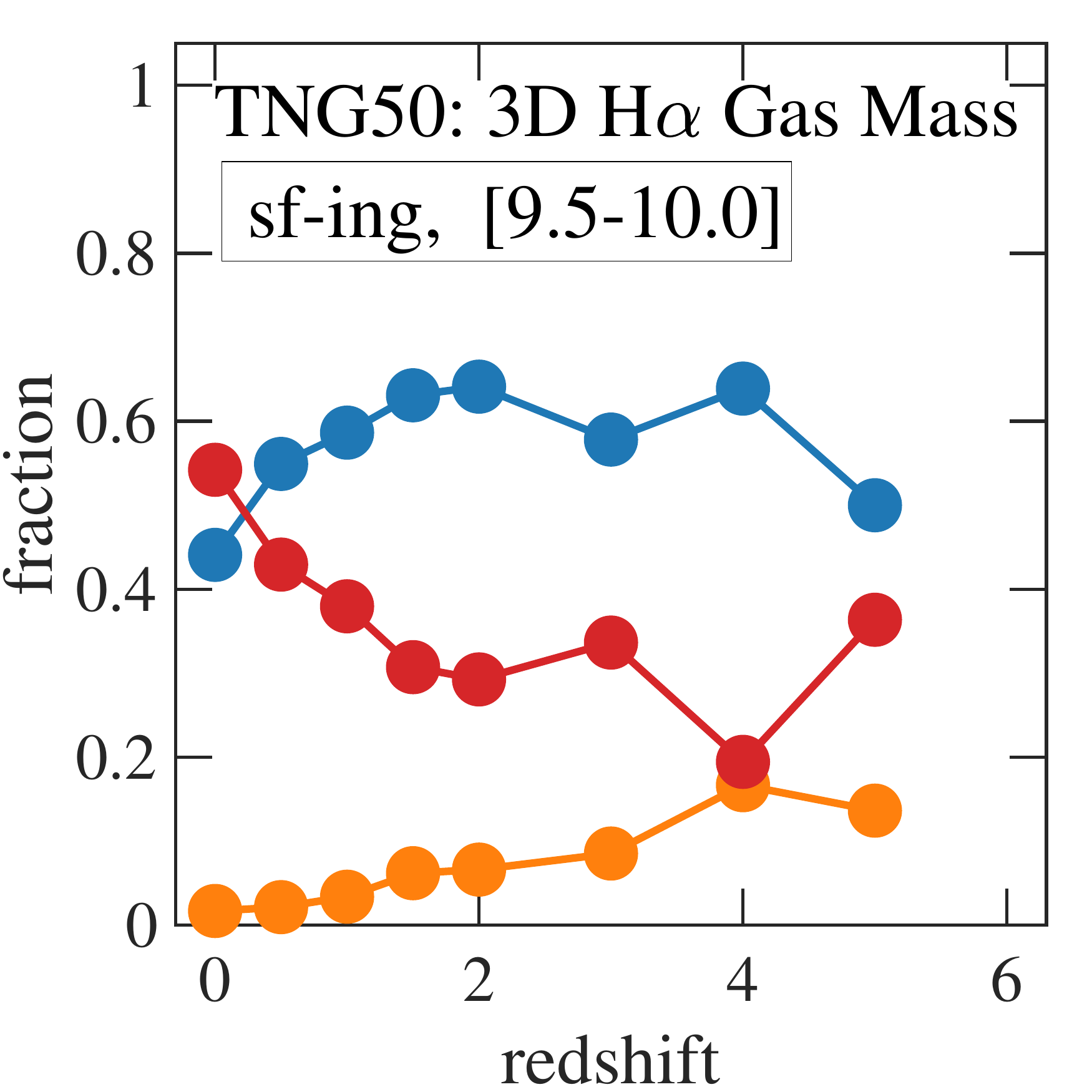}
\includegraphics[trim={2.6cm 0 0 0}, clip,height=3.8cm]{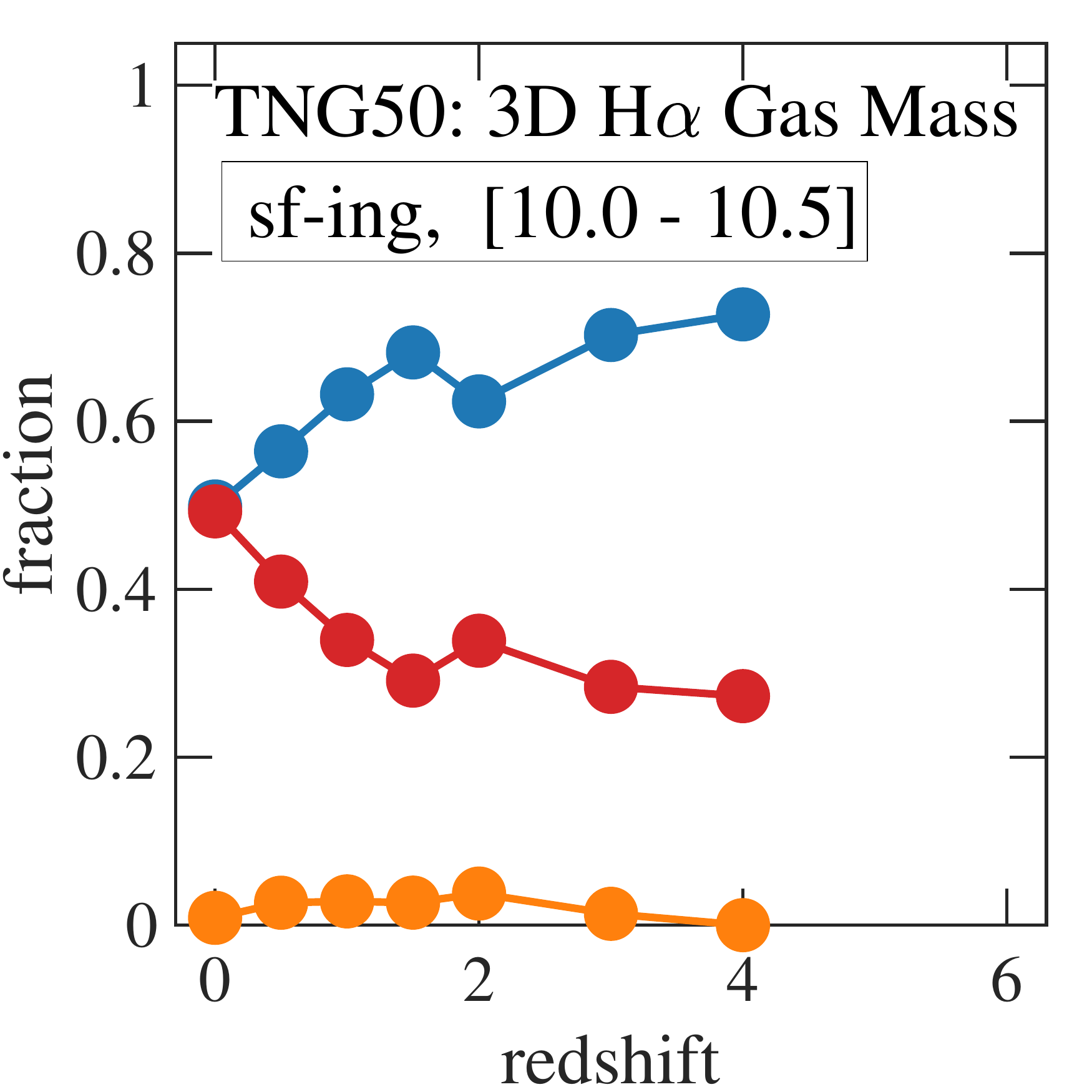}
\includegraphics[trim={2.5cm 0 0 0}, clip,height=3.8cm]{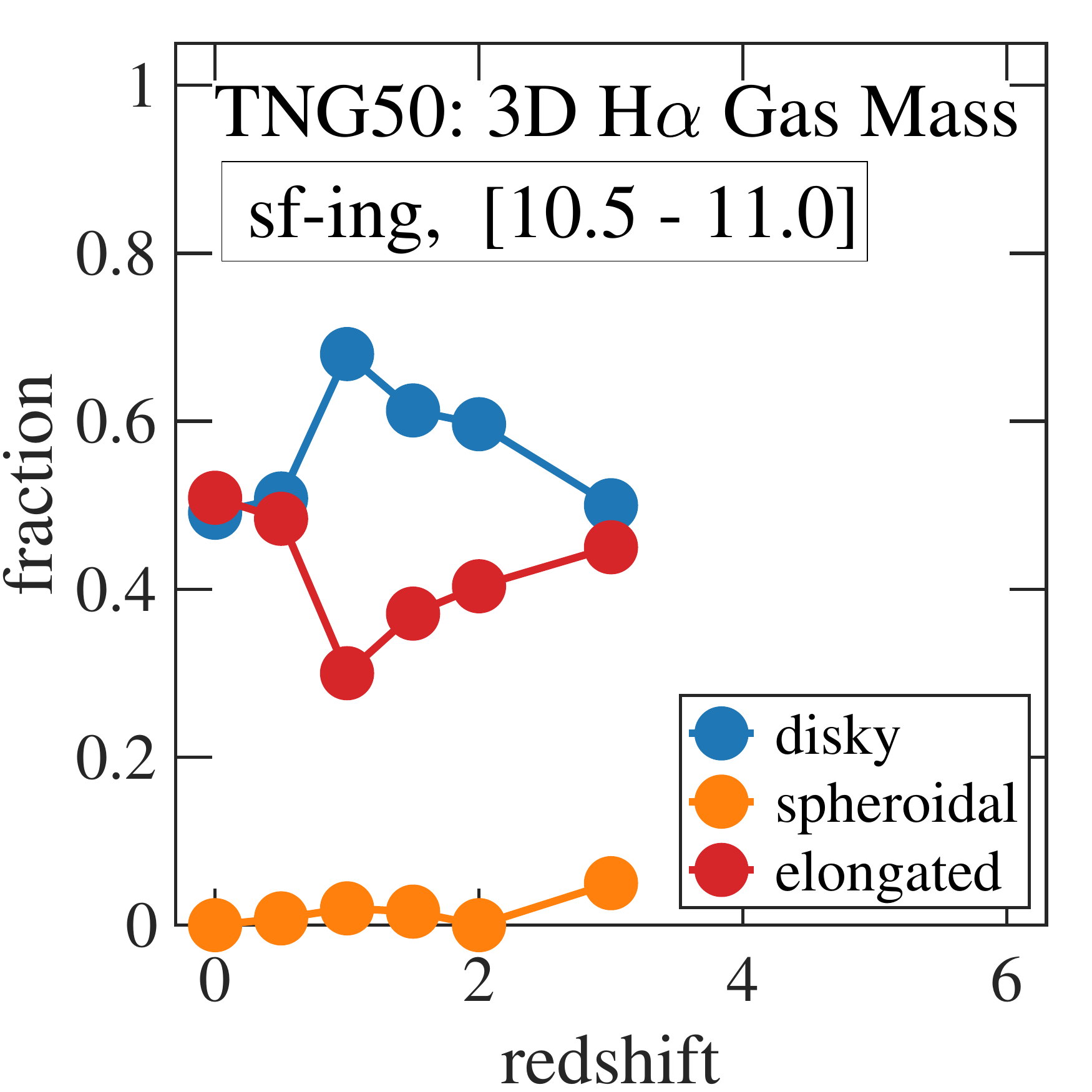}

\caption{\label{fig:shapesevolutions} The morphological fractions of TNG50 star-forming galaxies as a function of mass and time, according to the 3D mass shapes of their stellar component (top) and $\HA$-emitting gas (bottom). In both cases, the shapes are obtained by considering the mass distributions in a shell at twice the stellar half mass radius. Only bins with at least 15 galaxies are included. Solid and dotted faint blue curves show the observational constraints on the fraction of disks according to \citet{Zhang:2019} and \citet{VanderWel:2014}, both with CANDELS data at $z\gtrsim1$. While the fraction of disky galaxies always increases towards $z=0$ (consistently with observational findings), progressively more (less) elongated galaxies are present with shape measured in terms of their gaseous (stellar) components.} 
\end{figure*}

We quantify the relative balance among the three different morphological classes within the TNG50 galaxy population in Fig.~\ref{fig:shapesevolutions}, again for the stellar-mass and $\HA$-gas shapes (top vs. bottom respectively). In particular, we plot the fraction of galaxies classified as disks (blue), spheroids (orange), and elongated or prolate objects (red), as a function of redshift and in bins of increasing galaxy stellar mass (left to right). We adopt here mass bins similar to several found in the observational literature, in order to attempt a zeroth-order comparison. Specifically, the dashed and solid thick shaded curves represent the results by \citet{Zhang:2019} and \citet{VanderWel:2014} based on the same CANDELS data (but at the lowest redshifts, where the latter use SDSS data).

In terms of stellar mass, disk-like galaxies are more numerous towards lower redshift and higher mass: for example, for stellar masses of $10^{9.5-10}\MSUN$, disky galaxies increase rapidly from about 20-30 per cent of the total population at $z=4$ to $\gtrsim70$ per cent at $z\lesssim 0.5$; on the other hand, disk-like objects dominate the number of star-forming galaxies at all redshifts at the high-mass end ($\MS\gtrsim10^{10}\MSUN$). These findings are qualitatively compatible with those deduced from observations and are thus a confirmation of the TNG model as a whole, even though a more thorough comparison of methodologies and results is needed in the future. At the low-mass end ($\MS \lesssim 10^9\MSUN$) disk-like geometries are exceedingly rare, while elongated galaxies becomes progressively more frequent towards higher redshifts. Additionally we note that the CANDELS studies seem to find low fractions of `round' galaxies towards their low-mass end (not shown), whereas those becomes the dominant shape in TNG at $\MS < 10^{10}\MSUN$. In future analyses we will investigate this possible discrepancy, by focusing on spheroidal and elongated galaxies, by quantifying the consistency between observationally- and theoretically-derived 3D shapes and by checking how galaxy shapes change as a function of stellar ages.

In terms of 3D shapes of the $\HA$-emitting star-forming gas, the trends with redshift of the various morphological types are in general much weaker than for the stellar component. Towards the high-mass end ($\MS \gtrsim 10^{9.5}\MSUN$), the morphological mix changes only slightly within the studied redshift range, with disk-like $\HA$-detected galaxies representing 50-80 per cent of the star-forming population from $z\sim 0$ to $z\sim4-5$. It may be worth pointing out, on the other hand, that at the low mass end, in contrast to the stars, the fraction of elongated galaxies decreases steeply going back to high redshift, indicating a rather distinct morphological evolution across the population. Finally, although the majority of gaseous structures qualify as disks at most times and masses, they evolve towards lower sphericity with time, as can be noticed in the different panels of Fig.~\ref{fig:shapes}. In particular, within the populations of disk-like galaxies, gaseous structures become flatter at lower redshifts for fixed stellar mass.


\begin{figure*}
\centering   
\includegraphics[clip, width=16cm]{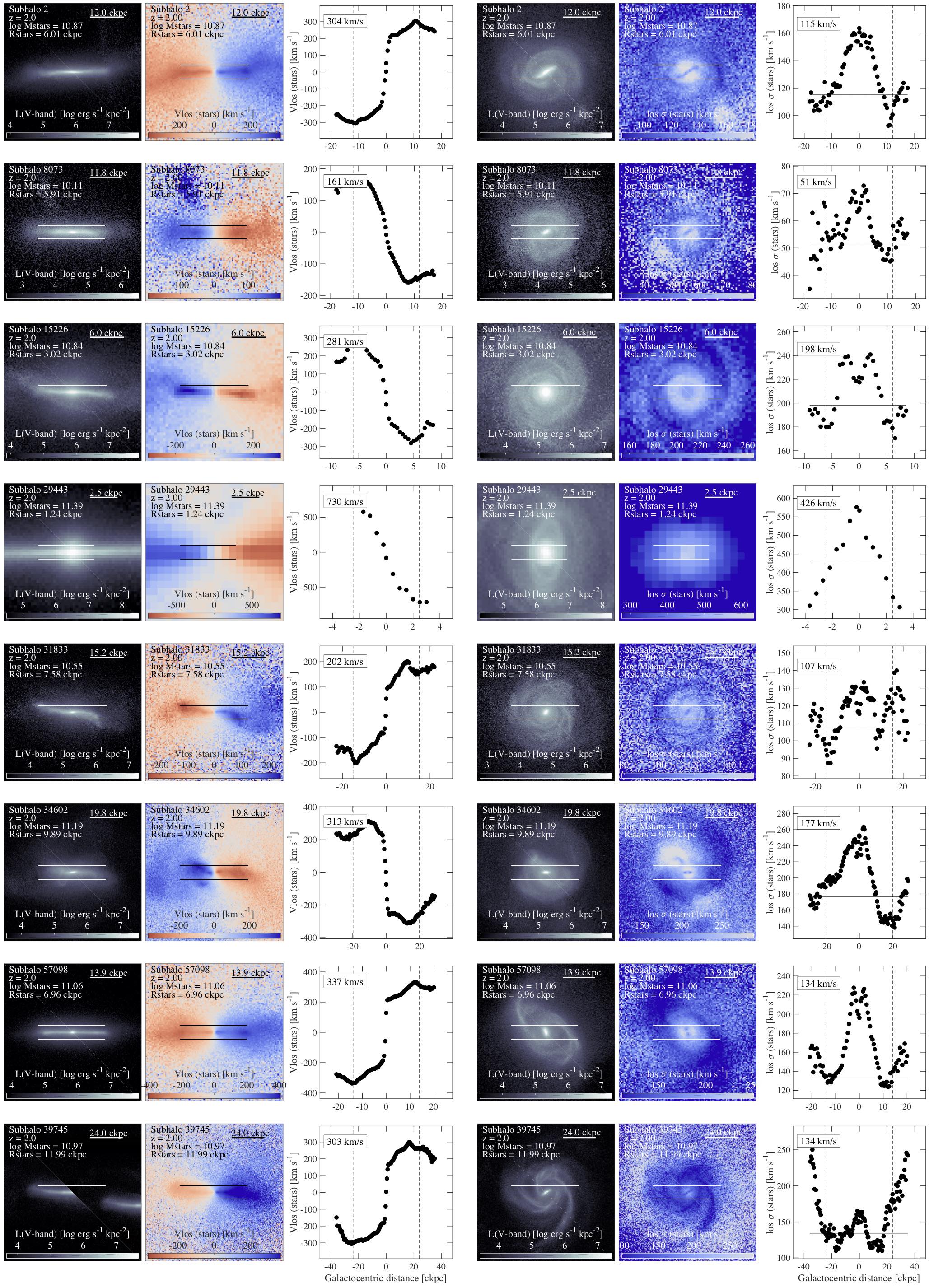}
\caption{\label{fig:kinematics1} V-band light maps, velocity maps and velocity profiles for the stellar component of a random selection of galaxies at $z=2$ from the TNG50 simulation. In the three leftmost columns, we show edge-on projections of the stellar light and mean line-of-sight velocity, in addition to the radial profiles of the mean line-of-sight velocity along the slit depicted by the black solid lines. In the three rightmost columns, we show the face-on projections of the stellar light and of the line-of-sight velocity dispersion, in addition to the average velocity dispersion profiles along the slit. For all kinematics, we pixelize using square pixels of 0.5 comoving kpc to a side. For each galaxy, two numbers are extracted as indicated in the panels: $\VMAX$ and $\sigma$ -- see Table~\ref{tab:velocities} for definitions.} 
\end{figure*}
\begin{figure*}
\centering   
\includegraphics[clip, width=16cm]{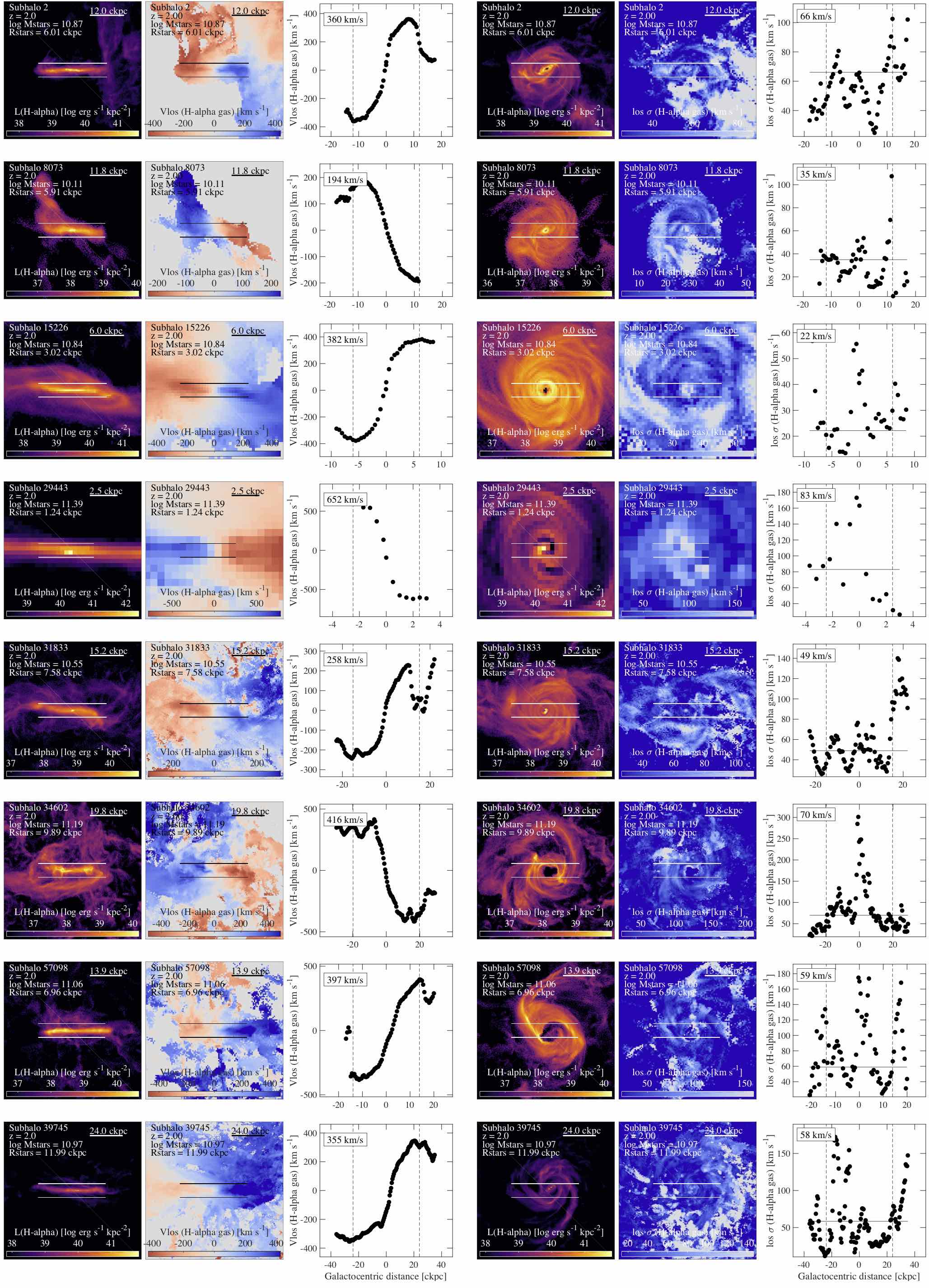}
\caption{\label{fig:kinematics2} $\HA$ light maps, velocity maps and velocity profiles for the star-forming and gaseous component of a random selection of galaxies at $z=2$ from TNG50. Galaxies and annotations are as in Fig.~\ref{fig:kinematics1}.} 
\end{figure*}

\section{Results on kinematical properties}
\label{sec:results_2}

One of the aims of this work is to contrast the information obtained through the structural analysis of galaxies (see previous Section) with properties deduced from their kinematics. In this Section, we explore the degree to which TNG50 star-forming galaxies are rotationally supported or dispersion dominated, performing our analysis on both the stellar and gaseous components, separately and then in comparison to one another.

\subsection{Light and kinematical maps}
\label{sec:kinematicmaps}

Figs.~\ref{fig:kinematics1} and \ref{fig:kinematics2} show maps of the light distribution and velocity fields of a random sample of TNG50 galaxies at $z=2$.\footnote{A more comprehensive set of examples is available on the IllustrisTNG website: \url{www.tng-project.org/explore/gallery/}.} In particular, Fig.~\ref{fig:kinematics1} focuses on the stellar and stellar light components, while Fig.~\ref{fig:kinematics2} provides information about the distribution of the $\HA$-emitting gas in the same galaxies. Each row depicts one galaxy, in edge-on projection (three leftmost panels) and face-on projection (three rightmost panels), where the determination of the projection is given by the mass distribution of the corresponding matter component. Here we do not apply any smoothing or adaptive binning, and the pixelization in the kinematical maps is meant to convey our underlying quantitative analysis procedure. Galaxy rotation velocity is obtained from the edge-on view (to maximize the signal from ordered rotation), while velocity dispersion is obtained from the face-on projection (to minimize any contamination from ordered motion). In both cases, only the line-of-sight component of the velocity field is used, and we directly measure average line-of-sight velocities in projected pixels or radial bins using the velocities of star particles (for stellar mass and V-band based measurements) and star-forming gas cells (for the $\HA$ measurements), as described in Section~\ref{sec:props}.
 
Our measurements are designed to capture the intrinsic dynamics of our simulated galaxies. These are are not a priori easily comparable to observation estimates, where for example the velocities of the gas are obtained via characterization of $\HA$ emission lines. 
However, in order to connect to the way observational measurements are carried out, we summarize the statistics of the velocity fields in similar fashions, e.g. by averaging velocities over radial bins of finite size or over projected pixels, taken here to be square for convenience, although one could also use equal S/N Voronoi tessellations as often done in integral field spectrograph (IFS) observations. We take 0.5 comoving kpc as our fiducial spatial resolution for the velocity maps and measurements, independent of redshift and galaxy properties. We do not weight the particle-averaged velocities by any observable and we do not impose signal-to-noise cuts to the mapped pixels/spaxels that contribute to the summary statistics, as would be the case in observations. The effects of these choices are quantified in Appendix~\ref{sec:app_sigma}. Instead we extract rotation-velocity and velocity-dispersion profiles exclusively from the inner, dense (i.e. most luminous) regions of galaxies. Specifically, we employ a virtual ``slit'' aligned with the structural major axis of the galaxy that extends from $[-2 \RSTARS, + 2 \RSTARS] $ and $[-1/5 \RSTARS, + 1/5 \RSTARS]$ along the major and minor (or middle) axes, respectively, as described in Section~\ref{sec:props}.

Figs.~\ref{fig:kinematics1} and \ref{fig:kinematics2} visualize the complexity and diversity of the galaxy population kinematics in TNG50. At $z=2$, the vast majority of star-forming massive galaxies exhibit clear rotation patterns, both in the stellar as well as in the gaseous bodies. Many galaxies present prominent stellar bulges (e.g. galaxies \#15226 and \#29443) and sometimes stellar bars are already present at early epochs (galaxy \#2). Both in the stellar and gaseous components, galaxy disks often exhibit warps at their edges (e.g. galaxy \#31833) that seem to correspond to cases where the rotation curves fall at distances slightly larger than the local maximum. 

One striking notion from the comparison between the stellar-based and $\HA$-based maps at these intermediate redshifts is the enhanced complexity of the latter: even through visual inspection of this small sample, it is evident that the line-of-sight velocity-dispersion field traced by the star-forming gas is less coherent, less organized and overall less spatially homogeneous than for the stars. A visual inspection of similar maps at lower redshifts (e.g. $z<1$) seems to point towards a decrease of such spatial inhomogeneity of the $\HA$ maps. 

\begin{table}
  \caption{\label{tab:velocities}Summary of the operational definitions of the rotation velocity ($\VMAX$) and velocity dispersion ($\sigma$) of a galaxy. These are adopted throughout, unless otherwise stated, and summarize the procedure described in Section~\ref{sec:props} and discussed in Section~\ref{sec:kinematicmaps}.}
\begin{tabular}{|p{1.1cm}|p{6.5cm} }
\hline
 Symbol & Notes \\
\hline
$\VMAX$ & Absolute maximum of a galaxy's rotation curve along its structural major axis, from its edge-on projection and within twice the stellar-half mass radius. The rotation curve is obtained by taking the unweighted mean along the edge-on line of sight of the resolution elements in bins of 0.5 comoving kpc. This estimate maximizes the contribution from ordered, rotational motions, although it could include contributions from any disordered motions along the plane of rotation.\\
&\\
$\sigma$ & Mean value of the velocity dispersion from a face-on projection of a galaxy, averaged in pixels of 0.5 comoving kpc a side that lie along the structural major axis and between one and two times the stellar-half mass radius. Represents an intrinsic measure, free from instrumental biases, of disordered motion along the galaxy minor axis. This estimate minimizes the ``contamination'' from ordered rotation, \rvvv{neglects the effects of thermal motions unless otherwise explicitly noted, }and captures all non-rotational (vertical and possibly extraplanar) motions, including those caused by self gravity, feedback (indirectly or directly), galaxy mergers and encounters, gas accretion, and so on. \\
\hline
\end{tabular}
\end{table}

In the profile panels of Figs.~\ref{fig:kinematics1} and \ref{fig:kinematics2} text annotations show the values we derive to characterize the velocity fields of each galaxy: $\VMAX$ and $\sigma$. The meaning and derivations of these two numbers are detailed in Table~\ref{tab:velocities}. 
Note that we do not fit the rotation curves and we do not take the asymptotic velocity value of the rotation curves as $\VMAX$. First, as noted above, many of the rotation curves exhibit signs of {\it falling} at some large distances, requiring fitting functions similar to those adopted e.g. by \citet{Wisnioski:2015} rather than arctangent-like functional shapes. Second, we have verified that our $\VMAX$ values typically occur well within the imposed radial distance maxima, and not where the curves still rise. As a result, our values for $\VMAX$ correctly represent where and to what degree kinematic disks are ``maximal''. 

Furthermore, we neglect any possible misalignment between the structural and kinetic major axes. This is an important issue in observations where the structural major axis is determined from the stellar light maps, while kinematics are extracted from the gaseous component. However, in our case the structural major axis and kinematical measures are both obtained from the spatial distribution of the same matter component/tracer\footnote{Cases of misalignment may not be negligible in certain regimes. For example, \citet{Krajnovic:2018} find an increased frequency of prolate galaxies which have their kinematic and structural position angles misaligned by 90 degrees at the high stellar mass end. However, these results refer to galaxies residing in the highest density environments in the local Universe and with stellar masses and sizes in excess of $10^{12}\MSUN$ and 10 kpc, respectively. This regime is not probed in this study.}. From a visual inspection of our TNG50 galaxies, structural and kinematics misalignments that are larger than 10-20 degrees are relatively rare, at least at $z\gtrsim0.5$: we have found just a few albeit interesting cases. For the $\HA$-emitting gas, these are clearly associated to complex large scale gas mass patterns or galaxy mergers, especially at $z\gtrsim1$; for the stellar components, misalignments are often connected to spheroidal-like stellar light distributions. As our quantitative results are based on thousands of galaxies, we can safely neglect these cases. Similarly, we neglect any possible issue associated to asymmetries in the velocity fields and profiles with respect to the minor axis. 

Finally, to determine velocity dispersion we prefer not to fit the profiles with a theoretical model because, especially for the $\HA$ gas, the actual profiles are often more complex than the base expectation of central peak with symmetric declining tails towards the outskirts. As described above, we have not pruned in any way our galaxy sample beyond residency on the SFMS. As a result, mergers, disturbed galaxies, galaxies with stellar spheroidal shapes, and both central and satellite objects are all analyzed in the same manner. Importantly, our velocity dispersions are by construction {\it not} the central values, and in fact exclude contributions from the central high-dispersion peak. Additionally, as a reminder, the velocity dispersions are measured in pixels and then the average of such pixels is taken, analogously to what is done in observations. Our principal goal is to assess to what extent galaxies are dominated by ordered or chaotic motions throughout their extended bodies, and what the contribution of turbulent motions is where kinematic disks are maximal (or where the rotation curves are flat). We have verified that in $\sim\,$95 per cent of the studied galaxies, the maximum of the rotation curve is reached between $1-2 R_{\rm stars}$, which is where we take average measures of the velocity dispersion.

\begin{figure*}
\centering         
\includegraphics[width=8.5cm]{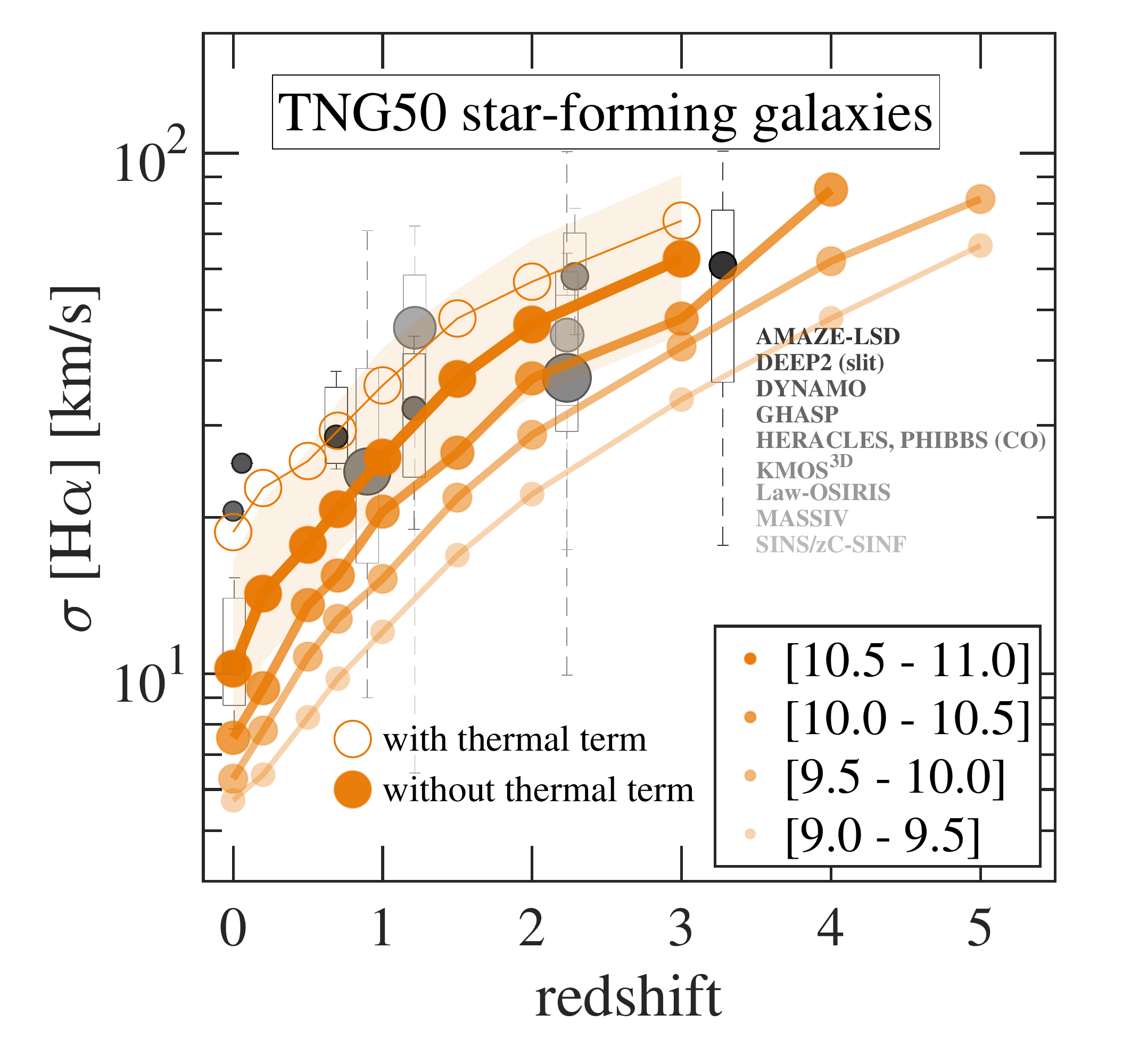}
\includegraphics[width=8.5cm]{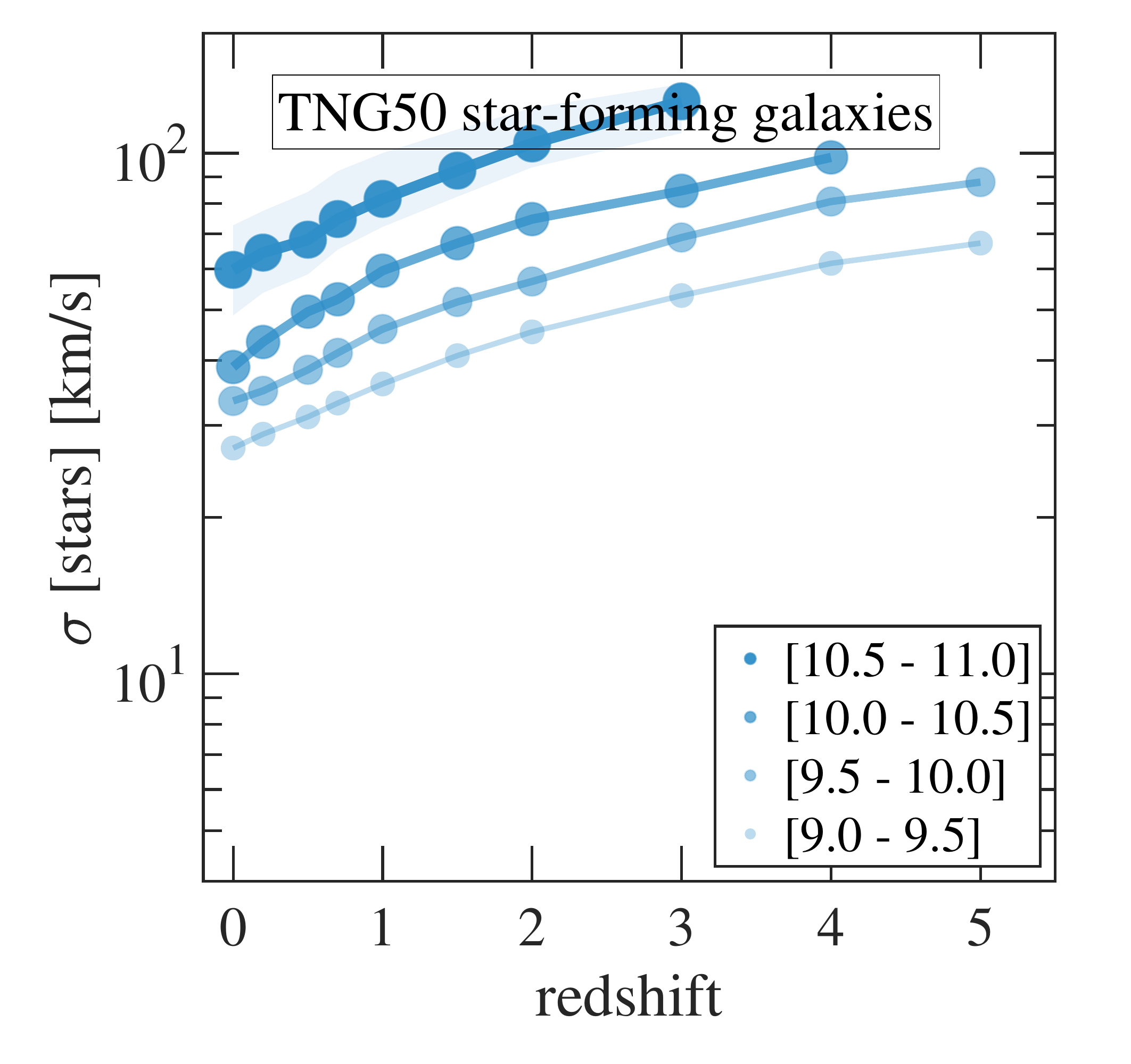}
\includegraphics[width=8.5cm]{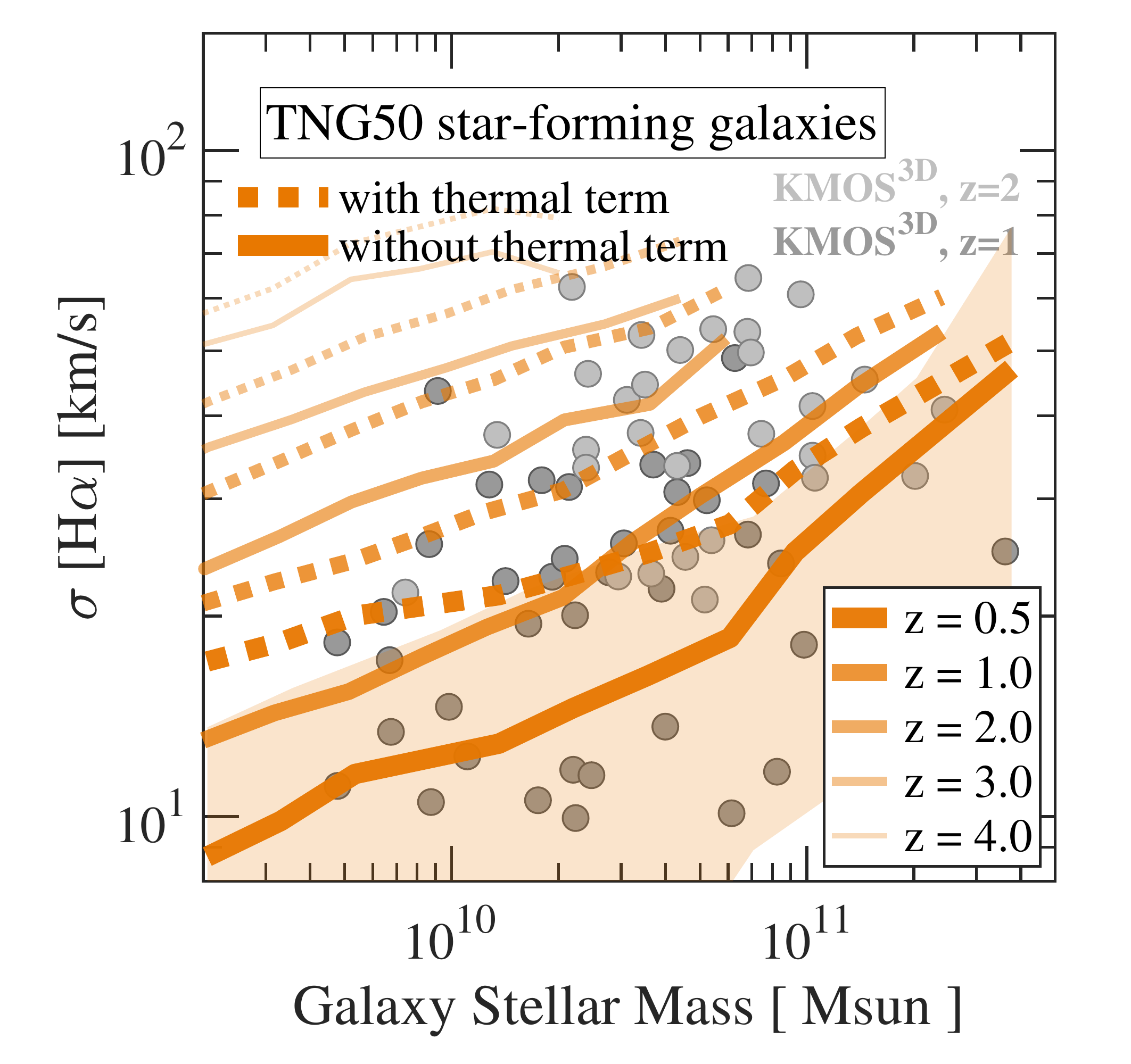}
\includegraphics[width=8.5cm]{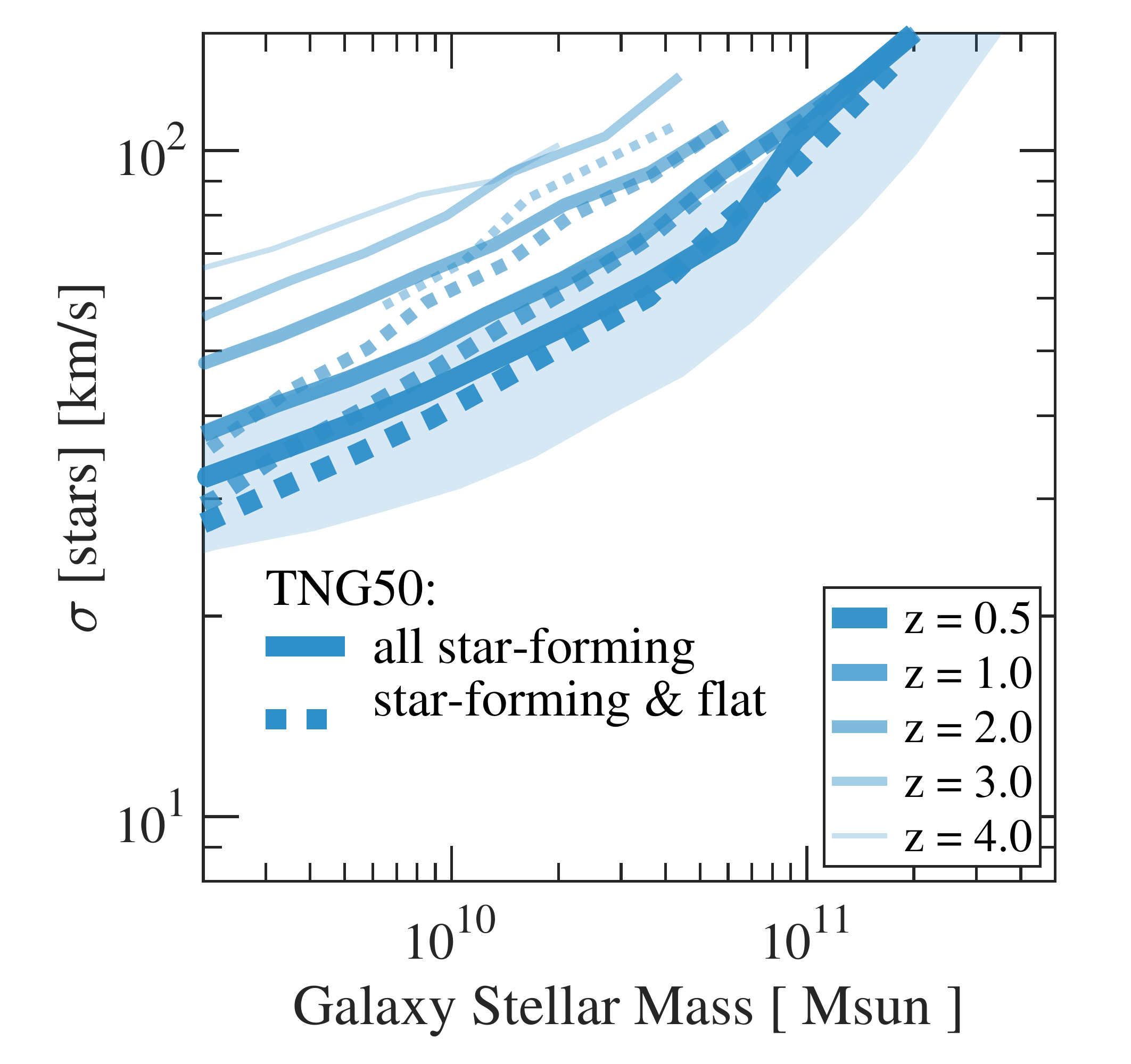}
\caption{\label{fig:sigma} Trends of the median velocity dispersions of TNG50 star-forming galaxies with redshift (top) and galaxy stellar mass (bottom), for the $\HA$-emitting gas (left, orange thick curves and markers) and the stellar component (right, blue thick curves and markers). $\sigma$ is measured where the rotation curve is flat or reaches its maximum, and it is obtained from the line-of-sight velocity field in face-on projections (see Table~\ref{tab:velocities}). \rvvv{In the top (bottom) panel, filled circles (solid curves) denote ``intrinsic" dispersions, i.e. measurements that do not account for the effects of thermal motions: they summarize } the velocity-dispersion maps of Figs.~\ref{fig:kinematics1} and \ref{fig:kinematics2} \rvvv{and represent the resolved outcome of the underlying galaxy formation model}. \rvvv{Empty circles and dashed curves in the top and bottom panels, respectively, include estimates for a gas thermal component (in the top, only for the most massive bin): see Appendix~\ref{sec:app_sigma} for details.} Gray symbols and annotations denote observational results for $10^{10.5}\MSUN$ galaxies from a compilation by \citet{Wisnioski:2015}, for reference (see text for details). In the bottom right panel, two sets of curves denote all star-forming galaxies (solid) vs. star-forming galaxies with ``flat'' stellar light distributions (dashed). At fixed galaxy stellar mass, galaxies are dominated by larger vertical incoherent motions at earlier epochs.}
\end{figure*}

\begin{figure*}
\centering                                           
\includegraphics[width=8.5cm]{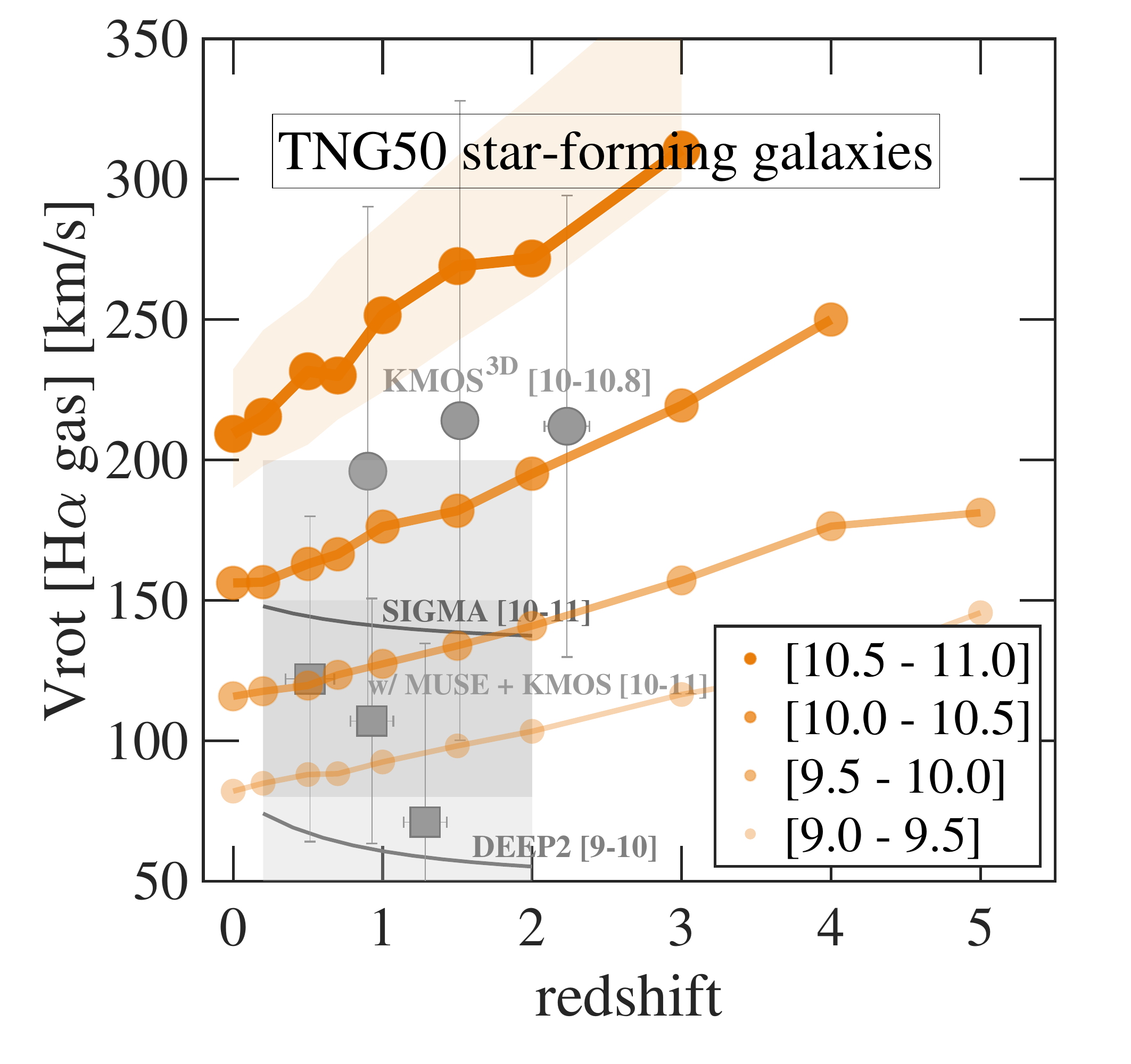}
\includegraphics[width=8.5cm]{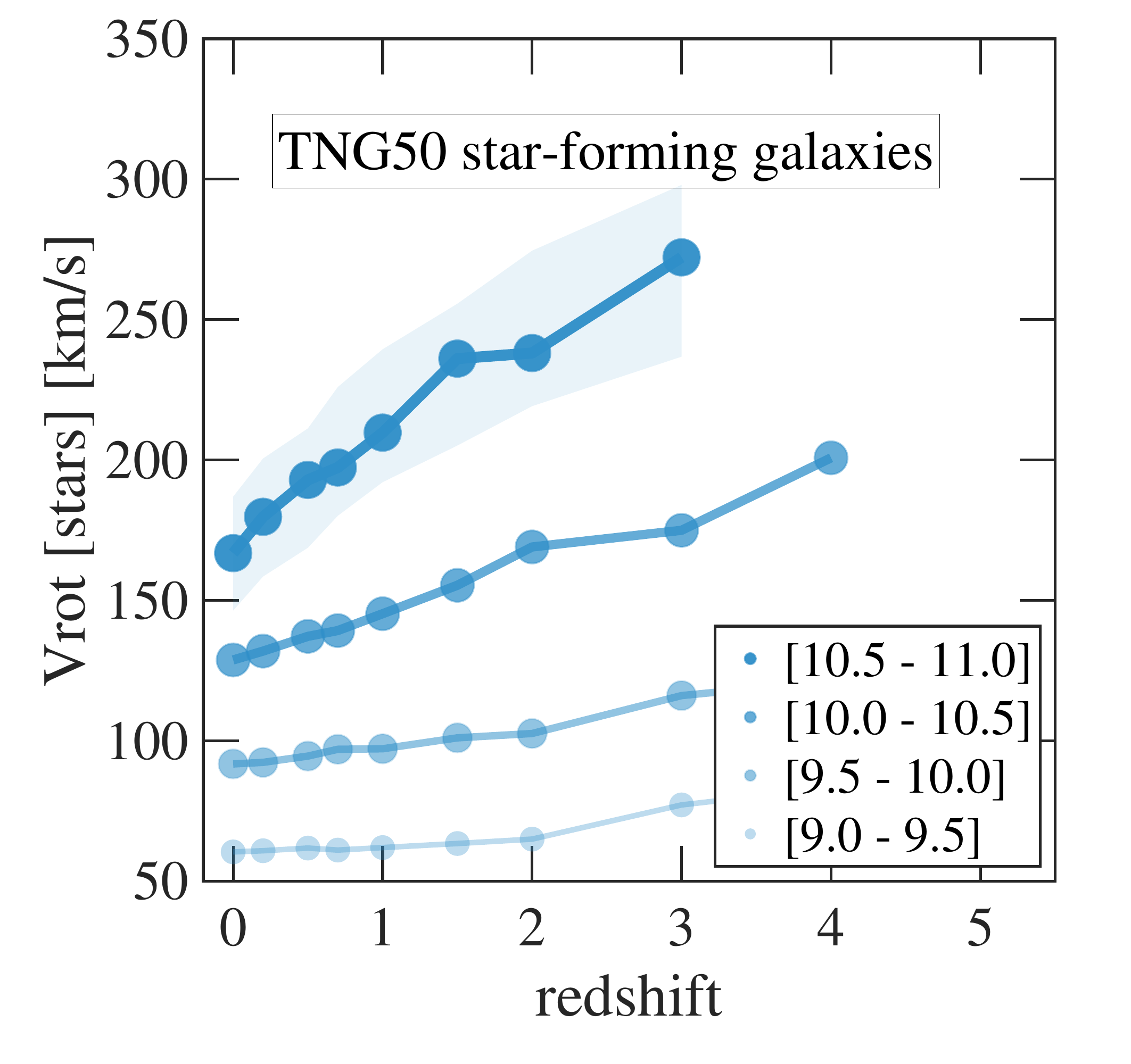}
\includegraphics[width=8.5cm]{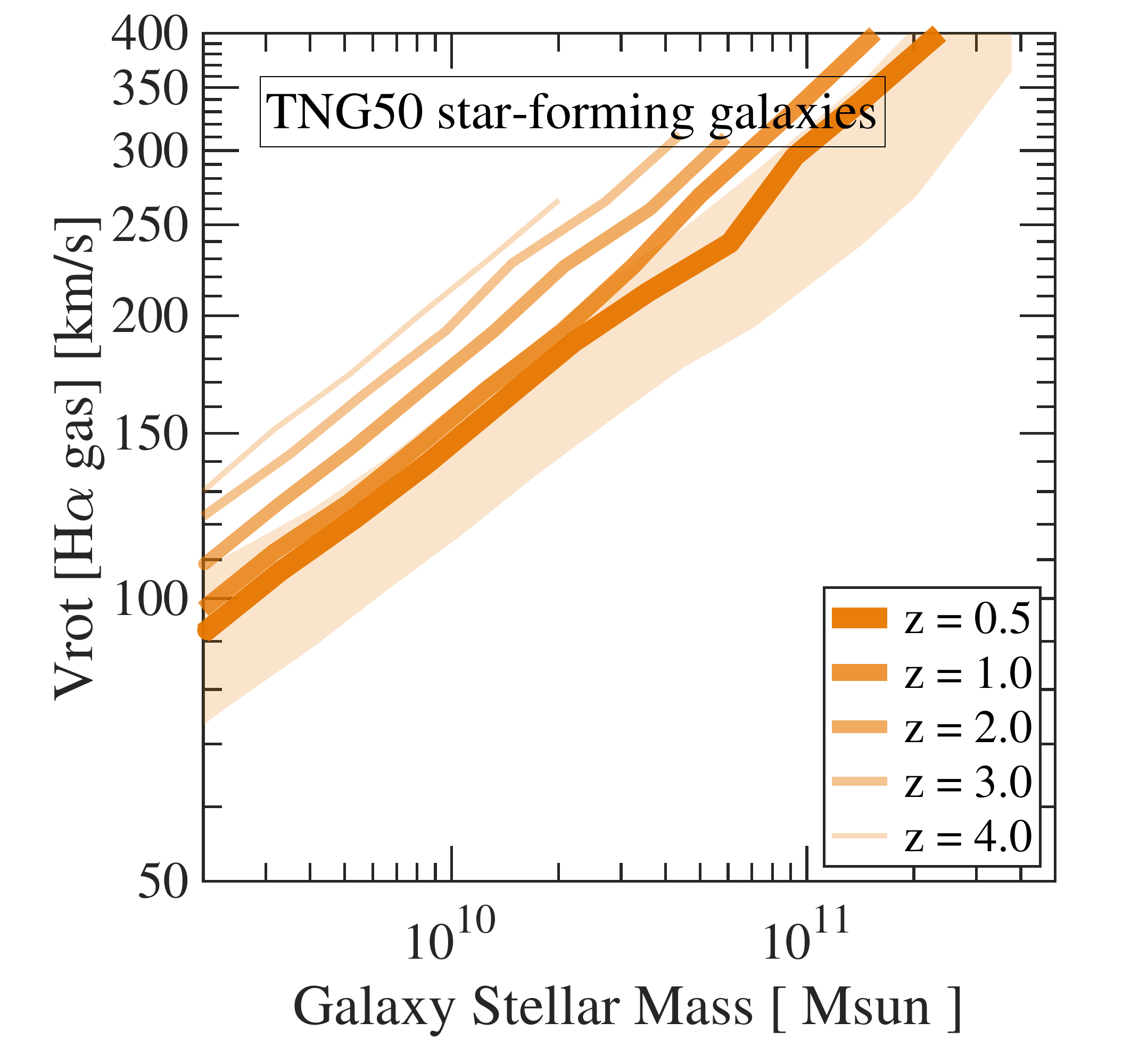}
\includegraphics[width=8.5cm]{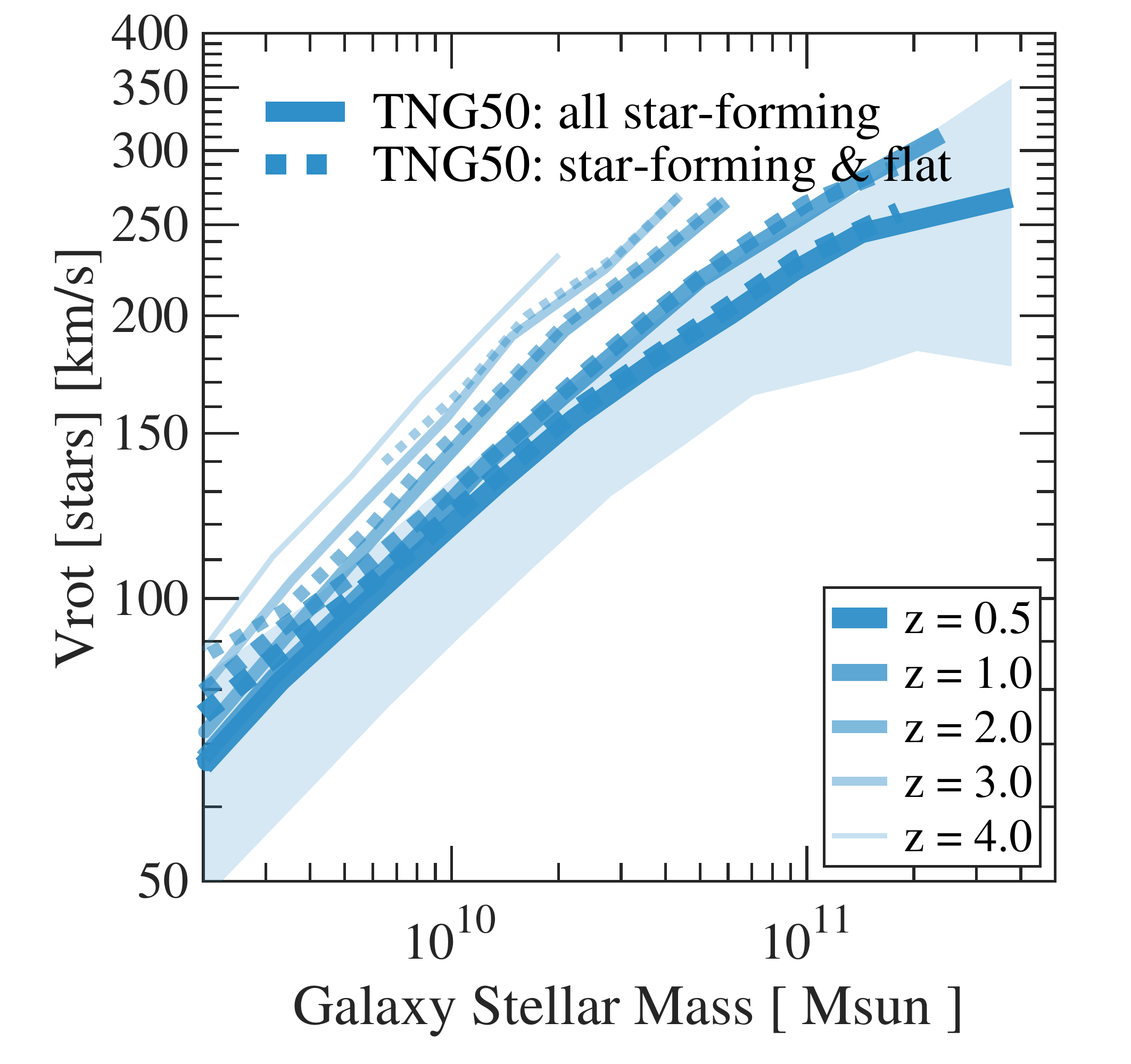}
\caption{\label{fig:vrot} Trends of the median rotational velocities of TNG50 star-forming galaxies with redshift (top) and galaxy stellar mass (bottom), for the $\HA$-emitting gas (left, orange thick curves and markers) and the stellar component (right, blue thick curves and markers). $\VMAX$ is the absolute maximum of the rotation curve in edge-on projection (see Table~\ref{tab:velocities}) and summarizes the line-of-sight-velocity maps of Figs.~\ref{fig:kinematics1} and \ref{fig:kinematics2}. In the bottom right panel, two sets of curves denote all star-forming galaxies (solid) vs. star-forming galaxies with ``flat'' stellar light distributions (dashed). In the top left panel, gray symbols and annotations denote observational results, from the analyses of \citet[][KMOS$^{\rm 3D}$]{Wisnioski:2015}, \citet[][SIGMA and DEEP2]{Simons:2017}, \rvvv{and  \citet[][with MUSE and KMOS]{Swinbank:2017}}, stellar mass ranges as indicated.}
\end{figure*}

\subsection{Rotation velocities and velocity dispersions}
\label{sec:kinematics}

The quantitative results of TNG50 kinematics are given in Figs.~\ref{fig:sigma} and \ref{fig:vrot}, for velocity dispersion and rotation velocity, respectively. In both cases, we quantify the median trends as a function of redshift, in bins of galaxy stellar mass (top panels), and as a function of galaxy stellar mass at different redshifts (bottom panels). We only focus on $10^9\MSUN$ galaxies and above, for two reasons. As typical kinematic measures are averaged over spatial scales of 0.5-1 kpc, we limit ourselves to study galaxies whose physical extent is comparatively similar or larger than the IFS-like pixel sizes adopted throughout. Secondly, we are certain that above such mass scale kinematic properties are converged (i.e. to better than a percent level) at the numerical resolution of TNG50 and throughout the studied redshift range (see Appendix~\ref{sec:app_res_kinematics}).

In both Figures, we provide $\HA$-based (left; orange curves: \rvvv{with and without thermal term}) and stellar V-band light (right; blue curves) based measurements. Several key results emerge. 1) Both velocity dispersion and rotation velocity decrease with time in galaxy populations of the same stellar mass, i.e. they are larger at larger redshifts. However, the redshift trends of the rotational velocities are considerably weaker than that of the velocity dispersion \rvvv{(Fig.~\ref{fig:sigma} vs. Fig.~\ref{fig:vrot}). The redshift trends are also weaker for the gas velocity dispersions that account for the effects of the gas thermal motions in comparison to those of the $\sigma$ intrinsic values (empty vs. filled symbols in the top left panel)}. 2) Rotational velocity is a stronger function of a galaxy stellar mass than dispersion. 3) At any given redshift and galaxy mass, and across stellar morphologies, the typical stellar velocity dispersions are higher than for $\HA$, by up to a factor of 3. \rvvv{In Figs.~\ref{fig:sigma}, bottom left panel, solid (dashed) curves represent results without (with) including the effects of thermal motions: these are relatively more important at low galaxy masses, with shifts by up to a factor of 3 (see also Appendix~\ref{sec:app_sigma}). } In terms of rotation velocity, stellar bodies fall behind the speed of the $\HA$-emitting dense gas. This incarnation of asymmetric drift exists also if we select only galaxies with 3D stellar morphologies consistent with disk-like shapes. In Figs.~\ref{fig:sigma} and \ref{fig:vrot}, on the right bottom panels, dashed curves represent the average trends of galaxies with disk-like, i.e. flat stellar morphologies ($h_{1/2} \leq 0.1 \times r_{1/2}$): for flat galaxies, median stellar dispersions are, as to be expected, smaller than those of non disky-shaped objects, but the difference in the median trends is more pronounced only towards higher redshifts and low masses i.e. when the galaxy populations are less dominated by disk-like stellar objects. Differences in the median rotation velocities are, in fact, negligible. To quote some numbers, across galaxy morphologies, for the average $10^{10}\MSUN$ galaxy at $z=1-2$ the gas dispersion is of order 20-40 $\KMPERS$ while its rotational velocity is closer to $\sim\,$150 $\KMPERS$. 

In order to provide an observational context, in the upper and lower left panels of Fig.~\ref{fig:sigma} we showcase the results of the kinematical analysis of galaxies across redshift, following the compilation by \citet{Wisnioski:2015}. Velocity dispersion data is given from a number of diverse observational programs, all indicated in shades of gray. Observational input includes $z>1$ IFS \citep[MASSIV, SINS/zC-SINF, OSIRIS, AMAZE, KMOS$^{\rm 3D}$:][]{Epinat:2012, Vergani:2012, ForsterSchreiber:2009, Law:2009, Gnerucci:2011, Wisnioski:2015} and long-slit \citep[DEEP2:][]{Kassin:2012} measurements from emission-lines, velocity dispersions derived from molecular gas at $z=1-2$ \citep[PHIBBS:][]{Tacconi:2013}, and low-redshift estimates \citep[GHASP, HERACLES, DYNAMO:][]{Epinat:2010, Leroy:2009, Green:2014}. All these refer to galaxies with stellar masses of roughly $\sim\,10^{10.5}\MSUN$. Despite the diversity of the underlying methodologies and the large scatter in the velocity dispersion at fixed mass and time, these observations suggest increasing velocity dispersion towards higher redshift at fixed mass, \rvvv{albeit with somewhat different strengths according to which low-redshift sample is favored. Very recently, \citet[][not shown]{Uebler:2019} have extended the KMOS$^{\rm 3D}$ results towards lower redshifts and reported a weaker velocity dispersion evolution of the gas than previously reported, when the same ISM phase is compared across cosmic epochs (mostly ionized and traced via $\HA$).} With respect to the redshift trends \rvvv{and with or without accounting for the thermal motions of the gas (filled vs. empty symbols)}, the TNG model and observations are overall in qualitative agreement. On the other hand, given the different tracers, selection functions and measurement techniques in the observational surveys compared to TNG50 galaxies, and given that small-scale turbulence in the dense ISM will be unresolved, the apparent reasonable match \rvvv{in normalization} between simulated and observed values of the velocity dispersions is encouraging although not conclusive. In Appendix~\ref{sec:app_sigma} we quantify that different measurement choices can affect the inferred values of the velocity dispersion by up to a factor of 2: within this uncertainty from both observational and simulation sides, we can conclude that TNG50 velocity dispersions fall well within the observational ball park.

For reference, with the caveat that no direct comparison is intended or possible, we also include selected observational results in the top left panel of Fig.~\ref{fig:vrot}: gray annotations. The large \rvvv{circular} data points represent the averages and standard deviations of the $\HA$ rotational velocities of KMOS$^{\rm3D}$ galaxies \citep[][Forster-Schreiber: priv. comm.]{Wisnioski:2015, Burkert:2016}, \rvvv{in the $10^{10-10.8}\MSUN$ range}. The shaded regions summarize the data by \citet{Simons:2017} from the SIGMA and DEEP2 surveys, with the thin solid curves denoting their best fit redshift trends \rvvv{(stellar mass ranges as indicated)}. \rvvv{Large square data points show the medians and galaxy-to-galaxy standard deviations of the results by \citet{Swinbank:2017}, obtained with MUSE and KMOS: these are in fact observed velocities at three times the scale disk length and here we include only their data from a relatively narrow mass bin: $10^{10-11}\MSUN$.} Again, these comparisons at face value are difficult to interpret. For example, the mass trend of the rotation velocity in TNG is much stronger than its redshift evolution (bottom left vs. top left panels of Fig.~\ref{fig:vrot}): observed redshift trends may be the result of an evolving mass selection, i.e. of different mass distributions at different redshift, and a comparison should be put forward only once the observed and simulated distributions of galaxy masses are matched as a function of redshift. \rvvv{Different operational definitions of rotational velocities can also complicate the comparison.} Keeping this in mind, here we note that the KMOS$^{\rm3D}$ measurements appear to be somewhat more consistent with TNG50 than those by \citet{Simons:2017} or \citet{Swinbank:2017}. In the Discussion session we expand on the redshift evolution of the rotational velocities.

While low redshift stellar-based kinematics are available in the observational literature, their derivations are too dissimilar from the procedure we adopt here and thus a comparison needs to be postponed. On the other hand, the stellar-based kinematical trends that we uncover with TNG50 at higher-redshifts, and their relation to the underlying gas motions, constitute a prediction of our numerical models. It will be possible to explicitly search for and test these expectations at intermediate and high redshifts using future observations with the James Webb Telescope (JWST) and Extremely Large Telescope (ELT). In fact, the LEGA-C survey \citep{VanDerWel2016} is providing the first systematic exploration of stellar dynamical structure of intermediate-redshift galaxies \citep[$z\lesssim1$,][]{Bezanson:2018} and the first results on star-forming galaxies are forthcoming.

\begin{figure*}
\centering                                      
\includegraphics[width=14cm]{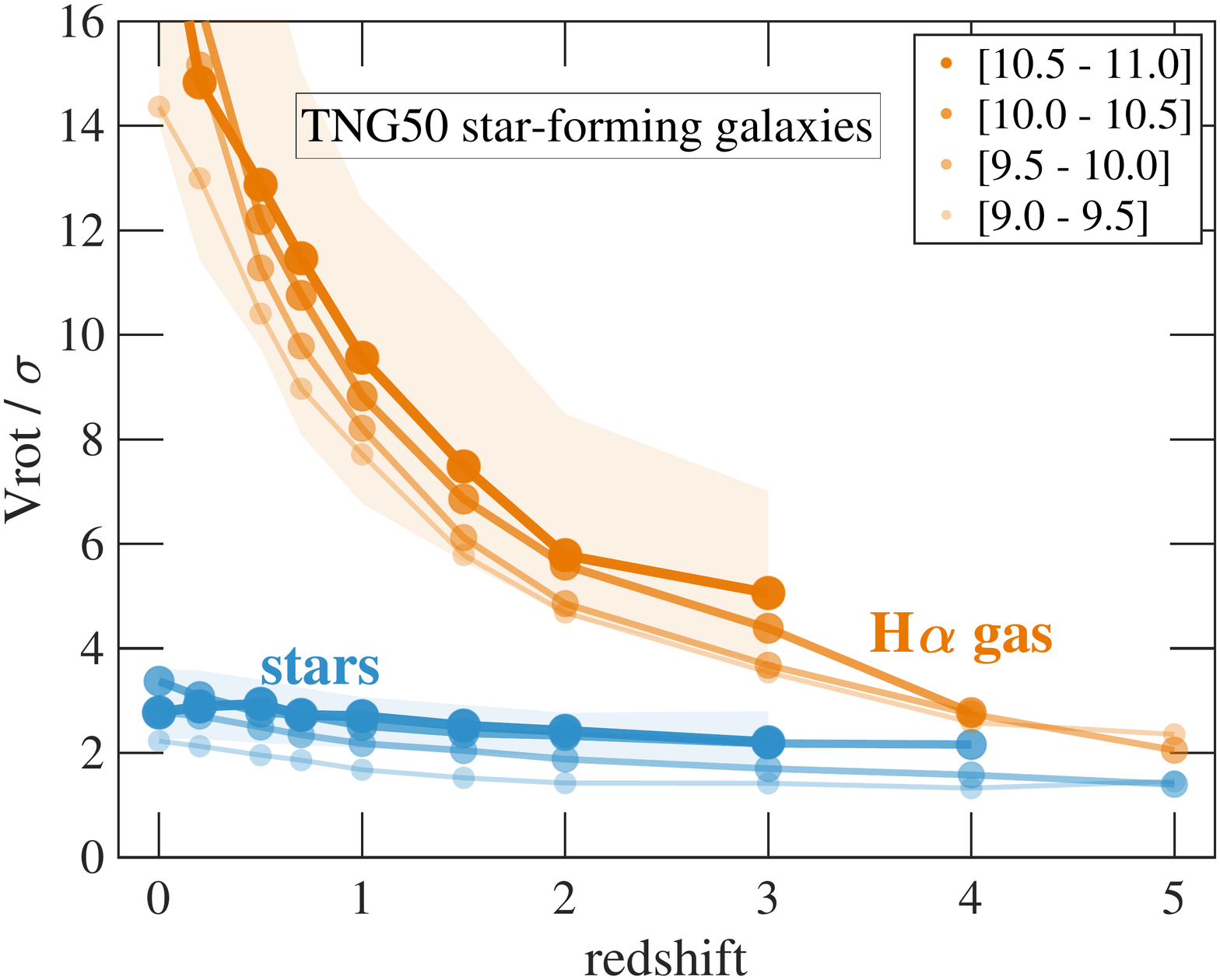}
\includegraphics[width=8.5cm]{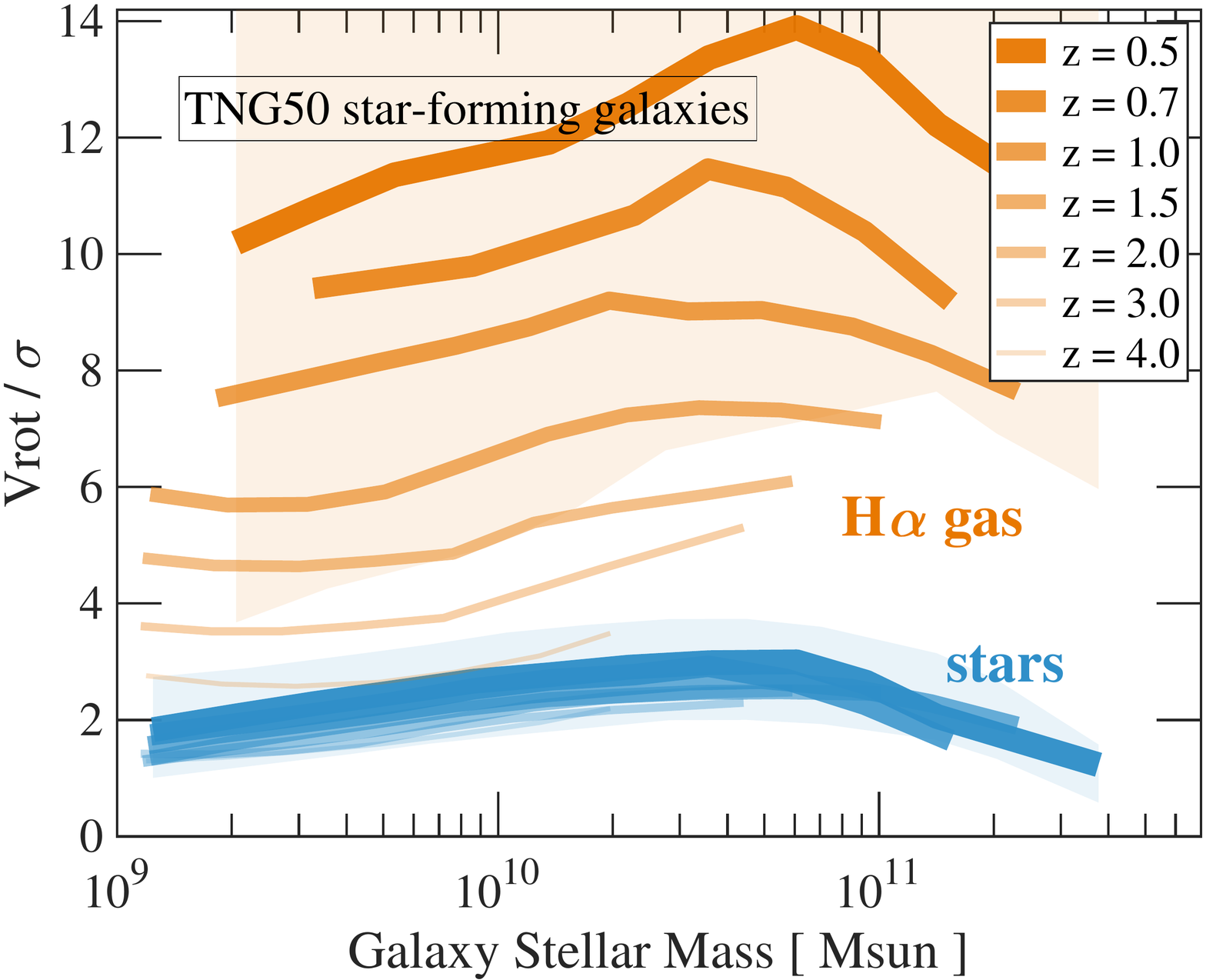}
\includegraphics[width=8.5cm]{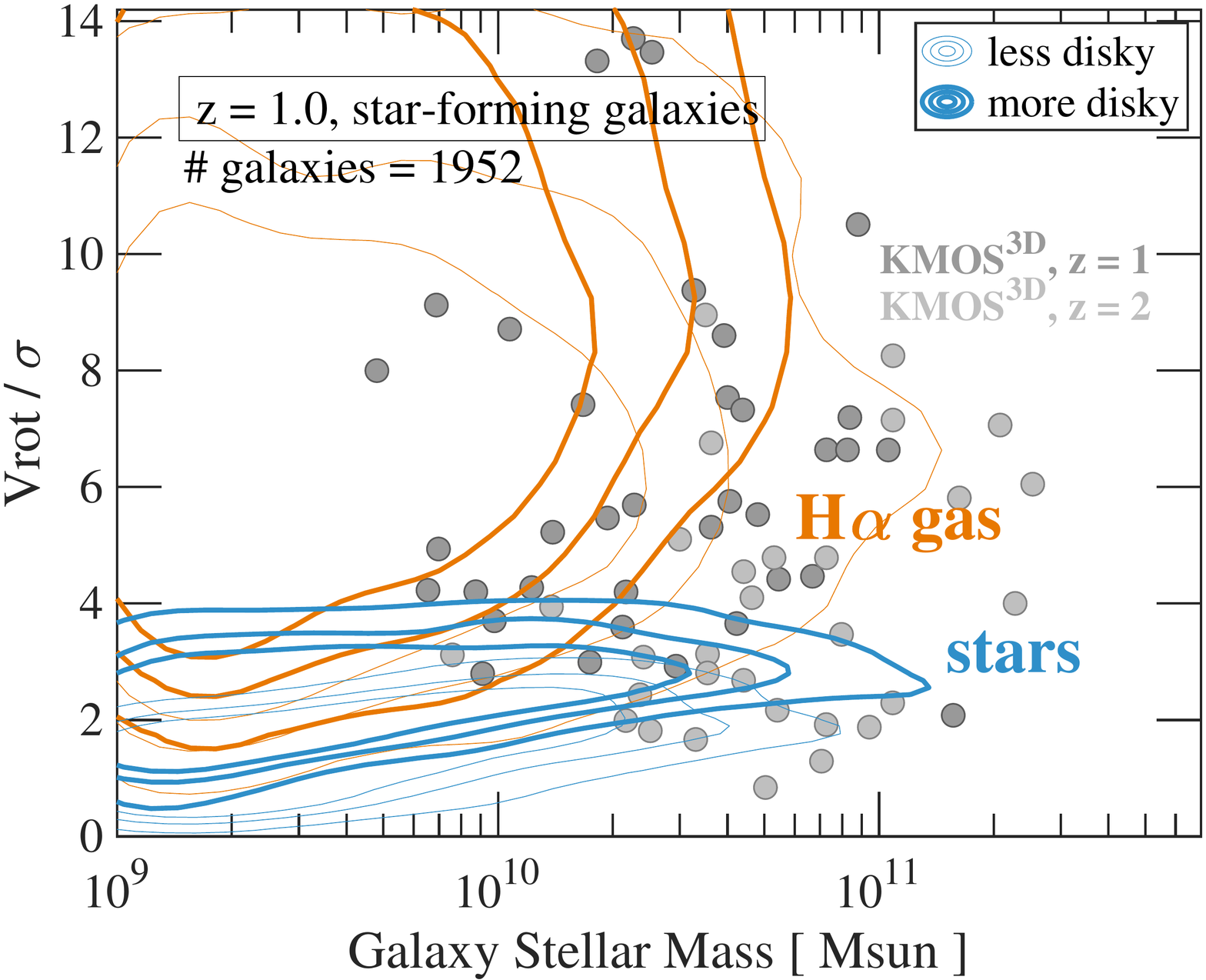}
\caption{\label{fig:Voversigma}  Degree of ordered vs. turbulent motion ($\VMAX/\sigma$) in TNG50 galaxies as a function of redshift in bins of galaxy stellar mass (top), as a function of galaxy stellar mass at various redshifts (bottom left) and at $z=1$ (bottom right). Solid curves and markers denote medians of the TNG50 $\VMAX/\sigma$, for the $\HA$-emitting gas (orange) and the stellar component (blue), separately. In the top and bottom left panels, {\it all} star-forming galaxies are included, independently of their structural, morphological type. In the bottom right panel, we visualize how galaxies with ``diskier'' or ``flatter'' structural morphologies (thicker contours, denoted as ``more disky'') exhibit larger values of $\VMAX/\sigma$ due to their higher degree of rotational support. Overall, the balance between ordered and disordered motions of the gaseous bodies increases substantially as the Universe evolves, and more so than the stellar counterparts.}
\end{figure*}

\subsection{V/$\sigma$ across times}
\label{sec:voversigma}

We quantify the balance between ordered and disordered (turbulent) motion in TNG50 star-forming galaxies in Fig.~\ref{fig:Voversigma}, by plotting the median values of $\VMAX/\sigma$ as a function of redshift and stellar mass. As before, orange curves and markers denote gas-based kinematics, while blue curves indicate stellar-based kinematics. 

In the main top panel, the time evolution of $\VMAX/\sigma$ is given in bins of galaxy stellar mass. For the first time, this plot unambiguously demonstrates that, at all times and masses, the dense gas component of star-forming galaxies is characterized by larger circular motions than the stellar material, with differences as large as a factor of several at low redshift and high mass. In fact, TNG50 star-forming galaxies host strongly rotating gaseous disks, more rotationally-supported at the highest mass-end and progressively so at more recent times: for $\sim 10^{10}\MSUN$ galaxies, the balance between ordered and chaotic motions increases from $\VMAX/\sigma \sim 4-5$ at $z\sim3$ to values exceeding 12 at $z\lesssim 0.5$. In other words, the TNG model is well aligned with recent observational claims whereby $z=1-2$ star-forming luminous galaxies show a larger component of random local gas motion than low-redshift counterparts of the same mass (see e.g. \citet{Wisnioski:2015} and gray markers from KMOS$^{\rm 3D}$ measurements in the bottom right panel). 

On the other hand, the stellar $\VMAX/\sigma$ of star-forming galaxies remains roughly constant across cosmic epochs, with lower levels of rotational support than the gas at $1-2$ stellar radii, and usually larger scale heights (but see Discussion). We reiterate that all main sequence galaxies are considered in the top and bottom left panels of Fig.~\ref{fig:Voversigma}, regardless of their morphology or shape, and that all stars contribute to the observables, regardless of age. In the bottom right panel of Fig.~\ref{fig:Voversigma} we therefore focus on $z=1$ and show how galaxies with ``diskier'' or ``flatter'' structural morphologies -- based on their relative sphericity compared to the average at a given stellar mass -- populate the $\VMAX/\sigma$--$\MS$ plane. We find that flatter, more disky morphologies (thicker contours) imply larger relative rotational motions, both for the stellar and gaseous components. If only disk-like morphologies were included in the main panel of Fig.\ref{fig:Voversigma}, the trends would be qualitatively similar, although the low-redshift enhancement would be more pronounced and $\VMAX/\sigma$ overall increased by up to a factor of $\sim 3$. Finally, the bottom left panel of Fig.~\ref{fig:Voversigma} demonstrates a rather mild galaxy-mass dependence for both stellar and gaseous $\VMAX/\sigma$ measures. The only exception is towards low redshift ($z\leq0.7$) where a peculiar non-monotonic mass trend starts to appear around $10^{10.5} - 10^{11}\MSUN$, particularly in the $\HA$-traced gas. The decline of $\VMAX/\sigma$ towards larger masses and lower redshifts could be related to the rapidly increasing importance of minor-merger activity, which would increase $\sigma$.

As a final remark, we note that the stellar values of $\VMAX/\sigma$ are lower than those of the gas and interpret this as a larger degree of vertical dispersion-supported motion in the extended stellar bodies of galaxies in comparison to the star-forming gas rotating disks. However, the stellar velocity dispersion fields are remarkably more coherent in space than for the gas, as noticed via inspection of Figs.~\ref{fig:kinematics1} and \ref{fig:kinematics2}. In practice, the $\VMAX/\sigma$ ratio, however measured, does not fully capture the spatial incoherence, complexity and chaotic nature of the gas kinematics across galaxy bodies. The richness of these kinematic fields and the interrelationship between gas-phase and stellar motion motivates more sophisticated observational diagnostics, more robust comparisons, and more theoretical explorations to better disentangle the kinematical evolution of galaxies across cosmic time.


\section{Discussion}
\label{sec:discussion}

\subsection{The impact of increased numerical resolution in the TNG galaxy formation model}
\label{sec:betterres}

The numerical resolution of the new TNG50 calculation allows us to reliably study galaxy structures to much lower masses than in previous uniform-volume calculations, to a level previously achieved only with zoom-in campaigns: down to $\MS=10^8\MSUN$ at low redshift, and to even smaller masses at earlier times. This fidelity across a broad range of masses enables us, in particular, to consider the time evolution of galaxy shapes and kinematics.

As we demonstrate in Appendix~\ref{sec:app_res}, the sizes and disk heights of TNG50 galaxies are not set by numerical softening choices for the gravitational forces. Stellar disks can be thinner (as thin as $\sim$100 pc) than the nominal softening lengths in the calculation. The vertical structure of the stellar and gaseous components appears to be the resolved outcome of an ensemble of physical ingredients -- above all, the depth and shape of the overall gravitational potential, which results from the interaction and orbital admixture of both collisional and collisionless matter components. 

With respect to sizes, the 2D and 3D extents of galaxies in TNG100 (the second-highest resolution realization of the TNG model) are well converged at the $\MS\sim10^{9-10}\MSUN$ mass scale. At larger galaxy masses and low redshifts ($z\lesssim$1), the high-resolution results of TNG50 move towards reconciling the previous 0.1-0.2 dex excess in effective radii in comparison to observations \citep{Genel:2018}. This improvement is suggested in Fig.~\ref{fig:sizes}, where gray symbols denote recent observational data.

TNG50 also produces galaxies with stellar and gaseous disks that are 2-3 times thinner than our previous simulations (i.e. TNG100), including at the massive end ($\MS\sim10^{10-11.5}\MSUN$). For example, the tails of the height distributions at $z=0.5$ reveal several massive galaxies ($\MS\sim1-5 \times 10^{10}\MSUN$) with stellar disk heights of only $\sim$100 physical parsec (in V-band light as well as actual stellar mass), even though their stellar disk sizes exceed 1.5-2 physical kpc. The origin, evolution, and nature of these ``razor-thin'' disks is a topic of exploration for future analyses. While a careful quantitative comparison to size observations is beyond the scope of this paper, we note that galaxies with thin stellar disks exist: for example, our own Milky Way's thin disk, with an exponential scale height of $300\pm60$ pc \citep[e.g.][]{Juric:2008}.

For the purposes of this paper, the enhanced resolution and the capability of modeling flatter galaxies produces more robust shape fractions across cosmic time (Figs.~\ref{fig:shapes} and \ref{fig:shapesevolutions}). Our tests with TNG50-2 return a qualitatively consistent picture but demonstrate that lower resolution generally underestimates the number of disky galaxies. Specifically, the growth in time of the disk fractions of low-mass galaxies is underestimated (by up to 20 percent), while it is overestimated at the highest mass end. This is consistent with lower resolution imposing thicker disk heights. The galaxy populations of TNG50-3 ($\sim$TNG300 resolution) are spheroidal-dominated at all but the $\MS\gtrsim10^{10.5}\MSUN$ regime, highlighting the difficulty of morphological analysis at this resolution.

In exploring the internal structural properties of galaxies with TNG50, one of the most important concerns (and caveats) is undoubtedly our relatively simple treatment of the unresolved dense interstellar medium (ISM). As in all Illustris and IllustrisTNG simulations \citep[as well as Auriga;][]{Grand:2017} we employ the effective treatment of the \citet{Springel:2003} subgrid model. This scheme treats the star-forming ISM as an unresolved multi-phase medium, and provides a pressure support which stabilizes gas against runaway cooling, fragmentation, and collapse. The impact on the vertical gaseous scale heights of disks is not immediately clear, and it could be the case that the ISM pressurization establishes in practice a minimum disk height, making them generally too `puffy' \citep[e.g. as demonstrated for EAGLE in][]{Trayford:2017}.

To explicitly test this issue, we have run a series of test simulations at TNG50-2 resolution, in smaller 25 Mpc/h volumes -- the `TNG model variation' series originally introduced in \citet{Pillepich:2018Method}. We have varied a parameter which controls the strength of the subgrid pressure term (the $q_{\rm EOS}$ parameter, as in \citet{Vogelsberger:2013b} Section 2.2, where $q=0$ represents a `weak' isothermal EOS at $10^4$ K, and $q=1$ represents the `maximal' strength EOS of the original model). Between the fiducial value of $q_{\rm EOS}=0.3$ and values of $0.5$ and $1.0$ we observe no systematic change in the stellar mass, V-band light, or $\HA$ disk heights, for any galaxy mass. Variations at the $20-30$ percent level arise, depending on galaxy mass, but have no obvious dependence on the pressurization strength. We conclude that the vertical structure of TNG50 star-forming disks negligibly depends on the underlying parameter choices of the ISM model.

Our multiphase ISM treatment also precludes more sophisticated modeling of some relevant observational signatures -- $\HA$, in particular. In this work we have used a canonical empirical scaling from SFR to $\HA$, neglecting any additional attenuation effects, as well as possible contributions from e.g. diffuse ionized extraplanar gas. Other model assumptions undoubtedly also play a role in some of the structural properties measured herein, including heights. For instance, our treatment of galactic-winds driven by supernovae is necessarily still simplified, and unresolved interactions of SN explosions with cold ISM phases could further modify gaseous disk structure. Despite the caveats, the TNG50 simulation is an important step towards capturing the dynamics of the stellar and gaseous components in disk galaxies, without compromising in statistics. 

\begin{figure*}
\centering 
\includegraphics[width=14cm]{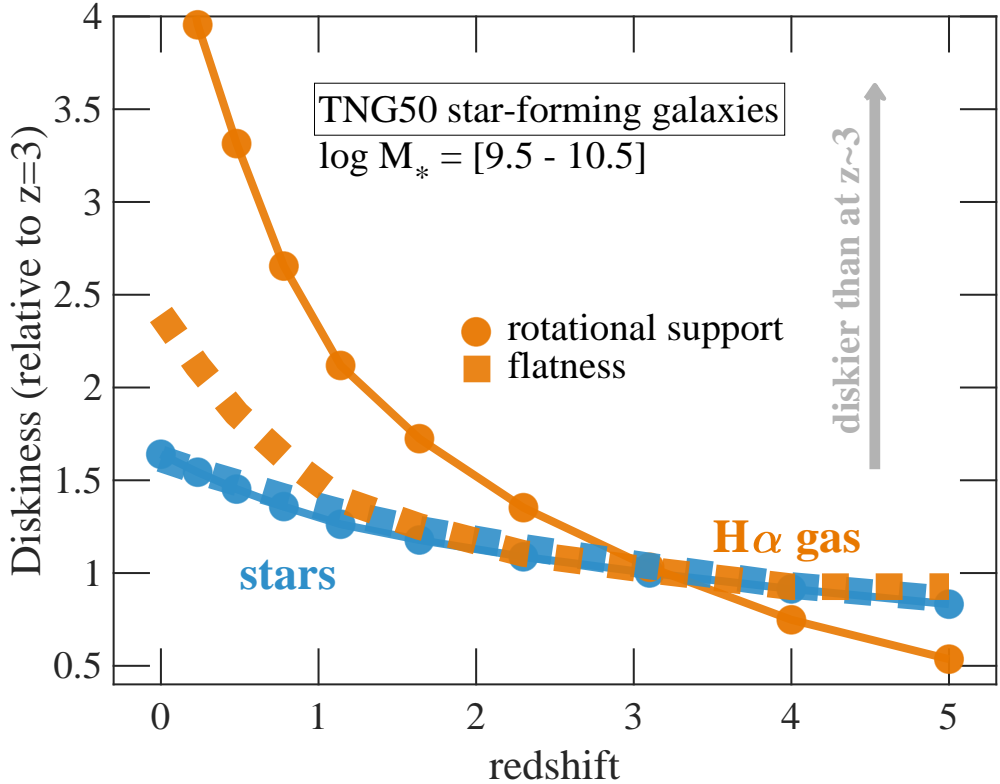}
\caption{\label{fig:diskiness}  Visual summary of the findings of this paper for the bulk of the TNG50 galaxy population. The level of ``diskiness'' of galaxies is given in comparison to their structures (dashed curves) and kinematics (solid dotted curves) quantified at early epochs: here we choose the comparison redshift around the cosmic noon, $z\sim3$. Curves denote median trends of flatness (the inverse of sphericity as per Section~\ref{sec:3Dshapes}) and degree of rotational support (i.e. $\VMAX/\sigma$, as per Section~\ref{sec:voversigma}). Galaxies connected across time at fixed stellar mass are progressively more {\it disky} as the Universe evolves.}
\end{figure*}

\subsection{On the settling of stellar and gaseous disks}
\label{sec:disc_disks}

In the previous Sections we have uncovered that the TNG model naturally predicts an increase in the abundance of disk-like galaxies towards lower redshifts within the star-forming population. We have demonstrated this to be the case in terms of stellar structural properties, based on the flatness or three-dimensional shapes of the stellar light and mass distributions (Figs.~\ref{fig:diskheights_ev} and \ref{fig:shapesevolutions}). We have also shown that this trend holds based on gas kinematics: the balance between ordered and chaotic motions of the $\HA$-emitting gas (Figs.~\ref{fig:Voversigma}). These findings are broadly consistent with observational claims, lending credibility to the overall functioning of the TNG model and to the broad range of its theoretical outcomes.

Interestingly, TNG50 galaxies favor a scenario where the cold star-forming gas in the simulation primarily arranges within configurations that qualify as disks (or elongated structures) for most regimes of mass and redshift (Fig.~\ref{fig:shapesevolutions}). Although the V-band versus $\HA$ (normalized) disk heights (Figs.~\ref{fig:diskheights_distribs} and \ref{fig:diskheights_ev}) appear, at all masses and redshifts, broadly similar, the star-forming gas mass in the simulation forms disk-like structures that are consistently flatter (i.e. thinner) than those of the stars, as is visible on the galaxy population basis (Fig.~\ref{fig:shapes}). 
This structural characterization is aligned with our kinematic finding that the $\HA$ gas is more rotationally supported (larger $\VMAX/\sigma$) than the stars. The gaseous component has a smaller (larger) velocity dispersion (rotation velocity) than the stars, at fixed galaxy stellar mass. 

From the structural view point, the change in time of the abundance of different morphological classes (Fig.~\ref{fig:shapesevolutions}) is quite different for stellar versus gaseous bodies, with the $\HA$-shape demographics being relatively stable across redshift, more so than their stellar counterparts. This contrast is somewhat exaggerated by the artificial classification into three separate classes: while $\HA$ morphologies qualify as disks in almost all cases, their shapes (at fixed galaxy stellar mass) evolve to become flatter (or more elongated) with time (Fig.~\ref{fig:diskheights_ev} and \ref{fig:shapes}), consistent with their increase in $\VMAX/\sigma$. 

In our model, the time evolution towards a larger contribution of rotation vs. dispersion-dominated motions results from a strong decrease of the average velocity dispersion from high to low redshifts (Fig.~\ref{fig:sigma}, top). For the typical $10^{10.5}\MSUN$ galaxy, the inner (vertical) velocity dispersion of the $\HA$ emitting gas decreases by a factor of a few between $z=2-3$ and $z\sim0.5$, a drop broadly consistent with observational constraints. The decline of $\sigma$ is more pronounced for gas than for stars, and is more pronounced than the decline of the rotational velocity with time. In fact, in contrast to  previous observational \citep[][at $z=0.2-1.2$]{Kassin:2012} and theoretical work \citep[for a handful of $z=0$ Milky Way-like galaxies]{Kassin:2014}, the $\HA$-probed $\VMAX/\sigma$ of TNG50 galaxies at fixed stellar mass is higher at more recent epochs even though the rotational velocities also decrease towards low redshift. At fixed stellar mass, the TNG50 rotational velocities increase with increasing redshift, approximately linearly. We have verified that the decline of $\VMAX$ towards recent epochs holds also with rotation curves weighted by $\HA$ luminosity, as would be the case in observations. 
It may be difficult to draw an unambiguous redshift trend from the $\VMAX$ compilation of data in Fig.12 of \citet{Kassin:2012} -- spanning from $z\sim0.2$ to 3.5 and including IFS results -- because of the large scatter and the mix of galaxies of different masses. More recently, \citet{Simons:2017} have found with DEEP2 and SIGMA data a mild increase with time of the average $\VMAX$ in galaxies with $10^{9-10}\MSUN$ stellar mass, but no time evolution at larger galaxy masses (see gray annotations in Fig.~\ref{fig:vrot}). Also the KMOS$^{\rm 3D}$ results seem consistent with no evolution, \rvvv{or rather, a marginal decrease towards low redshift}, while the results by \citet{Swinbank:2017} point towards \rvvv{an increase} of $\VMAX$ with time. \rvvv{Importantly, all these observational estimates span at least a factor of 2 in $\VMAX$ over comparable galaxy mass ranges.} In a dedicated study, we will address how the TNG50 rotational velocities presented in this paper map into circular velocities and what the simulation predicts in term of shape and time evolution of the Tully-Fisher relation.

The overall physical picture laid out in this paper is assembled in Fig.~\ref{fig:diskiness}, for a representative bulk of the TNG50 galaxy population across time ($\MS=10^{9.5-10.5}\MSUN$). There we show how the level of ``diskiness'' increases with time, from both a kinematics and structural perspective. In particular, we visualize the time trends of the average $\VMAX/\sigma$ (solid dotted curves) and flatness (dashed curves: the inverse of the 3D shape sphericity) normalized to their values at high redshift ($z\sim3$). As usual, stellar and gaseous kinematical and structural metrics are kept distinct (blue vs. orange). We omit the alternative flatness measure based on normalized disk heights as it gives indiscernible results from those already depicted.

Instead of defining disks based on a fixed threshold on any of our measures \citep[as previously proposed in the literature, e.g.][]{Kassin:2012,Newman:2013,Simons:2017}, we instead show the average trends, thereby avoiding possible biases because of choices in definition or measurement methodology. In fact, the majority of galaxies at the pivot redshift ($z=3$) can already be classified as gaseous disks, both from a kinematical ($\VMAX/\sigma\gtrsim 3$) and structural view point (axis ratios in the blue bottom right corner of the schemes in Fig.~\ref{fig:shapes}). Yet, their diskiness keeps increasing in time. The stellar components also evolve towards flatter and more rotationally-supported bodies, but more weakly with time than the gas. In addition, the stars continually straddle the boundary of what would reasonably be defined as a disk, or not. The rate of change in stellar flatness (blue square symbols) closely mirrors the rate of change in rotation (blue circles), while in the case of star-forming gas this relationship is less tightly coupled.

Note that the trends of Fig.~\ref{fig:diskiness} do not exhibit any particular dependence on galaxy mass (in the $10^{9-11}\MSUN$ stellar mass range). Yet, more massive galaxies have more disky structural and kinematical properties than lower mass galaxies, at all times, in agreement with the observational findings. On the other hand, the trends in Fig.~\ref{fig:diskiness} would be slightly shallower if we were to select only disk galaxies based on their 3D mass shapes, i.e. excluding elongated and spheroidal shapes. We postpone to future work a focused analysis of prolate configurations in both the gas and stellar components.

From the previous sections and Fig.~\ref{fig:diskiness} it is clear that stars and gas have different structural and kinematical behaviours within galaxy bodies, as expected given their different effective equations of motion. While stars are governed by collisionless dynamics, the gas is collisional and can dissipate energy through radiative cooling, for example. We observe that stars increase in their rotation at a similar rate to their flatness, whereas gas can have not only larger (lower) values of $\VMAX$ ($\sigma$), but also significant increases of rotational motion without this being reflected by similarly strong changes in flatness.

\subsection{Relating structural properties and kinematics}
\begin{figure}
\centering 
\includegraphics[trim={1.5cm 0 0 0}, height=4.2cm]{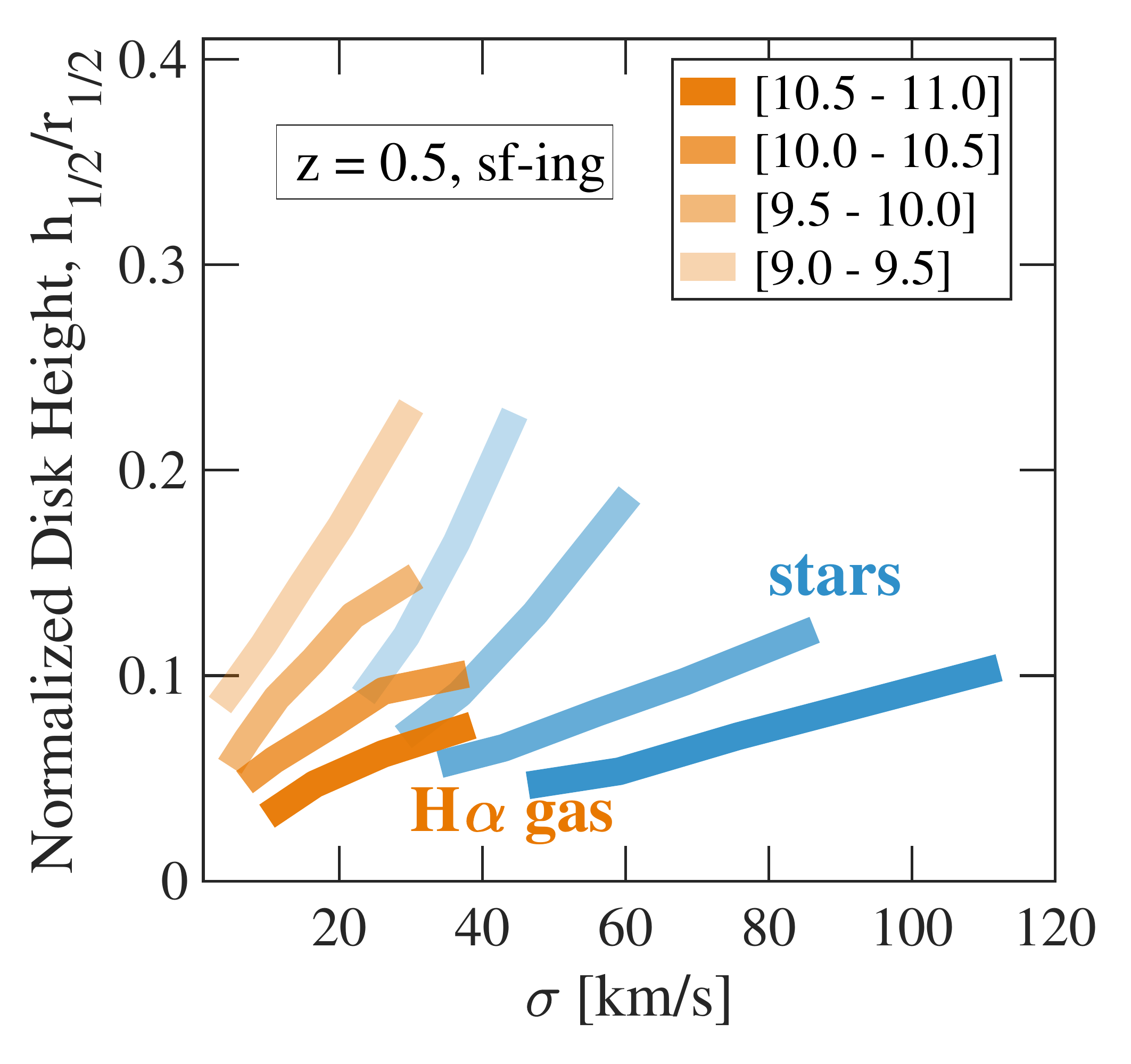}
\includegraphics[trim={1.5cm 0 0 0}, height=4.2cm]{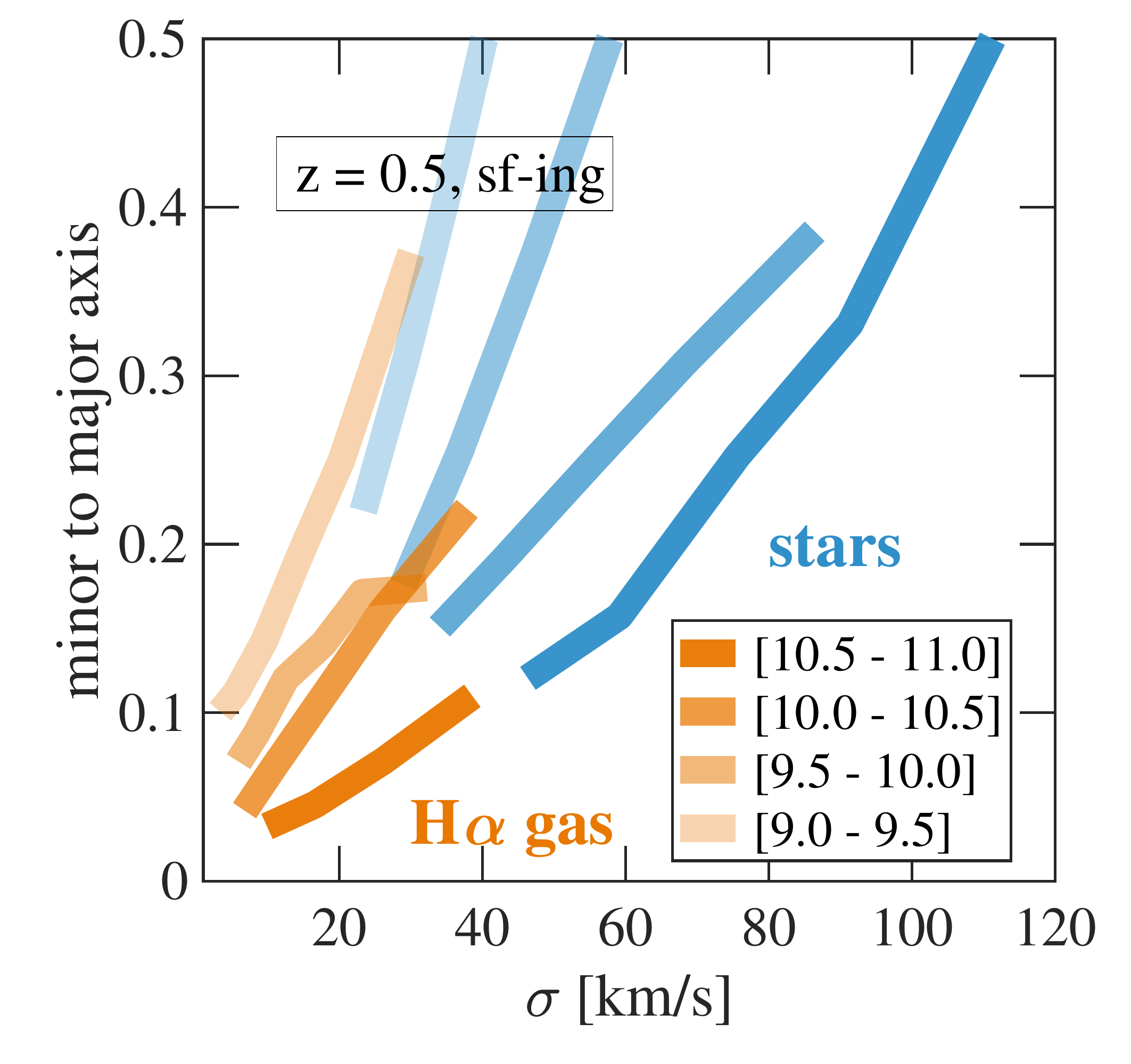}
\includegraphics[trim={1.5cm 0 0 0}, height=4.2cm]{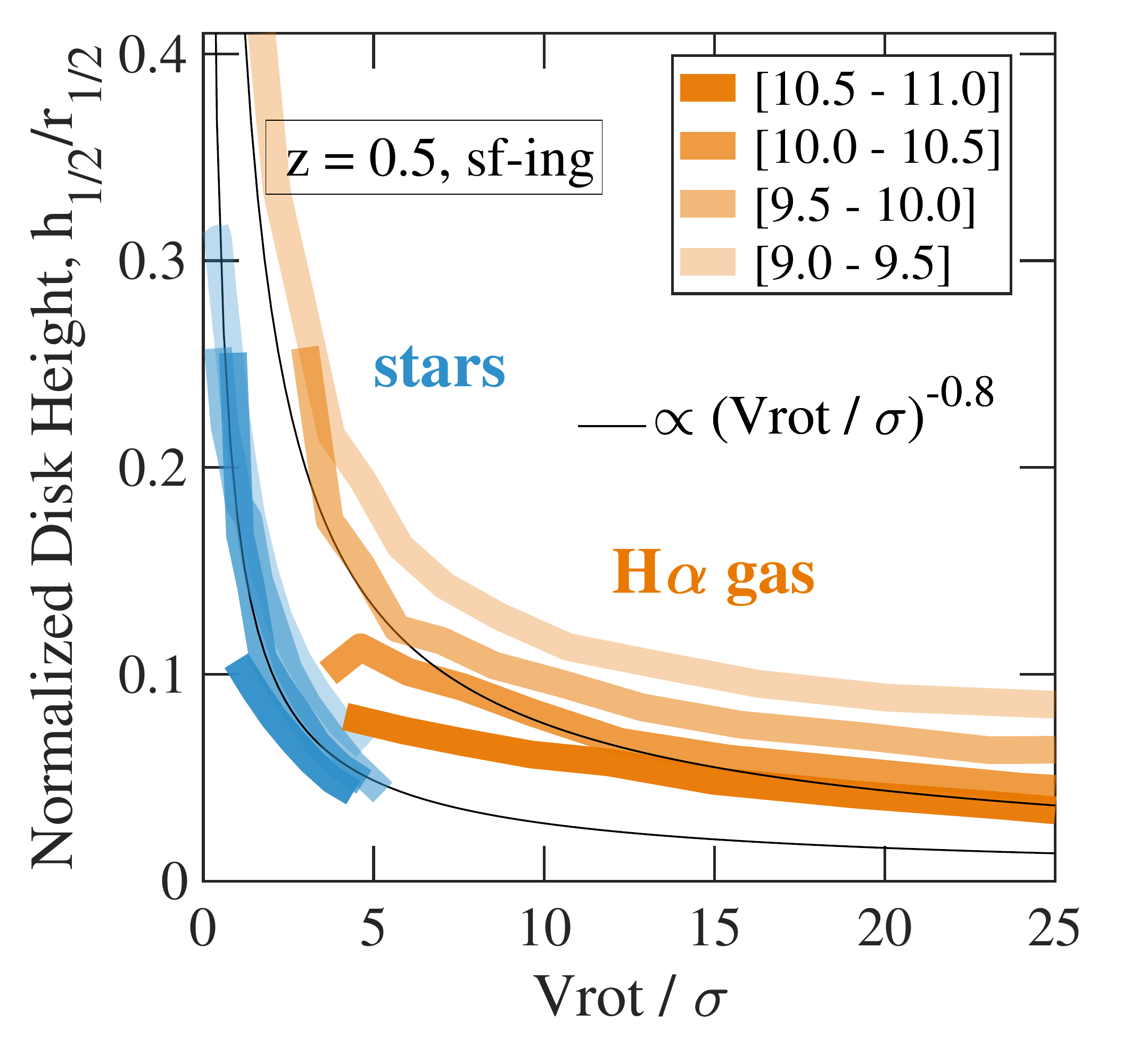}
\includegraphics[trim={1.5cm 0 0 0}, height=4.2cm]{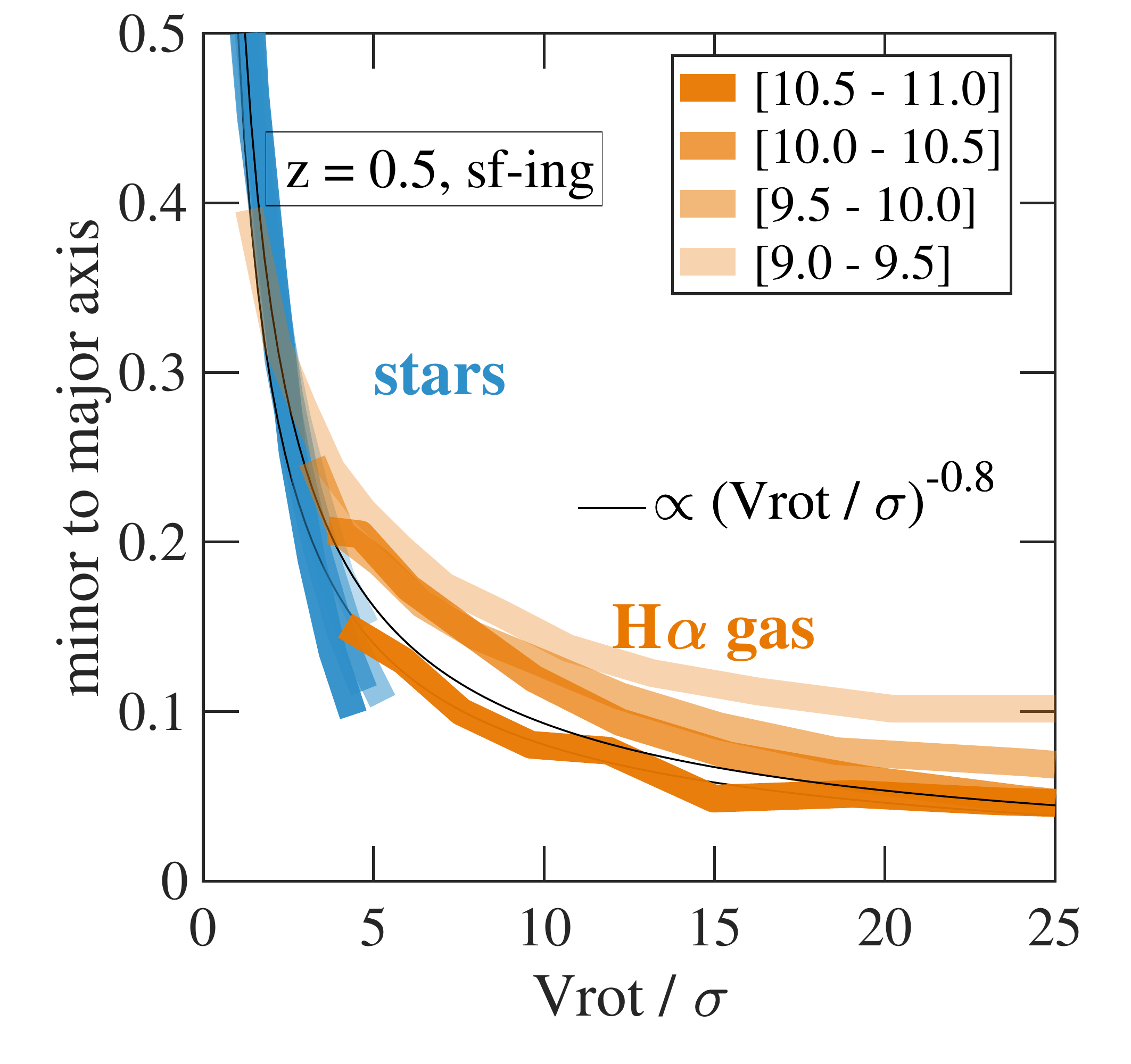}
\caption{\label{fig:strs_vs_kins} Connection between median structural and kinematical properties across the galaxy population of TNG50 at $z=0.5$. The top panels show the correlation of disk height and thickness with velocity dispersion, while the bottom panels show correlations with $\VMAX/\sigma$. Clear trends exist in all cases as a function of galaxy stellar mass, for both gas and stars.}
\end{figure}

In Fig.~\ref{fig:strs_vs_kins} we study in more detail the relation between structural and kinematic properties. We visualize the situation at low redshift ($z=0.5$), where the difference between gas and stars is significant. First, and expected, galaxies with larger (vertical) velocity dispersions are characterized by larger relative disk heights and larger minor to major axis ratios (top panels). In fact, we do not find a correlation between vertical dispersions and absolute physical disk height, for either stars or gas. Instead, we find that this relation depends on the average mass surface density  -- of gas and stars, respectively, or the combination of the two in addition to DM's --, as when normalizing by the 2D galaxy size (as shown). 

Interestingly, at fixed galaxy stellar mass, the linear relationships between vertical velocity dispersion and flatness are shallower for the gas than for the stars, except at the highest mass bin where they are comparable. As gas can cool, it seems plausible for the two relations to be different. However,  a two-fluid analysis is required to understand these trends and their inter-relationships beyond the intuitive expectation for which $\sigma^2 \propto h_{1/2} \Sigma$ (the latter symbol denoting mass surface density). We speculate that, while gas elements move along the tracks of Fig.~\ref{fig:strs_vs_kins}, the different relationships for the stellar component may reflect the effects of non-equilibrium heating and scattering processes. Furthermore, stellar and gaseous trends could plausibly be more similar if we were to select for only young stars.
It is clear, however, that the $\sigma$-$h_{\rm 1/2}/r_{\rm 1/2}$ correlation will also depend on other factors, such as gas surface density, gas fraction, and depth of the overall DM-dominated potential well, as suggested by the trends with mass. Importantly, at fixed galaxy stellar mass, smaller (more compact) galaxies have larger velocity dispersions, and particularly so towards lower redshift (not shown). This is in line with the findings of \citet{Newman:2013} with SINS/zC-SINF data at $z\sim2$ and translates into smaller galaxies being more dispersion-dominated (lower $\VMAX/\sigma$) at all times at fixed galaxy stellar mass. 

In the lower panels of Fig.~\ref{fig:strs_vs_kins} median relative galaxy heights are shown as a function of $\VMAX/\sigma$. Both stars and gas arrange on similar (although not identical) tracks with $h_{1/2}/r_{1/2} = q \propto (\VMAX / \sigma)^{\beta}$ with $\beta\sim -1$. A relation of this type is to be expected for the dynamics of a rotating, symmetric disk in a background potential, the exact shape depending on matter surface density, radial dependence of the rotation velocity, the functional form of the epicyclic frequency and the value of the Toomre Q-parameter. However and interestingly, stars and gas occupy completely different portions of this phase-space diagram. Particularly striking is the very steep inverse relation between galaxy thickness and rotation-to-dispersion budget for the stars and the fact that the stars are excluded from the regions with $\VMAX/\sigma$ larger than a few. This picture is also unchanged at higher redshifts ($z\gtrsim2$), except that the gas does not attain the largest values of $\VMAX/\sigma$.\\

\begin{figure*}
\centering                                      
\includegraphics[width=8cm]{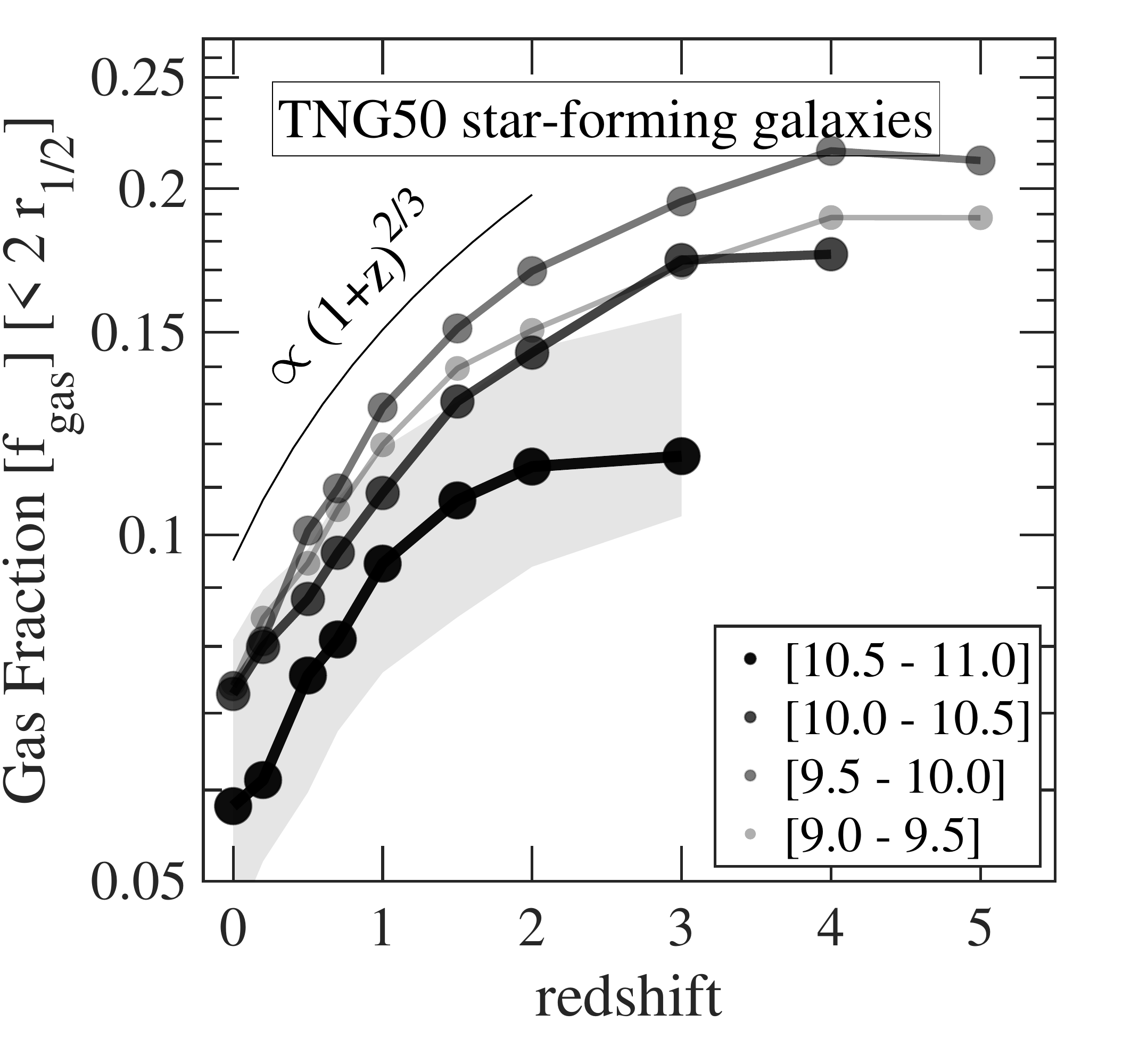}
\includegraphics[width=8cm]{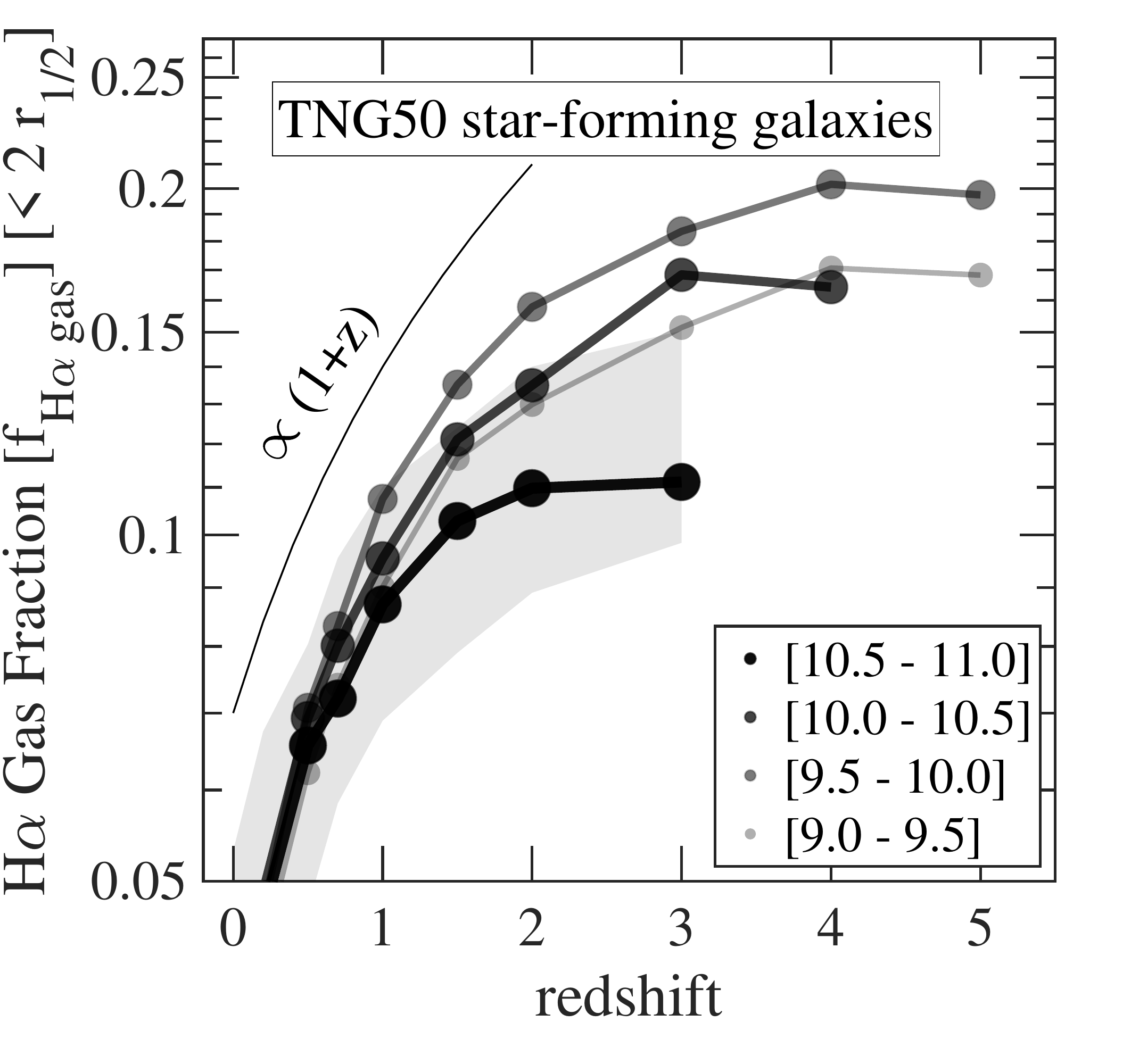}
\includegraphics[trim={0.5cm 1cm 1cm 1cm}, height=5.3cm]{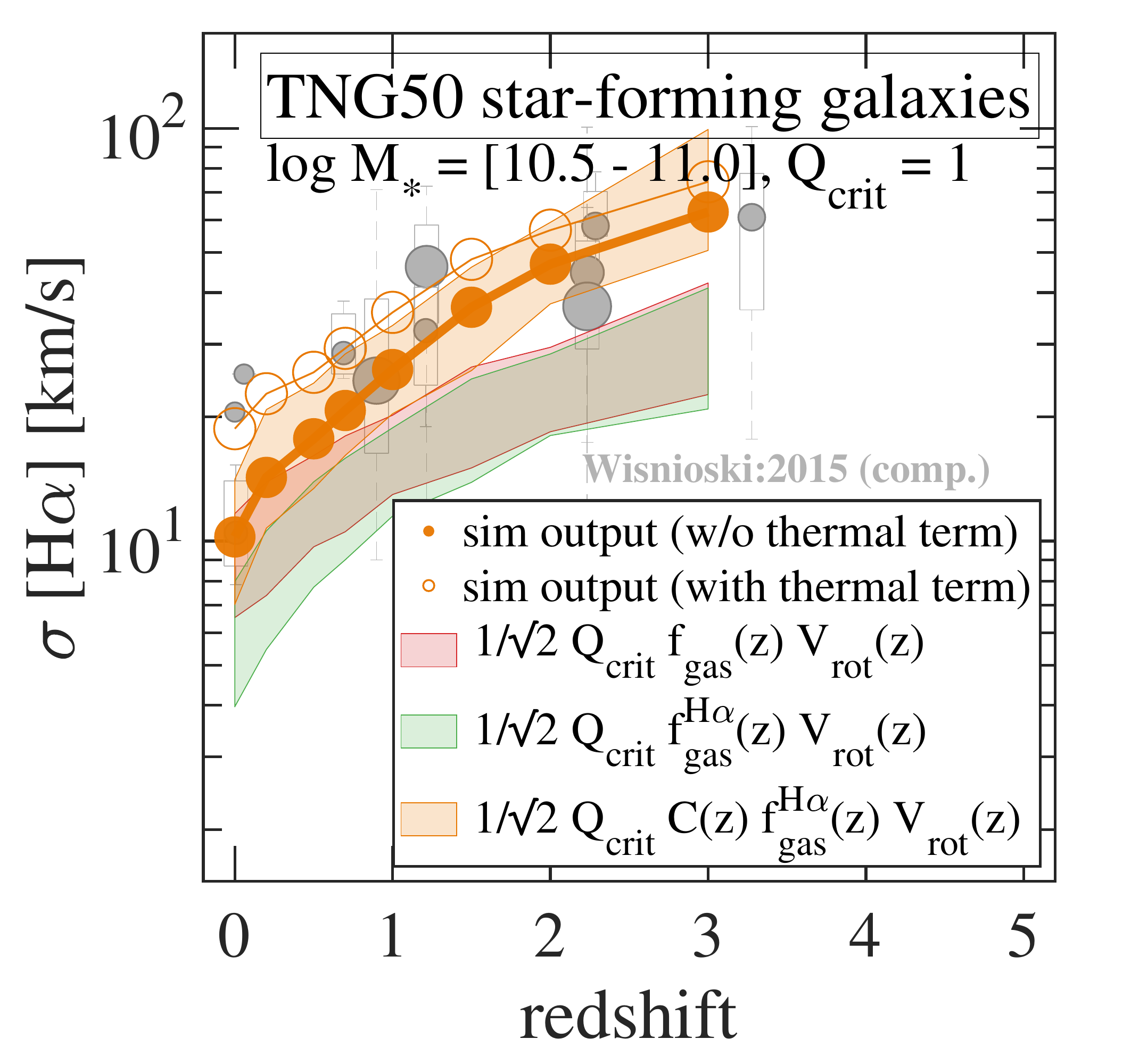}
\includegraphics[trim={0.5cm 1cm 1cm 1cm}, height=5.3cm]{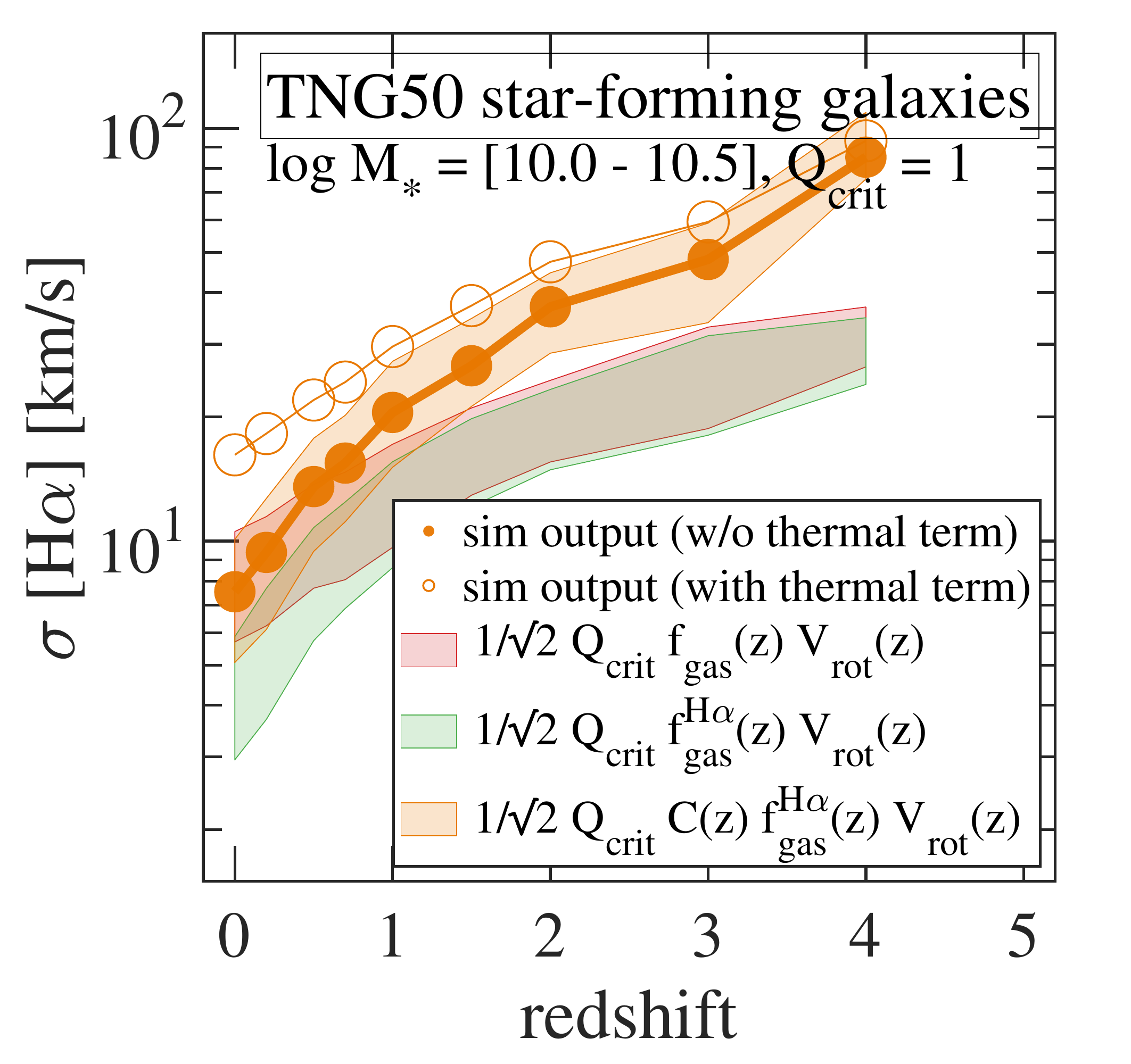}
\includegraphics[trim={0.5cm 1cm 1cm 1cm}, height=5.3cm]{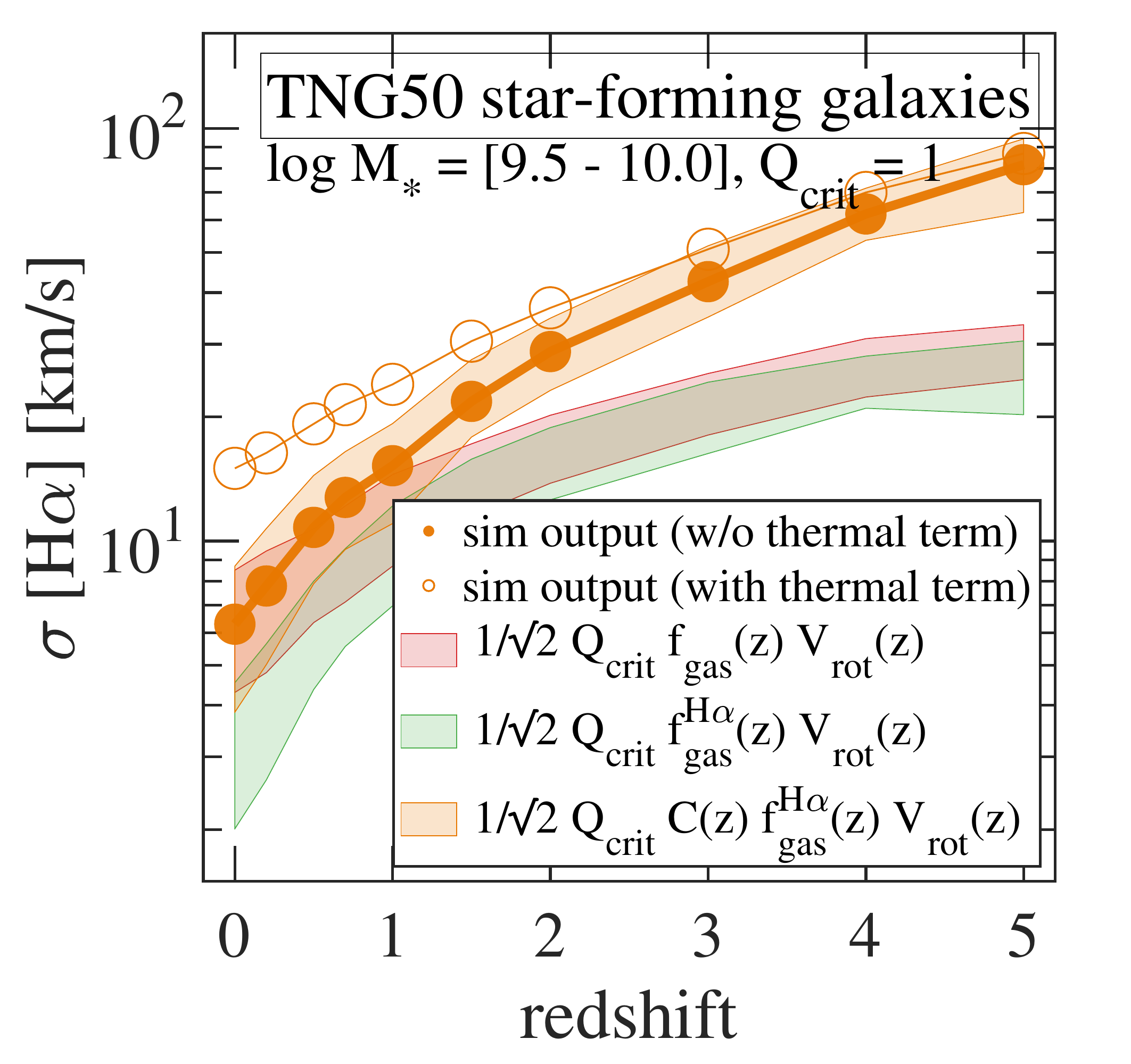}
\caption{\label{fig:ztrends}  Connection between the evolution of the gas fraction within galaxies and the evolution of the gas-probed velocity dispersion. Top panels: redshift evolution in TNG50 of the total gas mass (left) and the $\HA$-emitting gas (right), both normalized to the total dynamical mass of galaxies, all taken within twice the stellar half-mass radius. Circles and curves denote medians across galaxies in the specified stellar mass bins; the shaded gray areas indicate the 1-$\sigma$ galaxy-to-galaxy variation. Clearly, galaxy gas fractions are larger at higher redshifts, the trend being steeper for the $\HA$, i.e. star-forming, gas than for the total gas (at least at $z\lesssim 2-3$). Bottom panels: redshift evolution of the gas velocity dispersions in selected galaxy mass bins (decreasing from left to right): the orange curves \rvvv{and data points} are the outcome of the simulation, \rvvv{with and without accounting for the gas thermal motions: empty vs. filled symbols,} as in Fig.~\ref{fig:sigma}. Red and green shaded areas denote the gas velocity dispersions inferred, on a galaxy-by-galaxy basis, from other underlying physical quantities extracted from the simulations, like the gas fractions and the rotation velocities, following basic principles of disk instability theory \rvvv{(see text for details)}. We use throughout \rvvv{a constant and universal} $Q_{\rm crit}=1$. Orange shaded areas emphasize the need for an extended model \rvvv{when we want to reproduce the non-thermal only simulated results (filled orange symbols) via analytical arguments:} namely, one that allows for a redshift-dependent modulation, $C(z)$, including the effects of mechanisms that a) go beyond disk instability, b) that emerge in the full cosmological context, and c) that can directly alter the dynamics and phase-space properties of the gas: galaxy mergers, galaxy encounters, cosmic gas accretion, and large-scale feedback flows. \rvvv{Similar conclusions hold if we attempted to reproduce analytically the simulation results that account for the gas thermal motions (empty orange circles) in concert with the evolution of the ionized gas fraction: a redshift-dependent, albeit weaker, modulation would also be required.}}
\end{figure*}

\subsection{On the physical interpretation of the redshift trends}

The decided increase in the level of {\it diskiness} of star-forming galaxies at later cosmic epochs (Fig.~\ref{fig:diskiness}) seems to be predominantly driven by the rapid drop in velocity dispersion towards low redshifts: this decrease is much steeper than for the stars over the same time interval. 

Chaotic motions within star-forming galaxy disks are thought to be due to a number of physical mechanisms that affect both gas and, directly or indirectly, stars. These include thermal heating, small-scale turbulence in the ISM, galaxy mergers, galaxy encounters, gas inflows, and larger-scale outflows from feedback from stars and supermassive black holes. While our effective model of the ISM will underestimate small-scale turbulence, all other mechanisms are robustly captured in the TNG simulations. It is fair to speculate that a more `explicit' model of the ISM could produce higher velocity dispersions, although this remains to be demonstrated. In fact, larger dispersion could imply shorter turbulence dissipation times, hence hampering the intuitive enhancement. Further, it is currently unknown if, and to what degree, turbulence originating at scales of tens of parsecs and below contributes to observed gas velocity dispersions, averaged on 0.5-1 kpc scales.

We reiterate that our model for stellar feedback neglects small-scale interactions due to the hydrodynamically-decoupled wind particle scheme. In contrast, the AGN feedback is directly coupled, and energy injection from the central SMBH directly affect the coldest and densest gas in galaxies, producing multi-phase gas ejecta at thousands of km$\,$s$^{-1}$ (see \textcolor{blue}{Nelson et al. 2019}). Furthermore, even for SN-driven winds, our calculations do capture disordered motions indirectly induced by stellar feedback outflows, i.e. complex galactic-scale fountain flows. 

We can anticipate that the number and prominence of galaxy mergers, the rate of gas inflow, and the feedback-induced motions all decline as the Universe ages, the cosmic star formation rate stalls after the cosmic noon and the galactic gas fractions decrease. The curtailing of the physical processes that are capable of maintaining star-forming and dense gas dynamically hot against cooling is plausibly the reason why gaseous (and stellar) disks settle on rotation-dominated and thinner configurations towards recent epochs. In future analyses we will aim to quantitatively disentangle the physical drivers of the trends presented thus far. In \textcolor{blue}{Burkhart et al. in prep}, we compare the outcome of the Illustris and TNG simulations to a number of model incarnations for feedback-driven ISM turbulence and for ISM turbulence driven by gravitational instability, as summarized in \citet{Krumholz:2016}. Here, we expand upon our findings by focusing on the evolution of the gas velocity dispersion only and by connecting to what has been recently suggested in the literature in terms of gravitational instability.

\citet{Wisnioski:2015} have interpreted the observed decline in gas velocity dispersions at $z\lesssim 2-3$ within the theory of marginally stable disks coupled to a galaxy formation equilibrium model. In such a scenario,  star-forming galaxies are described in a steady equilibrium between gas inflows, star formation and outflows, and chaotic and turbulent motions in galactic disks are thought to be set by the balance between gas fuel and star formation. They find that the redshift evolution of $\sigma_{\rm gas}$ probed by currently-available data (e.g. gray annotations in the top left panel of Fig.~\ref{fig:sigma}) can be well reproduced adopting a Toomre $Q_{\rm crit}$ parameter equal to unity (a quasi stable disk) and by connecting a disk's velocity dispersion to its gas mass fraction \citep[see e.g. Eqs. 7 and 8 in][]{Wisnioski:2015}:
\begin{equation}
\sigma_{\rm gas}(z) = \frac{1}{\sqrt{2}} Q_{\rm crit} \VMAX f_{\rm gas}(z).
\label{eq:model}
\end{equation}
To this aim, \citet{Wisnioski:2015} deduced the evolution of the gas fraction in galaxies by invoking empirically-derived redshift trends for the gas depletion times and the star formation rates in galaxies. With the simulation, we can instead directly measure all involved physical quantities and further this line of thought within a fully self-consistent setup.

In Fig.~\ref{fig:ztrends}, upper panels, we quantify the redshift evolution of the gas fractions in galaxies according to our TNG50 model. In the top left panel, we follow the total gas mass within twice the stellar effective radius (all gas, irrespective of temperature, density, or phase), while in the right panel we focus on the $\HA$-emitting gas (i.e. here star-forming gas). In both cases we normalize by the total dynamical mass within the same aperture. Consistent with theoretical expectations and molecular-gas observations, star-forming high-redshift galaxies are characterized by larger gas masses than lower-redshift analogs at the same stellar mass. The redshift evolution is weaker for the total than the star-forming gas,  but in both cases the increase flattens at $z\gtrsim 3-4$.

We use these measurements, in concert with those obtained for the rotational velocities (Fig.~\ref{fig:vrot}), to infer the evolution of the velocity dispersion via Toomre disk instability theory and Eq.~\ref{eq:model}. Such model is applied on a galaxy-by-galaxy basis with the fixed assumption of marginally stable disks with $Q_{\rm crit}=1$ across cosmic time and across galaxy bodies. It should be noted from the onset that to apply the Toomre criterion globally to whole disks is debatable: here we follow the \citet{Wisnioski:2015} approach to show what it would return within the context of the TNG model. The exercise is visualized in the bottom panels of Fig.~\ref{fig:ztrends}, from high to lower mass galaxies from left to right. Orange big data points denote the outcome of the TNG50 simulation, as uncovered in Fig.~\ref{fig:sigma}: filled circles do not include the contributions from the gas thermal motions (we refer to them as intrinsic) while empty circles do. After the cosmic noon, the redshift evolution of the TNG50 $\HA$ velocity dispersion reads approximately $\sigma \propto (1+z)^{3/2}$ \rvvv{(no thermal term case)}, or slightly shallower depending on galaxy mass, \rvvv{but it is weaker than that when thermal motions are included, especially towards the low-mass end}. Note that, in the TNG model, also the rotational velocities evolve with time at fixed galaxy stellar mass: see top left panel of Fig.~\ref{fig:vrot} and discussion in Section~\ref{sec:disc_disks}.

Now, the red shaded areas denote the result of the model of Eq.~\ref{eq:model} applied to TNG50 galaxies by considering the whole gas fraction, while the green shaded areas are the results accounting only for the $\HA$-emitting gas. At least at low redshifts, the latter reproduce the trends of the \rvvv{intrinsic} velocity dispersions \rvvv{(green areas vs. orange filled circles)}, while in both cases the outcome is too shallow at $z\gtrsim1-2$ to reproduce the \rvvv{intrinsic} simulated data. \rvvv{Conversely, the same model reproduces somewhat the trends of the simulated velocity dispersions that account for thermal motions at intermediate and high redshifts (green areas vs. orange empty circles), while the evolution of the {\it total} gas fraction seems to better reproduce the total velocity dispersion across time (red areas vs. orange empty circles)}. It should be noted that, even within the framework of such a model, it is not at all obvious what gas {\it phase} should be included (and within what effective physical aperture). In fact, a single-fluid model for galactic disks is most probably highly unrealistic, not just because of the contribution of the stars -- differently important at different epochs and masses, but also because of the interplay between the different gaseous components. 
\rvvv{Furthermore}, our measurements do account for extra-planar and asymmetric motions, that are certainly not within the scope of the invoked disk models. Finally, as pointed out above, a priori $Q_{\rm crit}$ cannot be treated as a global parameter, as it describes the stability at a certain position in a disk, based on local quantities. All these caveats need to be kept in mind whenever interpreting the preliminary comparison between simulated and modeled data of Fig.~\ref{fig:ztrends}, lower panels.

Interestingly, the low-redshift simulation data would seem to favor $Q_{\rm crit} \sim 2$: namely, the green \rvvv{(red)} area multiplied by a factor of 2 would match the simulated data points \rvvv{without (with) thermal term} at $z\lesssim1-2$ and $\MS \gtrsim 10^{10.5}\MSUN$. It would therefore be consistent with the physical interpretation of \citet{Wisnioski:2015}, modulo a normalization factor. With the caveat that $\sigma_{\rm gas}$ measurements can be systematically uncertain by up to a factor of 2 (see Appendix~\ref{sec:app_sigma}), it could be argued that the `observed' \rvvv{low-redshift} $\sigma_{\rm gas}$ can be interpreted as due to gravitational motions within more-or-less steady state disks, with no need for contributions from turbulence, thermal motions, outflows, etc. Furthermore, this result could be interpreted as a verification that $Q_{\rm crit}$ can in fact be treated as a global parameter derived from global measurements, at least at $0 < z \lesssim1-2$ and high galaxy masses.

However, even by allowing $Q_{\rm crit}>1$, the proposed models cannot justify the continuous increase in velocity dispersion at $z\gtrsim1-2$, in agreement with the conclusions by \citet{Girard:2018}: \rvvv{for TNG50, this is the case for the intrinsic simulated velocity dispersions, i.e. without accounting for thermal motions}. Based on all the findings within this work and others, it is expected that low-redshift star-forming galactic disks are more stable, with larger values of the Toomre parameter, than high-redshift counterparts. High-redshift galaxies are also observed to exhibit more irregular morphologies and kinematics and to be more gas rich -- this probably implying lower $Q_{\rm crit}$ values. Therefore a redshift-dependent $Q_{\rm crit}$ (with lower values at higher redshifts) would not be useful to solve the inconsistency between \rvvv{intrinsic} simulated and ``analytically''-modeled data at high redshifts.

At last, in Fig.~\ref{fig:ztrends}, orange shaded areas emphasize that an extended model is required: one that allows for a redshift-dependent fudge factor, $C(z)$. \rvvv{We explicitly show this for the simulated data that does not account for the gas thermal motions.} This correction clearly increases with increasing redshift and is somewhat larger for progressively smaller galaxies. We interpret this factor $C(z)$ to represent the effects of those physical mechanisms that go beyond simple disk instability and that emerge in the full cosmological context -- hence the choice for the symbol. As mentioned above, such mechanisms can include galaxy mergers, galaxy encounters, cosmic gas accretion, and large-scale feedback flows, that are all expected to be more frequent or more important at high redshift. Plausibly, their effects manifest not only in regulating the gas content of galaxies across cosmic epochs but also in directly determining non-equilibrium dynamics of the gas (and thus its velocity field and phase-space properties). 
\rvvv{Similar conclusions hold if we attempted to reproduce analytically the simulation results that account for the gas thermal motions in concert with the evolution of the ionized gas fraction: a redshift-dependent modulation would also be required, albeit weaker and targeted to better fit the low rather than the high redshift results. This could indeed be interpreted as a requirement for a larger $Q_{\rm crit}$ at lower than higher redshifts, consistently with expectations and with the lines of thoughts put forward by \citet{Uebler:2019}. In practice, we expect that all these cosmic time-dependent physical mechanisms play in concert: their relative contributions and redshift evolution will be the topic of future work.}


\section{Summary and Conclusions} 
\label{sec:summary}

In this work, and together with the companion paper by \textcolor{blue}{Nelson et al. 2019}, we have introduced a new cosmological magneto-hydrodynamical simulation for galaxy physics: TNG50, the third, final, and most demanding run of the IllustrisTNG project. 

Within the current landscape of galaxy simulations, TNG50 provides a unique combination of statistics and resolution. It realizes a uniform volume of 50 comoving Mpc on a side at ``zoom''-like resolution. In practice, at $z=0$, the TNG50 simulation simultaneously samples both a $10^{14}\MSUN$ Virgo-like cluster as well as about one hundred Milky Way-mass haloes and thousands of lower-mass dwarf galaxies. This provides generous statistics of massive galaxies, even at high and intermediate redshifts, with $\sim\,$70 galaxies more massive than $\MS = 10^{11}\MSUN$ at $z=1$ and 380 galaxies with $\MS\geq10^{10}\MSUN$ at $z=2$. All these are modeled with a uniform baryonic mass resolution of $8.5\times10^4\MSUN$, a collisionless gravitational softening of 288 parsecs, and a typical cell size of $70-140$ parsecs within the star-forming regions of galaxies (see Table~\ref{tab:sims} and Fig.~\ref{fig:cellsizes} for resolution details).

Leveraging this new numerical laboratory we have investigated the structural and kinematics properties of TNG50 star-forming galaxies across cosmic time. In particular, we have selected galaxies on the star-forming main sequence above $10^7\MSUN$ at $0.5 \lesssim z \lesssim 5$ (Fig.~\ref{fig:ssfr}) and measured their sizes, disk heights, 3D shapes, maximum rotation velocities, and intrinsic velocity dispersions (see Section~\ref{sec:props} for methodological details). Our findings are summarized as follows:

\begin{itemize}

\item The stellar sizes of our simulated galaxies range from a median of 10 physical kpc for $10^{11}\MSUN$ objects at $z\sim 0.5$ to $0.5-2$ physical kpc for $10^{8-9}\MSUN$ galaxies at $z\sim 4-5$. More massive galaxies are more extended in both V-band and $\HA$ light but this relation flattens considerably at high redshifts ($z\gtrsim2$ and $\MS\lesssim10^{10}\MSUN$): see Fig.~\ref{fig:sizes}. According to our model, intermediate and low mass galaxies, $\MS < 10^{10}\MSUN$, exhibit very weak redshift evolution, with their stellar sizes increasing by a factor of two at most between $z=5$ and $0.5$. We find that V-band and $\HA$ half-light radii trace each other closely, within a factor of 1.5 for galaxies with $10^{7-11.5}\MSUN$ at $0.5 < z < 5$. \\

\item The bulk of the $\MS = 10^{9-11}\MSUN$ main-sequence galaxy population has a typical edge-on ``thickness'' of $200-400$ physical pc, independent of morphological type and redshift. These values are similar for V-band and $\HA$ light profiles, and are converged to better than 20-40 percent at the resolution of TNG50. Massive star-forming galaxies exhibit progressively flatter morphologies at lower redshift, based on their edge-on and face-on light profiles (Figs.~\ref{fig:diskheights_distribs} and \ref{fig:diskheights_ev}). In particular, galaxy ``flatness'' decreases by up to a factor of a few between $z=4-5$ and $z=0.5-1$, where the typical star-forming $10^{10-11}\MSUN$ galaxy has $h_{\rm 1/2} \lesssim 0.1 r_{\rm 1/2}$ at recent epochs ($z\lesssim 1$). The high-mass redshift trend is driven by both height and size evolution, although the latter may dominate.\\

\item We analyze the 3D intrinsic shapes of the stellar as well as $\HA$-gas mass distributions (Figs.~\ref{fig:shapes} and \ref{fig:shapesevolutions}). Based on stellar-mass geometry, the fraction of galaxies with disk-like morphologies increases with time (i.e. more disks at lower redshifts, for all $\MS$) and is higher for larger stellar mass at fixed redshift. For example, above $\MS\gtrsim 10^{10.5}\MSUN$, more than 60 per cent of star-forming galaxies exhibit disk-like stellar morphologies at all times. We also find that elongated, `cigar-like' stellar mass distributions are more frequent towards high redshift and low mass. On the other hand, the redshift trends of $\HA$-based morphologies are weaker than those derived from stellar mass. This star-forming gas settles into disk-like (or elongated) morphologies for the majority (60-100 per cent) of $>10^9\MSUN$ galaxies at all times ($z\lesssim5$).\\
 
\item The kinematic fields of TNG50 galaxies (Figs.~\ref{fig:kinematics1} and \ref{fig:kinematics2}) display a rich phenomenology and wide diversity across the simulated galaxy population. The vertical, intrinsic gas velocity dispersion $\sigma$ is typically more complex and less spatially coherent than the analog traced by star light. Yet, at any given redshift and mass, and without selecting on stellar morphology, the typical velocity $\sigma_{\rm stars}$ is larger than $\sigma_{\rm gas}$. For both tracers velocity dispersion decreases with time; the $\HA$-probed dispersion declines by a factor of a few between $z\sim2$ and $z\sim0.5$ independent of mass (Fig.~\ref{fig:sigma}). Nevertheless, clear rotation patterns are identified in both the stellar and gaseous components in all star-forming galaxies also at high redshifts ($z\gtrsim1$, $\MS\gtrsim 10^9\MSUN$, Fig.~\ref{fig:vrot}). \\


\item In practice, our model predicts that the vast majority of $\MS > 10^{9}\MSUN$ star-forming galaxies are rotationally-supported gas disks at all times, with values of $\VMAX/\sigma$ in excess of 2-3 ($z\lesssim5$, Fig.~\ref{fig:Voversigma}). These gaseous structures become dynamically colder towards lower redshift. On the other hand, due to their collisionless dynamics, stars in the same galaxies are always dynamically hotter (lower $\VMAX/\sigma$ values) than the gas out of which they form, with little variation in redshift or mass. In the TNG model, the increasing gas $\VMAX/\sigma$ with time proceeds despite the pronounced decrease of rotational velocities towards low redshift, a trend which may be in tension with some observational findings, yet is qualitatively consistent with others.

\end{itemize}

In conclusion, the TNG50 simulation produces a main-sequence galaxy population whose stellar and gas structural properties and kinematics evolve across mass and time, and are qualitatively consistent with observational findings. Namely, the fraction of disk-like galaxies based on 3D stellar shapes rises with both cosmic time and galaxy stellar mass, and the vast majority of $10^{9-11.5}\,\MSUN$ star-forming galaxies are rotationally-supported gaseous disks at all times, although dynamically hotter at earlier epochs. This concordance is a non-trivial confirmation of the outcome of the numerical model and the plausibility of the underlying physical assumptions. It encourages future analyses that we will undertake in order to identify and further disentangle the physical drivers of the trends presented here, and the validity of our general approach, i.e. despite the unavoidable simplifications of the underlying galaxy physics model, particularly in the treatment of the star formation, of the cold dense gas, and of the feedback from stars on small-scales. In a preliminary attempt to interpret the redshift trends of the gas velocity dispersions, we find that disk instability theory coupled with the time evolution of the ionized gas fraction in galaxies can describe the simulation outcome only at $z\lesssim1-2$, but may require additional ingredients at higher redshifts.

Our analysis of TNG50 has uncovered novel predictions for the relation between stellar and $\HA$-traced galaxy properties and between structural and kinematical measurements. These insights will help guide the interpretation of future observational results. Notably, despite the presence of larger velocity dispersions at high redshift, cold and dense gas principally arranges in disky or elongated shapes at {\it all} times and masses. Interestingly, we find that the fraction of ``elongated'' $\HA$ shapes dominates at low masses at all times ($\MS \lesssim 10^{9.5}\MSUN{}$ and $z\lesssim6$) and is not negligible also at larger masses. In other words, in the TNG model, star-formation occurs in relatively thin disk-like or slab configurations of dense gas, in accordance with the basic principles of galaxy formation theory. At the same time rotationally-dominant motions in gaseous galactic bodies, as represented by high values of gas $\VMAX/\sigma$, increase with time and correlate with an analog, albeit weaker, flattening of gaseous disks towards low redshift. In contrast, gas kinematics ehibit degrees of rotationally-supported motions that far exceed those of their stellar counterparts, even in galaxies with very flat, disk-like stellar morphologies. This apparent dichotomy reflects the different nature of stars and gas: as the latter can dissipate energy, it cannot remain in high-dispersion equilibrium states with crossing orbits, in contrast to the stars.

In conclusion, with this paper we put forward quantitative predictions for the evolution of galaxy internal kinematics at intermediate and high-redshifts, particularly in as of yet unexplored stellar light tracers, that will be possible to test with ambitious future observatories such as the James Webb Space Telescope (JWST), Giant Magellan Telescope (GMT), and Extremely Large Telescope (ELT).


\section*{Acknowledgements}

The authors thank the anonymous reviewer for useful discussions and suggestions, that have contributed to expand upon the paper messages. AP thanks Philipp Lang, Ryan Leaman, Emily Wisnioski, Sandy Faber, Joel Primack, Ortwin Gerhard, Caroline Straatman, Bhawna Motwani, Gandhali Joshi, Sedona Price, Hannah Uebler, and Natascha Foerster Schreiber for insightful conversations and input. 
SG, through the Flatiron Institute, is supported by the Simons Foundation.
FM acknowledges support through the Program ``Rita Levi Montalcini'' of the Italian MIUR. AvdW acknowledges funding through the H2020 ERC Consolidator Grant 683184.
The primary TNG simulations were realized with compute time granted by the Gauss Centre for Supercomputing (GCS): TNG50 under GCS Large-Scale Project GCS-DWAR (2016; PIs Nelson/Pillepich), and TNG100 and TNG300 under GCS-ILLU (2014; PI Springel) on the GCS share of the supercomputer Hazel Hen at the High Performance Computing Center Stuttgart (HLRS). GCS is the alliance of the three national supercomputing centres HLRS (Universit{\"a}t Stuttgart), JSC (Forschungszentrum J{\"u}lich), and LRZ (Bayerische Akademie der Wissenschaften), funded by the German Federal Ministry of Education and Research (BMBF) and the German State Ministries for Research of Baden-W{\"u}rttemberg (MWK), Bayern (StMWFK) and Nordrhein-Westfalen (MIWF). Additional simulations for this paper were carried out on the Draco and Cobra supercomputers at the Max Planck Computing and Data Facility (MPCDF).

\bibliographystyle{mnras}
\bibliography{TNGPapers_TNG50_1}


\appendix
\label{sec:appendix}

\section{Measurements of the velocity dispersion}
\label{sec:app_sigma}

In this Appendix we compare different operational definitions of a galaxy velocity dispersion. Given a 2D map of a galaxy with finite spatial resolution, throughout the paper our fiducial $\sigma$ is the average of the pixel-based velocity dispersion ($\sigma_{\rm pixel}$) of its stellar or gas resolution elements:
\begin{equation}
\sigma = \frac{1}{N_{\rm pixels}} \sum_{\rm pixels} \sigma_{\rm pixel}.
\label{eq:pixel_averages}
\end{equation}
The sum is performed over the number of pixels that, in our fiducial setup, are found along an imaginary long slit between one and two times the stellar half-mass radius of a galaxy (the long slit being aligned with the structural major axis of the mass or light distribution). The reference pixel size is 0.5 comoving kpc. Here $\sigma_{\rm pixel}$ reads as follows:
\begin{equation}
 \sigma_{\rm pixel} = \sqrt{\frac{\sum_i w_i v_i^2}{\sum_i w_i}},
 \label{eq:sigmapixel_all}
\end{equation}
where $v_i$ denotes the line of sight velocity of the stellar particles or (star-forming) gas cells within a pixel. The weights $w_i$ are taken to be unity in our fiducial setup. However, observers can commonly access, by construction, luminosity-weighted averages, so that the weights above can be e.g. the V-band or the $\HA$ luminosity for the stellar and gaseous components, respectively. 

\begin{figure}
\centering                                      
\includegraphics[width=8cm]{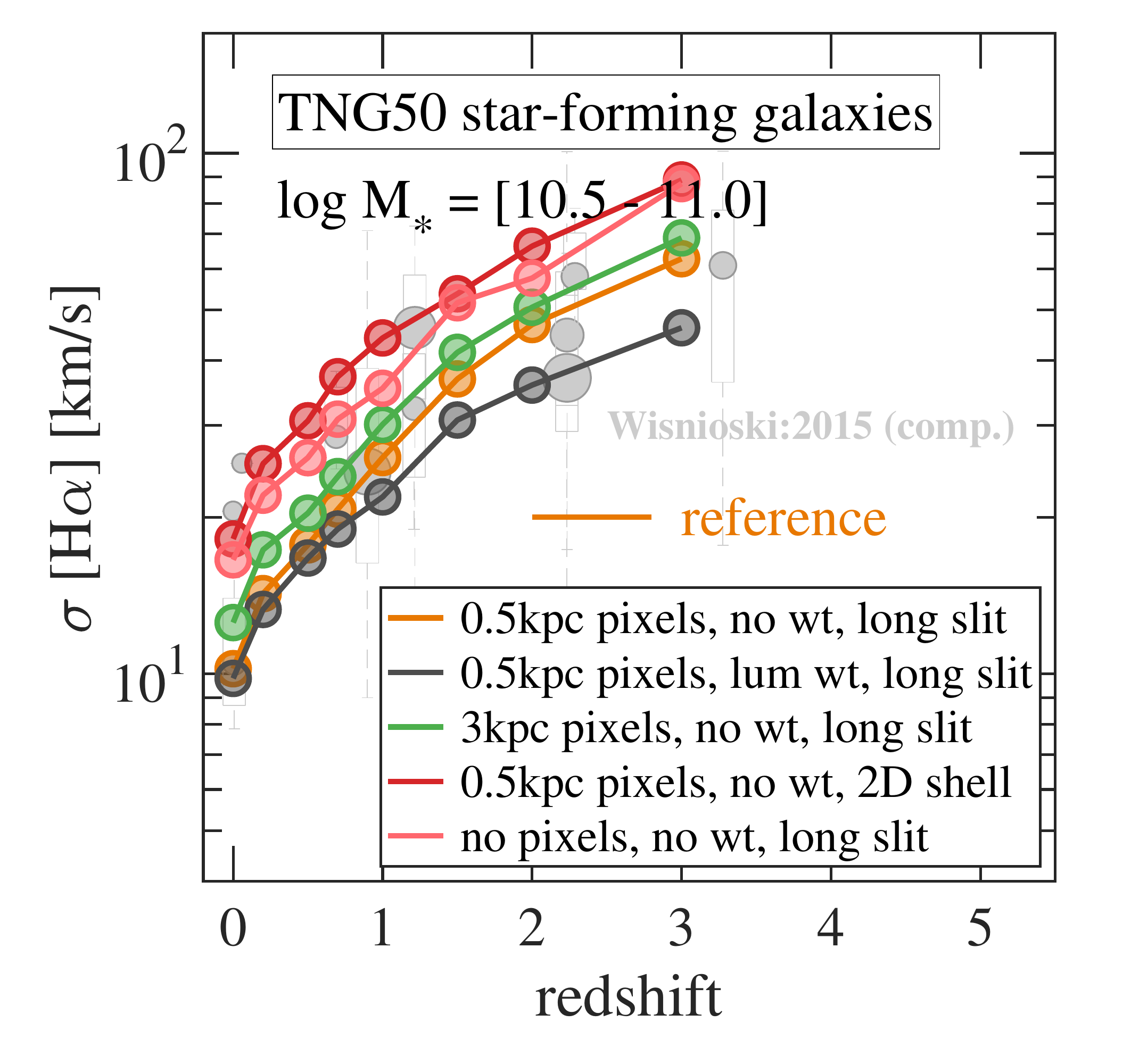}
\includegraphics[width=8cm]{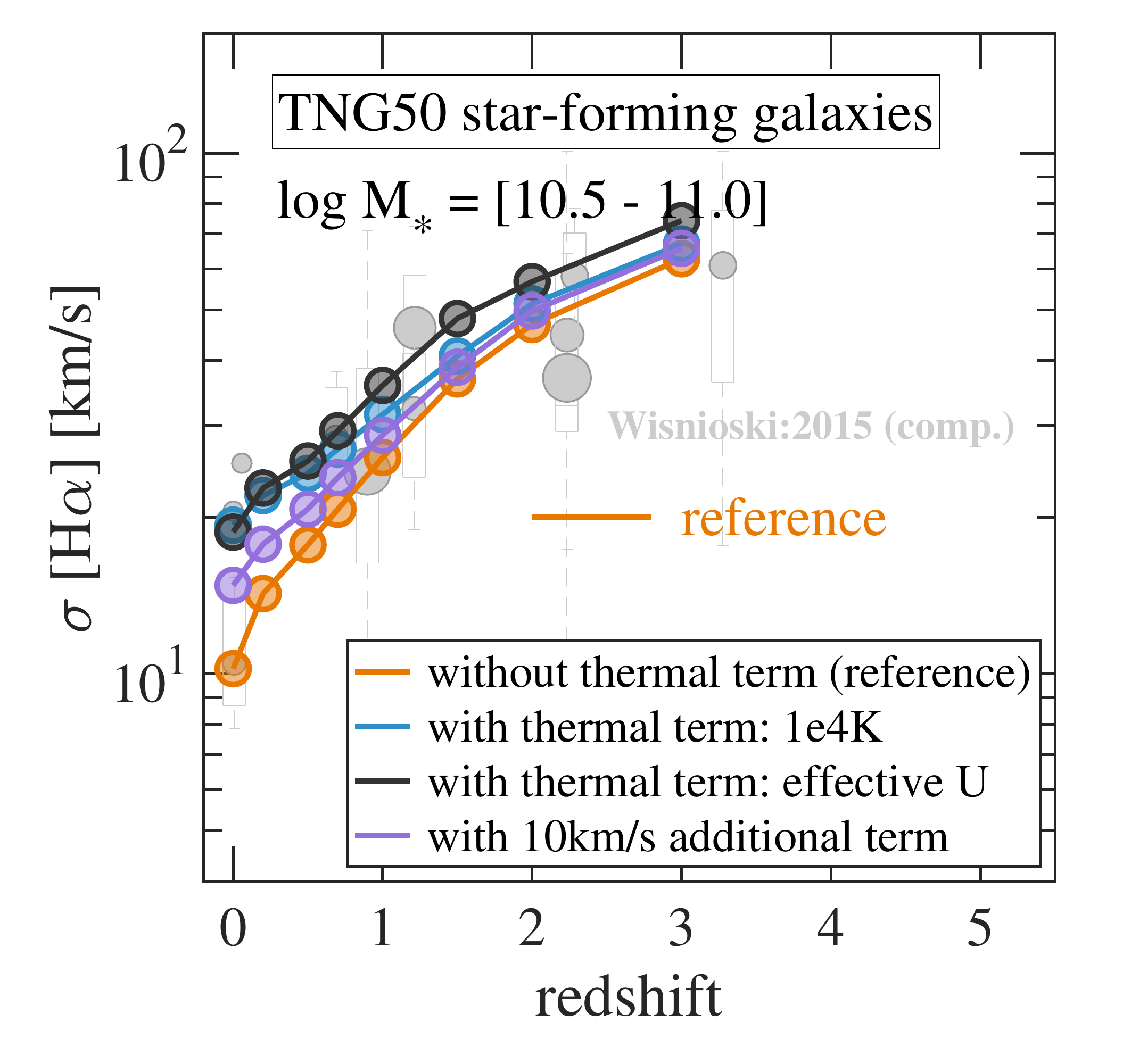}
\caption{\label{fig:sigma_measurements} Effects of different measurement choices on the redshift evolution of the average $\HA$ velocity dispersions for $10^{10.5-11}\MSUN$ TNG50 galaxies. In both panels, the orange curves denote our \rvvv{reference} choice and are identical to those discussed in the main body of the text \rvvv{(e.g., filled circles in Fig.~\ref{fig:sigma}, top left panel)}. In the top panel here, we evaluate different choices in the averages: with and without weights, for different pixel sizes, or averaging across different regions of a galaxy body. \rvvv{All curves in the top panels neglect the contributions from a thermal component to the velocities}. In the bottom panels, we quantify the effects of thermal motions. All options fall within the ball park of currently-available observational constraints, yet implying slightly different redshift trends.}
\end{figure}

In Fig.~\ref{fig:sigma_measurements}, top panel, we hence demonstrate how alternative {\it averaging} choices affect the measurements. We focus here on the redshift evolution of the $\HA$-gas velocity dispersion, in analogy to the top left panel of Fig.~\ref{fig:sigma} \rvvv{(filled symbols, neglecting thermal motions)}. We concentrate on the high-mass end of the galaxy population so that we can use currently-available observational constraints as a benchmark (gray data points and annotations): similar results hold also at lower galaxy masses. The reference measures are in orange. $\HA$ luminosity weighted dispersions are somewhat lower than the unweighted ones (black vs. orange data), while larger pixel sizes tend to somewhat overestimate the velocity dispersions (green vs. orange). The red curve denotes the typical velocity dispersions obtained by averaging the gas parcels not within a long slit along the galaxy major axis but along a circular shell at $1-2\times R_{\rm stars}$ (always in face-on projections). On the other hand, the pink results show what would be obtained with a more theoretically-oriented approach, one that is not based on a pixelized map and where the first average of Equation~\ref{eq:pixel_averages} is waived (yet by accounting for the contribution of only those simulation resolution elements at the fiducial location within the galaxy). Importantly, none of these alternatives imply substantial changes in the redshift evolution: galaxy gas velocity dispersions are much higher at higher redshifts than at $z\lesssim1$ and the redshift trends are consistent across the options analyzed here. On the other hand, the overall normalizations can vary by up to a factor of 2 but are still within the ball park that is currently probed by observations.

In the case of gas, thermal motions can contribute non negligibly to the velocity dispersion measured in a galaxy. In practice, the pixel-based velocity dispersion can be generalized as follows (see also e.g. Equation 8 of \citealt{Diemer:2018}):
\begin{equation}
 \sigma_{\rm pixel, gas} = \sqrt{\frac{\sum_i w_i (v_i^2 + \sigma^2_{\rm thermal})}{\sum_i w_i}}.
 \label{eq:sigmapixel_gas}
\end{equation}
The thermal-motion term can be written as $\sigma^2_{\rm thermal} = P/\rho = u (\gamma -1)$, i.e. as a function of pressure and density or via the internal energy per unit mass, $u$, and the adiabiatic index, $\gamma = 5/3$.

In Fig.~\ref{fig:sigma_measurements}, bottom panel, we quantify the contribution of thermal chaotic motions to galaxies' gas velocity dispersions. However, in this paper we mostly focus on $\HA$-emitting gas that, by construction in our model, coincides with star-forming gas. Because of the two-phase ISM approximation adopted within the TNG model \citep{Springel:2003}, for star-forming gas it is difficult to assign a physically-consistent velocity dispersion that reflects its temperature: the temperatures or internal energies for the star-forming gas cells in the TNG model are {\it effective}, i.e. dictated by the effective equation of state for the ISM invoked by \citet{Springel:2003}. Therefore, we explore three options as follows:
\begin{itemize}
\item we neglect the conceptual and practical inconsistency highlighted above and simply compute the gas velocity dispersions of TNG50 galaxies with  $\sigma^2_{\rm thermal} = u (\gamma -1)$;
\item as we are interested in the kinematics of $\HA$ gas and this is expected to have a temperature of $10^4$K, we take $\sigma^2_{\rm thermal} = \sqrt{\frac{3 k_B T}{m_H}}$, namely the width of a Maxwellian velocity distribution for Hydrogen (hence $m_H$) with $T = 10^4$K  and $k_B$ denoting the Boltzmann constant;
\item we add a fixed additional term of $\sigma_{\rm thermal} = 10$ km s$^{-1}$ in order to have a basic quantitative benchmark and by invoking the typical, observationally-inferred, zeroth-order estimate of the velocity dispersion of the cold phase of the ISM.
\end{itemize}
The results are shown in the bottom panel of Fig.~\ref{fig:sigma_measurements} in black, blue, and purple, respectively, as they compare to the \rvvv{reference, intrinsic} measurement in orange, with $\sigma^2_{\rm thermal} \equiv 0$. The relative effects of thermal motions are larger at lower redshift, where they can boost the gas velocity dispersions by up to a factor of $\sim2$ for $\sim 10^{10}\MSUN$ galaxies (and up to a factor of $\sim3$ for lower mass galaxies, e.g. at the $10^9\MSUN$ scale -- not shown).  In practice, by including the thermal motions, the increase of the galaxies' gas velocity dispersions with increasing redshift is somewhat weaker. Yet, overall the effects at fixed redshift are similar or smaller than those induced by simply adopting different averaging choices (bottom vs. top panel of Fig.~\ref{fig:sigma_measurements}). Furthermore, in all cases here explored, the TNG50 estimates fall within the ball park of observational findings.

\rvvv{Unless otherwise explicitly stated, results in the main body of the paper} shall be intended for gas velocity dispersions that do {\it not} account for thermal motions: these should be more robustly compared to current or future observational estimates where the effects of thermal broadening are suitably modeled and subtracted. \rvvv{However, because to separate thermal from other motions via the $\HA$ line alone is almost impractical from an observational perspective, we include in relevant instances also the TNG50 results that account for a thermal component: we do so in Figs.~\ref{fig:sigma} and \ref{fig:ztrends} (e.g. empty circles and dashed curves) by reporting the estimates based on the effective internal energy of the star-forming gas cells (black data points in Fig.~\ref{fig:sigma_measurements}, bottom panel).}

\section{Resolution studies of the diagnostics presented in this paper}
\label{sec:app_res}

\begin{table*}
  \caption{Table of physical and numerical parameters for the four resolution levels of the TNG50 simulation presented in this paper. The physical parameters are: the box volume, the box side-length, the initial number of gas cells, dark matter particles, and Monte Carlo tracer particles. The target baryon mass, the dark matter particle mass, the $z$\,=\,0 Plummer equivalent gravitational softening of the collisionless component, the same value in comoving units, the maximum softening applied to blackholes, and the minimum comoving value of the adaptive gas gravitational softenings. Additional characterizations of the gas 
  resolution, measured at redshift zero: the minimum physical gas cell radius, the median gas cell radius, the mean radius of star-forming gas cells
  The numerical parameters are: the number of high time frequency subbox volumes, the number of snapshots in each subbox, the total number of timesteps to $z$\,=\,0, the total run time including substructure identification in millions of CPU core hours, and the number of compute cores used.}
  
  \label{simTableBig}
  \begin{center}
    \begin{tabular}{lcllll}
    
     \hline\hline
     
 Run Name & & TNG50-1 & TNG50-2 & TNG50-3 & TNG50-4 \\ \hline\hline
 Volume & [\,cMpc$^3$\,] & $51.7^3$ & $51.7^3$ & $51.7^3$ & $51.7^3$  \\
 $L_{\rm box}$ & [\,cMpc/$h$\,] & 35 & 35 & 35 & 35 \\
 $N_{\rm GAS}$ & - & $2160^3$ & $1080^3$ & $540^3$ & $270^3$  \\
 $N_{\rm DM}$ & - & $2160^3$ & $1080^3$ & $540^3$ & $270^3$  \\
 $N_{\rm TRACER}$ & - & $2 \times 2160^3$ & $2 \times 1080^3$ & $2 \times 540^3$ & $1 \times 270^3$ \\
 $m_{\rm baryon}$ & [\,M$_\odot$/$h$\,] & $5.7 \times 10^4$ & $4.6 \times 10^5$ & $3.7 \times 10^6$ & $2.9 \times 10^7$ \\
 $m_{\rm DM}$ & [\,M$_\odot / h$\,] & $3.1 \times 10^5$ & $2.5 \times 10^6$ & $2.0 \times 10^7$ & $1.6 \times 10^8$  \\
 $m_{\rm baryon}$ & [\,10$^5$\,M$_\odot$\,] & 0.85 & 6.8 & 54.2 & 433.8 \\
 $m_{\rm DM}$ & [\,10$^5$\,M$_\odot$\,] & 4.5 & 36.3 & 290.4 & 2323.2 \\
 $\epsilon_{\rm DM,stars}^{z=0}$ & [\,pc\,] & 288 & 576 & 1152 & 2303  \\
 $\epsilon_{\rm DM,stars}$ & [\,ckpc/$h$\,] & 0.39 $\rightarrow$ 0.195 & 0.78 $\rightarrow$ 0.39 & 1.56 $\rightarrow$ 0.78 & 3.12 $\rightarrow$ 1.56 \\
 $\epsilon_{\rm BH,max}$ & [\,ckpc/$h$\,] & 2.0 & 4.0 & 4.60 & 5.36 \\
 $\epsilon_{\rm gas,min}$ & [\,cpc/$h$\,] & 50 & 100 & 200 & 400 \\ \hline
 $r_{\rm cell,min}$ & [\,pc\,] & 6 & 18.6 & 64.6 & 168 \\
 $\bar{r}_{\rm cell}$ & [\,kpc\,] & 6.2 & 12.9 & 25.1 & 50.1  \\ 
 $\bar{r}_{\rm cell,SF}$ & [\,pc\,] & 140 & 280 & 562 & 1080 \\ 
 $N_{\rm subboxes}$ & - & 3 & 3 & 3 & 3 \\
 $N_{\rm snaps,sub}$ & - & 3829 & 1985 & 4006 & 2333 \\
 $\Delta t$ & - & 11606566 & 7691447 & 1324711 & 239399 \\
 CPU Time & [\,Mh\,] & $\sim$130 & 4.8 & 0.170 & 0.006  \\
 $N_{\rm cores}$ & - & 16320 & 5120 & 1056 & 256\\
 \hline

    \end{tabular}
  \end{center}
\end{table*}

In this Appendix we comment on resolution convergence. As a reminder, we do not adjust in any way our model parameters as a function of numerical resolution, but for the choices on the softening lengths \citep[see][for a discussion]{Pillepich:2018Method}. Different galaxy properties have different convergence trends and a priori it is not given that the quantitative outcomes of our simulations are identical across resolutions. For example, as demonstrated in Appendix A of \cite{Pillepich:2018} for TNG100 and TNG300, a galaxy's stellar mass is in general larger at better resolutions, and this resolution-dependent shift is smaller at higher redshifts and larger masses. Similarly, galaxy star formation rates tend to be higher at improved resolution, as denser gas is better captured. This mostly affects massive and quenched galaxies and, in fact, the locus of the star-forming main sequence is indistinguishable at $0\leq z \leq 2$ between TNG100 and TNG300 \citep{Donnari:2019}. We focus here on galaxy sizes, disk heights, and kinematic summary statistics like $\VMAX$ and $\sigma$: for this purpose, we contrast the results from the resolution variations of the TNG50 simulation. 

The numerical parameters of the TNG50 resolution series are given in Table~\ref{simTableBig}. The runs TNG50-2, TNG50-3, and TNG50-4 have 8, 64, and 512 times lower mass resolution than the flagship run, corresponding to factors of 2, 4, 8 larger gravitational softening lengths. As a reminder, TNG100 (TNG300) has a particle resolution that is a factor of 2 worse than TNG50-2 (TNG50-3), and a factor of 16.4 (132) worse than TNG50-1.

\subsection{Galaxy Sizes}

\begin{figure*}
\centering                                      
\includegraphics[width=5.8cm]{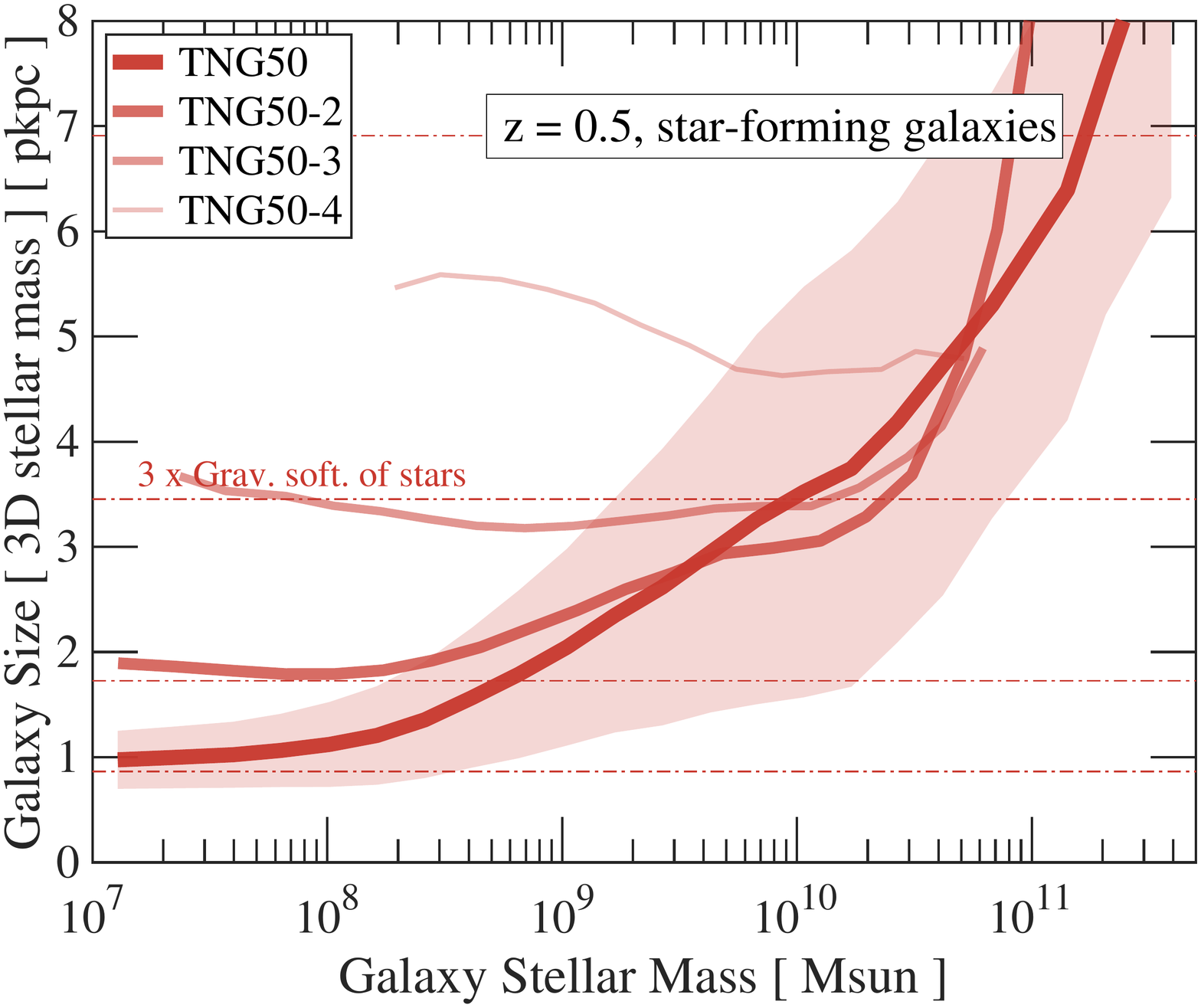}
\includegraphics[width=5.8cm]{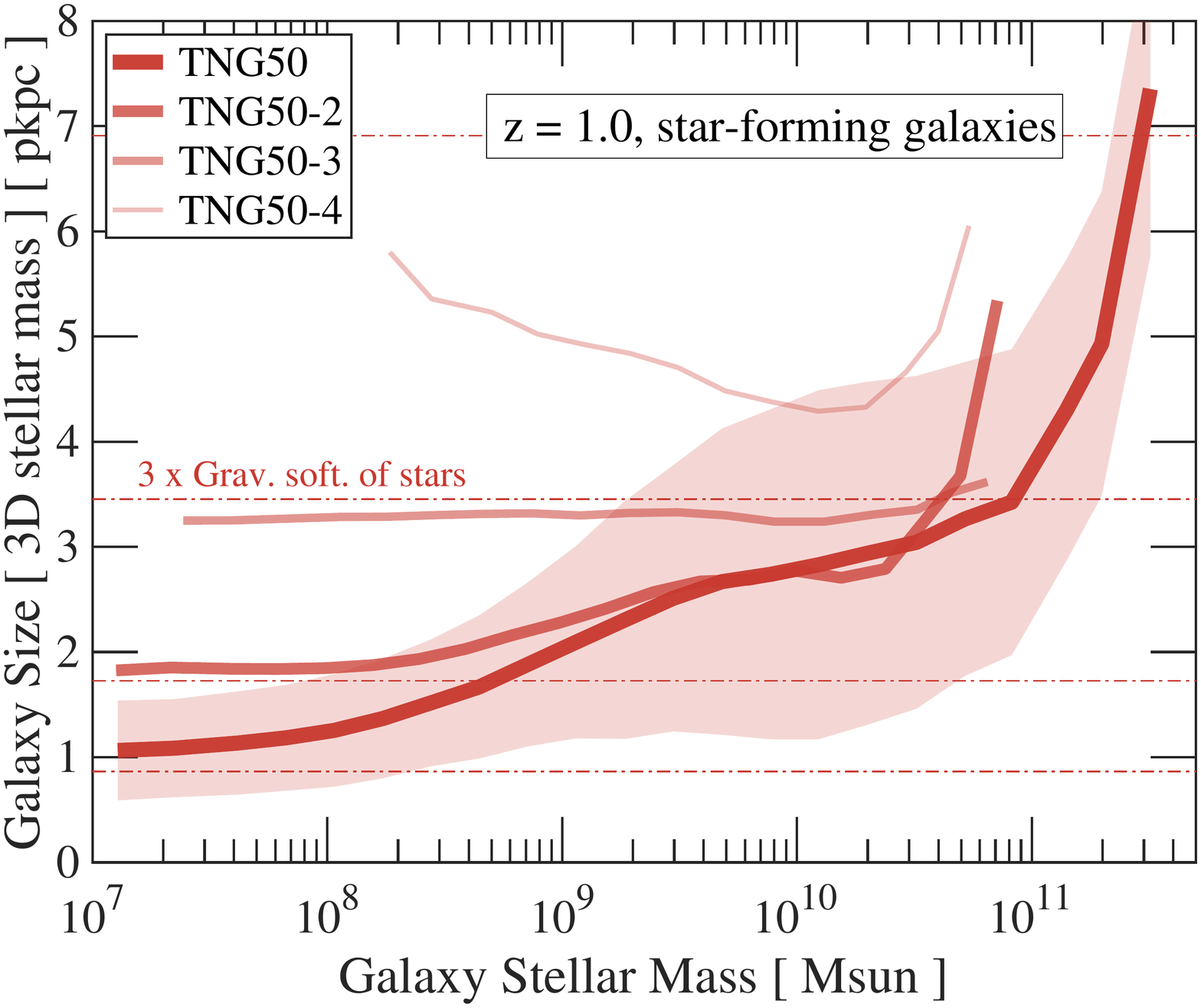}
\includegraphics[width=5.8cm]{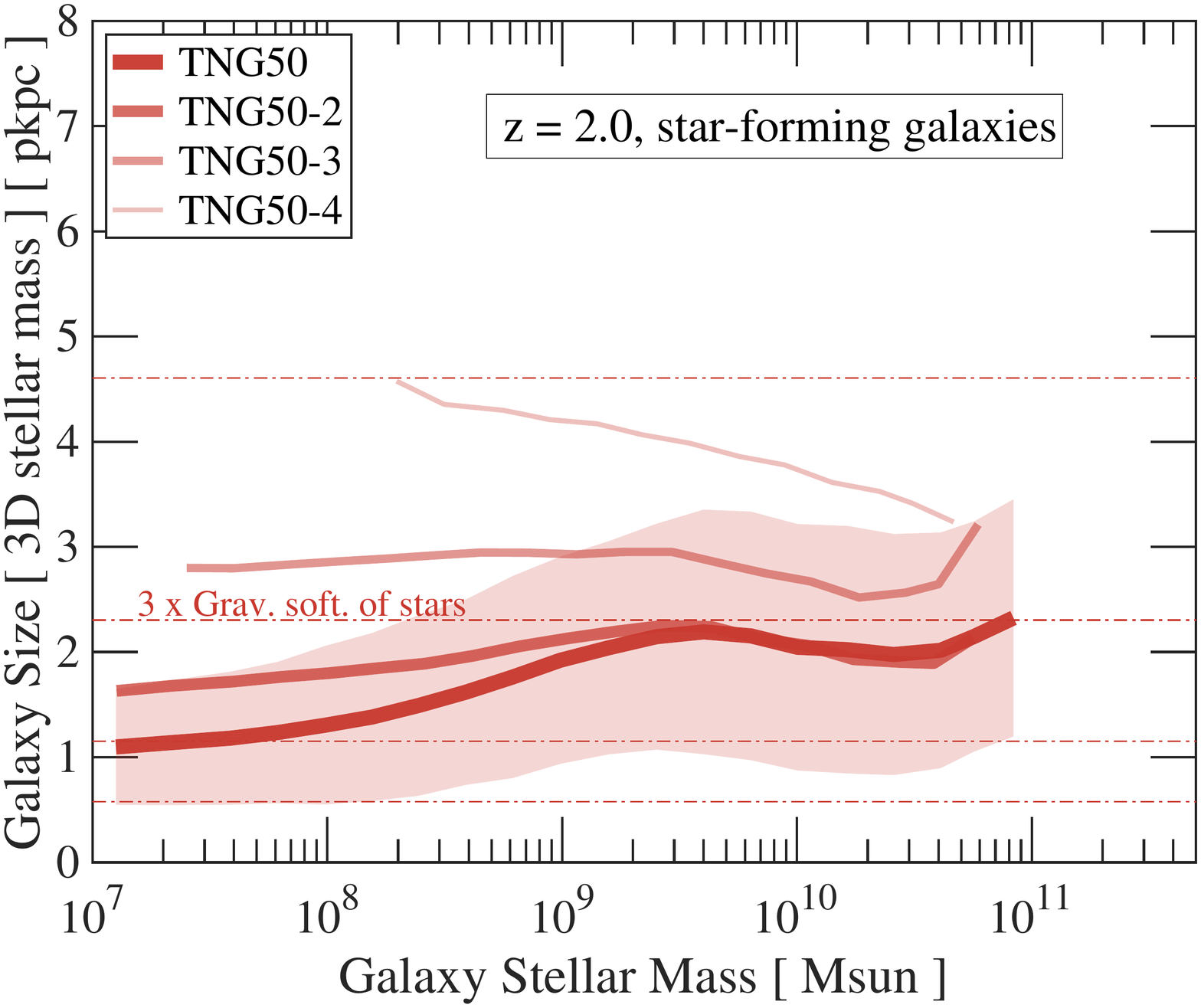}
\includegraphics[width=5.8cm]{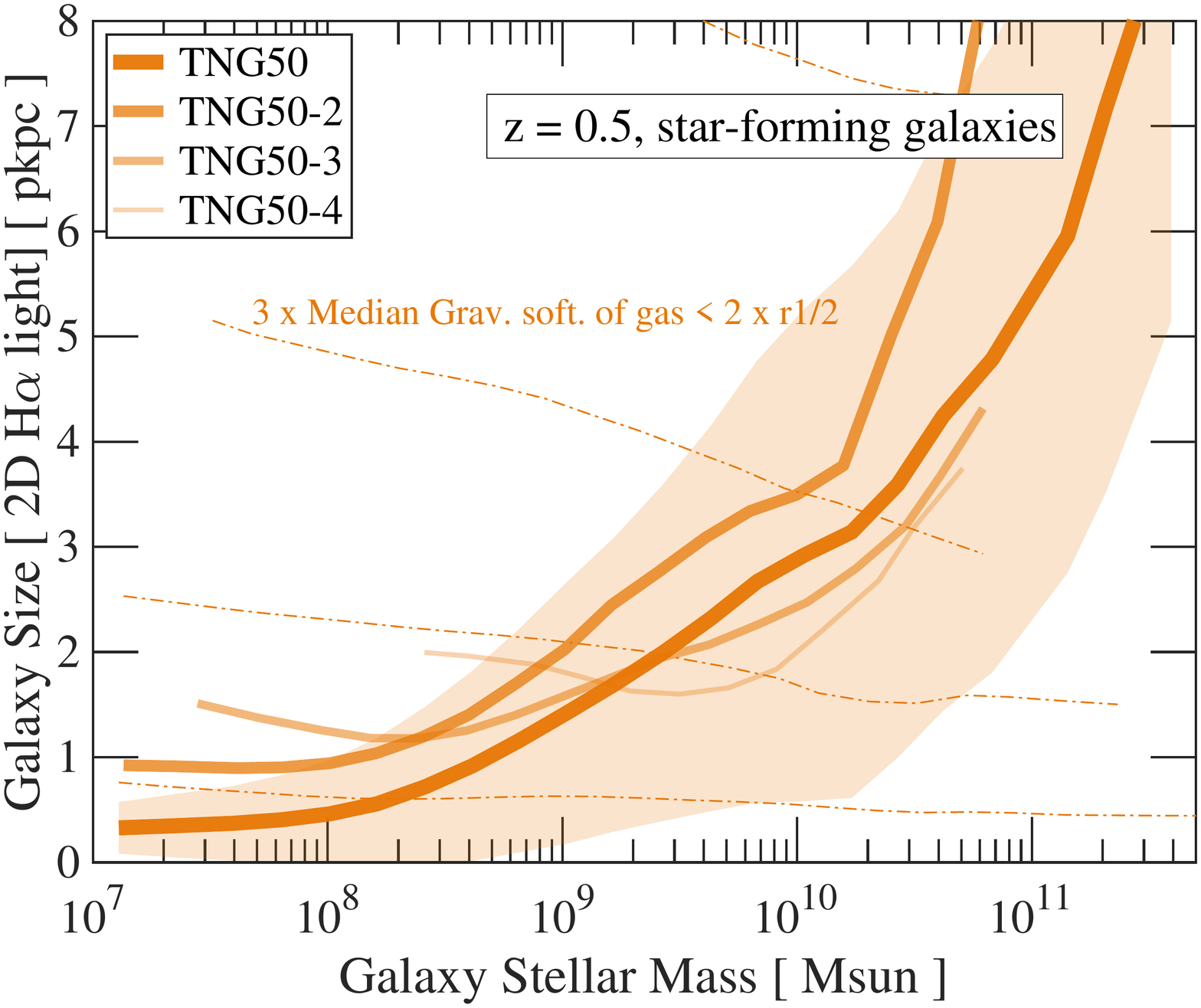}
\includegraphics[width=5.8cm]{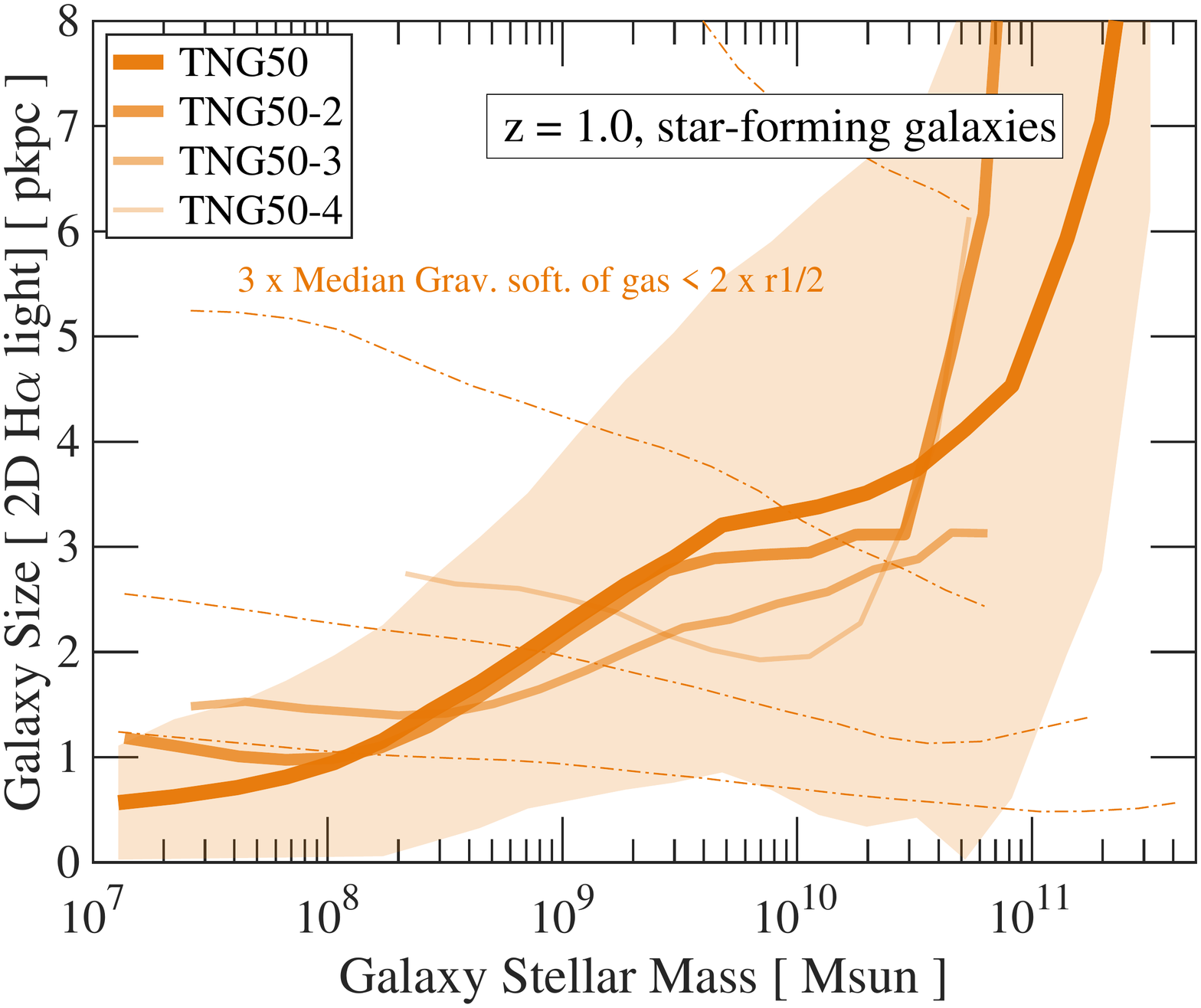}
\includegraphics[width=5.8cm]{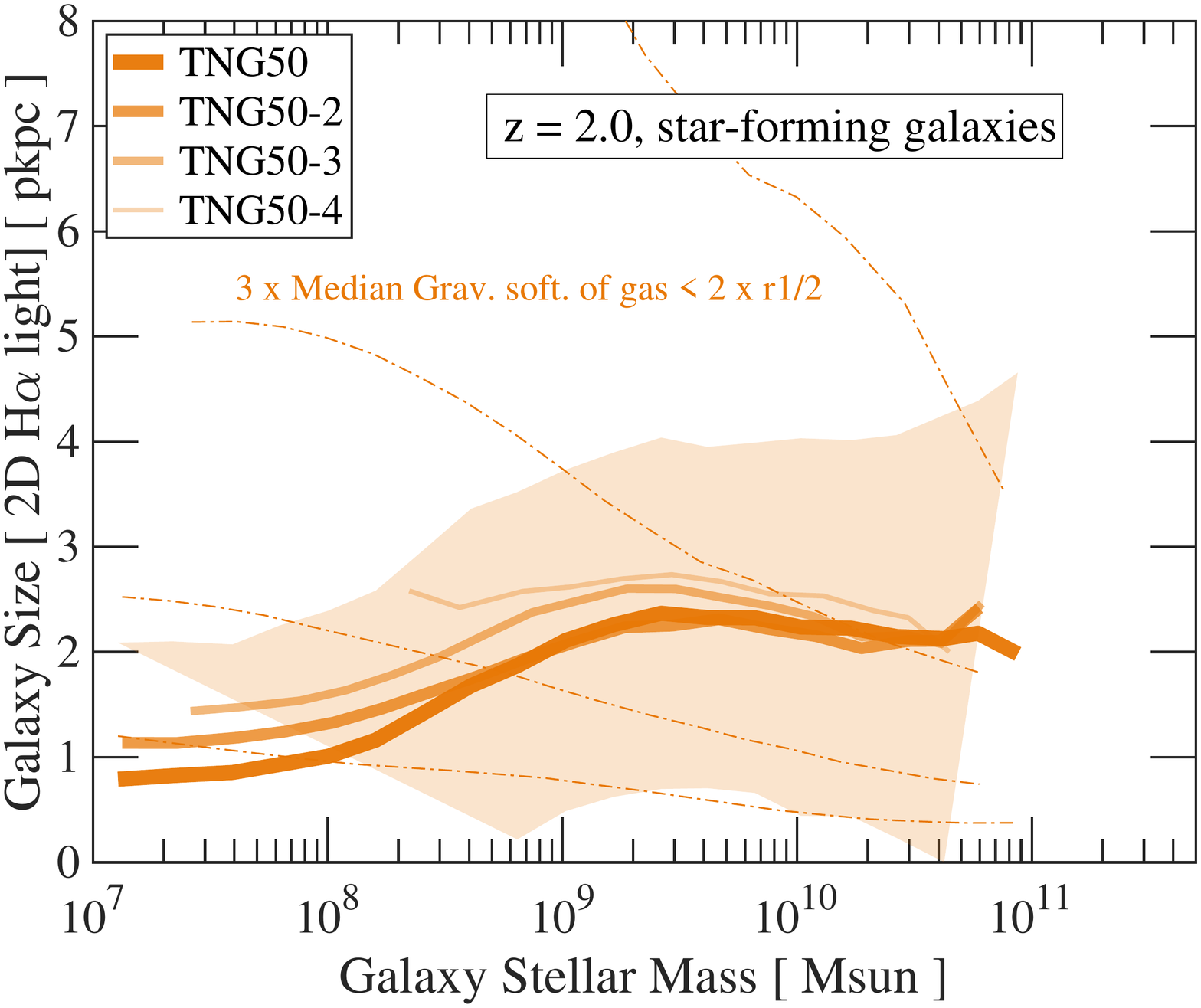}
\caption{\label{fig:sizes_res} Median galaxy sizes as a function of galaxy stellar mass for different resolution realizations of the same cosmological volume: thicker and darker curves denote progressively better numerical resolution. We show 3D stellar half-mass radii on the top and 2D $\HA$ half-light radii on the bottom, at three representative redshifts. Shaded areas indicate the 1-$\sigma$ galaxy-to-galaxy variation at fixed stellar mass: only for TNG50. Dash-dotted lines mark the locus of three times the typical gravitational softening lengths of the stellar particles and gas cells in the different simulations (see text). In the TNG model, galaxy sizes are converg{\it ing} at all mass scales and times, albeit with some complexity at the $10^{10}\MSUN$ knee and above, at low redshifts. Size convergence is reached faster (i.e. also with limited resolution) at higher redshifts. Galaxy sizes do {\it not} trivially depend on the underlying choices of the softening lengths, at least above a certain stellar mass limit.} 
\end{figure*}

In Fig.~\ref{fig:sizes_res} we compare the size-mass relation across different resolution realizations of the same halo population: darker and thicker curves denote progressively better resolution. We explicitly show results at three representative points in time, $z=0.5$, $z=1$, and $z=2$ (from left to right), and for the stellar sizes we study mass rather than light distributions (top). In contrast, for the gas, we compute $\HA$ 2D sizes (bottom).

At fixed galaxy stellar mass, the median relations show that better resolution generally gives smaller galaxy sizes in both stars and star-forming gas -- at $z\lesssim1$ this is the case below $10^{9-10}\MSUN$ galaxies. Yet, in the TNG model, galaxy sizes are converg{\it ing} at all mass scales and times. By this we mean that the difference between TNG50 and TNG50-2 is smaller than that between TNG50-2 and TNG50-3, and so on (in line with the results already shown, but for $z=0$ only, in Appendix A of \citet{Pillepich:2018Method}). This convergence is faster at higher redshifts: across a range of 512 (8) in particle mass (spatial) resolution, sizes are more similar at higher redshifts (e.g. $z=2$) than below e.g. $z\lesssim1$.

At later epochs, the knee in the size-mass relation can be more or less pronounced, based on the available resolution, likely due to two effects. First, different resolutions produce slightly different stellar-to-halo mass relations (see e.g. Appendix A of \citet{Pillepich:2018Method} and \citet{Pillepich:2018}). Second, towards the low-mass end, limited numerical resolution produces more extended galaxies. This effect is stronger, and extends to larger stellar mass, with lower resolution. Furthermore, at the high-mass end, galaxies are more extended at lower resolution: both because their in-situ intrinsic sizes are larger, as well as because they are the result of mergers of lower-mass galaxies, that are likewise larger at lower resolutions. This difference also manifests if we study trends with halo mass rather than galaxy stellar mass, the former being largely invariant to resolution changes.

By comparison to the outcome of TNG50, on the other hand, TNG100 galaxies in the $10^{9-10}\MSUN$ range have median stellar sizes that are converged to better than 20-40 per cent at recent epochs (e.g. below $z\sim2$). This statement provides quantitative confirmation to the results in this regime of \citet{Genel:2018}, while the findings described above at the higher mass end possibly reconcile the 0.1-0.2 dex overestimation identified by \citet{Genel:2018} in comparison to observations.

Overall, the improved numerical resolution of TNG50 allows us to reliably study galaxy structures to even smaller masses. From the comparison across resolutions of Fig.~\ref{fig:sizes_res}, we infer that the median sizes of TNG50 represent the resolution-independent outcome of the TNG model down to galaxy stellar masses of $\sim10^8\MSUN$, i.e. they are resolved to better than 20-30 per cent accuracy. On the other hand, judging from the flattening of the size-mass relations below this mass limit below $z\sim1$, and by comparison to the analog but enhanced flattening of the sizes at lower resolutions, TNG50 $10^7\MSUN$ galaxies may still be larger than what our model would predict at even better resolution.

In Fig.~\ref{fig:sizes_res}, dash-dotted curves mark the locus of three times the typical gravitational softening lengths of the stellar particles and gas cells in the different simulations and galaxies ($\sim2.8$ times our softening lengths indicating the boundary between softened and fully Newtonian forces). For example, in TNG50-3, this boundary is roughly 3.4 physical kpc at $z\leq1$. For the gas component, we indicate the median softening of the gas cells enclosed within the stellar galaxy body ($< 2\RSTARS$) for the typical galaxy at a given stellar mass. As described in Section~\ref{sec:res}, the gravitational softening of the gas cells is tied to their spatial size, in turn depending on ambient density. As a result, cells at the centers of more massive galaxies are smaller than those in less massive galaxies, this trend being steeper at worse resolutions: see orange dash-dotted curves. 

It is manifest from the comparison between median galaxy sizes and typical softening lengths as a function of galaxy mass at any given time that the latter are not responsible for setting galaxy sizes in our calculations, at least above a minimum mass that depends on numerical resolution. This is particularly evident for the star-forming gas sizes (lower panels), where the typical softening scale decreases with galaxy mass whereas galaxy sizes increase with mass. It can also be appreciated as a function of time. As noted in the main text, the extent of $10^{8-9}\MSUN$ galaxies remains constant at about 1-2 physical kpc essentially at all times. This invariance occurs despite the different redshift evolution of the underlying numerical softening lengths. Given the different trends of galaxy sizes and softening lengths with time, it is clear that the softening per se' is {\it not} responsible for setting the physical extent of TNG50 galaxies. An exception to this fact arises only, as our resolution tests suggest, at small galaxy masses and low redshifts (in TNG50, $\MS\lesssim10^8\MSUN$ and $z\lesssim1$). We expand on this in the next subsection, by focusing on galaxy disk heights, an even more stringent test of numerical resolution.

\subsection{Galaxy Disk Heights}
\begin{figure*}
\centering                                      
\includegraphics[height=4.7cm]{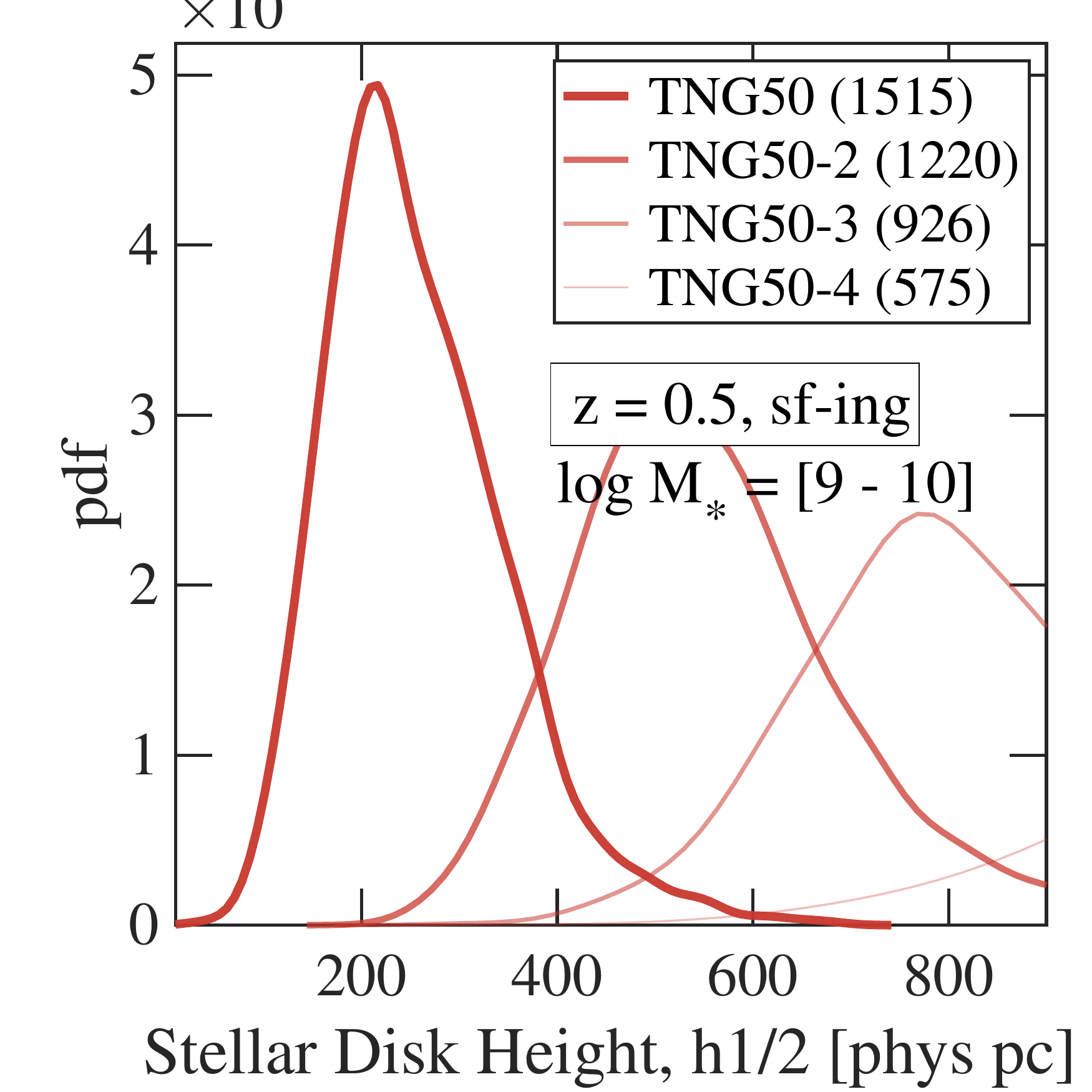}
\includegraphics[height=4.7cm]{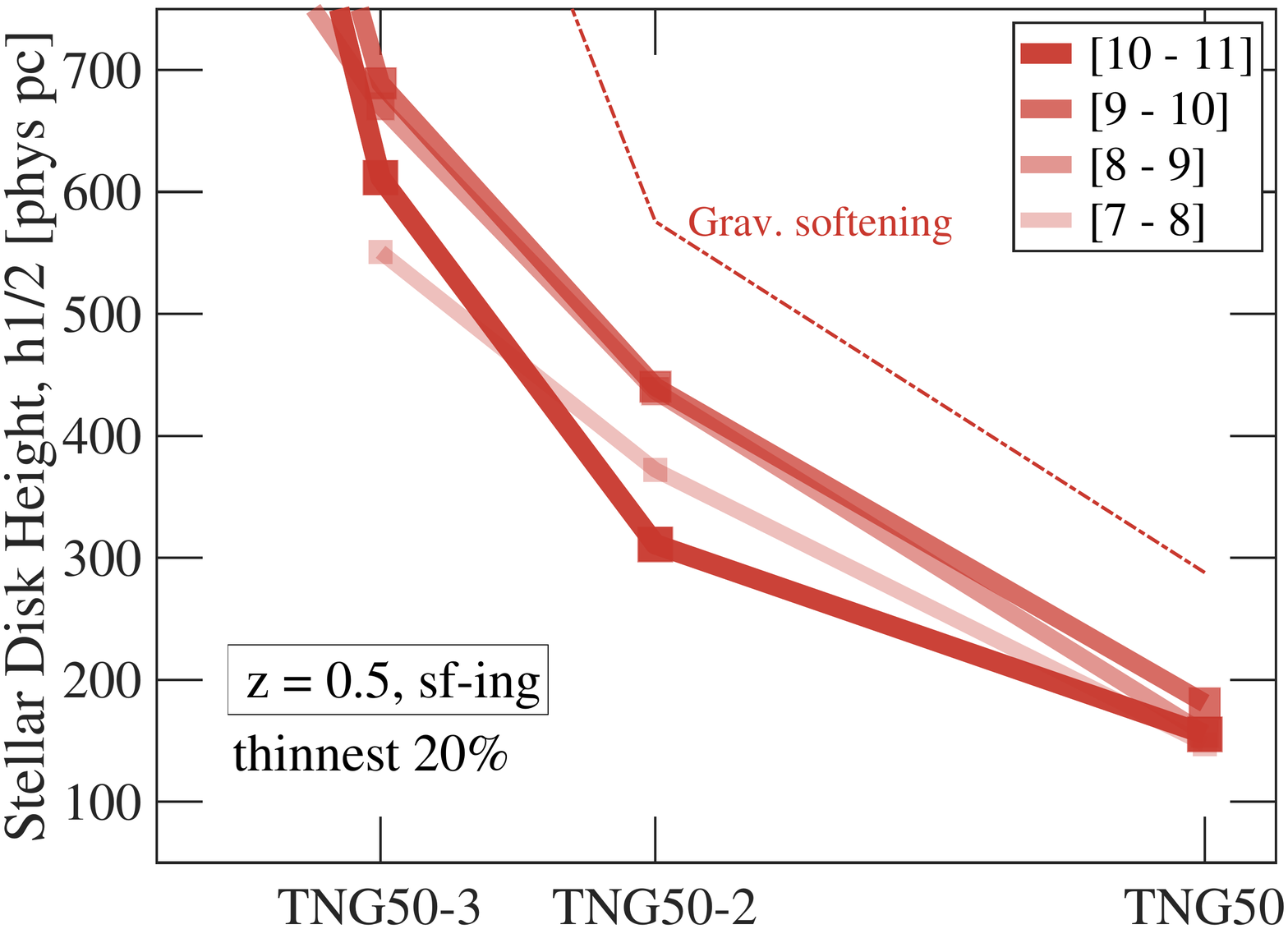}
\includegraphics[height=4.7cm]{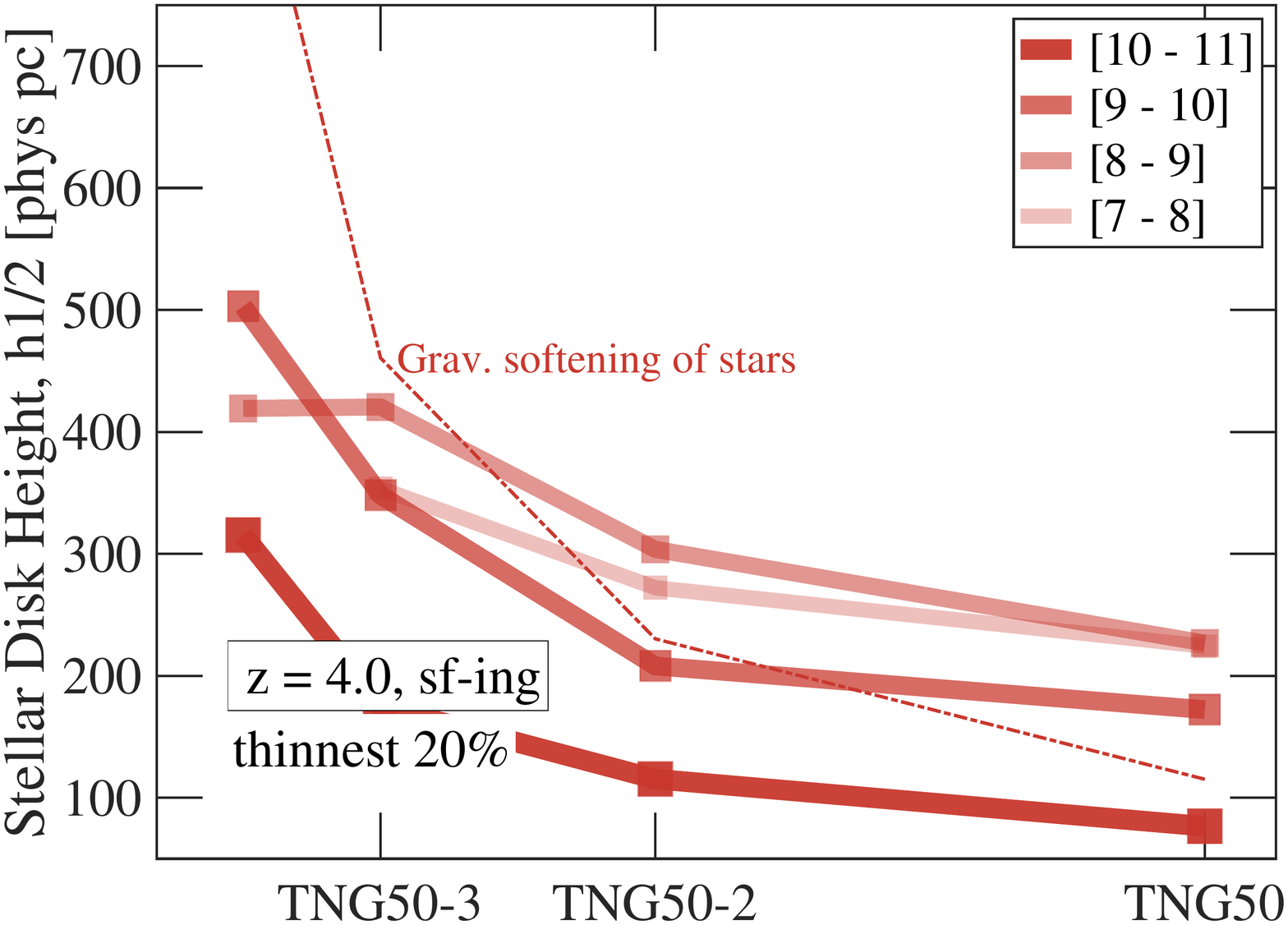}
\includegraphics[height=4.7cm]{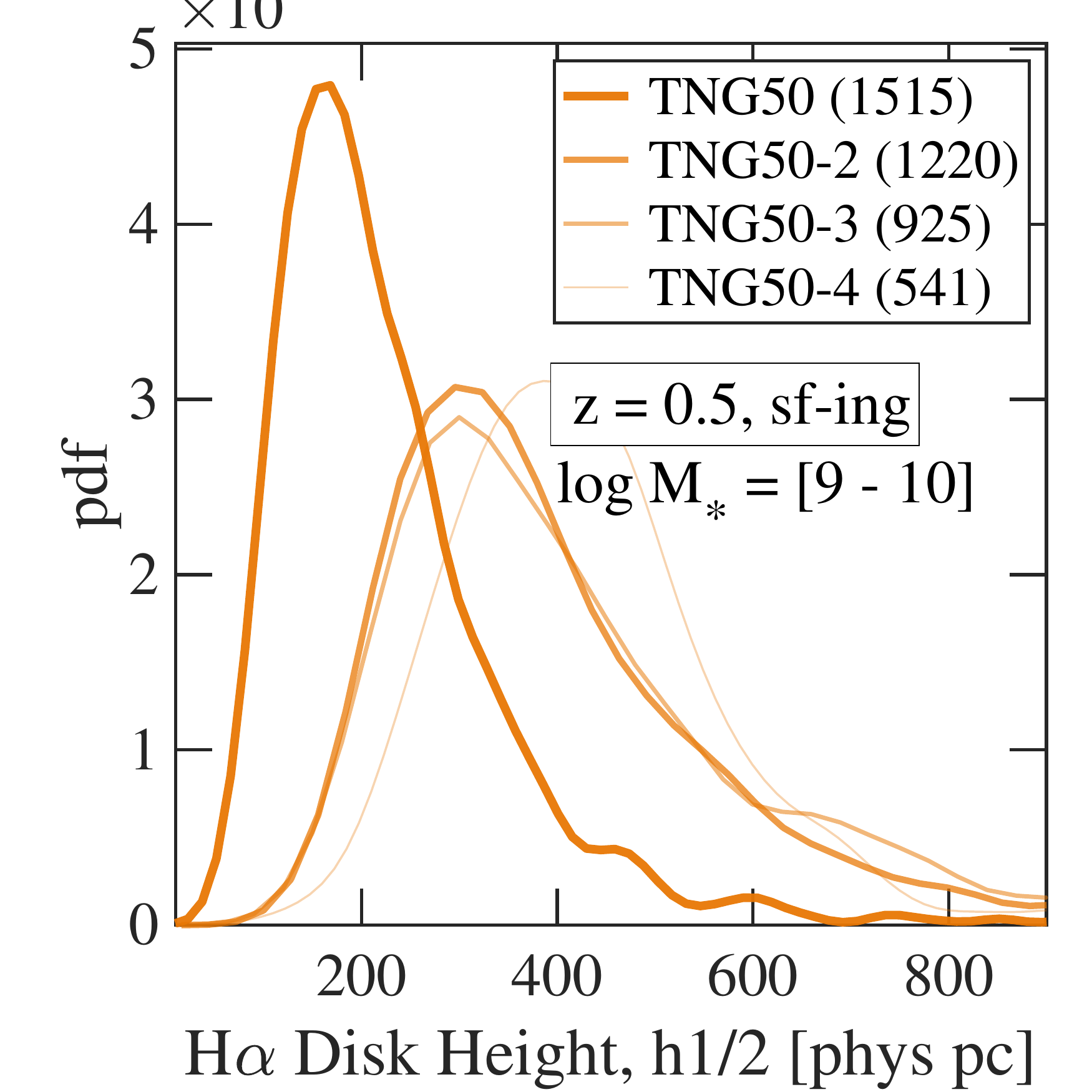}
\includegraphics[height=4.7cm]{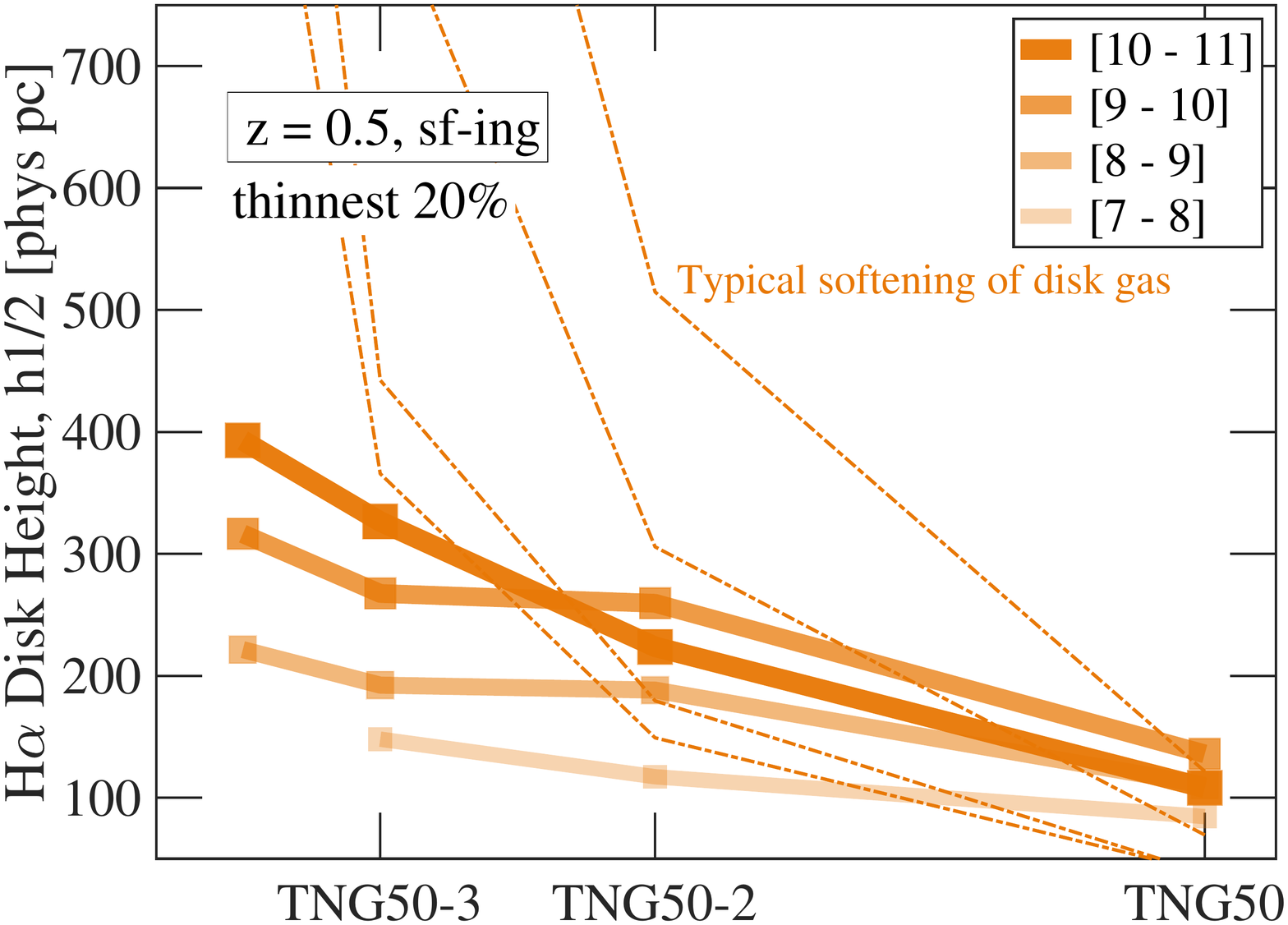}
\includegraphics[height=4.7cm]{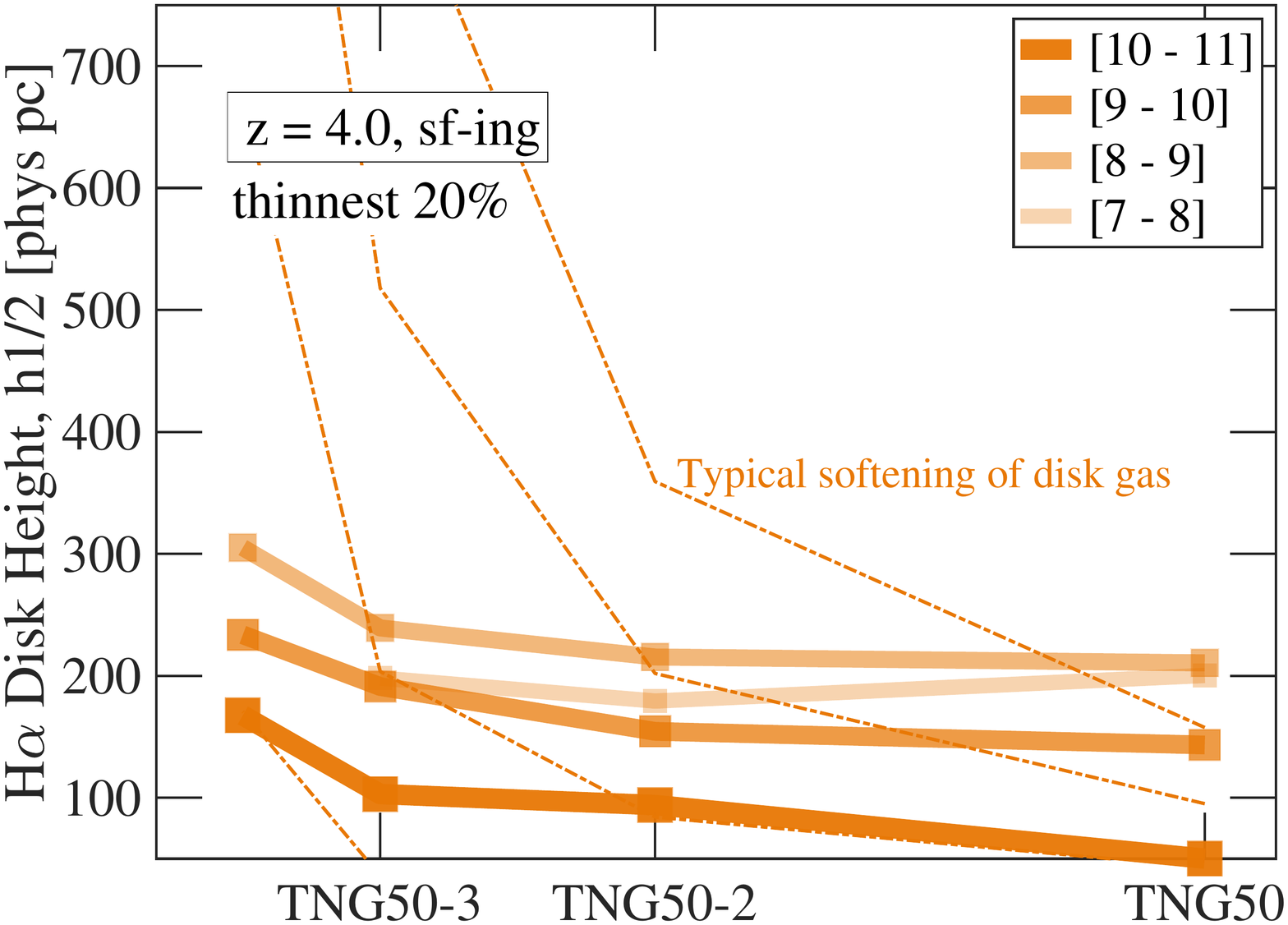}
\caption{\label{fig:heights_res} Galaxy heights for different resolution realizations of the same cosmological volume: thicker and darker curves denote progressively better numerical resolution. Stellar-mass and $\HA$-light disk heights are shown on the top and bottom, respectively. On the leftmost column, we show the distributions across galaxy populations in a given mass bin at $z=0.5$. On the right, the typical heights of the 20 per cent thinnest galaxies are shown as a function of numerical resolution, with changes in both mass and spatial resolution, at $z=0.5$ (center), and $z=4$ (right). At lower redshifts, the enhanced resolution of TNG50 allows us to reach disk heights up to factors of 2 thinner than at the next resolution level, TNG50-2, similar to TNG100. Within our model, galaxy heights are converg{\it ing}, and better levels of convergence are reached faster at higher redshifts. } 
\end{figure*}

In Section~\ref{sec:heights}, we have demonstrated that the bulk of TNG50 galaxies have typical disk heights as small as 100-300 physical parsecs across masses and times. This is the case for the vertical distributions of stellar mass, V-band light and $\HA$ light. The dependence of such heights on numerical resolution is non trivial. Importantly, we see how galaxy sizes and heights (i.e. half-mass or half-light heights) can be smaller than any nominal gravitational softening length. 

In Fig.~\ref{fig:heights_res}, we compare disk heights in simulations with different mass and spatial numerical resolutions. Stellar mass-based measures are shown on the top, $\HA$ heights on the bottom. Darker and thicker curves denote better resolution. We assess convergence at two representative redshifts: $z=0.5$ (two leftmost columns) and $z=4$ (rightmost column). In the leftmost panels, distributions of galaxy disk heights are shown at recent times for $10^{9-10}\MSUN$ galaxies for TNG50, TNG50-2, TNG50-3, and TNG50-4 (the latter almost outside the plotted range for the stellar heights).  
In this mass range, differently than for the 3D or 2D galaxy sizes, the enhanced numerical resolution of TNG50 is fundamental: TNG50 galaxies are on average two times thinner than those in TNG50-2 (and hence in TNG100). The resolution dependence of disk heights is smaller at higher redshifts ($z>0.5$). This resolution dependence is less (similarly) pronounced for more (less) massive galaxies than the ones depicted in Fig.~\ref{fig:heights_res}. 

Now, the distributions of thicknesses may reflect two simultaneous effects: numerical ``heating'' due to limited resolution, as well as a different mix of galaxy morphologies due to the different functioning of the TNG model at different resolutions. To focus on the former, in the two rightmost columns of Fig.~\ref{fig:heights_res}, we show the resolution trends of the median disk heights of the 20 per cent thinnest galaxies in different mass bins. Despite the non-negligible effects of resolution on heights, our model is numerically robust and well-behaved in the sense that a better numerical resolution implies thinner galaxies and, in most regime, heights are converg{\it ing}. In other words, the curves in Fig.~\ref{fig:heights_res} exhibit shallower derivatives at progressively better resolutions. In fact, $\HA$ heights converge much faster than the stellar heights at any given time, as do V-band heights, albeit to a lesser degree (not shown). All heights converge faster at higher redshifts and, usually, at larger galaxy stellar masses. Given the trends in Fig.~\ref{fig:heights_res}, TNG50 disk heights can be considered converged to better than 20-40 per cent across all studied masses and redshifts. Conversely, degraded resolution would return somewhat imprecise redshift evolutions of disk heights at fixed stellar mass. 

For reference, red dash-dotted curves denote the typical gravitational softening lengths of the stellar particles. These are, by design, smaller in physical units at higher redshift (for $z>1$). Importantly, across simulations, disk heights do not necessarily decrease by half when the underlying softening lengths do so. This in fact appears to be the case at lower redshifts for the stellar component of intermediate-mass galaxies between TNG50 and TNG50-2 ($z\lesssim1$, top left and  middle panels in Fig.~\ref{fig:heights_res}), but is not the case across resolution variations, for all mass bins, and, importantly, at higher redshifts. At high redshifts ($z\gtrsim1$), the resolution trends of disk heights appear almost insensitive to the changes in gravitational softening (rightmost panels of Fig.~\ref{fig:heights_res}, solid vs. dash-dotted lines). This fact is more pronounced for the heights of the $\HA$ light: orange dash-dotted curves indicate the gravitational softening lengths of the 5th-percentile smallest gas cells within a galaxy's body ($< 2\RSTARS$), for the average galaxy in each mass bin. This threshold provides an approximate metric for the typical gravitational softening of the gaseous component in the star-forming disks. Even in this case, the resolution trends of the $\HA$ disk heights do not reflect the changes in gravitational softening of the relevant matter component. We conclude that the vertical structure of the stellar and gaseous components is determined by an ensemble of factors (starting with the depth of the overall potential) and is not directly tied to numerical choices of spatial resolution (see Section~\ref{sec:betterres} for a discussion on the possible dependence of heights on other modeling choices, chiefly the ISM effective equation of state).

\begin{figure}
\centering                                      
\includegraphics[width=8cm]{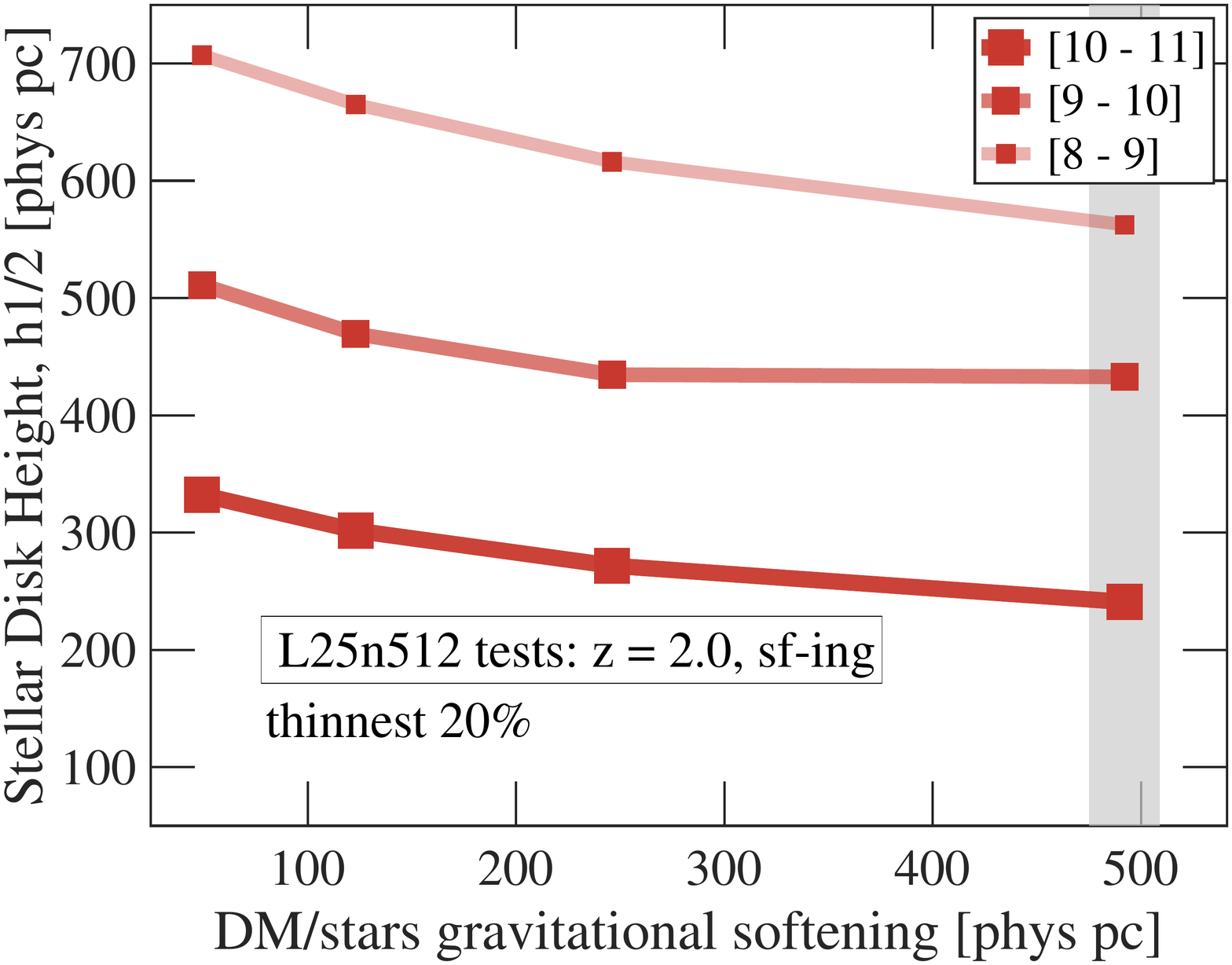}
\caption{\label{fig:heights_runtests} A smaller gravitational softening does not necessarily imply thinner or smaller galaxies. Here the median stellar disk heights of disk-like (thin) galaxies are shown for different simulations that differ exclusively for the underlying choice of the DM and stellar particle gravitational softening. Darker and thicker symbols denote larger galaxy stellar masses. For the problem at hand, the gray band denotes the fiducial choice.} 
\end{figure}

\begin{figure*}
\centering                                      
\includegraphics[width=4.3cm]{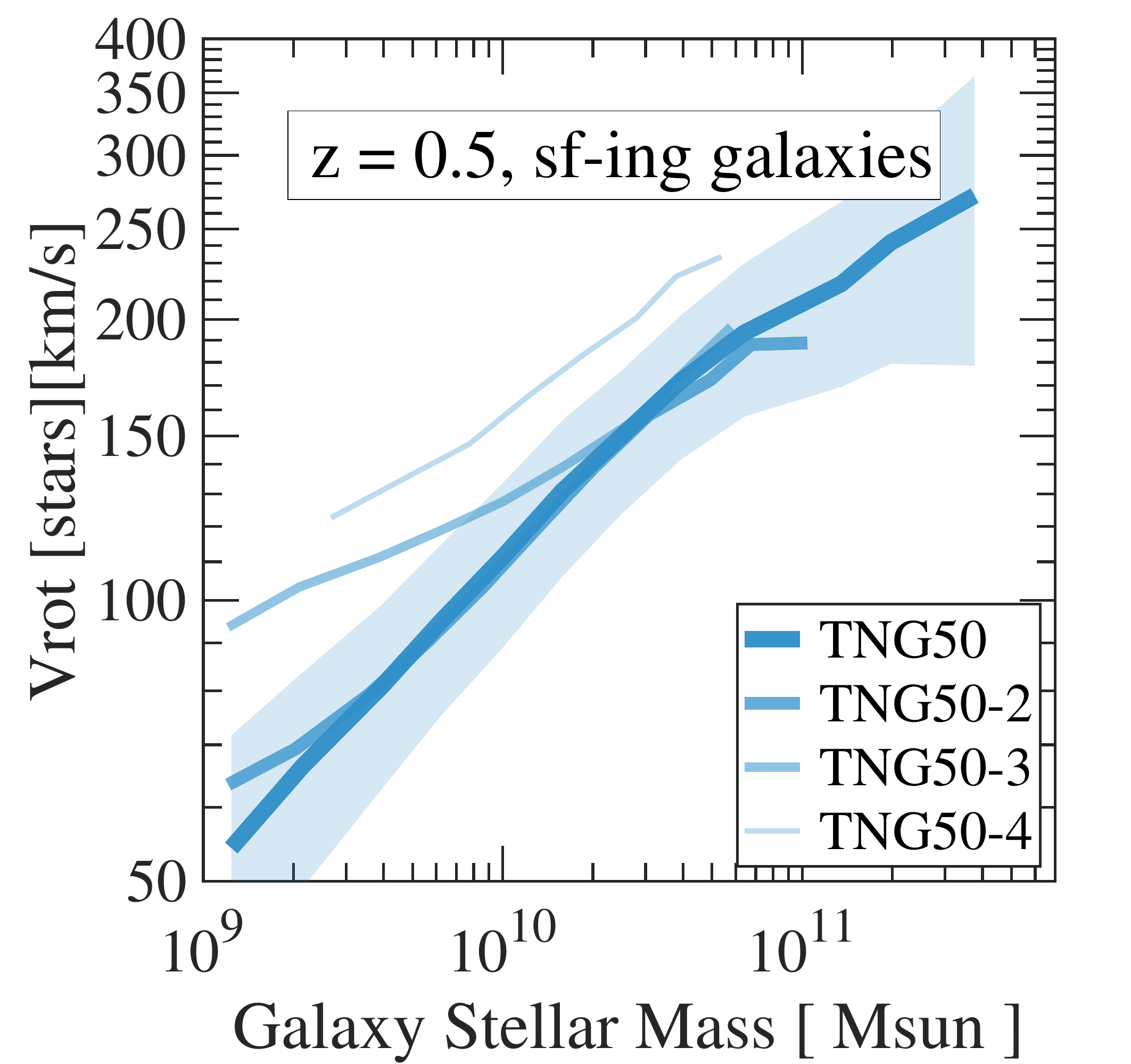}
\includegraphics[width=4.3cm]{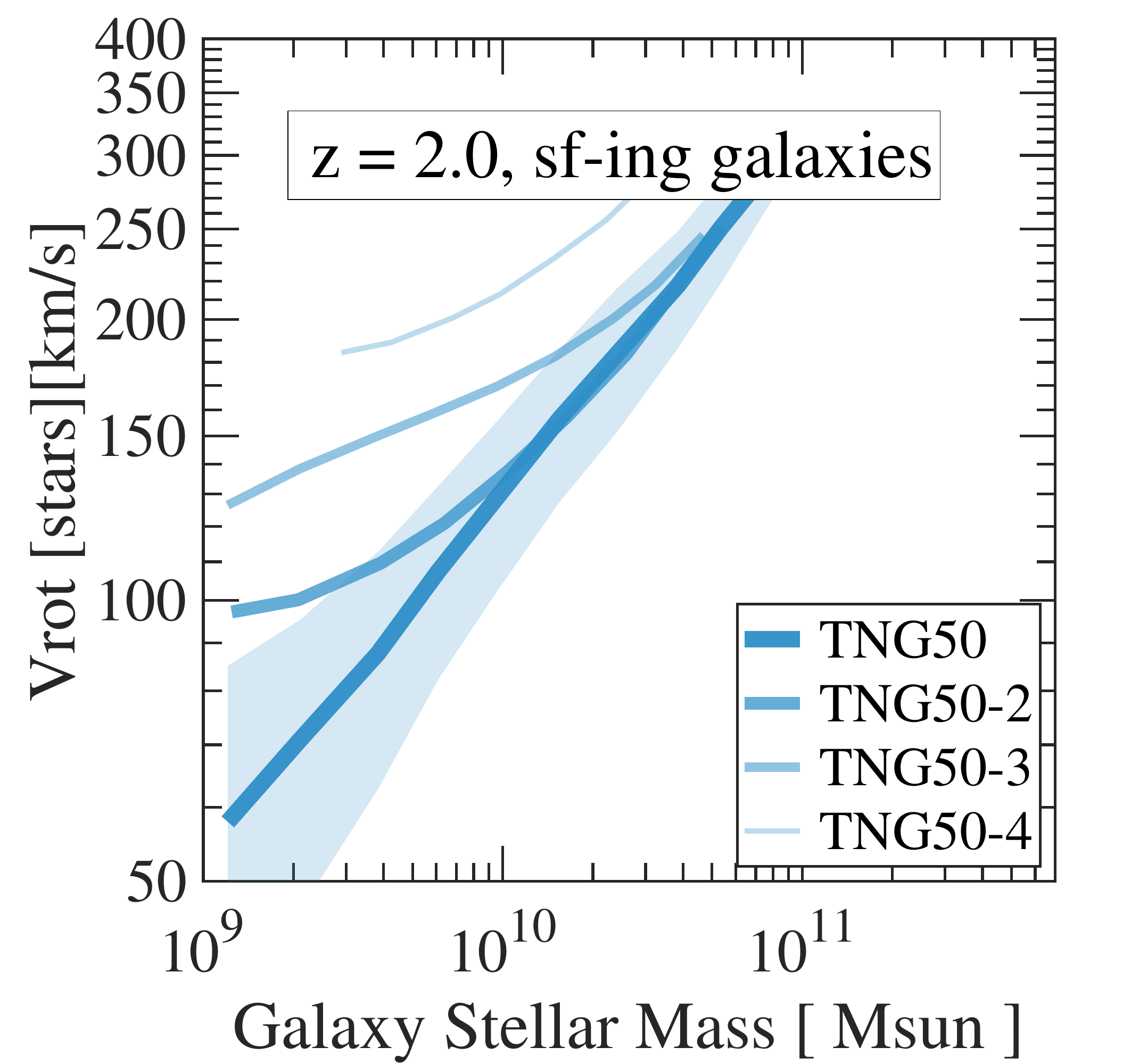}
\includegraphics[width=4.3cm]{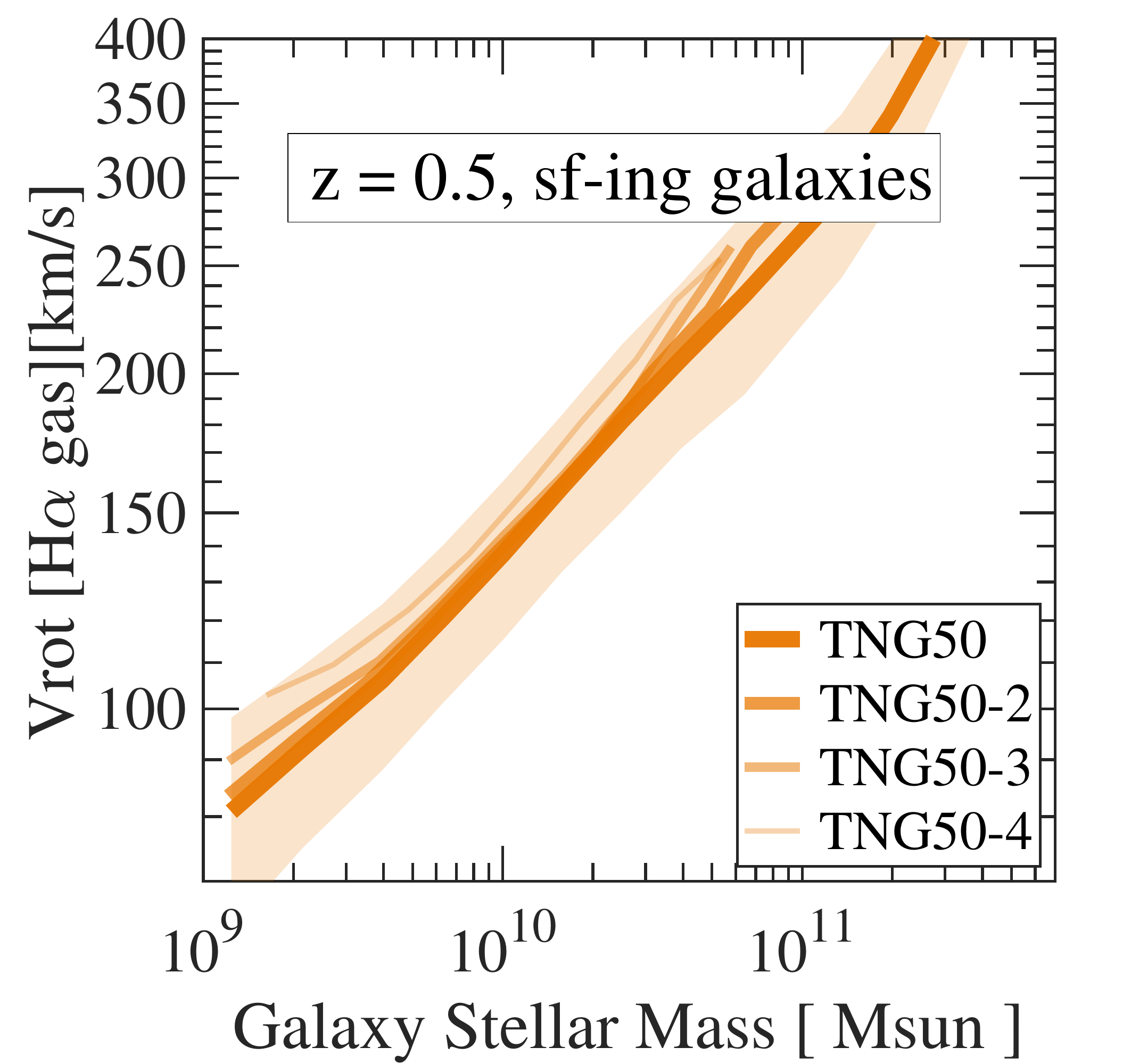}
\includegraphics[width=4.3cm]{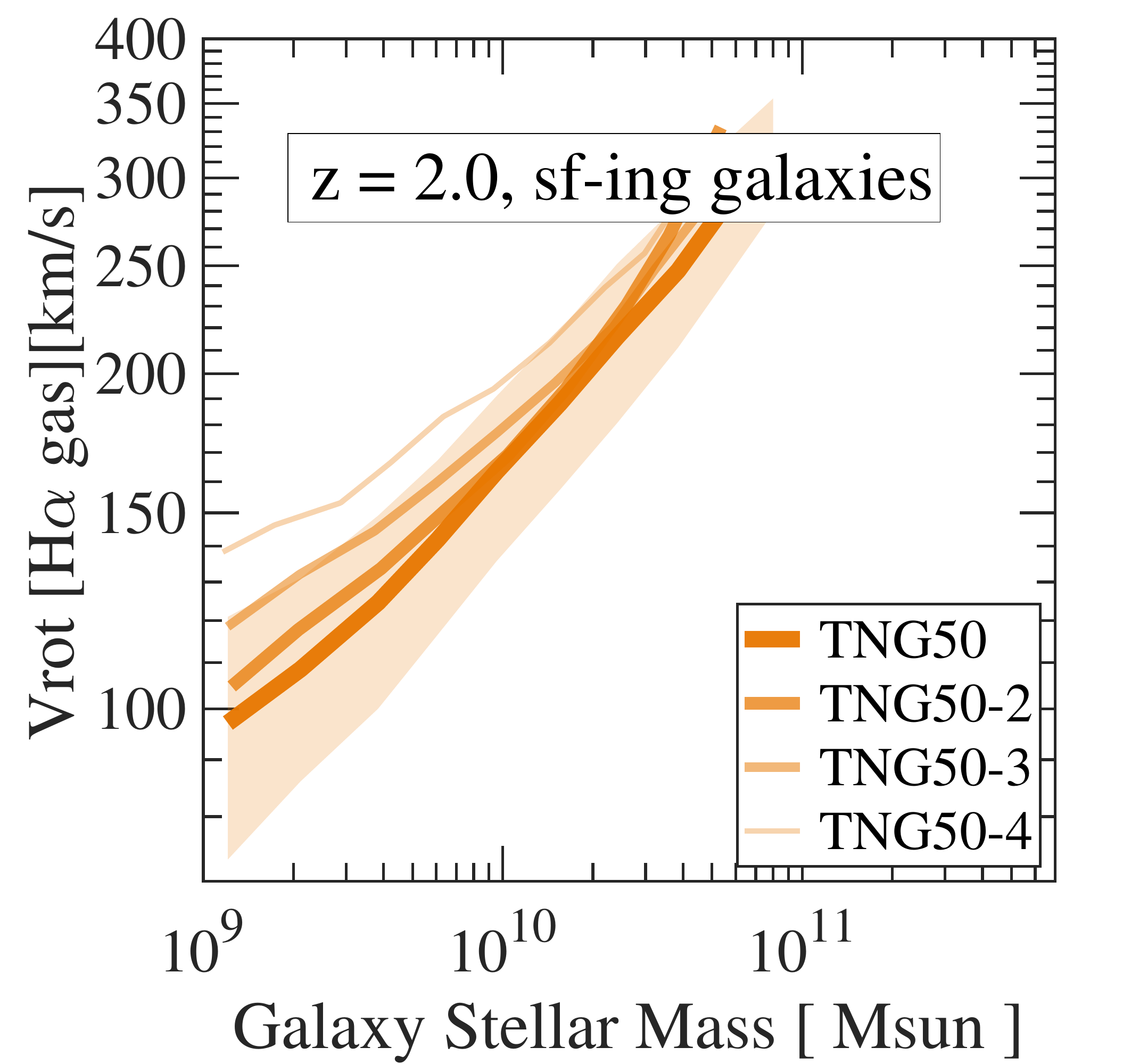}
\includegraphics[width=4.3cm]{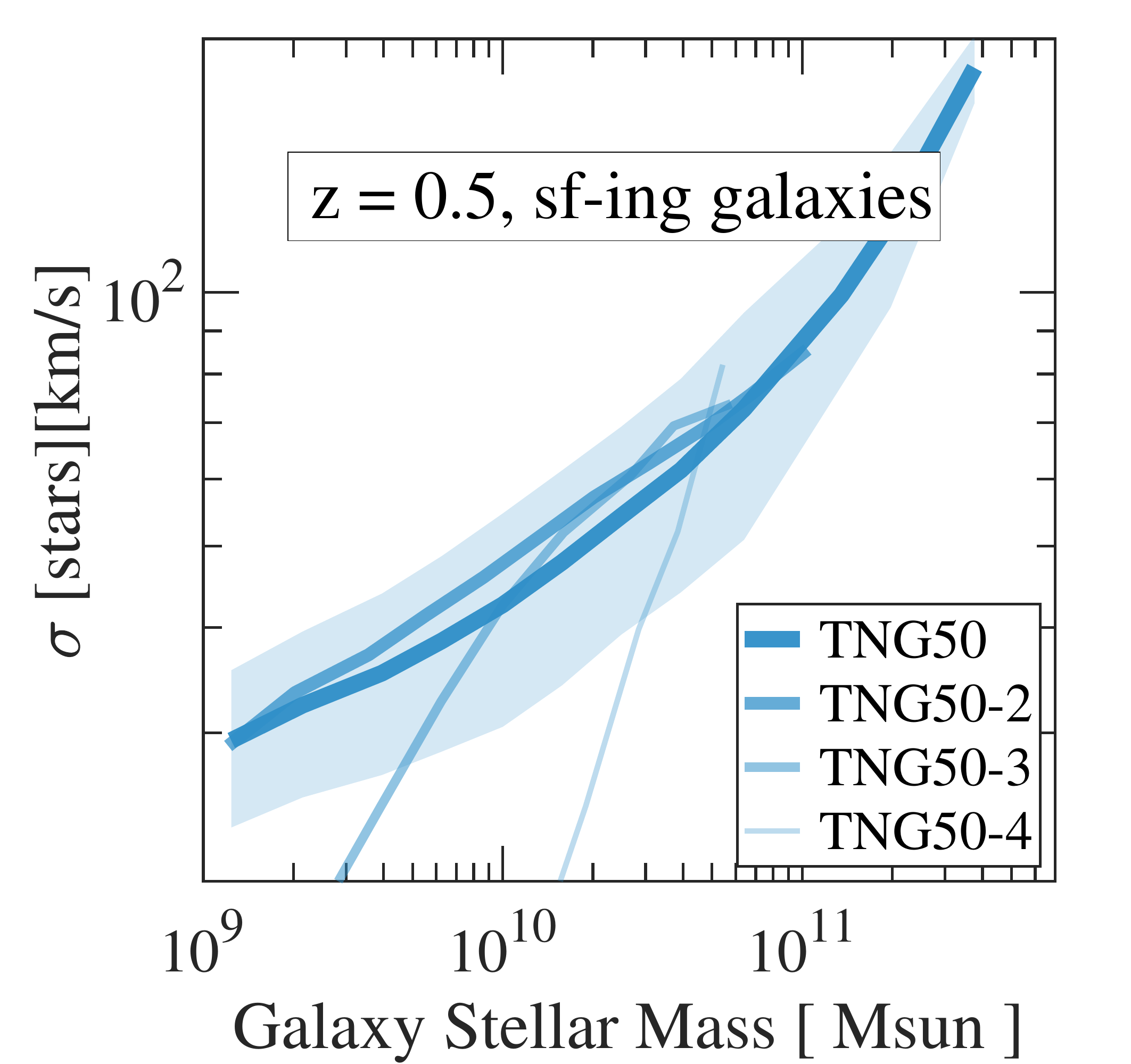}
\includegraphics[width=4.3cm]{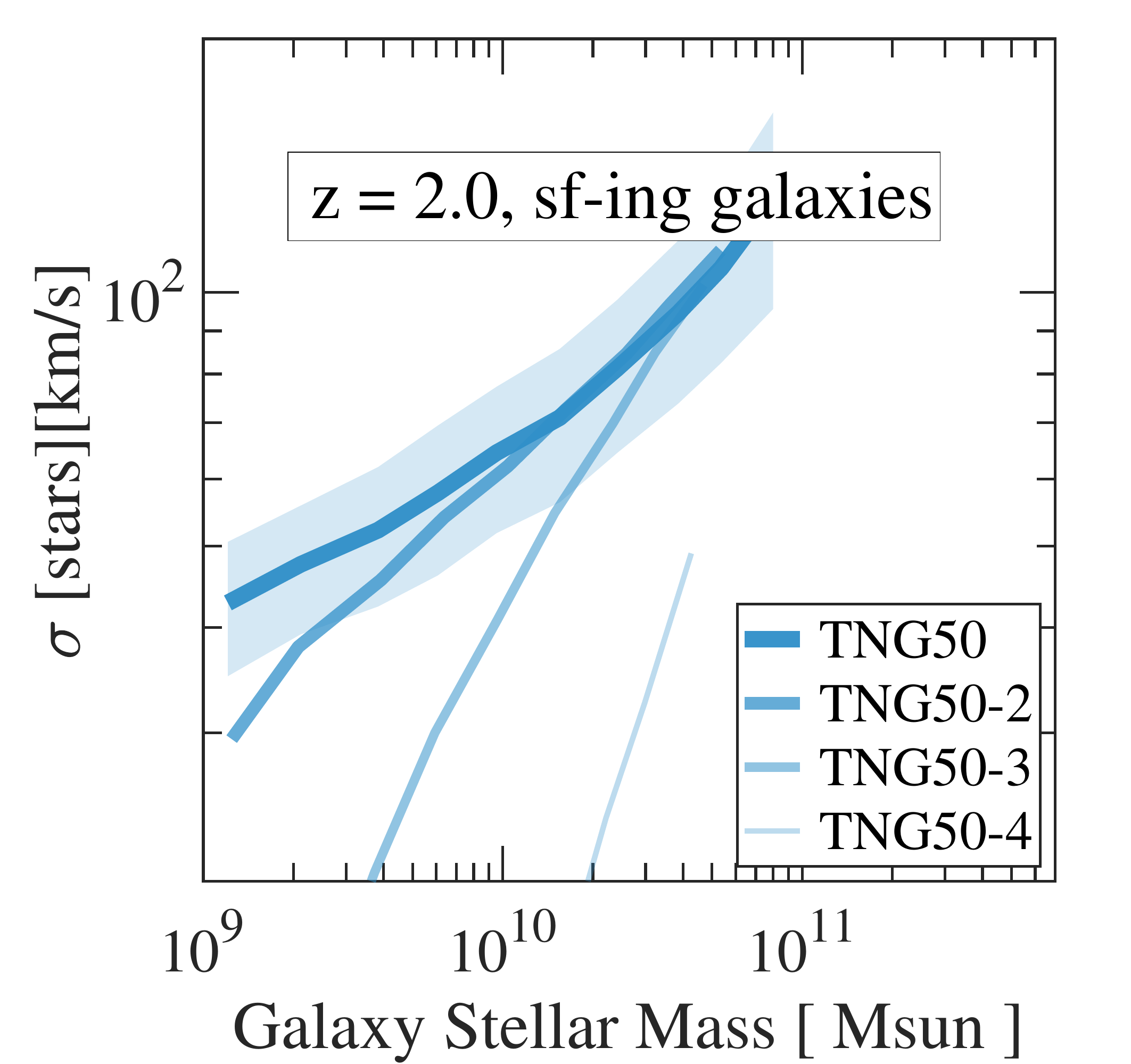}
\includegraphics[width=4.3cm]{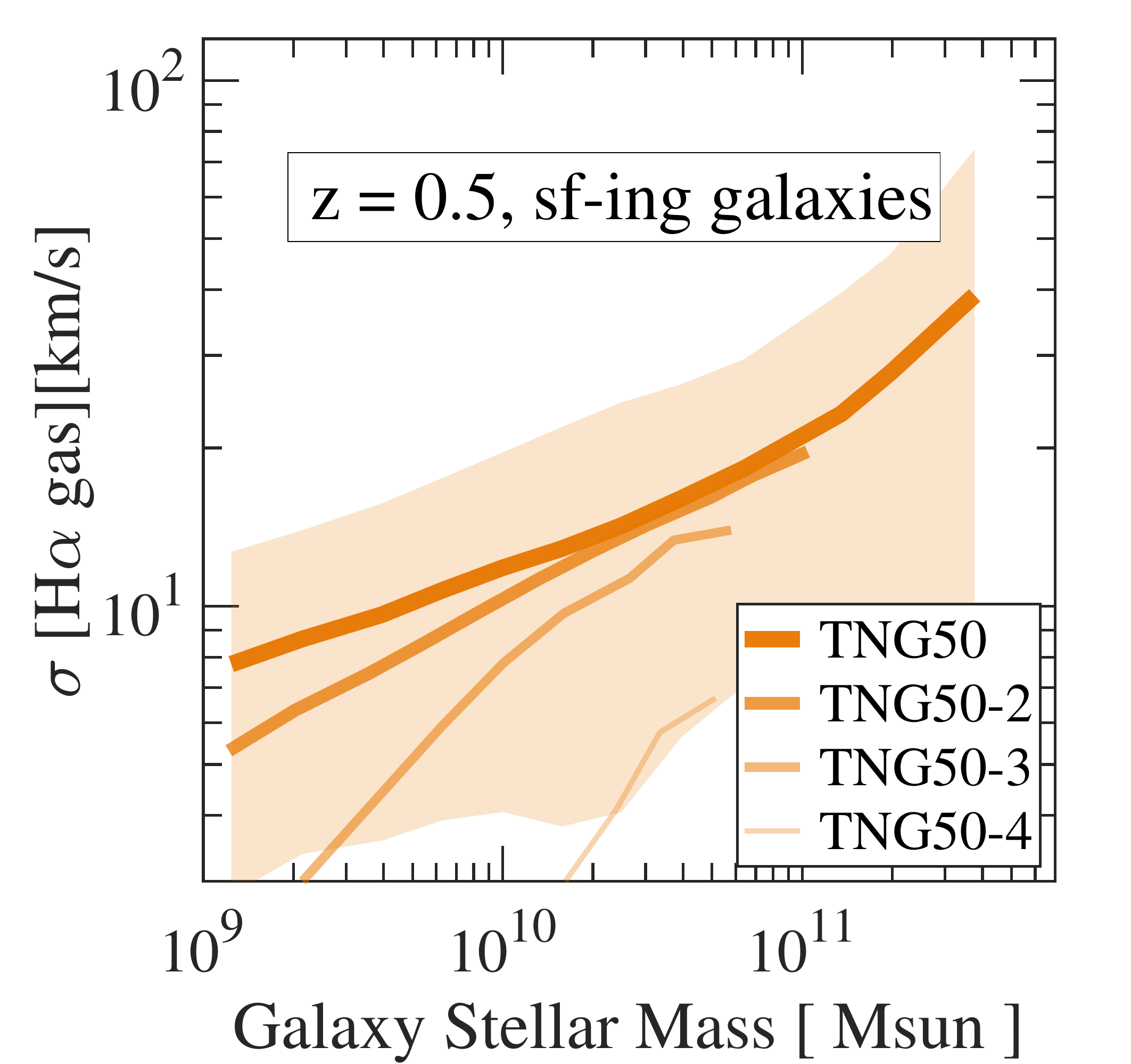}
\includegraphics[width=4.3cm]{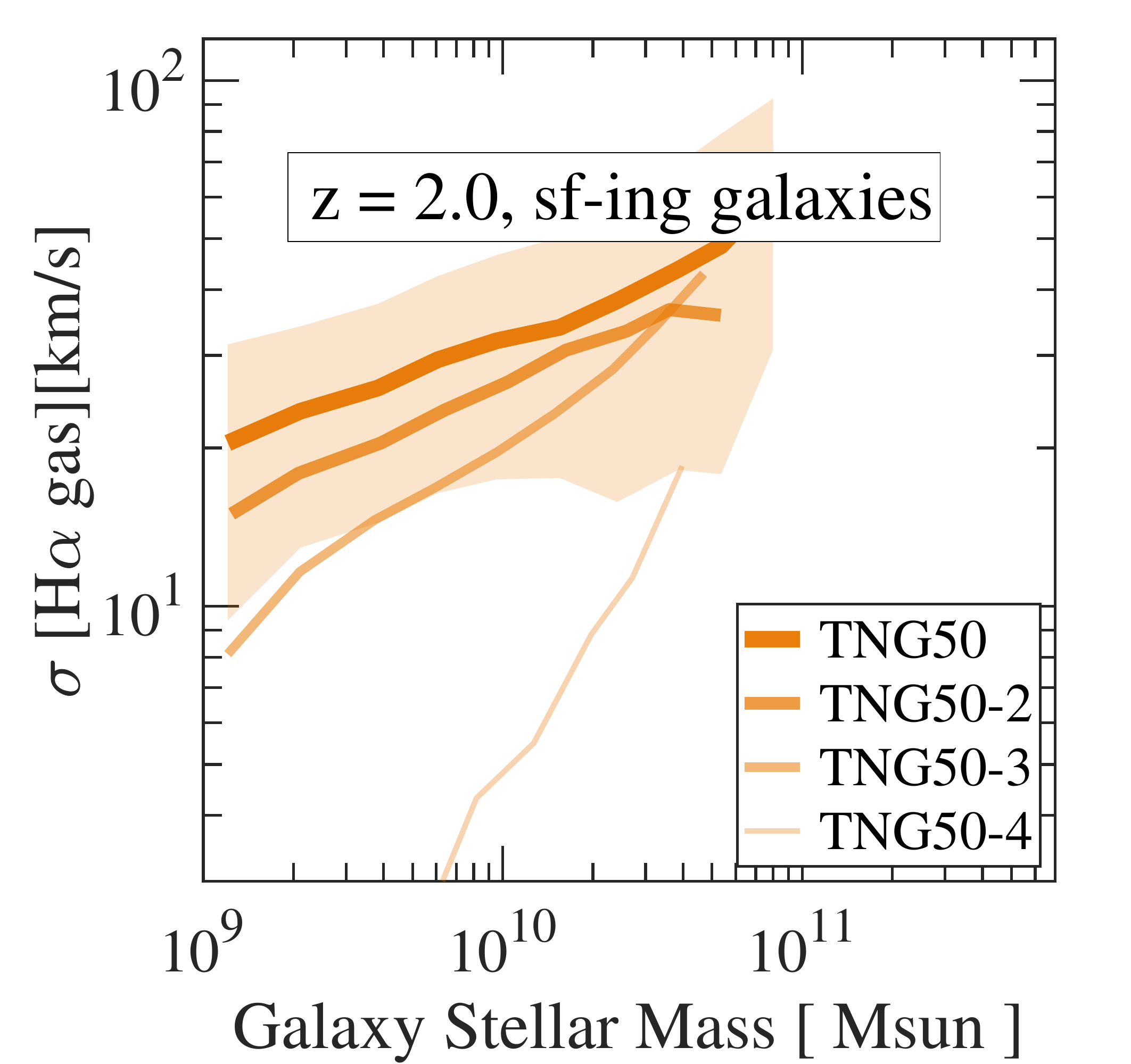}
\caption{\label{fig:kinematics_res} Median galaxy kinematic properties as a function of galaxy stellar mass for different resolution realizations of the same cosmological volume: as so far, thicker and darker curves denote progressively better numerical resolution. We show the rotational velocity (top) and velocity dispersions (bottom) of both the stellar and gaseous components (left and right, respectively). We bracket the redshift range by showing $z=0.5$ and 2 results. In the TNG model, convergence of the galaxy kinematics is reached faster (i.e. also with limited resolution) at lower redshifts, and for the gas rather than the stellar material. At TNG50 resolution, $\VMAX$ and $\sigma$ are quantitatively robust all the way down to galaxy stellar masses of $10^9\MSUN$.}
\end{figure*}

The perceived impact of the numerical softening on galaxy structure is often overestimated \citep[but see][]{Grand:2017, Hopkins:2018}, especially within complex multi-physics and multi-matter galaxy models like TNG. We demonstrate this with the results of the experiment depicted in Figure \ref{fig:heights_runtests}. There, we show the typical heights of thin galaxies obtained in a series of four test simulations, all of the same 25 Mpc h$^{-1}$ box at about TNG100 (i.e. with 512$^3$ DM particles and gas cells) -- these are part of the `model variant' series introduced in \citet{Pillepich:2018Method}. In the tests, {\it all} model parameters are unchanged except for the choice of the gravitational softening, which is chosen to be 2, 4, and 10 times smaller than its fiducial value. 

It is usually expected that smaller softening produce thinner galaxies. In fact, the effect is minimal, if not reversed. From these tests, it is clear that the dependence of galaxy structure (both height and size) on the gravitational softening is more complex than a simply imposed numerical floor. Indeed, at fixed particle resolution, improving the spatial resolution for DM and stars produces a slight height \textit{increase} of galaxies. The same effect is found for galaxy sizes. Reducing too aggressively the softening can produce a counter-intuitive slight enlargement of the galaxy heights and sizes, as a result of increased two-body heating. 


\subsection{Galaxy Kinematic Measures}
\label{sec:app_res_kinematics}
We conclude this study of the convergence properties of the TNG model by analyzing galaxy kinematics, specifically the rotation velocity and velocity dispersions of both the stellar and $\HA$ components. It has been shown that, for example in the context of the EAGLE model, kinematic properties such as the stellar angular momentum and the stellar spin parameters converge more slowly as a function of number of resolution elements per galaxy than structural properties like stellar galaxy sizes \citep{Lagos:2017, Lagos:2018b}. We recover similar qualitative findings. 

In Fig.~\ref{fig:kinematics_res} we focus on two representative redshifts ($z=0.5$ and $2$) and study how the median rotational velocities (top) and velocity dispersions (bottom, \rvvv{without thermal component}) vary as a function of stellar mass across the different-resolution runs of the TNG50 series: stellar kinematics are given on the left, $\HA$-gas properties on the right. Firstly, as is the case for the structural properties, kinematic properties of the gas component depend less strongly on numerical resolution than their stellar counterparts, also for very low resolutions (as in TNG50-3 and -4). Importantly, TNG50 and TNG50-2 return $\HA$ kinematic measures that are consistent at the 20-30 per cent accuracy across the studied redshift range and for galaxy stellar masses above $10^9\MSUN$. The trends in the four right panels of Fig.~\ref{fig:kinematics_res} suggest that, at TNG50 resolution, we can trust the gas $\VMAX$ and $\sigma$ to the $20-30$ per cent level of accuracy all the way down to $1-2\times10^8\MSUN$ in stars, and hence to much better accuracy for all the mass range we have explored in Sections~\ref{sec:results_2} and \ref{sec:discussion}.

Furthermore, we see that poorer numerical resolution implies larger rotational velocities and lower velocity dispersions. Differently than for the structural properties like disk heights and galaxy sizes, the effects of poor numerical resolution are larger at larger redshifts. Finally, for the stellar kinematics (four left panels), properties are converged once a minimum number of stellar particles samples the galaxies. At low redshift ($z\lesssim0.5$), TNG100 kinematics (similar to TNG50-2) are converged, and in fact indistinguishable from those of TNG50, for galaxies of $10^9\MSUN$ and above i.e. at least 1000 particles -- this minimum stellar mass is at least a factor of 8 smaller for TNG50. Namely, we find that at low redshifts, e.g. $z\sim0.5$, galaxy stellar kinematics in TNG50 are robust and converged down to $\sim10^8\MSUN$. Higher redshift results are, on the other hand, more constraining: TNG50 kinematics are converged at the percent level also at $z\gtrsim1$ as long as galaxies are sampled with at least 10,000 stellar particles. We are thus confident of the robustness of the quantitative results and trends uncovered in the main text of the paper, concerning the kinematics of galaxies of $10^9\MSUN$ and above.

\end{document}